\documentclass[10pt,letterpaper]{article}
\usepackage[top=0.85in,left=2.75in,footskip=0.75in]{geometry}

\usepackage{amsmath,amssymb}
\usepackage{mathrsfs}
\usepackage{mathbbol}

\usepackage{changepage}

\usepackage[utf8x]{inputenc}
\usepackage{textcomp,marvosym}
\usepackage{cite}

\usepackage{caption}

\usepackage{nameref,hyperref}

\usepackage[right]{lineno}
\usepackage{microtype}
\DisableLigatures[f]{encoding = *, family = * }

\usepackage[table]{xcolor}

\usepackage{array}

\newcolumntype{+}{!{\vrule width 2pt}}

\usepackage{rotating}

\newcommand{\blue}[1]{\textcolor[rgb]{0,0,0}{#1}}

\newlength\savedwidth

\usepackage{graphicx}
\usepackage{scrextend}   % supuestamente para hacer multiples referencias a footnotes: \footnote{\label{jondemor} note on jondemor } ... \footref{jondemor} y para itemizes con texto (labeling)
\usepackage{url}
\usepackage{lipsum}
\usepackage{cite}
\usepackage[all]{xy}
\usepackage{xcolor}
\usepackage{authblk}

\usepackage{amsmath}
\usepackage{dsfont}
\usepackage{cancel}

\usepackage{pbox}
\usepackage{bbm}
\usepackage{blindtext}  % Para itemizes con texto
\addtokomafont{labelinglabel}{\sffamily}
\usepackage[normalem]{ulem}
\definecolor{marcelo}{rgb}{0., 0, 1}
\definecolor{marina}{rgb}{0.1, 0.4, 0.1}
\definecolor{jesus}{rgb}{1, 0., 0.}
\definecolor{thomas}{rgb}{1, 0.5, 0.}
\definecolor{praveen}{rgb}{0, 0, 0}

\newcommand{\chngP}[1]{\textcolor{praveen}{{\textbf{#1}}}}

\newcommand{\vect}[1]{\boldsymbol{#1}}

\urldef{\mailsa}\path|{marina.martinez, jesus.malo}@uv.es|

\usepackage[aboveskip=1pt,labelfont=bf,labelsep=period,justification=justified,singlelinecheck=off]{caption}

\makeatletter
\renewcommand{\@biblabel}[1]{\quad#1.}
\makeatother

\date{}

\usepackage{lastpage,fancyhdr,graphicx}
\usepackage{epstopdf}
\fancyheadoffset[L]{2.25in}
\fancyfootoffset[L]{2.25in}
\newcommand{\hiddensubsubsection}[1]{
    \stepcounter{subsubsection}
    \subsubsection*{\arabic{section}.\arabic{subsection}.\arabic{subsubsection}\hspace{1em}{#1}}
}

\begin{document}
\vspace*{-0.9in}

\begin{flushleft}
{\huge
\textbf\newline{Derivatives and Inverse of Cascaded \\[0.3cm] Linear+Nonlinear Neural Models} % Please use "title case" (capitalize all terms in the title except conjunctions, prepositions, and articles).
}
\newline
\\
M. Martinez-Garcia\textsuperscript{1,2},
P. Cyriac\textsuperscript{3},
T. Batard\textsuperscript{3},
M. Bertalm\'{i}o\textsuperscript{3},
J. Malo\textsuperscript{1*}
\\
%\bigskip
\vspace{0.2cm}
\textbf{1} \small{Image Processing Lab., Univ. Val\`encia, Spain}\\
\textbf{2} \small{Instituto de Neurociencias, CSIC, Alicante, Spain}\\
\textbf{3} \small{Information and Communication Technologies Dept., Univ. Pompeu Fabra, Barcelona, Spain}\\
%\bigskip
\vspace{-0.0cm}

% Insert additional author notes using the symbols described below. Insert symbol callouts after author names as necessary.
%
% Remove or comment out the author notes below if they aren't used.
%
% Primary Equal Contribution Note
%\Yinyang These authors contributed equally to this work.

% Additional Equal Contribution Note
% Also use this double-dagger symbol for special authorship notes, such as senior authorship.
%\ddag These authors also contributed equally to this work.

%\textcurrency Current Address: Dept/Program/Center, Institution Name, City, State, Country % change symbol to "\textcurrency a" if more than one current address note
% \textcurrency b Insert second current address
% \textcurrency c Insert third current address


% Group/Consortium Author Note
%\textpilcrow Membership list can be found in the Acknowledgments section.

% Use the asterisk to denote corresponding authorship and provide email address in note below.
* \small{jesus.malo@uv.es}

\end{flushleft}
\section*{Abstract}

In vision science, cascades of \emph{Linear}+\emph{Nonlinear} transforms are very successful in modeling a number of perceptual experiences~\cite{Carandini12}.
However, the conventional literature is usually too focused on only describing the \emph{forward} input-output transform.
%, many times expressed in terms of filters.

Instead, in this work we present the mathematics of such cascades beyond the forward transform, namely the Jacobian matrices and the inverse.
%These analytic results are important for three reasons:
\blue{The fundamental reason for this analytical treatment is that it offers useful analytical insight into the
% behavior of the system.
% Here we include examples of such insight into
\emph{psychophysics}, the \emph{physiology}, and the \emph{function} of the visual system.
For instance, we show how the trends of the sensitivity (volume of the discrimination regions) and the adaptation of the
receptive fields can be identified in the expression of the Jacobian w.r.t. the stimulus.
This matrix also tells us which regions of the stimulus space are encoded more efficiently in multi-information terms.
The Jacobian w.r.t. the parameters shows which aspects of the model have bigger impact in the response, and hence their relative relevance.
The analytic inverse implies conditions for the response and model parameters to ensure appropriate decoding.
From the experimental and applied perspective,}
(a)~the Jacobian w.r.t. the stimulus is necessary in new experimental methods based on the synthesis of visual stimuli with interesting geometrical properties,
(b)~the Jacobian matrices w.r.t. the parameters are convenient to learn the model from classical experiments or alternative goal optimization, and
(c)~the inverse is a promising model-based alternative to blind machine-learning methods for neural decoding \blue{that do not include meaningful biological information}.
% Moreover, \blue{a number of} properties of the neural model are more intuitive by using this kind of vector formulation.

The theory is checked by building and testing a vision model that actually follows the modular program suggested in~\cite{Carandini12}.
% of our approach to cascaded  \emph{Linear}+\emph{Nonlinear} models,
Our \blue{illustrative} \emph{derivable} and \emph{invertible} model consists of a cascade of modules that account for brightness, contrast, energy masking, and wavelet masking.
To stress the generality of this modular setting we show examples where some of the canonical \emph{Divisive Normalization} modules are substituted by equivalent modules such as the \emph{Wilson-Cowan} interaction model~\cite{Wilson72,Cowan16} (at the V1 cortex) or a tone-mapping model~\cite{Cyriac15} (at the retina).
\small{.}

% \linenumbers

\newpage
\tableofcontents
%\begin{figure}[b!]
%  \centering
%  \vspace{-0.5cm}
%  \begin{tabular}{c}
%   \hspace{-0cm}\includegraphics[height=8.7cm,width=10cm]{scheme2.JPG}
%  \end{tabular}
%  %\captionsetup{labelformat=empty}
%  %\caption{\scriptsize{\textbf{Analytical results enable new applications}.}}
%\end{figure}
\newpage
%--------------------------------------------
\section{Introduction}
%--------------------------------------------
%\input{1_intro_2nd.tex}
\label{intro}

\vspace{-0.2cm}
The mathematics of \emph{Linear}+\emph{Nonlinear} (L+NL) transforms is interesting in neuroscience because cascades of such modules are key in explaining a number of perceptual experiences~\cite{Carandini12}.
For instance, in \emph{visual neuroscience}, perceptions of color, motion and spatial texture are tightly related
to L+NL models of similar functional form~\cite{Brainard05,Simoncelli98,Watson97}.
The literature is usually focused on describing the behavior, i.e. setting the parameters of the \emph{forward} input-output transform.
% so as to obtain the responses from arbitrary stimuli, which is the obvious goal of predictive science.
However, understanding the transform computed by the sensory system, $S$, goes beyond predicting the output from the input.
The mathematical properties of the model (namely the \emph{derivatives}, $\nabla S$, and the \emph{inverse}, $S^{-1}$),
are also relevant.
\blue{Here we show that the Jacobian matrices and the inverse provide analytical insight into fundamental
aspects of the \emph{psychophysics} of the visual system, its \emph{physiology}, and its \emph{function}.
% (e.g. information transmission).
Additionally, the Jacobian matrices and the inverse enable new experimental designs, data analysis and applications
in \emph{visual neuroscience}. Finally, related applied disciplines like \emph{image processing} that require computable and interpretable
models of visual perception may also benefit from this formulation.}
\vspace{-0.35cm}

\paragraph*{Derivatives are relevant.}
The Jacobian, $\nabla S$, represents a local linear approximation of the nonlinear system, $S$.
\blue{From a fundamental perspective}, the analytical expressions of the Jacobian matrices have a variety of interests in visual neurosicence.
\blue{In \textbf{\emph{physiology}}}, the dot product definition of \emph{receptive field} introduced for linear systems \cite{Olshausen96,Ringach02}
can be extended to nonlinear systems using the Jacobian matrix with regard to the stimulus.
Therefore, this Jacobian is convenient to properly formulate concepts such as adaptive (stimulus dependent) receptive
fields or adaptive features.
% Here we show how the context-dependence of the receptive fields of the sensors can be seen
% in the expression of the Jacobian w.r.t the stimulus.
On the other hand, the Jacobian of the response w.r.t the parameters allows to assess the impact of the different aspects of the model
on the response, and hence the relative relevance of these aspects.
% For instance, as shown below, the impact of different receptive fields may be different over the stimulus space, and may depend
% on the sensor. Interestingly, these trends can be seen in the expression of the Jacobian wrt the parameters.
\blue{In \textbf{\emph{psychophysics}}}, the sensitivity of the system is characterized by its discrimination abilities
(inverse of the volume of the regions determined by the \emph{just noticeable differences} -JNDs- \cite{Graham89,Regan91}).
Discrimination depends on models to compute perceptual differences from the internal representation~\cite{Epifanio03,Malo06a,Laparra10a}
or on models of noise at the internal representation \cite{Series09,Fonseca16,Laparra17}.
In any of these cases, the way the sensory system, $S$, deforms the stimulus space is critical to understand
how the discrimination regions in the internal representation transform back into the image space. It does not matter that
these internal JNDs are implied by internal noise or by an assumed internal metric.
The change of variable theorem~\cite{Spivak,Dubrovin82} implies that the Jacobian w.r.t. the stimulus controls how the
volume element is enlarged or compressed in the deformations suffered by the representation
along the neural pathway. That is the key to describe how the metric matrices change under nonlinear
transforms in Riemannian geometry~\cite{Dubrovin82}.
For the same reason, this Jacobian wrt the stimulus is also the key to characterize the propagation of noise throughout the system \cite{Ahumada87}.
% Here we show how the general trends of the sensitivity over the stimulus space can be seen from the determinant of the metric based on
% the Jacobian wrt the stimulus.
\blue{In analyzing the \textbf{\emph{function}} of the sensory system in information-theoretic terms} the relation between the information and the volume
of the signal manifold is crucial~\cite{Cover06,Studeny98}.
According to this, for the same geometrical reasons stated above~\cite{Spivak,Dubrovin82}, the Jacobian wrt the stimulus plays an important role in
determining the amount of information lost (or neglected) along the neural pathway.
More specifically~\cite{Studeny98}, the Jacobian wrt the stimulus determines the multi-information shared by the different sensors of
the neural representation.
% Below we show that the Jacobian implies different efficiency (bigger or lower multi-information reduction) in different regions of the stimulus space.

%As a result, subjective distances also depend on the Jacobian w.r.t. the stimulus~\cite{Epifanio03,Malo06a,Laparra10a},
%and these distances characterize the discrimination ability (or threshold psychophysics).
%Therefore when discrimination is related to the noise at the internal representation \cite{Series09,Fonseca16,Laparra17},
%thresholds at the input space are also characterized by the Jacobian w.r.t. the input.
%% Moreover, in minimization procedures... Relation with synthesis in NN.

\blue{From an experimental and applied perspective, the Jacobian matrices also have relevance in visual neuroscience}.
Novel psychophysical techniques such as Maximum Differentiation~\cite{Wang08,Malo15,Wang16,Malo16}
synthesize stimuli for the experiments through the gradient of the perceptual distance,
and it depends on the Jacobian w.r.t. the input.
On the other hand, characterizing the Jacobian w.r.t. the parameters is also important.
First, it is relevant in order to learn the L+NL cascade that better reproduces classical experiments (e.g. physiological responses or psychophysical judgements), as opposed to approaches that rely on exhaustive search (as in \cite{Watson02,Laparra10a,Winawer13,Bertalmio17}).
Second, an explicit expression for this Jacobian is important to understand the optimization for alternative goals such as optimal coding, as opposed to approaches that rely on implicit automatic differentiation (as in \cite{LaparraICLR17}).

\blue{Finally, related disciplines such as \emph{image processing} may benefit from analytically interpretable models}.
Reliable subjective image distances (and hence the Jacobian w.r.t. stimulus) have paramount relevance in image processing applications judged by human viewers~\cite{Sakrison77,Watson93,Wang09}. Examples include tone mapping and contrast enhancement~\cite{Mantiuk06}, image coding \cite{Wallace91,Epifanio03,Malo06a}, motion estimation and video coding~\cite{LeGall91,Malo01a,Watson02}, denoising~\cite{Gutierrez06,Laparra10b}, visual pattern recognition~\cite{Coen13}, or search in image databases~\cite{DelBimbo99}. In all these cases, either the subjective distance between the original and the processed image has to be minimized, or the distance used to find image matches has to be perceptually meaningful.
\vspace{-0.15cm}

\paragraph*{Inverse is relevant.} In neuroscience, visual brain decoding~\cite{Kay08,Kamitani05,Marre15} may benefit from the analytic inverse, $S^{-1}$, because it may lead to improvements of the current techniques based on blind regression \cite{Marina16}.
Interestingly, the benefits of the inverse may not only be limited to straightforward improvements in decoding:
the inverse may also give rise to more accurate methods to estimate the model. For instance,
the best parameters of $S$ would be those that lead to better reconstructions through the corresponding $S^{-1}$.
Note that another relevant point of $\nabla S$ is its relation to $S^{-1}$:
according to the theorem of the inverse function \cite{Spivak}, the non-singularity of the Jacobian is the necessary condition for the existence of the inverse.

In the image processing side, the relevance of the inverse is obvious in perceptual image/video coding where the signal is transformed to the perceptual representation prior to quantization~\cite{Wallace91,LeGall91,Malo01a,Malo06a}: decompression implies the inverse to reconstruct the image. Another example is white balance based on human color constancy (or chromatic adaptation): in general, adaptation may be understood as a transform to an invariant representation which is insensitive to irrelevant changes (as for instance the nature of the illumination)~\cite{Laparra12,Gutmann14,Laparra15}. Models of this class of invariant representations could be easily applied for color constancy if the transform is invertible.

\vspace{0.35cm}
In this paper we derive three analytic results for neural models consisting on cascades of canonical Linear+Nonlinear modules:
(i) the Jacobian with regard to the stimulus, (ii) the Jacobian with regard to the parameters, and (iii) the inverse.

We discuss the use of the above results in the context of illustrative
%
% As an example of the general approach introduced to work with these cascaded neural models, we propose a
%
\emph{derivable} and \emph{invertible} vision models made of cascades of L+NL modules.
% Toy example.
% Large scale.
\blue{This kind of models is used to illustrate both (a) the fundamental insight that can be obtained from the analytical expressions
as well as (b) their usefulness in designing new experiments and applications in visual neuroscience.}

\blue{Regarding the insight obtained from analytical expressions, in \textbf{\emph{physiology}},
(a.1) we show how the context-dependence of the receptive fields of the sensors can be
explicitly seen in the expression of the Jacobian w.r.t the stimulus.
Likewise, (a.2) we show that the expression of the Jacobian wrt the parameters reveals that the impact in the response of uncertainty
at the filters (or synaptic weights) may vary over the stimulus space, and this trend may depend on the sensor.
In \textbf{\emph{psychophysics}}, (a.3) we show how the general trends of the sensitivity over the stimulus space can be seen from the
determinant of the metric based on the Jacobian wrt the stimulus.
Finally, in studying the \textbf{\emph{function}} of the system in coding terms, (a.4) we show that the Jacobian wrt the stimulus implies different
efficiency (different multi-information reduction) in different regions of the stimulus space.}

Regarding the experimental and applied interest of the expressions, we address three examples:
\blue{(b.1)} the Jacobian wrt the image is used for stimuli generation in \emph{geometry}-based psychophysics as in \cite{Malo15};
\blue{(b.2)} the Jacobian wrt the parameters is used to maximize the alignment with subjective distortion measures,
improving the brute-force approaches in \cite{Watson02,Laparra10a,Bertalmio17};
and finally, \blue{(b.3)} we discuss how the analytic inverse may be a successful alternative to decoding techniques based on blind linear regression \cite{Stanley99} or nonlinear kernel-ridge regression \cite{Marre15}.

To stress the generality of the modular L+NL cascade
we show examples where some of the canonical Divisive Normalization modules~\cite{Carandini12} are substituted by equivalent
modules such as the Wilson-Cowan interaction~\cite{Wilson72,Cowan16} (at the V1 cortex) or a
tone-mapping model~\cite{Cyriac15} (at the retina).
\blue{Of course, these were selected just as illustrative nonlinearities in an active field in which new alternatives are being explored \cite{Cui13,Simoncelli15,Cui16,Antolik16}.}

\vspace{0.35cm}
Despite the relevance of these ubiquitous neural models, the above mathematical issues have not been addressed in detail in the experimental literature.
%Interestingly, even though the machine learning literature deals with similar (deep) architectures \cite{Goodfellow16}, these details are not made explicit either because of the increasing popularity of automatic differentiation \cite{Baydin17}.
Interestingly, although the machine learning literature deals with similar architectures \cite{Goodfellow16}, these details are also not made explicit due to the growing popularity of automatic differentiation \cite{Baydin17}.
%For instance, the Jacobian w.r.t. the parameters used to optimize biologically plausible L+NL cascades according to psychophysical data or to efficient coding principles is hidden behind automatic differentiation in \cite{LaparraICLR17,LaparraJOSA17}.
For instance, in \cite{Antolik16,Laparra17,LaparraICLR17,LaparraJOSA17,Bethge17}
biologically plausible L+NL architectures are optimized according to
physiological data, psychophysical data or to efficient coding principles.
Unfortunately, the Jacobian w.r.t. the parameters was hidden behind automatic differentiation.

On the contrary, \blue{here we show how the explicit expressions provide intuition on the role of biologically relevant parameters}.

%%%%%%%%%%%%%%%%%%%%%%%%%%%%%%%%%%%%%%%%%%%%%%%%%%%%%%%%%%%%%%%%%%%%%%%%%  END OF 1. INTRO
%%%%%%%%%%%%%%%%%%%%%%%%%%%%%%%%%%%%%%%%%%%%%%%%%%%%%%%%%%%%%%%%%%%%%%%%%%%%%%%%%%%%%%%%%%%%%%%%%%%%%%%%%%%%%%%%%%%%%%%%%%%%%%%%%%%%%%%%%%%%%%%%%%
%%%%%%%%%%%%%%%%%%%%%%%%%%%%%%%%%%%%%%%%%%%%%%%%%%%%%%%%%%%%%%%%%%%%%%%%%%%%%%%%%%%%%%%%%%%%%%%%%%%%%%%%%%%%%%%%%%%%%%%%%%%%%%%%%%%%%%%%%%%%%%%%%%
%%%%%%%%%%%%%%%%%%%%%%%%%%%%%%%%%%%%%%%%%%%%%%%%%%%%%%%%%%%%%%%%%%%%%%%%%%%%%%%%%%%%%%%%%%%%%%%%%%%%%%%%%%%%%%%%%%%%%%%%%%%%%%%%%%%%%%%%%%%%%%%%%%
%%%%%%%%%%%%%%%%%%%%%%%%%%%%%%%%%%%%%%%%%%%%%%%%%%%%%%%%%%%%%%%%%%%%%%%%%%%%%%%%%%%%%%%%%%%%%%%%%%%%%%%%%%%%%%%%%%%%%%%%%%%%%%%%%%%%%%%%%%%%%%%%%%

%%--------------------------------------------
\section{Results}
%%--------------------------------------------

\subsection{Notation and general considerations}
\label{general_cons}

%-------------------------------------------------------------
\subsection*{Stimuli as vectors}
%-------------------------------------------------------------
\addcontentsline{toc}{subsubsection}{Stimuli as vectors}

An \emph{image} in the retina, $\vect{x}^0(\vect{p},\lambda)$, is a function describing the spectral irradiance in each spatial location, $\vect{p}$, and wavelength, $\lambda$.
\blue{Here regular, bold, and capital letters will represent scalars, vectors and matrices (or multivariate applications) respectively.}
Assuming a dense enough sampling, the continuous input can be represented by a discrete spectral array with no information loss, regardless of the specific sampling pattern \cite{Tekalp95}. Here we will assume Cartesian sampling in space and wavelength.
This implies that the spectral cube consists of $b$ matrices of size $h \times w$, where the $l$-th matrix represents the discrete spatial distribution of the energy of $l$-th discrete wavelength ($l = 1,\ldots, b$).
In vision models the spatio-spectral resolution of viewers should determine the sampling frequencies. Given the cut-off frequencies of the contrast sensitivities \cite{Campbell68,Mullen85}, and given the smoothness of the achromatic and opponent spectral sensitivities \cite{Stiles82,Fairchild13}, the spatial dimensions may be sampled at about 80 samples/deg (cpd) and the spectral dimension at about 0.1 samples/nm \cite{Wandell95}.

Using an appropriate rearrangement of the spectral array, the input image can be thought as a \emph{vector} in a $d_0$-dimensional space,
\vspace{-0.4cm}
\begin{equation}
   \vect{x}^0(\vect{p},\lambda) \,\,\,\, \xrightarrow{\,\,\,\,\,\,\,\,\,\, \text{vect} \,\,\,\,\,\,\,\,\,\,} \,\,\,\, \vect{x}^0 = \left(\begin{array}{c}
   %$\xrightarrow{\text{Yes, there is! As long as you like\ldots}}$
                                             x^0_1 \\
                                             x^0_2 \\
                                             \vdots \\
                                             x^0_k \\
                                             \vdots \\
                                             x^0_{d_0} \\
                                         \end{array}
                                         \right)
  \label{the_image}
\end{equation}
i.e. the input {stimuli}, $\vect{x}^0(\vect{p},\lambda)$, which are functions defined in a discrete 3-dimensional domain, are rearranged as a $d_0$-dimensional \emph{column vectors}, $\vect{x}^0 = \text{vect}\left(\vect{x}^0(\vect{p},\lambda)\right) \in \mathbb{R}^{d_0 \times 1}$, where $d_0 = h \times w \times b$. Note that with the considered sampling frequencies, the dimension of the input stimuli is \emph{huge} even for moderate image sizes (small angular field in the visible spectral range).
% For example, a scene subtending 1 degree (in the horizontal and vertical directions) in the [400,700] nm visible range with the referred sampling frequencies implies input vectors of dimension 80*80*30=192000.

The particular scanning pattern in the rearrangement function, $\text{vect}(\cdot)$, has no major relevance as long as it can be inverted back to the original spatio-spectral domain. Here we will use the  \emph{last-dimension-first} convention used in the \texttt{Matlab} functions \texttt{im2col.m} and \texttt{col2im.m}. The selected rearrangement pattern has no fundamental effect, but it has to be taken into account to make sense of the structure of the matrices of the model acting on the input vector.

This rearrangement function, $\text{vect}(\cdot)$, will be also a convenient choice when computing derivatives with regard to the elements of the matrices involved in the model.

%----------------------------------------
\subsection*{The visual pathway: modular L+NL architecture.}
%----------------------------------------
\addcontentsline{toc}{subsubsection}{The visual pathway: modular L+NL architecture}

The visual system may be thought as an operator, $S$, transforming the input $d_0$-dimensional vectors (stimuli) into $d_n$-dimensional output vectors (or sets of $d_n$ responses),
\vspace{-0.25cm}
\begin{equation}
  \xymatrixcolsep{2pc}
  %\xymatrix{ x^0  \,\,\,\, \ar[r]^{\scalebox{1.2}{\emph{S}}} & \,\,\,\, \vect{x}^n
  \xymatrix{ \vect{x}^0  \,\,\,\, \ar@/^1.1pc/[r]^{\scalebox{0.85}{$S(\vect{x}^0,\vect{\Theta})$}} & \,\,\,\, \vect{x}^n
  }
  \label{global_response}
\end{equation}
where $\vect{x}^n \in \mathbb{R}^{d_n \times 1}$ is the response vector, and $\vect{\Theta}$ is the set of parameters of the model. Vectorial output is equivalent to considering $d_n$ separate sensors (or mechanisms) acting on the stimulus, $\vect{x}^0$, leading to the corresponding individual responses, $x^n_k$, where $k = 1, 2, \ldots d_n$.
In this view, the $k$-th sensor would be responsible for the $k$-th dimension of the response vector, $\vect{x}^n$.
The number of separate sensors analyzing the signal may not be the same as the input dimension, so in general $d_n \neq d_0$.
The number of parameters of the model, $d_{\vect{\Theta}}$, depends on the specific functional form of the considered transform.

%\subsubsection*{Multi-layer structure.}
As suggested in \cite{Carandini12}, the global response described above may be decomposed as a series of feed-forward elementary operations, or a cascade of $n$ modules (stages or layers), $S^{(i)}$, where $i = 1,2,\cdots,n$,
\vspace{-0.2cm}
\begin{equation}
  \xymatrixcolsep{2pc}
  %\ar[r]^{1}
  %\ar@[blue][r]^{\textcolor{blue}{\lambda^{-1/2}B^T}}
  \xymatrix{ \vect{x}^0 \ar@/_1pc/[r]_{S^{(1)}} \ar@/^2pc/[rrrrrr]^{\scalebox{0.9}{$S(\vect{x}^0,\vect{\Theta})$}} & \vect{x}^1  \ar@/_1pc/[r]_{S^{(2)}}  & \vect{x}^2 \cdots \!\!\!\!\!\!\!\!\!\!\!\!\!\! & \vect{x}^{i-1} \ar@/_1pc/[r]_{S^{(i)}}  & \vect{x}^i  \cdots \!\!\!\!\!\!\!\!\!\!\!\!\!\! &  \vect{x}^{n-1} \ar@/_1pc/[r]_{S^{(n)}}  & \vect{x}^n
  % \ar@{..>}@/^3pc/[lll]^{M_{V1}^+}
  }
  \label{modular}
\end{equation}
\vspace{-0.0cm}
i.e. the global response is the composition of the elementary responses:
\begin{equation}
    S = S^{(n)} \circ S^{(n-1)} \circ \cdots \circ S^{(2)} \circ S^{(1)}
    \nonumber
\end{equation}

The intermediate representations of the signal along this response path may have different dimension, i.e. $\vect{x}^i \in \mathbb{R}^{d_i \times 1}$, because the number of mechanisms in stage $S^{(i)}$ may be different from the number of mechanisms in $S^{(i-1)}$.
%--------------------------------------------------------
%\subsubsection*{Linear-nonlinear modules.}
%--------------------------------------------------------
%\addcontentsline{toc}{subsubsection}{Linear-nonlinear modules}
Each layer in the above deep network architecture has its own parameters, $\vect{x}^{i+1} = S^{(i)}(\vect{x}^{i},\vect{\Theta}^i)$. Again, $d_{\vect{\Theta}^i}$, depends on the specific functional form of the $i$-th layer.
Each layer performs a linear+nonlinear (L+NL) operation:
\vspace{-0.0cm}
\begin{equation}
  \xymatrixcolsep{2pc}
  \xymatrix{ \cdots \vect{x}^{i-1} \ar@/_2pc/[rr]_{S^{(i)}} \ar[r]^{\,\, \mathcal{L}^{(i)}} & \vect{y}^i \ar[r]^{\mathcal{N}^{(i)}} & \vect{x}^i \cdots}
  \label{module}
  \vspace{-0.0cm}
\end{equation}
i.e. each layer is a composition of two operations: $S^{(i)} = \mathcal{N}^{(i)} \circ \mathcal{L}^{(i)}$.
Let us briefly note that, while in most models the linear operation is followed by the nonlinear one, which is why we use this formulation here, in some instances an inverted scheme of a nonlinear+linear model might be more suitable \cite{Schwartz11, Schwartz12}. That scenario can be handled by our framework as well, after some trivial modification (e.g. choosing the linear operation in the first layer of the L+NL model to be the identity, so that the first layer becomes in practice the nonlinear operation of the first layer followed by the linear operation of the second layer, and the whole cascade gets shifted into a NL+L form).
\vspace{0.1cm}

The linear operation, $\mathcal{L}^{(i)}$, is represented by a matrix $L^i \in \mathbb{R}^{d_i \times d_{i-1}}$. The number of rows in the matrix $L^i$ corresponds to the number of \emph{linear} sensors in layer $S^{(i)}$. This number of mechanisms determines the dimension of the linear output, $\vect{y}^i \in \mathbb{R}^{d_i \times 1}$,
% where the subscript $L$ stands for \emph{linear} (this comment comes from our previous notation x_L),
\begin{equation}
       \vect{y}^i = L^i \cdot \vect{x}^{i-1}
       %\,\,\,\,\, \textrm{where}  \,\,\,\,\,  \vect{x}^i_{L l} = \sum_{m=1}^{d_{i-1}} L^{(i)}_{l m} x^{i-1}_j
       \label{linear}
\end{equation}
In the nonlinear operation, $\mathcal{N}^{(i)}$, each output of the previous linear operation undergoes a saturation transform. Phenomena such as \emph{masking} or \emph{lateral inhibition} imply that the saturation of $y^i_{k}$ should depend on the neighbors $y^i_{k'}$ with $k' \neq k$. This saturation is usually formalized using \emph{divisive normalization} \cite{Carandini12}.
This adaptive saturation is a canonical neural operation and it is at the core of models for color \cite{Brainard05}, motion \cite{Simoncelli98}, and spatial texture vision \cite{Watson97}. Nevertheless, other alternative nonlinearities may be considered as discussed below.
% which modifies the moduli, (as a function of neighbor moduli), but preserves the sign.
In general, this saturating interaction will depend on certain parameters $\vect{\theta}^i$,
\begin{equation}
     \vect{x}^{i} = \mathcal{N}^{(i)}(\vect{y}^i,\vect{\theta}^i)
\end{equation}

Summarizing, in this cascaded setting, the parameters of the $i$-th layer are
(1) the weights of the bank of \emph{linear} sensors represented in $\mathcal{L}^{(i)}$ (the rows of the matrix $L^i$), and
(2) the parameters of the nonlinear saturating interaction, $\mathcal{N}^{(i)}$, i.e.
\begin{equation}
       \vect{\Theta}^i = \{ L^i, \vect{\theta}^i \}
\end{equation}
 Note that according to Eq. \ref{linear}, the rows of the $L^i$ play the same scalar-product role as standard linear receptive fields \cite{Olshausen96,Ringach02}). The only difference is that the rows of $L^i$ are defined in the space of vectors $\vect{x}^{i-1}$ instead of being defined in the input image space (of vectors $\vect{x}^{0}$).

%-------------------------------------------------------------
\subsection*{Canonical and alternative nonlinearities}
%-------------------------------------------------------------
\addcontentsline{toc}{subsubsection}{Canonical and alternative nonlinearities}

\paragraph*{Divisive Normalization in matrix notation.}
%\addcontentsline{toc}{subsubsection}{\red{Divisive Normalization}}
The conventional expressions of the \emph{canonical} divisive normalization saturation use an element-wise
formulation \cite{Carandini12},
\begin{equation}
       x^i_k =  \mathcal{N}^{(i)}(\vect{y}^i,\vect{\theta}^i)_k = \textrm{sign}(y^i_{k}) \,\, \frac{ |y^i_{k}|^{\gamma^i}  }{b^i_k + \sum_{k'} H^i_{k k'} |y^i_{ k'}|^{\gamma^i}} = \textrm{sign}(y^i_{k}) \,\, \frac{|y^i_{k}|^{\gamma^i}}{\mathcal{D}^{(i)}(|\vect{y}^i)|)_k}
       = \textrm{sign}(y^i_{k}) \,\, \frac{e^i_{k}}{\mathcal{D}^{(i)}(\vect{e}^i)_k}
       \label{conventionalDN}
\end{equation}

%\begin{equation}
%       x^i_k =  \mathcal{N}^{(i)}(\vect{y}^i,\vect{\Theta}^i)_k =  \frac{ \sum_{k'} K^i_{k k'} |y^i_{k'}|^{\gamma^i}  }{b^i_k + \sum_{k'} H^i_{k k'} |y^i_{ k'}|^{\gamma^i}} = \frac{\sum_{k'} K^i_{k k'} |y^i_{k'}|^{\gamma^i}}{\mathcal{D}^{(i)}(|\vect{y}^i)|)_k}
%       \label{conventionalDNK}
%\end{equation}
This expression, in which the \emph{energy} of each linear response is $e^i_k = |y^i_k|^{\gamma^i}$, combines conventional \emph{matrix-on-vector} operations (such as the product $H \cdot \vect{e}$ in the denominator)
% effect of the activity of the neighbor mechanisms, $k'$, in the $k$-th mechanism through the matrix $H^{i}$)
with a number of \emph{element-wise} operations: the division of each coefficient of the vector in the numerator by the corresponding coefficient of an inhibitory \emph{denominator vector}, $\mathcal{D}^{(i)}$; the element-wise absolute value (or rectification) to compute the \emph{energy}; the element-wise exponentiation; the element-wise computation of sign, and its preservation in the response through an element-wise product. Therefore, the parameters of this divisive normalization are:
% the interaction matrix in the numerator, $K$;
the excitation and inhibition exponent, $\gamma$; the semisaturation constants in the vector, $\vect{b}$; and the interaction matrix in the denominator, $H$,
\begin{equation}
       \vect{\theta}^i =  \{ \gamma^i, \vect{b}^i, H^i \}
       \nonumber
\end{equation}

%\begin{equation}
%       \vect{\theta}^i =  \{ K^i, \gamma^i, b^i, H^i \}
%       \nonumber
%\end{equation}

%While the interaction in the numerator is usually taken to be the identity matrix since the input to the nonlinear stage is usually taken to be the \emph{energies} of the linear responses (the rectified responses raised to $\gamma$), here we included $K$ because it is convenient in a specific stage (the contrast computation stage) of the specific model used in the illustrations of the discussion. See Supplementary Material S1 for the details of the model.
%More importantly,
The matrix-on-vector operation in the denominator is key in understanding \emph{masking} and \emph{adaptation}.
This is because the $k$-th row of $H^i$ describes how the neighbor activities $|y^i_{k'}|^{\gamma_i}$ saturate (or mask) the response of the $k$-th nonlinear response. The effect of these parameters are extensively analyzed elsewhere \cite{Carandini12}.

From a formal perspective, the combination of element-wise and matrix-on-vector operations in the conventional expression makes differentiation and inversion from Eq. \ref{conventionalDN} extremely cumbersome. This can be alleviated by a matrix-vector expression where the individual coefficients, $k$, are not explicitly present. Incidentally, this matrix expression will imply more efficient code in matrix-oriented environments such as \texttt{Matlab}.

In order to get such matrix-vector form, it is convenient to recall the equivalence between the element-wise (or Hadamard) product and the operation with diagonal matrices \cite{Minka00}.
% Thomas P. Minka. Old and New Matrix Algebra Useful for Statistics. 2000 (cited in the MatrixCookbook2012)
Given two vectors $\vect{a}$ and $\vect{b}$, their Hadamard product is:
\begin{equation}
      \vect{a} \odot \, \vect{b} = \mathds{D}_{\vect{a}} \cdot \vect{b} = \mathds{D}_{\vect{b}} \cdot \vect{a}
      % \nonumber
      \label{haddamard}
\end{equation}
where $\mathds{D}_{\vect{a}}$ is the diagonal matrix with vector $\vect{a}$ in the diagonal.

Using the matrix form of the Hadamard product and the definitions of \emph{energy}, $\vect{e}^i = |\vect{y}^i|^{\gamma^i}$ and
\emph{denominator vector}, $\mathcal{D}^{(i)}( \vect{e}^i ) = \vect{b}^i + H^i \cdot \vect{e}^i$, the conventional Divisive Normalization, Eq. \ref{conventionalDN}, can be re-written with diagonal matrices without referring to the individual components of the vectors:

\begin{equation}
 \vect{x}^i \,\, = \,\, \mathcal{N}^{(i)}(\vect{y}^i,\vect{\theta}^i) \,\, = \,\, \mathds{D}_{\textrm{sign}(\vect{y}^i)} \cdot \mathds{D}_{\left( \vect{b}^i + H^i \cdot \vect{e}^i \right)}^{-1} \cdot \vect{e}^i \,\, = \,\, \mathds{D}_{\textrm{sign}(\vect{y}^i)} \cdot \mathds{D}_{\mathcal{D}^{(i)}(\vect{e}^i)}^{-1} \cdot \vect{e}^i
 \label{divisive_norm2}
\end{equation}

%\begin{equation}
% \begin{aligned}
% \vect{x}^i = \mathcal{N}^{(i)}(\vect{y}^i,\vect{\theta}^i) = & \mathds{D}_{\textrm{sign}(\vect{y}^i)} \cdot \mathds{D}_{\mathcal{D}^{(i)}(\vect{e}^i)}^{-1} \cdot \vect{e}^i \\
% \textrm{where} \,\,\,\, & \vect{e}^i = |\vect{y}^i|^{\gamma^i} \\
%                \,\,\,\, & \mathcal{D}^{(i)}( \vect{e}^i ) = \vect{b}^i + H^i \cdot \vect{e}^i
% \end{aligned}
% \label{divisive_norm2}
%\end{equation}

%\begin{equation}
% \vect{x}^i = \mathcal{N}^{(i)}(\vect{y}^i,\vect{\Theta}^i) = \mathds{D}_{\mathcal{D}^{(i)}(\vect{e}^i)}^{-1} \cdot K^i \cdot \vect{e}^i
% \label{divisive_norm2K}
%\end{equation}
%\vspace{0.2cm}

\noindent where the model parameters are $\vect{\theta}^i =  \{ \gamma^i, \vect{b}^i, H^i \}$.
%Here the normalization only acts on the absolute values, but the matrix $D_{\textrm{sign}(\vect{y}^i)}$ acting on the vector $\mathcal{N}^{(i)}(a)$ preserves the sign of each $y^i_{k}$.
%Note that all the operations in $\mathcal{N}^{(i)}(a)$ are element-wise except the matrix-on-vector, $H^{(i)} \cdot a$, within the denominator vector $\mathcal{D}^{(1)}(a)$.
Similarly to Von-Kries adaptation \cite{Fairchild05}, this matrix form of Divisive Normalization is nonlinear because the diagonal of the matrix depends on the signal.
The derivation of the results (proofs given in the Supplementary Materials) shows that the above matrix version of Divisive Normalization is extremely convenient to avoid cumbersome individual element-wise partial derivatives and to compute the analytic inverse.

%Of course variations exist (\emph{Note!}): \red{Naka-Rushton, different excitation/inhibition exponents, different regularized versions (e.g. see \cite{Carandini12})}, but the important concept is \emph{saturation} and \emph{interaction} through $H$. This leads to a neighbor-dependent adaptive saturation. Simplified dimension-wise saturation (logarithm, exponents $<1$,...) neglect
%the interactions. \red{(\emph{Note!}) Figure and connections with deep-networks.} Throughout the document the text (\emph{Note!}) means that the corresponding paragraph requires additional elaboration.

\paragraph*{Alternative nonlinearities: Wilson-Cowan equations and tone-mapping.}
%\addcontentsline{toc}{subsubsection}{\red{Wilson-Cowan as nonlinear transform}}
Even though all the elementary L+NL layers of the deep network in Eq. \ref{modular} could be implemented by a composition of Eqs. \ref{linear} and \ref{divisive_norm2} (as suggested in \cite{Carandini12}), here we also consider particular alternatives for
the nonlinearities that have been proposed to account for the response at specific stages in the visual pathway.
Namely, the Wilson-Cowan equations \cite{Wilson72,Cowan16}, which could account for the masking between local-oriented sensors \cite{Bertalmio17};
and nonlinear models of brightness perception such as the ones used in tone mapping \cite{Cyriac15,Cyriac16,KaneBertalmio16}.
The consideration of these alternatives for specific stages stresses the generality of the proposed framework since, as shown in the examples of the Discussion, the network equations can be applied no matter the specific functional form of each stage (provided the elementary derivatives and inverses are known).

The Wilson-Cowan equations \cite{Wilson72,Cowan16} describe the temporal evolution of the mean activity of a population of neurons at the V1 cortex. In what follows, we consider the following form of the Wilson-Cowan equations
\begin{equation}
\label{Wilson-Cowan equation}
\overset{.}{\vect{x}^i}(t) = - \alpha \, \vect{x}^i(t) + \mu \, W \cdot f ( \vect{x}^i(t)) \, + \lambda \vect{y}^i
%\overset{.}{\vect{x}^i}_{\!\!\! k}(t) = - \alpha \, \vect{x}^i_{\!k}(t) + \mu \sum_{k' \in \{1,\cdots,d_i\}} W_{k,k'} \, f ( \vect{x}^i_{\!k'}(t)) \, + \lambda \vect{y}^i_k
\end{equation}
where $\alpha,\mu,\lambda$ are coupling coefficients, $W=W_{k,k'}$ is a kernel which decays with the difference $|k-k'|$, $f$ is a sigmoid function and $t$ is time.\\
\\
The steady-state equation of the evolution equation (\ref{Wilson-Cowan equation}) is
\begin{equation}
\label{steady-state Wilson-Cowan equation}
%0 = - \alpha \, \vect{x}^i_{\!k} + \mu \sum_{k' \in \{1,\cdots,d_i\}} W_{k,k'} \, f ( \vect{x}^i_{\!k'}) \, + \lambda \vect{y}^i_k
0 = - \alpha \, \vect{x}^i + \mu \, W \cdot \, f ( \vect{x}^i) \, + \lambda \vect{y}^i
\end{equation}
Existence and uniqueness of the solution of the steady-state equation (\ref{Wilson-Cowan equation}) are not guaranteed in the general case. We refer the reader to \cite{Palma-Amestoy09} for some conditions on the coefficients \chngP{$\alpha,\mu,\lambda$} and the sigmoid $f$ for which the existence and uniqueness of the solution is guaranteed.\\
\\
From now on, we assume that we are in a case where we have existence and uniqueness of the solution of the steady-state equation. Then, we define the Wilson-Cowan transform $N^{(i)}(\vect{y}^i)$ of $\vect{y}^i$ as the unique solution $\vect{x}^i$ of the steady-state equation.

\vspace{0.2cm}

While the Wilson-Cowan equations are sensible for populations of cortical neurons, brightness-from-luminance models may account for nonlinearities at earlier stages of the visual pathway (e.g. in the retina).
An illustrative example of these specific nonlinearities which is connected to image enhancement applications through tone mapping is the two-gamma model in \cite{Cyriac16}.
In this model the nonlinear saturation is a simple exponential function with no interaction between neighbor dimensions,
\begin{equation}
    \vect{x} = \textrm{sign}(\vect{y}) \odot |\vect{y}|^{\gamma(\,|\vect{y}|\,)}
    \label{two-gamma}
\end{equation}
where all operations (sign, rectification, exponentiation) are dimension-wise.
However, note that the exponent is a function of the magnitude of the input tristimulus value. Specifically,
\begin{equation}
      \gamma(|\vect{y}|)  =  \gamma_H - (\gamma_H-\gamma_L)\cdot \dfrac{\mu_1^m}{(\mu_1^m + |\vect{y}|^m)}
      \label{exponente}
\end{equation}
The exponent has different values for low and high inputs, $\gamma_L$ and $\gamma_H$ respectively (hence the two-gamma name).
The transition of $\gamma$ between $\gamma_L$ and $\gamma_H$ happens around the value \chngP{$|\vect{y}| = \mu_1$}.
This transition is smooth, and its sharpness is controlled by the exponent $m$.
% \red{(\emph{Note!}) Figure on the nonlinear response ($\gamma$ on inset).}

This expression for $\gamma$ has statistical grounds since the resulting nonlinearity approximately equalizes the \blue{\emph{probability density function}} (PDF) of luminance values in natural scenes \cite{HuangMumford99,KaneBertalmio16},
which is a sensible goal in the information maximization context \cite{Laughlin83}.
This nonlinearity can be applied both to linear luminance values \cite{Cyriac16,Cyriac16b} as well as to linear opponent color channels \cite{MacLeod03b,Laparra12}.
Therefore, this specific nonlinearity could be applied after a linear stage where the spectrum in each spatial location is transformed into opponent tristimulus values.
Special modification of the nonlinearity around zero is required to address the singularity of the derivative in zero. We will be more specific on this point when we address
the Jacobian of this two-gamma model below.

%--------------------------------------------------------
\subsection*{Jacobian matrices of L+NL cascades}
%--------------------------------------------------------
\addcontentsline{toc}{subsubsection}{Jacobian matrices of L+NL cascades}

In the modular setting outlined above, variation of the responses may come either from variations of the stimulus, $\vect{x}^0$, or from variations of the parameters, $\vect{\Theta}$.
On the one hand, for a given set of fixed parameters, many properties of the sensory system depend on how the output depends on the stimuli, i.e. many properties depend on the Jacobian of the transform with regard to the image, $\nabla_{\!\!\vect{x}^0} S$ (where the subindex at the derivative operator indicates the derivation variable).
In particular, this Jacobian is critical to decode the neural representation (existence of inverse), and to describe perceptual distance between stimuli.
As an example, the Discussion shows how this Jacobian is key in the generation of stimuli fulfilling certain geometric requirements involved in recent psychophysics.
% \red{Supplementary Material S3 gives a more exhaustive list of properties that depend on $\nabla_{\!\!\vect{x}^0} S$}.
On the other hand, when looking for the parameters that better explain certain experimental behavior, it is necessary to know how the response depends on the parameters,
i.e. the key is the Jacobian with regard to the parameters, $\nabla_{\!\!\vect{\Theta}} S$.
As an example, the Discussion shows how this Jacobian can be used to maximize the correlation with subjective opinion in visual distortion psychophysics.

In these notation preliminaries we address the general properties of these Jacobian matrices (both $\nabla_{\!\!\vect{x}^0} S$ and $\nabla_{\!\!\vect{\Theta}} S$) in the context of the modular network outlined above.
% and its effect on the invertibility of the response transform (existence of $S^{-1}$).
The interest of these preliminaries is that we show that the problem of computing $\nabla_{\!\!\vect{x}^0} S$ and $\nabla_{\!\!\vect{\Theta}} S$ reduces to the computation of the Jacobian matrices of the elementary
nonlinearities ($\nabla_{\!\!\vect{y}^i} \mathcal{N}^{(i)}$ and $\nabla_{\!\!\vect{\theta}^i} \mathcal{N}^{(i)}$ respectively).
These elementary Jacobians, $\nabla_{\!\!\vect{y}^i} \mathcal{N}^{(i)}$ and $\nabla_{\!\!\vect{\theta}^i} \mathcal{N}^{(i)}$, and the inverse, ${\mathcal{N}^{(i)}}^{-1}$ (whose existence is related to $\nabla_{\!\!\vect{y}^i} \mathcal{N}^{(i)}$), are the three analytical results of the paper, and will be addressed in the next subsections. Specifically, for the divisive normalization, in Eq. \ref{deriv_DN} (result I), Eqs. \ref{jacobian2_global}-\ref{dNdc} (result II), and Eq. \ref{inv_DN} (result III).
%The expressions presented in the remaining of this notation subsection come from
%straightforward application of the properties of (1) the derivatives of composition of functions (the chain rule),
%and (2) the integrability of initial value problems \cite{Logan94}.

\paragraph*{Local-linear approximation.}
The response function, $S$, can be seen as a nonlinear change of coordinates depending on the (independent) variables $\vect{x}^0$ and $\vect{\Theta}$.
Therefore, around certain $(\vect{x}^0_A, \vect{\Theta}_A)$, this function can be expanded in Taylor series and its properties depend
on the matrices of derivatives with regard to these variables \cite{Dubrovin82,Spivak}, in this case, the Jacobian matrices $\nabla_{\!\!\vect{x}^0} S$ and $\nabla_{\!\!\vect{\Theta}} S$,
\begin{equation}
      \Delta \vect{x}^n = \nabla_{\!\!\vect{x}^0} S \cdot \Delta \vect{x}^0 + \nabla_{\!\!\vect{\Theta}} S \cdot \Delta \vect{\Theta}
      \label{linear_approx}
\end{equation}
This is the \emph{local-linear approximation} of the nonlinear response for small perturbations of the stimulus or the parameters.
In Eq. \ref{linear_approx} the derivatives are computed at $(\vect{x}^0_A,\vect{\Theta}_A)$, the vector $\Delta \vect{x}^0 \in \mathbb{R}^{d_0 \times 1}$ is the variation of the stimulus; and $\Delta \vect{\Theta} \in \mathbb{R}^{d_{\vect{\Theta}} \times 1}$
is a vector with a perturbation of the $d_{\vect{\Theta}}$ parameters in the model.
%As will be clear below, by concatenating the parameters of the different layers and vectorizing when necessary, it is possible to build a single %vector of parameters $\vect{\Theta}$ of dimension $d_{\vect{\Theta}}$. The first point addressed in the Methods section \ref{dLineardL} shows an %explicit example of the convenience of the vectorization of the parameters.
Note that the column vector of model parameters (of dimension $d_\Theta$) is obtained
simply by concatenating the parameters of the different layers.

The Jacobian with regard to the parameters necessarily has variables from different layers, so it makes an extensive use of the chain rule.
Therefore, lets start with the Jacobian with regard to the stimulus and then, let's introduce the chain rule for this simpler case.

\paragraph*{Global Jacobian with regard to the stimulus.}
At certain point $\vect{x}^0_A$, one may make independent variations in all the dimensions of the input. Note that statistical independence of the dimensions of the stimuli is a different issue (different from formal mathematical independence in the expression). Actually, in general, the dimensions of natural stimuli are not statistically independent \cite{Barlow01,Laparra15}.
Omitting the (fixed) parameters, $\vect{\Theta}_A$, for the sake of clarity, the Jacobian with regard to the input is the following concatenation
(independent variables imply concatenation of derivatives \cite{Dubrovin82,Spivak}),
%\begin{eqnarray}
%      \nabla_{\!\!\vect{x}^0} S(\vect{x}^0_A) &=&
%      \Bigg[\frac{\partial \vect{x}^n(\vect{x}^0_A)}{\partial x^0_1}, \cdots, \frac{\partial \vect{x}^n(\vect{x}^0_A)}{\partial x^0_j}, \cdots ,\frac{\partial \vect{x}^n(\vect{x}^0_A)}{\partial x^0_{d_0}} \Bigg]
%      \nonumber
%      %\label{global_jacobian}
%\end{eqnarray}
\begin{eqnarray}
      \nabla_{\!\!\vect{x}^0} S(\vect{x}^0_A) &=&
      \Bigg[\frac{\partial S(\vect{x}^0_A)}{\partial x^0_1}, \cdots, \frac{\partial S(\vect{x}^0_A)}{\partial x^0_j}, \cdots ,\frac{\partial S(\vect{x}^0_A)}{\partial x^0_{d_0}} \Bigg]
      \nonumber
      %\label{global_jacobian}
\end{eqnarray}
%where $\frac{\partial \vect{x}^n(\vect{x}^0_A)}{\partial x^0_j} \in \mathbb{R}^{d_n \times 1} \,\,\,\, \forall j$.
where $\frac{\partial S(\vect{x}^0_A)}{\partial x^0_j} \in \mathbb{R}^{d_n \times 1} \,\,\,\, \forall j$.
Expanding these column vectors, we see that
$\nabla_{\vect{x}^0} S \in \mathbb{R}^{d_n \times d_0}$:
\begin{eqnarray}
      \nabla_{\!\!\vect{x}^0} S(\vect{x}^0_A)
          &=&
      \begin{bmatrix}
      \dfrac{\partial S(\vect{x}^0_A)_1}{\partial x_1^0} &        & \dfrac{\partial S(\vect{x}^0_A)_1}{\partial x_j^0} &        & \dfrac{\partial S(\vect{x}^0_A)_1}{\partial x_{d_0}^0}\\
      \vdots&       & \vdots &    &\vdots\\
      \dfrac{\partial S(\vect{x}^0_A)_k}{\partial x_1^0} & \cdots &\dfrac{\partial S(\vect{x}^0_A)_k}{\partial x_j^0}  & \cdots &  \dfrac{\partial S(\vect{x}^0_A)_k}{\partial x_{d_0}^0}\\
       \vdots&      & \vdots &    &\vdots\\
       \dfrac{\partial S(\vect{x}^0_A)_{d_n}}{\partial x_1^0} &     &\dfrac{\partial S(\vect{x}^0_A)_{d_n}}{\partial x_j^0}  &     &  \dfrac{\partial S(\vect{x}^0_A)_{d_n}}{\partial x_{d_0}^0}
      \end{bmatrix}
      \label{global_jacobian1}
\end{eqnarray}
%\begin{eqnarray}
%      \nabla_{\!\!\vect{x}^0} S(\vect{x}^0_A) &=&
%      \Bigg[\frac{\partial \vect{x}^n(\vect{x}^0_A)}{\partial x^0_1}, \cdots, \frac{\partial \vect{x}^n(\vect{x}^0_A)}{\partial x^0_j}, \cdots ,\frac{\partial \vect{x}^n(\vect{x}^0_A)}{\partial x^0_{d_0}} \Bigg] \\[0.5cm]
%          &=&
%      \begin{bmatrix}
%      \dfrac{\partial x^n(\vect{x}^0_A)_1}{\partial x_1^0} & \cdots & \dfrac{\partial x^n(\vect{x}^0_A)_1}{\partial x_j^0} & \cdots & \dfrac{\partial x^n(\vect{x}^0_A)_1}{\partial x_{d_0}^0}\\
%      \vdots& \ddots& \vdots &\ddots &\vdots\\
%      \dfrac{\partial x^n(\vect{x}^0_A)_k}{\partial x_1^0} & \cdots &\dfrac{\partial x^n(\vect{x}^0_A)_k}{\partial x_j^0}  & \cdots &  \dfrac{\partial x^n(\vect{x}^0_A)_k}{\partial x_{d_0}^0}\\
%       \vdots& \ddots& \vdots &\ddots &\vdots\\
%       \dfrac{\partial x^n(\vect{x}^0_A)_{d_n}}{\partial x_1^0} & \cdots &\dfrac{\partial x^n(\vect{x}^0_A)_{d_n}}{\partial x_j^0}  & \cdots &  \dfrac{\partial x^n(\vect{x}^0_A)_{d_n}}{\partial x_{d_0}^0}
%      \end{bmatrix}
%      \nonumber
%      %\label{global_jacobian}
%\end{eqnarray}

Note that this Jacobian may depend on the input, $\vect{x}^0$, because the \emph{slope} of the response
(the behavior of the system) may be different in different points of the stimulus space.

Note also that, for fixed parameters, according to Eq. \ref{linear_approx}, the global nonlinear behavior of the system can be linearly approximated in a neighborhood of some stimulus, $\vect{x}^0_A$,
using the Jacobian with regard to the stimulus, i.e. variations of the response linearly depend on variations of the input for small distortions $\Delta \vect{x}^0$.
%\begin{eqnarray}
%      \nonumber S(\vect{x}^0_A + \Delta \vect{x}^0) & \approx & S(\vect{x}^0_A) + \nabla S(\vect{x}^0_A) \cdot \Delta \vect{x}^0 \\[2mm]
%                 \Delta \vect{x}^n & \approx & \nabla S(\vect{x}^0_A) \cdot \Delta \vect{x}^0
%      \label{linear_approx}
%\end{eqnarray}

\paragraph*{Chain rule: global Jacobian in terms of the Jacobians of the layers.}
The Jacobian of the composition of functions (e.g. the multi-layer architecture we have here), can be decomposed as the product of the individual Jacobian matrices.
For example, given the composition, $f \circ g \circ h =  f\Big(g\big(h(\vect{x})\big)\Big)$, the application of the \emph{chain rule} leads to:
\begin{equation}
     \nabla_{\!\!\vect{x}} \, f \Big(g\big(h(\vect{x})\big)\Big) = \frac{\partial f}{\partial \vect{x}} = \frac{\partial f}{\partial g} \cdot \frac{\partial g}{\partial h} \cdot \frac{\partial h}{\partial \vect{x}} = \nabla_{\!\!g} \, f \cdot \nabla_{\!\!h} \, g \cdot \nabla_{\!\!\vect{x}} \, h
     \nonumber
\end{equation}
% where the derivation variable is explicitly indicated as a subindex at each derivative symbol.
Note that when inputs and outputs are  multidimensional (\emph{matrix chain-rule}) the order of the product of Jacobians is important for obvious reasons.
Following the above, the Jacobian of the cascade can be expressed in terms of the Jacobian of each layer:
\begin{equation}
      \nabla_{\!\!\vect{x}^0} S = \nabla_{\!\!\vect{x}^{n-1}} S^{(n)} \cdot \nabla_{\!\!\vect{x}^{n-2}} S^{(n-1)} \cdot \cdots \cdot \nabla_{\!\!\vect{x}^{i-1}} S^{(i)} \cdot \cdots \cdot \nabla_{\!\!\vect{x}^{1}} S^{(2)} \cdot \nabla_{\!\!\vect{x}^{0}} S^{(1)} = \prod_{i = n}^{1} \nabla_{\!\!\vect{x}^{i-1}} S^{(i)}
      \label{chain_rule1}
\end{equation}
Similarly to $\nabla_{\!\!\vect{x}^{0}} S$, in general $\nabla_{\!\!\vect{x}^{i-1}} S^{(i)}$ \blue{depends on the input} and is rectangular. Note that $\nabla_{\!\!\vect{x}^{i-1}} S^{(i)} \in \mathbb{R}^{d_{i} \times d_{i-1}}$.
Given the L+NL structure of each layer, $S^{(i)} = \mathcal{N}^{(i)} \circ \mathcal{L}^{(i)}$, we can also apply the chain rule inside each layer,
\begin{equation}
      \nabla_{\!\!\vect{x}^{i-1}} S^{(i)} = \nabla_{\!\!\vect{y}^{i}} \mathcal{N}^{(i)} \cdot L^i
      \label{chain_rule2}
\end{equation}
where we used the trivial derivative of a linear function \cite{Matrix12}: $\nabla_{\!\!\vect{x}^{i-1}} \mathcal{L}^{(i)} = \nabla_{\!\!\vect{x}^{i-1}} L^i \cdot \vect{x}^{i-1} = L^i$.

Note that assuming we know the parameters of the system (the linear weights, $L^i$, in each layer, and the parameters of the nonlinearities, $\vect{\theta}^i$),
after Eqs. \ref{chain_rule1} and \ref{chain_rule2} the final piece to compute the Jacobian of the system with regard to the stimulus
is the Jacobian of the specific nonlinearities, $\nabla_{\!\!\vect{y}^{i}} \mathcal{N}^{(i)}$.
Solving this remaining unknown will be the first analytical result of the paper (Result I), namely Eq. \ref{deriv_DN}.

\paragraph*{Jacobian with regard to the parameters.}
For a given set of parameters, $\vect{\Theta}_A$, one may introduce independent perturbations in the parameters of each layer.
Therefore, the Jacobian with regard to the parameters is the following concatenation,
\begin{eqnarray}
      \nabla_{\!\!\vect{\Theta}} S &=&
      \Bigg[\nabla_{\!\!\vect{\Theta}^1} S \,\,\,\, \nabla_{\!\!\vect{\Theta}^2} S \,\,\,\, \cdots \,\,\,\, \nabla_{\!\!\vect{\Theta}^n} S \Bigg]
      \label{global_jacobian2}
\end{eqnarray}
where each $\nabla_{\!\!\vect{\Theta}^i} S \in \mathbb{R}^{d_n \times d_{\vect{\Theta}^i}}$ is a rectangular matrix with $d_{\vect{\Theta}^i}$ being the dimension of $\vect{\Theta}^i$;
and the input $(\vect{x}_A, \vect{\Theta}_A)$ was omitted for the sake of clarity.
Note that actual independence among the different parameters is different from formal mathematical independence in the expression. In fact, certain interaction between layers can be required to get certain computational goal.

Applying the chain rule for the Jacobian with regard to the parameters of the $i$-th layer,
\begin{eqnarray}
      \nabla_{\!\!\vect{\Theta}^i} S &=& \frac{\partial \vect{x}^n}{\partial \vect{x}^{n-1}} \cdot \frac{\partial \vect{x}^{n-1}}{\partial \vect{x}^{n-2}} \cdot \,\, \cdots \,\, \cdot \frac{\partial \vect{x}^{i+1}}{\partial \vect{x}^{i}} \cdot \frac{\partial \vect{x}^{i}}{\partial \vect{\Theta}^{i}} \nonumber \\
                           &=& \left[\prod_{l=n}^{i+1} \nabla_{\!\!\vect{x}^{l-1}} S^{(l)}\right]  \cdot  \nabla_{\!\!\vect{\Theta}^i} S^{(i)}
      \label{chain_rule3}
\end{eqnarray}
Note how Eq. \ref{chain_rule3} makes sense, both dimensionally and qualitatively. First, note that $\prod_{l=n}^{i+1} \nabla_{\!\!\vect{x}^{l-1}} S^{(l)} \in \mathbb{R}^{d_n \times d_i}$ and
$\nabla_{\!\!\vect{\Theta}^i} S^{(i)} \in \mathbb{R}^{d_i \times d_{\vect{\Theta}^i}}$. Second, it makes sense that the effect of changing the parameters in the $i$-th layer has two terms: one describing
how the change affects the response of this layer (given by $\nabla_{\!\!\vect{\Theta}^i} S^{(i)}$), and other describing the propagation of the perturbation through the remaining layers of the network
(given by the product of the other Jacobians -with regard to the stimulus!-, $\prod_{l=n}^{i+1} \nabla_{\!\!\vect{x}^{l-1}} S^{(l)}$).

\vspace{0.3cm}
Now, taking into account that in each layer the parameters come from the linear and the nonlinear parts, and these could be varied independently, we obtain:
\begin{equation}
      \nabla_{\!\!\vect{\Theta}^i} S^{(i)} = \Bigg[ \nabla_{\!\!L^i} S^{(i)} \,\,\,\,\,\,\,\,\, \nabla_{\!\!\vect{\theta}^i} S^{(i)} \Bigg] = \Bigg[ \frac{\partial \vect{x}^i}{\partial \vect{y}^i} \cdot \frac{\partial \vect{y}^i}{\partial L^i} \,\,\,\,\,\,\,\,\, \nabla_{\!\!\vect{\theta}^i} \mathcal{N}^{(i)} \Bigg]
      \nonumber
\end{equation}
where we applied the chain rule in the Jacobian with regard to the matrix $L^i$, and the fact that, by definition, $\nabla_{\!\!\vect{\theta}^i} S^{(i)} = \nabla_{\!\!\vect{\theta}^i} \mathcal{N}^{(i)}$.

Further development of the first term requires the use of the derivative of a linear function with regard to the elements in the matrix $L^i$. This technical issue is addressed in the Supplementary Material \ref{dLineardL}. Using the result derived there, namely Eq. \ref{deriv_linear_param}, the above equation reduces to:
\begin{equation}
      \nabla_{\!\!\vect{\Theta}^i} S^{(i)} = \Bigg[ \nabla_{\!\!\vect{y}^i} \mathcal{N}^{(i)} \cdot \mathds{B}^{d_i}_{({\vect{x}^{i-1}}^\top)} \,\,\,\,\,\,\,\,\, \nabla_{\!\!\vect{\theta}^i} \mathcal{N}^{(i)} \Bigg]
      \label{elementary_jacobian2}
\end{equation}
where, as stated in Eq. \ref{deriv_linear_param}, $\mathds{B}^{d_i}_{({\vect{x}^{i-1}}^\top)}$ is just a block diagonal matrix made from $d_i$ replications of the (known) vector $\vect{x}^{i-1}$,
and this expression assumes that the elements of the perturbations $\Delta L^i$ are vector-arranged row-wise, e.g. using $vect({\Delta L^i}^\top)$.
Note that in Eq. \ref{elementary_jacobian2}, the only unknown terms are the Jacobian of the nonlinearity: $\nabla_{\!\!\vect{y}^i} \mathcal{N}^{(i)}$, already referred to as the first analytical result of this work (Eq. \ref{deriv_DN}), and
$\nabla_{\!\!\vect{\theta}^i} \mathcal{N}^{(i)}$, which will be the second analytical result of the work (Result II), namely Eqs. \ref{jacobian2_global}-\ref{dNdc}.

%--------------------------------------------------------
\subsection*{Jacobian and perceptual distance}
%--------------------------------------------------------
\addcontentsline{toc}{subsubsection}{Jacobian and perceptual distance}

In the input-output setting represented by $S$, perceptual decisions (e.g. discrimination between stimuli) will be made on the basis of the information available in the response (output) space and not in the input space.
This role of the response space in stimulus discrimination is consistent with
(i) the psychophysical practice that assumes uniform just noticeable differences in the response domain to derive
the slope of the response from experimental thresholds \cite{Watson97,Brainard05,Laparra12}, and
(ii) the formulation of subjective distortion metrics as Euclidean measures in the response domain \cite{Teo94a,Epifanio03,Malo06a,Laparra10a}.

\paragraph*{Perceptual distance: general expression.} The perceptual distance, $\mathbb{d}_p$, between two images, $\vect{x}^0_A$ and $\vect{x}^0_B$, can be defined as the \emph{Euclidean distance in the response domain}:
\begin{equation}
      \mathbb{d}^2_p(\vect{x}^0_A, \vect{x}^0_B) = \left|\vect{x}^n_B - \vect{x}^n_A\right|^2_2 = \left(\vect{x}^n_B - \vect{x}^n_A\right)^\top \cdot \left(\vect{x}^n_B - \vect{x}^n_A\right) =  \Delta {\vect{x}^n}^\top \cdot \Delta {\vect{x}^n}
      \label{distance}
\end{equation}

An Euclidean distance in the response domain implies a non-Euclidean measure in the input
image domain \cite{Dubrovin82,Pons99,Epifanio03,Laparra10a,Laparra17}.
One may imagine that, for nontrivial $S^{-1}$, the inverse of the points in the sphere of radius $|\Delta \vect{x}^n|_2$ around the point $\vect{x}^n_A$ will no longer be a sphere (not even a convex region!) in the input space.
The size and orientation of these \emph{discrimination regions} determine the visibility of distortions $\Delta \vect{x}^0$
on top of certain background image, $\vect{x}^0_A$.
Different Euclidean lengths in the image space (different $|\Delta \vect{x}^0|_2$) will be required in different directions in order to lead to the same perceptual distance $ \mathbb{d}_p$.
The variety of orientations and sizes of the well-known Brown-MacAdam color discrimination regions \cite{Brown49} is an intuitive
(just three-dimensional) example of the above concepts.
% Brown, Walter R.J.; MacAdam, David L. (October 1949). "Visual sensitivities to combined chromaticity and luminance differences" (abstract). JOSA 39 (10): 808–834. doi:10.1364/JOSA.39.000808

\paragraph*{Perceptual distance: 2nd-order approximation.} Assuming the local-linear approximation of the response around the reference image, Eq. \ref{linear_approx}, we have $\Delta \vect{x}^n = \nabla_{\vect{x}} S(\vect{x}^0_A) \cdot \Delta \vect{x}^0$. Under this approximation, the perceptual distance from the reference image reduces to:
\begin{equation}
  \mathbb{d}^2_p(\vect{x}^0_A, \vect{x}^0_B) = \Delta {\vect{x}^0}^\top \cdot \nabla_x S(\vect{x}^0_A)^\top \cdot \nabla_x S(\vect{x}^0_A) \cdot \Delta \vect{x}^0,
  \label{distance2}
\end{equation}
with $\vect{x}^0_B=\vect{x}^0_A+\Delta {\vect{x}^0}$. Therefore, the matrix $M(\vect{x}^0_A) = \nabla_x S(\vect{x}^0_A)^\top \cdot \nabla_x S(\vect{x}^0_A)$ plays the role of a non-Euclidean metric matrix induced by the sensory system.
This is a 2nd-order approximation because in this way, perceived distortion only depends on the interaction between the deviations in \emph{pairs} of locations: $M(\vect{x}^0_A)_{ij} \Delta x^0_i \Delta x^0_j$.

Note that a constant value for the distance in Eq. \ref{distance2} defines an ellipsoid oriented and scaled according to the metric matrix $M(\vect{x}^0_A)$.
In this 2nd-order approximation, the \emph{discrimination regions} reduce to \emph{discrimination ellipsoids}. The properties of these ellipsoids depend on the metric and hence on
the Jacobian of the response w.r.t. the stimulus (i.e. on Result~I below). In particular, the orientation depends on the eigenvectors of $M$ and the scaling depends on the eigenvalues.

The simplicity of Eq.\ref{distance2} depends on the assumption of quadratic norm in Eq. \ref{distance} (as opposed to other possible summation exponents in Minkowski metrics \cite{Spivak}).
Note that using other norms would prevent writing the distance in the response domain through the dot product of $\Delta \vect{x}^n$.
Therefore, the linear approximation would not be that easy.
With non-quadratic summation the distance would still depend on the elements of the Jacobian (and hence on Result~I),
but the expression would be more complicated, and the reasoning through Jacobian-related eigenvectors would not be as intuitive.

%%%%%%%%%%%%%%%%%%%%%%%%%%%%%%%%%%%%%%%%%%%%%%%%%%%%%%%%%%%%%%%%%%%%%%%%%  END OF 2.1 NOTATION
%%%%%%%%%%%%%%%%%%%%%%%%%%%%%%%%%%%%%%%%%%%%%%%%%%%%%%%%%%%%%%%%%%%%%%%%%%%%%%%%%%%%%%%%%%%%%%%%%%%%%%%%%%%%%%%%%%%%%%%%%%%%%%%%%%%%%%%%%%%%%%%%%%
%%%%%%%%%%%%%%%%%%%%%%%%%%%%%%%%%%%%%%%%%%%%%%%%%%%%%%%%%%%%%%%%%%%%%%%%%%%%%%%%%%%%%%%%%%%%%%%%%%%%%%%%%%%%%%%%%%%%%%%%%%%%%%%%%%%%%%%%%%%%%%%%%%
%%%%%%%%%%%%%%%%%%%%%%%%%%%%%%%%%%%%%%%%%%%%%%%%%%%%%%%%%%%%%%%%%%%%%%%%%%%%%%%%%%%%%%%%%%%%%%%%%%%%%%%%%%%%%%%%%%%%%%%%%%%%%%%%%%%%%%%%%%%%%%%%%%
%%%%%%%%%%%%%%%%%%%%%%%%%%%%%%%%%%%%%%%%%%%%%%%%%%%%%%%%%%%%%%%%%%%%%%%%%%%%%%%%%%%%%%%%%%%%%%%%%%%%%%%%%%%%%%%%%%%%%%%%%%%%%%%%%%%%%%%%%%%%%%%%%%

\subsection{Result I: Jacobian with regard to the stimulus}
\label{result1}

The problem of computing the Jacobian with regard to the stimulus in the cascade of L+NL modules, $\nabla_{\!\!\vect{x}^0} S$,
reduces, according to Eqs. \ref{chain_rule1} and  \ref{chain_rule2}, to the computation of the Jacobian of
the nonlinearity with regard to the stimulus in every layer, $\nabla_{\!\!\vect{y}^i} \mathcal{N}^{(i)}$.
In this section we give the analytical result of the required Jacobian, $\nabla_{\!\!\vect{y}^i} \mathcal{N}^{(i)}$, in the
canonical divisive normalization case, and for two alternative nonlinearities. Proofs of this first set of analytical results are
given in the Supplementary Material \ref{J_wrt_stimul}.
The role of this analytical result in generating stimuli for novel psychophysics is illustrated in the Discussion, Section \ref{MAD_section}.

%----------------------------------------------------------
\paragraph*{Jacobian of the canonical nonlinearity with regard to the stimulus.}
%----------------------------------------------------------
%\addcontentsline{toc}{subsubsection}{Jacobian of the Divisive Normalization}
The matrix form of the divisive normalization, Eq. \ref{divisive_norm2}, based on the diagonal matrix notation for the Hadamard products, is convenient to easily compute the Jacobian
(see the explicit derivation in the Supplementary Material \ref{J_wrt_stimul}), which
leads to,
\begin{equation}
      \hspace{-1cm}
      \nabla_{\!\!\vect{y}^i} \mathcal{N}^{(i)} =  \mathds{D}_{\textrm{sign}(\vect{y}^i)} \cdot
      \mathds{D}^{-1}_{\mathcal{D}^{(i)}(\vect{e}^i)} \cdot \left[ I - \mathds{D}_{\left(\frac{\vect{e}^i}{{\mathcal{D}^{(i)}(\vect{e}^i)}}\right)} \cdot H^{i}
      \right]  \cdot \mathds{D}_{\left(\gamma^i |\vect{y}^i|^{\gamma^i -1}\right)} \cdot \mathds{D}_{\textrm{sign}(\vect{y}^i)}
      \label{deriv_DN}
\end{equation}

%\begin{equation}
%      \hspace{-1.4cm}
%      \nabla_{\!\!\vect{y}^i} \mathcal{N}^{(i)} =
%      \left[ \mathds{D}^{-1}_{\mathcal{D}^{(i)}(\vect{e}^i)} \cdot K^{i} - \mathds{D}_{\left(\frac{\vect{e}^i}{{\mathcal{D}^{(i)}(\vect{e}^i)}^2}\right)} \cdot H^{i}
%      \right]  \cdot \mathds{D}_{\left(\gamma^i |\vect{y}^i|^{\gamma^i -1}\right)}
%      \label{deriv_DNK}
%\end{equation}

 Eq. \ref{deriv_DN} shows that the Jacobian, $\nabla_{\!\!\vect{y}^i} \mathcal{N}^{(i)}$, depends on the subtraction of two matrices,
 where the first one is \emph{diagonal}
 % if the linear gain $K^{i}$ is scalar or point-wise,
 and the second one depends on $H^{i}$, the matrix describing the interaction between the intermediate linear responses.
 Note that the role of the interaction is \emph{subtractive}, i.e. it reduces the slope (for positive $H^{i}$).
In situations where there is no interaction between the different coefficients of $\vect{y}^i$, $H^{i}_{kl}=0 \,\,\, \forall k\neq l$, the resulting $\nabla_{\!\!\vect{y}^i} \mathcal{N}^{(i)}$ is point-dependent, but diagonal.

Eq. \ref{deriv_DN} also shows that the sign of the linear coefficients has to be
considered \emph{twice} (through the multiplication by the diagonal matrices at the left and right).
This detail in the sign (which is crucial to set the direction in gradient descent), was not properly addressed in previous reports of this Jacobian (e.g. in \cite{Malo06a,Laparra10a,Malo10}) because this literature was focused on properties which are independent of the sign (diagonal nature, effect on the metric, and determinant respectively).

%----------------------------------------------------------
\paragraph*{Jacobian of alternative nonlinearities with regard to the stimulus.}
%----------------------------------------------------------
%\addcontentsline{toc}{subsubsection}{\red{Jacobian of the Wilson-Cowan transform}}
The forward Wilson-Cowan transform does not have an explicit expression since the solution evolves from a differential equation. As a result, there is no analytic solution of the Jacobian either. However its inverse is analytical (as detailed in the next section, Eq. \ref{Inv W-C transform}). Therefore, given the relation between the Jacobian matrices of inverse functions, namely $\nabla_{\!\!\vect{y}^i} {\mathcal{N}^{(i)}} = \left( \nabla_{\!\!\vect{x}^i} {\mathcal{N}^{(i)}}^{-1} \right)^{-1}$, we can compute the Jacobian of the \emph{forward} Wilson-Cowan transform from the Jacobian of its inverse.

Specifically, derivation with regard to the response in the analytic inverse given in Eq. \ref{Inv W-C transform} is straightforward, and it leads to:
\begin{equation}
\nabla_{\!\!\vect{x}^i} {\mathcal{N}^{(i)}}^{-1}(\vect{x}^i)  = \dfrac{1}{\lambda} (\alpha I - \mu \, W \cdot \mathds{D}_{f'(\vect{x}^i)})
\label{Jacobian inverse W-C}
\end{equation}
As a result, the Jacobian of the \emph{forward} Wilson-Cowan nonlinearities at the point $\vect{y}^i_A$ is,
\begin{equation}
\nabla_{\!\!\vect{y}^i} \mathcal{N}^{(i)}(\vect{y}^i_A) = \left(\nabla_{\!\!\vect{x}^i}{\mathcal{N}^{(i)}}^{-1} (\vect{x}^i_A) \right)^{-1}
\label{Jacobian W-C}
\end{equation}
assuming that $\nabla_{\!\!\vect{x}^i}{\mathcal{N}^{(i)}}^{-1}$ is nonsingular at $\vect{x}^i_A = \mathcal{N}^{(i)}(\vect{y}^i_A)$.
Note that, in general, this Jacobian matrix will be nondiagonal because of the inhibitory interactions between sensors expressed in the (nondiagonal) matrix W.

\vspace{0.2cm}

For the other example of alternative nonlinearity, the two-gamma saturation model,
the Jacobian with regard to the stimulus is a diagonal matrix since this special nonlinearity is a point-wise operation.
From Eq. \ref{two-gamma}, according to the derivation given in the Supplementary Material \ref{J_wrt_stimul}, the Jacobian of the two-gamma model is:
%\begin{equation}
%      \nabla_{\!\!\vect{y}} \mathcal{N} = \mathds{D}_{|\vect{y}|^{\gamma(|\vect{y}|)}} \cdot \Big[  \frac{\partial \gamma(|\vect{y}|)}{\partial |\vect{y}|} \cdot \mathds{D}_{\log |\vect{y}|} + \mathds{D}_{\left(\frac{\gamma(|\vect{y}|)}{|\vect{y}|}\right)} \Big]
%      \label{jacobian_two_gamma}
%\end{equation}
\begin{equation}
      \nabla_{\!\!\vect{y}} \mathcal{N} =
      \mathds{D}_{|\vect{y}|^{\gamma(|\vect{y}|)}} \cdot \Big[
      \mathds{D}_{\Big( (\gamma_H-\gamma_L)\cdot \dfrac{m \, |\vect{y}|^{ (m-1) }\cdot  \mu_1^m}{ (\mu_1^m + |\vect{y}|^{m})^2 } \Big)}
      \cdot \mathds{D}_{\log |\vect{y}|} + \mathds{D}_{\left(\frac{\gamma(|\vect{y}|)}{|\vect{y}|}\right)} \Big]
      \label{jacobian_two_gamma}
\end{equation}
%where, given Eq. \ref{exponente}, the straightforward derivative of $\gamma$ is:
%\chngP{
%\begin{equation}
%      \dfrac{\partial \gamma(|\vect{y}|)}{\partial |\vect{y}|} = \mathds{D}_{\Big( (\gamma_H-\gamma_L)\cdot \dfrac{m \, |\vect{y}|^{ (m-1) }\cdot  \mu_1^m}{ (\mu_1^m + |\vect{y}|^{m})^2 } \Big)}
%      \nonumber
%\end{equation} }
Note that the logarithm and the division by $|\vect{y}|$ imply a singularity in zero. Then, in order to guarantee the differentiability of the nonlinear transform,
we propose a modification of the nonlinearity in a small neighborhood of 0. By choosing an arbitrarily small, $\epsilon$, so that $0<\epsilon<<1$,
we modify Eq. \ref{two-gamma} for small inputs in this way,
\begin{equation}
\vect{x} = \left \{
\begin{array}{lc}
\textrm{sign}(\vect{y}) \odot |\vect{y}|^{\gamma(\,|\vect{y}|\,)} & \textrm{if} \,\, | \vect{y} | \geq \epsilon \\
\\
\textrm{sign}(\vect{y}) \odot (a_1 |\vect{y}|^2 + a_2 |\vect{y}|) & \textrm{if} \,\, |\vect{y}| \leq \epsilon
\end{array}
\right.
\label{two_gamma differentiable}
\end{equation}
where,
%$$
%a_1 = \dfrac{\epsilon \,  \dfrac{\partial \,  |\vect{y}|^{\gamma(\,|\vect{y}|\,)}}{\partial |\vect{y}|}(\epsilon) - \epsilon  %^{\gamma(\epsilon)}}{\epsilon^2}
%$$
%and
%$$
%a_2= \dfrac{2 \,  \epsilon ^{\gamma(\epsilon)}   - \epsilon \,  \dfrac{\partial \,  %|\vect{y}|^{\gamma(\,|\vect{y}|\,)}}{\partial |\vect{y}|}(\epsilon)}{\epsilon}
%$$

$
\,\,\,\,\,\,\,\,\,\,\,\,\,\,\,\,\,\,\,\,\,\,\,\, a_1 = \dfrac{\epsilon \,  \dfrac{\partial \,  |\vect{y}|^{\gamma(\,|\vect{y}|\,)}}{\partial |\vect{y}|}(\epsilon) - \epsilon  ^{\gamma(\epsilon)}}{\epsilon^2}
\,\,\,\,\,\,\,\,
\textrm{and}
\,\,\,\,\,\,\,\,
a_2= \dfrac{2 \,  \epsilon ^{\gamma(\epsilon)}   - \epsilon \,  \dfrac{\partial \,  |\vect{y}|^{\gamma(\,|\vect{y}|\,)}}{\partial |\vect{y}|}(\epsilon)}{\epsilon}
$
\vspace{0.5cm}

With this modification around zero the two-gamma nonlinearity and its derivative are continuous and well defined everywhere:
the Jacobian for $|\vect{y}|>\epsilon$ would be given by  Eq. \ref{jacobian_two_gamma}, and for smaller inputs $\nabla_{\!\!\vect{y}} \mathcal{N} = 2 a_1 |\vect{y}| + a_2$,
which is well defined at zero.

%%%%%%%%%%%%%%%%%%%%%%%%%%%%%%%%%%%%%%%%%%%%%%%%%%%%%%%%%%%%%%%%%%%%%%%%%  END OF 2.2 JACOBIAN WRT STIMULUS
%%%%%%%%%%%%%%%%%%%%%%%%%%%%%%%%%%%%%%%%%%%%%%%%%%%%%%%%%%%%%%%%%%%%%%%%%%%%%%%%%%%%%%%%%%%%%%%%%%%%%%%%%%%%%%%%%%%%%%%%%%%%%%%%%%%%%%%%%%%%%%%%%%
%%%%%%%%%%%%%%%%%%%%%%%%%%%%%%%%%%%%%%%%%%%%%%%%%%%%%%%%%%%%%%%%%%%%%%%%%%%%%%%%%%%%%%%%%%%%%%%%%%%%%%%%%%%%%%%%%%%%%%%%%%%%%%%%%%%%%%%%%%%%%%%%%%
%%%%%%%%%%%%%%%%%%%%%%%%%%%%%%%%%%%%%%%%%%%%%%%%%%%%%%%%%%%%%%%%%%%%%%%%%%%%%%%%%%%%%%%%%%%%%%%%%%%%%%%%%%%%%%%%%%%%%%%%%%%%%%%%%%%%%%%%%%%%%%%%%%
%%%%%%%%%%%%%%%%%%%%%%%%%%%%%%%%%%%%%%%%%%%%%%%%%%%%%%%%%%%%%%%%%%%%%%%%%%%%%%%%%%%%%%%%%%%%%%%%%%%%%%%%%%%%%%%%%%%%%%%%%%%%%%%%%%%%%%%%%%%%%%%%%%

\subsection{Result II: Jacobian with regard to the parameters}
\label{result2}

The problem of computing the Jacobian with regard to the parameters in the cascade of L+NL modules, $\nabla_{\!\!\vect{\Theta}} S$,
reduces, according to Eqs. \ref{global_jacobian2} - \ref{elementary_jacobian2}, to the computation of the Jacobian of the nonlinearity with regard to the parameters in every layer, $\nabla_{\!\!\vect{\theta}^i} \mathcal{N}^{(i)}$.
In this section we give the analytical result of the required Jacobian in the
canonical divisive normalization case. Proofs of this second analytical result are given in the Supplementary Material \ref{J_wrt_param}.
The role of this analytical result in getting optimal models from classical psychophysics is illustrated in the Discussion, Section \ref{use_jacobian_param}.

\paragraph*{Jacobian w.r.t. parameters: general equations}
The parameters of the divisive normalization of the $i$-th layer that may be independently modified are $\vect{\theta}^i =  \{\gamma^i, \vect{b}^i, H^i \}$. Therefore, $\nabla_{\!\!\vect{\theta}^i} \mathcal{N}^{(i)}$ is given by this concatenation:
\begin{equation}
      \nabla_{\!\!\vect{\theta}^i} \mathcal{N}^{(i)} = \Bigg[ \nabla_{\!\!\gamma^i} \mathcal{N}^{(i)}   \,\,\,\,\nabla_{\!\!\vect{b}^i} \mathcal{N}^{(i)}   \,\,\,\,\nabla_{\!\!H^i} \mathcal{N}^{(i)} \Bigg]
      \label{jacobian2_global}
\end{equation}
%\begin{equation}
%      \nabla_{\!\!\theta^i} \mathcal{N}^{(i)} = \Bigg[ \nabla_{\!\!K^i} \mathcal{N}^{(i)}   \,\,\,\,  \nabla_{\!\!\gamma^i} \mathcal{N}^{(i)}   \,\,\,\,\nabla_{\!\!b^i} \mathcal{N}^{(i)}   \,\,\,\,\nabla_{\!\!H^i} \mathcal{N}^{(i)} \Bigg]
%      \label{jacobian2_globalK}
%\end{equation}
where, according to the derivation given in the Supplementary Material \ref{J_wrt_param}, we have,
\begin{eqnarray}
      \nabla_{\!\!\gamma^i} \mathcal{N}^{(i)} \!&=&\!  \mathds{D}_{\textrm{sign}(\vect{y}^i)} \!\cdot\! \mathds{D}^{-1}_{\mathcal{D}^{(i)}(\vect{e}^i)} \!\cdot\! \Bigg[ \mathds{D}_{log |\vect{y}^i|}
      - \mathds{D}^{-1}_{\mathcal{D}^{(i)}(\vect{e}^i)} \!\cdot\! \mathds{D}_{( H^i \cdot \mathds{D}_{\vect{e}^i} \cdot log |\vect{y}^i| )} \Bigg] \!\cdot\! \vect{e}^i \label{jacobian2_gamma} \\
      \nabla_{\!\!\vect{b}^i} \mathcal{N}^{(i)} \!&=&\! - \mathds{D}_{\textrm{sign}(\vect{y}^i)} \cdot \mathds{D}_{\vect{e}^i} \cdot \mathds{D}^{-2}_{\mathcal{D}^{(i)}(\vect{e}^i)}  \label{jacobian2_b} \\
      \nabla_{\!\!H^i} \mathcal{N}^{(i)} \!&=&\!  - \mathds{D}_{\textrm{sign}(\vect{y}^i)} \cdot  \mathds{D}_{\vect{e}^i} \cdot \mathds{D}^{-2}_{\mathcal{D}^{(i)}(\vect{e}^i)} \cdot \mathds{B}^{d_i}_{({\vect{e}^i}^\top)} \label{jacobian2_H}
\end{eqnarray}
%\begin{eqnarray}
%      \nabla_{\!\!K^i} \mathcal{N}^{(i)} \!&=&\! \mathds{D}^{-1}_{\mathcal{D}^{(i)}(\vect{e}^i)} \cdot \mathds{B}^{d_i}_{({\vect{e}^i}^\top)} \label{jacobian2_K} \\
%      \nabla_{\!\!\gamma^i} \mathcal{N}^{(i)} \!&=&\!  \mathds{D}^{-1}_{\mathcal{D}^{(i)}(\vect{e}^i)} \!\cdot\! \Bigg[ K^i \!\cdot\! \mathds{D}_{log |\vect{y}^i|}
%      - \mathds{D}^{-1}_{\mathcal{D}^{(i)}(\vect{e}^i)} \!\cdot\! \mathds{D}_{( H^i \cdot \mathds{D}_{\vect{e}^i} \cdot log |\vect{y}^i| )} \!\cdot\! K^i \Bigg] \!\cdot\! \vect{e}^i \label{jacobian2_gammaK} \\
%      \nabla_{\!\!b^i} \mathcal{N}^{(i)} \!&=&\! - \mathds{D}_{(K^i \cdot \vect{e}^i)} \cdot \mathds{D}^{-2}_{\mathcal{D}^{(i)}(\vect{e}^i)}  \label{jacobian2_bK} \\
%      \nabla_{\!\!H^i} \mathcal{N}^{(i)} \!&=&\!  - \mathds{D}_{( K^i \cdot \vect{e}^i)} \cdot \mathds{D}^{-2}_{\mathcal{D}^{(i)}(\vect{e}^i)} \cdot \mathds{B}^{d_i}_{({\vect{e}^i}^\top)} \label{jacobian2_HK}
%\end{eqnarray}
where $\mathds{D}_{\vect{v}}$ stands for a diagonal matrix with vector $\vect{v}$ in the diagonal as stated in Eq. \ref{haddamard}, and $\mathds{B}^d_{\vect{v}}$ stands for a block diagonal matrix built by $d$-times replication of the matrix (or vector) $\vect{v}$ as stated in the Supplementary Material \ref{dLineardL} (in Eqs. \ref{block_diag} and \ref{deriv_linear_param}).
Note also that, consistently with the derivative of a linear function w.r.t. its parameters (in Suppl. Material \ref{dLineardL}),
in order to apply the Jacobian in Eq.~\ref{jacobian2_H} on small perturbations of the matrix, $H^i$, the corresponding perturbation should undergo row-wise vectorization.
For instance, imagine $H^i$ is perturbed so that $H^i_{*} = H^i + \Delta H$. Then, the perturbation in the response should be computed as
$\Delta \vect{x}^i = \nabla_{\!\!H^i} \mathcal{N}^{(i)} \cdot vect(\Delta H^\top)$.
\vspace{0.2cm}

\paragraph*{Jacobian w.r.t. parameters: specific equations for Gaussian kernels}
The qualitative meaning of $H^i$ (interaction between neighboring neurons) naturally leads to propose specific structures in the rows of these matrices.
For instance, stronger interaction between closer neurons naturally leads to the idea of Gaussian kernels \cite{Watson97}.
This functional parametrization implies a dramatic reduction in the number of unknowns because each row, $H^i_{k \star}$, with dimension $d_i$, could be described by a Gaussian defined by with only two parameters: amplitude and width.
In the considered retina-V1 pathway the identity of the sensors is characterized by its 2D spatial location or by its 4D spatio-frequency location.
In the most general case the index, $k$, of the sensor has spatio-frequency meaning:
\begin{equation}
k \,\, \textrm{denotes a wavelet-like index} \,\,\,\, \Rightarrow  \,\,\,\, k \equiv (p_{k1},p_{k2},f_k,\phi_k) \nonumber
\end{equation}

\noindent where $\vect{p}_k = (p_{k1},p_{k2})$ is the optimal 2D location, $f_k$ is the optimal spatial frequency, and $\phi_k$
is the optimal orientation of the $k$-th sensor.
In V1, the interaction between the linear response $y^i_k$ and the neighbors $y^i_{k'}$
decreases with the distance between $k$ and $k'$ in space, frequency and orientation \cite{Watson97}.
Restricting ourselves to intra-subband interactions (which incidentally are the most relevant \cite{Malo10,Laparra10a}) one has:
\begin{equation}
H^i_{k k'} = \left \{
\begin{array}{lc}
0 & \,\, \forall  \,\, k' \notin \textrm{subband} \,\,\, k \\
\\
c^i_k \, \frac{dp_{k1} dp_{k2}}{2 \pi \,\, {\sigma^i_k}^2} \,\, e^{-\frac{ \Delta^2_{k k'} }{2 \,\, {\sigma^i_k}^2}} & \forall \,\,  k' \in \textrm{subband} \,\,\, k
\end{array}
\right.
\label{GaussianH}
\end{equation}

\noindent where the relevant parameters are $c^i_k$ and $\sigma^i_k$ which respectively stand for the amplitude and width of the Gaussian centered in the $k$-th sensor. $\Delta^2_{k k'} = (\vect{p}_k-\vect{p}_{k'})^\top \cdot (\vect{p}_k-\vect{p}_{k'})$
is the squared distance between the sensors,
and $dp_{k1} dp_{k2}$ is just the spatial area of the discrete grid of sensors that sample the visual space in this subband.
This implies that the pool of all interactions is $\sum_{k'} H^i_{k k'} = c^i_k$.

In the case of different interactions per sensor (different Gaussian in each row, $H^i_{k \star}$), derivatives with regard to the independent widths are,
\begin{equation}
      \nabla_{\!\!\vect{\sigma}^i}  \mathcal{N}^{(i)} = \Big[ \nabla_{\!\!\sigma^i_1}  \mathcal{N}^{(i)} \,\,\, \nabla_{\!\!\sigma^i_2}  \mathcal{N}^{(i)}\,\, \cdots \,\, \nabla_{\!\!\sigma^i_k}  \mathcal{N}^{(i)}\,\, \cdots \,\, \nabla_{\!\!\sigma^i_{d_i}} \mathcal{N}^{(i)} \Big]
      \label{jacobian2_H_gauss_sig1}
\end{equation}
With this parametrization of $H$ we can develop Eq. \ref{jacobian2_H} further: the dependence on individual widths can be obtained by using $\nabla_{\sigma^i_k} \mathcal{N}^{(i)} = \nabla_{H^i} \mathcal{N}^{(i)} \cdot \nabla_{H^i_{k \star}} H^i \cdot \nabla_{\sigma^i_k} H^i_{k \star}$, and the final result (see the Supplementary Material \ref{J_wrt_param}) is:
\begin{equation}
     \nabla_{\vect{\sigma}^i} \mathcal{N}^{(i)} = -\textrm{diag} \left[ \mathds{D}_{\textrm{sign}(\vect{y}^i)} \cdot \mathds{D}_{\vect{e}^i} \cdot \mathds{D}^{-2}_{\mathcal{D}^{(i)}(\vect{e}^i)} \cdot
     \left(
%       \begin{array}{c}
%                   {\vect{e}^i}^\top  \\
%                   {\vect{e}^i}^\top  \\
%                   \vdots  \\
%                   {\vect{e}^i}^\top
%       \end{array}
       \begin{array}{ccc}
                  \hspace{0.5cm} & {\vect{e}^i}^\top & \hspace{0.5cm} \\
                  \hspace{0.5cm} & {\vect{e}^i}^\top & \hspace{0.5cm} \\
                  \hspace{0.5cm} & \vdots & \hspace{0.5cm} \\
                  \hspace{0.5cm} & {\vect{e}^i}^\top & \hspace{0.5cm}
       \end{array}
     \right) \cdot F^i
     \right]
     \label{dNdsigma}
\end{equation}
where,
\begin{equation}
F^i_{k k'} = \left \{
\begin{array}{lc}
0 & \,\, \forall  \,\, k' \notin \textrm{subband} \,\,\, k \\
\\
c^i_k \, \frac{dp_{k1} dp_{k2}}{2 \pi \,\, {\sigma^i_k}^5} \,\, \left(  \Delta^2_{k k'} - 2 \, {\sigma^i_k}^2 \right) \,\, e^{-\frac{  \Delta^2_{k k'} }{2 \,\, {\sigma^i_k}^2}} & \forall \,\,  k' \in \textrm{subband} \,\,\, k
\end{array}
\right.
\nonumber
\end{equation}

A diagonal matrix for $\nabla_{\vect{\sigma}^i} \mathcal{N}^{(i)}$ makes sense because the modification of the interaction width of a sensor only affects the nonlinear response of this sensor (similarly to the diagonal nature of $\nabla_{\vect{b}^i} \mathcal{N}^{(i)}$ in Eq. \ref{jacobian2_b}).

The derivative with regard to the vector of amplitudes of the Gaussian interactions,
$\nabla_{\vect{c}^i} \mathcal{N}^{(i)}$, is a concatenation of columns (similarly to Eq. \ref{jacobian2_H_gauss_sig1}). It can also be computed from the chain rule and from the derivative w.r.t the corresponding variables. The result is:

\begin{equation}
     \nabla_{\vect{c}^i} \mathcal{N}^{(i)} = -\textrm{diag} \left[ \mathds{D}_{\textrm{sign}(\vect{y}^i)} \cdot \mathds{D}_{\vect{e}^i} \cdot \mathds{D}^{-2}_{\mathcal{D}^{(i)}(\vect{e}^i)} \cdot
     \left(
       \begin{array}{ccc}
                  \hspace{1cm} & {\vect{e}^i}^\top & \hspace{1cm} \\
                  \hspace{1cm} & {\vect{e}^i}^\top & \hspace{1cm} \\
                  \hspace{1cm} & \vdots & \hspace{0.5cm} \\
                  \hspace{1cm} & {\vect{e}^i}^\top & \hspace{1cm}
       \end{array}
     \right) \cdot G^i
     \right]
     \label{dNdc}
\end{equation}
where,
\begin{equation}
G^i_{k k'} = \left \{
\begin{array}{lc}
0 & \,\, \forall  \,\, k' \notin \textrm{subband} \,\,\, k \\
\\
\frac{dp_{k1} dp_{k2}}{2 \pi \,\, {\sigma^i_k}^2} \,\, e^{-\frac{  \Delta^2_{k k'} }{2 \,\, {\sigma^i_k}^2}} & \forall \,\,  k' \in \textrm{subband} \,\,\, k
\end{array}
\right.
\nonumber
\end{equation}

The number of free parameters can be further reduced if one assumes that the values of the semisaturation, $b^i_k$, or the parameters of the Gaussians, $c^i_k$ and $\sigma^i_k$, have certain structure (e.g. constant along the visual space in each subband). One may impose this structure in Eqs. \ref{jacobian2_b}, \ref{dNdsigma} and \ref{dNdc} by right-multiplication of the jacobian by a binary matrix that describes the structure of the considered vector. For instance, assuming the same width all over each scale in a two-scales image representation, one only has two independent parameters. In that case:
\begin{equation}
      \nabla_{\vect{\sigma}^i_{\textrm{struct}}} \mathcal{N}^{(i)} = \nabla_{\vect{\sigma}^i} \mathcal{N}^{(i)} \cdot M_{\textrm{struct}}
      \nonumber
\end{equation}
where, the structure matrix selects which coefficients belong to each scale:
\begin{equation}
M_{\textrm{struct}} =      \left(
       \begin{array}{cc}
                  1 & 0 \\
                  1 & 0 \\
                  \vdots & \vdots \\
                  1 & 0 \\
                  0 & 1 \\
                  0 & 1 \\
                  \vdots & \vdots \\
                  0 & 1 \\
       \end{array}
     \right)
\nonumber
\end{equation}

%%%%%%%%%%%%%%%%%%%%%%%%%%%%%%%%%%%%%%%%%%%%%%%%%%%%%%%%%%%%%%%%%%%%%%%%%  END OF 2.3 JACOBIAN WRT PARAMETERS
%%%%%%%%%%%%%%%%%%%%%%%%%%%%%%%%%%%%%%%%%%%%%%%%%%%%%%%%%%%%%%%%%%%%%%%%%%%%%%%%%%%%%%%%%%%%%%%%%%%%%%%%%%%%%%%%%%%%%%%%%%%%%%%%%%%%%%%%%%%%%%%%%%
%%%%%%%%%%%%%%%%%%%%%%%%%%%%%%%%%%%%%%%%%%%%%%%%%%%%%%%%%%%%%%%%%%%%%%%%%%%%%%%%%%%%%%%%%%%%%%%%%%%%%%%%%%%%%%%%%%%%%%%%%%%%%%%%%%%%%%%%%%%%%%%%%%
%%%%%%%%%%%%%%%%%%%%%%%%%%%%%%%%%%%%%%%%%%%%%%%%%%%%%%%%%%%%%%%%%%%%%%%%%%%%%%%%%%%%%%%%%%%%%%%%%%%%%%%%%%%%%%%%%%%%%%%%%%%%%%%%%%%%%%%%%%%%%%%%%%
%%%%%%%%%%%%%%%%%%%%%%%%%%%%%%%%%%%%%%%%%%%%%%%%%%%%%%%%%%%%%%%%%%%%%%%%%%%%%%%%%%%%%%%%%%%%%%%%%%%%%%%%%%%%%%%%%%%%%%%%%%%%%%%%%%%%%%%%%%%%%%%%%%

\subsection{Result III: Analytic inverse}
\label{result3}

The inverse of the global transform can be obtained inverting each individual L+NL layer in turn,
\begin{equation}
      S^{-1} = {S^{(1)}}^{-1} \circ {S^{(2)}}^{-1} \circ \cdots \circ {S^{(n-1)}}^{-1} \circ {S^{(n)}}^{-1}
      \label{inverse_network}
\end{equation}
where,
\begin{equation}
      \vect{x}^{i-1} ={S^{(i)}}^{-1}(\vect{x}^i) = {L^i}^\dag \cdot {\mathcal{N}^{(i)}}^{-1}(\vect{x}^i)
      \label{inverse_layer}
\end{equation}
Here we will focus on the ${\mathcal{N}^{(i)}}^{-1}$ part because the linear part can be addressed by
standard matrix inversion.

Here we present the analytical inverse of the canonical divisive normalization and of the Wilson-Cowan alternative. The inverse of the two-gamma nonlinearity is not addressed here but in the Supplementary Material \ref{example_model} because, given the coupling between the input and the exponent, it has no analytical inverse. Nevertheless, a simple and efficient iterative method is proposed there to compute the inverse.
The role of the analytical inverse in improving conventional decoding of visual signals is illustrated in the Discussion, Section \ref{use_inverse}.

A note on the linear part: the eventual rectangular nature of $L^i$ (different number of outputs than inputs in the $i$-th layer) requires standard pseudoinverse, $(\cdot)^\dag$, instead of the regular square-matrix inversion, $(\cdot)^{-1}$;
and it may be regularized through standard methods \cite{Numerical92,Golub} in case $L^i$ is ill-conditioned.
Information loss in the pseudoinverse due to strong dimensionality reduction in $L^i$ is not serious in the central region of the visual field due to mild undersampling of the fovea throughout
the neural pathway \cite{Kandel}.
The only aspect of the input that definitely cannot be recovered from the
responses is the spectral distribution in each location.
In color perception models the first stage is linear spectral integration
to give opponent tristimulus values in each spatial location \cite{Fairchild13}.
This very first linear stage is represented by a extremely fat
rectangular matrix, $L^1 \in \mathbb{R}^{3 \times 300}$, in each location (300 wavelengths in the spectral visible region reduce to 3 tristimulus values),
which definitely is not invertible though standard regularized pseudoinversion.
Therefore, the inversion of a standard \emph{retina-V1} model such as the one used in the Discussion may recover the tristimulus images but not the whole hyperspectral array.

\blue{The \emph{metamerism} concept (the many-to-one transform) can be generalized beyond the
spectral integration. In higher levels of processing, it has been suggested that stimuli may be not be represented by the specific responses of a population of neurons, but by their statistical properties \cite{Freeman11}.
These statistical summaries could be thought as a stronger nonlinear dimensionality reduction which cannot be decoded through regular pseudoinversion. Therefore, the proposed inverse is applicable only to the (early) stages in which the information is still encoded in the responses of the population and not in summarized descriptions of these responses.
}

\paragraph*{Analytic inverse of the Divisive Normalization.}
%\addcontentsline{toc}{subsection}{Analytic inverse of the Divisive Normalization}
Analytic inversion of standard divisive normalization, Eq. \ref{conventionalDN}, is not obvious.
However, using the diagonal matrix notation for the Hadamard product, the inverse is (see Supplementary Material \ref{proof_inverse}),
\begin{equation}
       \vect{y}^i = {\mathcal{N}^{(i)}}^{-1}(\vect{x}^i) = \mathds{D}_{\text{sign}(\vect{x}^i)} \cdot \left[ \left( I -  \mathds{D}_{|\vect{x}^i|} \cdot H^{i}  \right)^{-1} \cdot \mathds{D}_{\vect{b}^i} \cdot |\vect{x}^i| \right]^{\frac{1}{\gamma^i}}
       \label{inv_DN}
\end{equation}
%
%\begin{equation}
%       \vect{y}^i = {\mathcal{N}^{(i)}}^{-1}(\vect{x}^i) = \left[ \left( K^{i} - \mathds{D}_{\vect{x}^i} \cdot H^{i}  \right)^{-1} \cdot \mathds{D}_{b^i} \cdot \vect{x}^i \right]^{\frac{1}{\gamma^i}}
%       \label{inv_DNK}
%\end{equation}
where $[v]^{\frac{1}{\gamma^i}}$ is element-wise exponentiation of elements of the vector $v$.
%Finally, the signal representation at the previous layer can be obtained from the linear responses
%through the inverse of the linear transform: $\vect{x}^{i-1} = {L^{(i)}}^+ \cdot \vect{y}^i$.

Consistently with generic inverse-through-integration approaches based on $\nabla_{\vect{x}} S^{-1}$ \cite{Epifanio00},
here Eq. \ref{inv_DN} shows more specifically that in this linear-nonlinear architecture, \emph{inversion} reduces to \emph{matrix inversion}.
While the linear filtering operations, $L^i$, may be inverted without the need of an explicit matrix inversion through surrogate signal representations (deconvolution in the Fourier or Wavelet domains),
there is no way to avoid the inverse $\left( I - \varepsilon \right)^{-1}$ in Eq. \ref{inv_DN}.
This may pose severe computational problems in high-dimensional situations (e.g. in redundant wavelet representations).
A series expansion alternative for that matrix inversion was proposed in \cite{Malo06a}, where it is substituted by a (more affordable) series of matrix-on-vector operations.

%In any case, note the consistency between the invertibility condition that can be drawn from Eq. \ref{inv_DN} and Eq. \ref{deriv_DN}: in both cases the inverse exists only if
%$\left( I -  \mathds{D}_{|\vect{x}^i|} \cdot H^{i}  \right)$ is non-singular.

\paragraph*{Inverse of the Wilson-Cowan equations.}
%\addcontentsline{toc}{subsection}{\red{Inverse of the Wilson-Cowan transform}}
The expression of the inverse of the Wilson-Cowan transform is straightforward: by reordering the terms in the steady-state equation, Eq. \ref{steady-state Wilson-Cowan equation}, it follows,
\begin{equation}
 \vect{y}^i = {N^{(i)}}^{-1}(\vect{x}^i) = \dfrac{1}{\lambda} (\alpha \vect{x}^i - \mu \, W \cdot f(\vect{x}^i))
 \label{Inv W-C transform}
\end{equation}
Note that this inverse function is easily derivable w.r.t $\vect{x}^i$, which is required to obtain the corresponding Jacobian of the \emph{forward} transform Eq. \ref{Jacobian W-C}.

\paragraph*{Relation between Result I and Result III.}
Result III (inverse) is obviously related to Result I (Jacobian with regard to the stimulus) because
a sufficient condition for invertibility is that the Jacobian with regard to the stimulus is nonsingular
for every image. Note that if the Jacobian is non singular, the inverse of the Jacobian can be integrated and hence,
the global inverse can be obtained from the local-linear approximations as in other \emph{local-to-global}
methods, e.g. \cite{Epifanio00,Malo06b,Laparra12,Laparra15}.

This general statement is perfectly illustrated by the similarity between Eqs. \ref{inv_DN} and \ref{deriv_DN}.
According to Eq. \ref{inv_DN}, inverting the divisive normalization reduces to inverting $\left( I -  \mathds{D}_{|\vect{x}^i|} \cdot H^{i} \right)$. Similarly, according to Eq. \ref{deriv_DN}, the singularity of the Jacobian depends on the very same matrix.
\blue{As a result, specific interest on invertible models would imply restrictions to the response \emph{and} the parameters of $H$: the eigenvalues of $\mathds{D}_{|\vect{x}^i|} \cdot H^{i}$ have to be smaller than 1 \cite{Malo06a}.}

%%%%%%%%%%%%%%%%%%%%%%%%%%%%%%%%%%%%%%%%%%%%%%%%%%%%%%%%%%%%%%%%%%%%%%%%%  END OF 2.4 INVERSE
%%%%%%%%%%%%%%%%%%%%%%%%%%%%%%%%%%%%%%%%%%%%%%%%%%%%%%%%%%%%%%%%%%%%%%%%%%%%%%%%%%%%%%%%%%%%%%%%%%%%%%%%%%%%%%%%%%%%%%%%%%%%%%%%%%%%%%%%%%%%%%%%%%
%%%%%%%%%%%%%%%%%%%%%%%%%%%%%%%%%%%%%%%%%%%%%%%%%%%%%%%%%%%%%%%%%%%%%%%%%%%%%%%%%%%%%%%%%%%%%%%%%%%%%%%%%%%%%%%%%%%%%%%%%%%%%%%%%%%%%%%%%%%%%%%%%%
%%%%%%%%%%%%%%%%%%%%%%%%%%%%%%%%%%%%%%%%%%%%%%%%%%%%%%%%%%%%%%%%%%%%%%%%%%%%%%%%%%%%%%%%%%%%%%%%%%%%%%%%%%%%%%%%%%%%%%%%%%%%%%%%%%%%%%%%%%%%%%%%%%
%%%%%%%%%%%%%%%%%%%%%%%%%%%%%%%%%%%%%%%%%%%%%%%%%%%%%%%%%%%%%%%%%%%%%%%%%%%%%%%%%%%%%%%%%%%%%%%%%%%%%%%%%%%%%%%%%%%%%%%%%%%%%%%%%%%%%%%%%%%%%%%%%%

%%--------------------------------------------
\section{Discussion}
%%--------------------------------------------

\blue{In this section we consider illustrative vision models based on cascades of L+NL stages to point out (a) the fundamental insight into the system behavior that can be obtained from the analytic expressions, and (b) the usefulness of the expressions to develop new experiments and methods in visual neuroscience.
The first consist on using the analytical expressions to identify basic trends in physiology, in psychophysics and in the function of the sensory system.
Specifically, (a.1)~we show how the context-dependence of the receptive fields of the sensors can be
explicitly seen in the expression of the Jacobian w.r.t. the stimulus.
(a.2)~We show that the expression of the Jacobian w.r.t. the parameters reveals that the impact in the response of uncertainty at the filters, or synaptic weights, varies over the stimulus space, and this trend is different for different sensors.
(a.3)~We show how the general trends of the sensitivity over the stimulus space can be seen from the
determinant of the metric based on the Jacobian w.r.t. the stimulus.
(a.4)~We show that this Jacobian also implies different
efficiency (different multi-information reduction) in different regions of the stimulus space.}
\blue{The second includes (b.1)~stimulus design in novel psychophysics,
(b.2)~more accurate model fitting in classical physiology and psychophysics,
and (b.3)~new proposals for decoding of visual signals.}

\begin{figure}[!t]
	\centering
    \small
    \setlength{\tabcolsep}{2pt}
    \vspace{-1.80cm}
    \begin{tabular}{c}
    %\hspace{-1.0cm}  \includegraphics[height=0.92\textheight]{general_scheme_bbb.JPG} \\
    \hspace{-5.15cm}  \includegraphics[height=0.90\textheight]{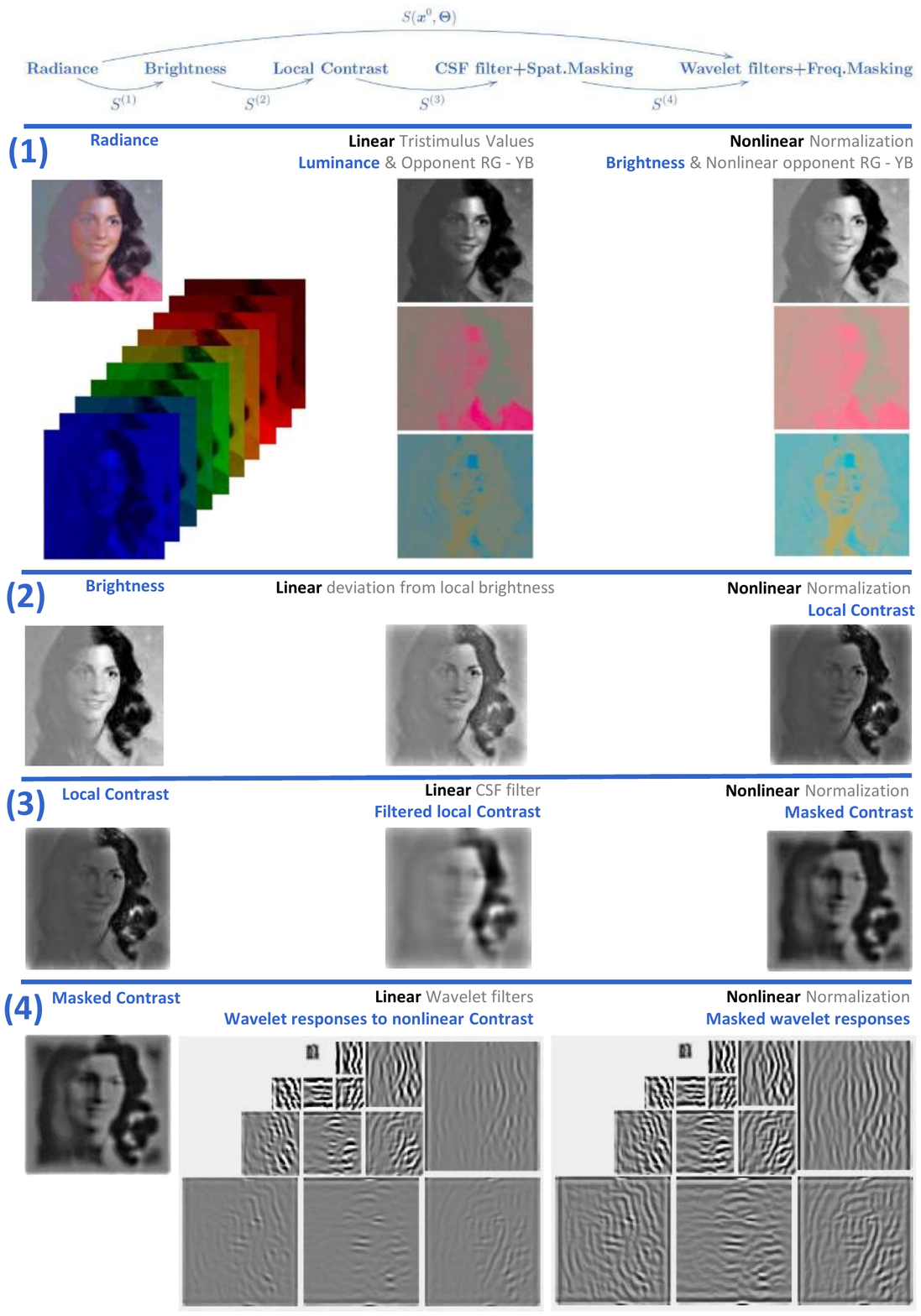} \\
    \vspace{-0.0cm}
    \hspace{-5.3cm}\pbox{1.4\textwidth}{\textbf{Fig 1.} \scriptsize{\textbf{A cascade of isomorphic L+NL modules based on canonical Divisive Normalization.}
                         The input is the spatial distribution
                         of the \emph{spectral irradiance} at the retina.
                         %(the hyperspectral cube $\vect{x}^0$).
                         For this illustration we modified an image from the USC-SIPI Database \cite{USCdatabase}
                         to reduce the contrast at the left part of the stimulus.
                         (1)~The linear part of the first layer consist of three positive LMS spectral sensitivities and a linear recombination of the LMS values with positive/negative weights. This leads to three tristimulus values in each spatial location:
                         one of them is proportional to the luminance, and the other two have opponent chromatic meaning (red-green and yellow-blue).
                         %(as for instance in \cite{Ingling77}).
                         These linear tristimulus responses undergo adaptive saturation transforms.
                         %Restricting ourselves to the luminance distribution,
                         %$\vect{y}^1$,
                         Perception of \emph{brightness} is mediated by an adaptive Weber-like nonlinearity applied to the luminance at each location. This nonlinearity enhances the response in the regions with small linear input (low luminance).
                         %$\vect{x}^2$.
                         (2)~The linear part of the second layer computes the deviation of the brightness at each location from the local brightness. Then, this deviation is nonlinearly normalized by the local brightness
                         % (the small blurred image on top)
                         to give the local nonlinear contrast.
                         % (the small image below the local brightness).
                         % A weighted sum of these two terms leads to a signal containing both brightness and contrast information.
                         (3)~The responses to local contrast are convolved by center surround receptive fields (or filtered by the Contrast Sensitivity Function). Then the linearly filtered contrast is nonlinearly normalized by the local contrast. Again normalization increases the response in the regions with small input (low contrast).
                         (4)~After linear wavelet transform, each response is normalized by the activity of the neurons in the surround. Again, the activity relatively increases in the regions with low input. The common effect of the nonlinear modules throughout the network is response equalization. Supplementary Material \ref{toolbox} (Fig. \ref{demo1}) shows PDFs of the responses along the network
                         which are consistent with previous reports of the predictive effect of Divisive Normalization.}
                         }
                         %\label{FMmodel}
                         \end{tabular}
    \vspace{-0.5cm}
\end{figure}

% \blue{Before going into the details of the specific illustrations of the proposed expressions,
\blue{Let's briefly describe the kind of vision model used as example throughout the discussion.}
The case-study model follows the program suggested in \cite{Carandini12}: a cascade of four isomorphic canonica L+NL modules addressing brightness, contrast, frequency filtered contrast masked in the spatial domain, and orientation/scale masking.
\blue{The general architecture is certainly not new, but the proposed expressions were very helpful to tune psychophysically these specific modules to work together for the first time.}
%%%% Strictly speaking, the model could have been determined using automatic differentiation as oposed to the use of the explicit expressions (both for MAD and experiments for the reproduction of distortion perception, however this would have prevented the intuition obtained from the expressions.
The response of the model on an image is illustrated in Fig.~1.
% \label{FMmodel}

\blue{Before going into the many details of the full 4-layer model (given in the Supplementary Material \ref{example_model}, and in the code available in} {\footnotesize \verb"http://isp.uv.es/docs/BioMultiLayer_L_NL.zip"}), \blue{let's look at a cartoon version for a better interpretation of the analytical expressions.}

\blue{Consider a system with only three sensors acting on three-pixel images. Consider it is a cascade of just two L+NL layers,
one for brightness and the next for spatial frequency analysis:}
\begin{itemize}
       \item \blue{Layer 1: brightness from radiance,}
       \vspace{-0.4cm}
                 \blue{
                 \begin{eqnarray}
                       \vect{y}^1 &=& V_\lambda \cdot \vect{x}^0 \nonumber \\
                       \vect{x}^1 &=& (\vect{y}^1)^{\gamma_1} \nonumber
                 \end{eqnarray}}
                 \vspace{-0.4cm}
       \item \blue{Layer 2: spatial frequency analyzers and contrast response,}
                 \blue{
                 \begin{eqnarray}
                       \vect{y}^2 &=& G \cdot F \cdot \vect{x}^1 =\left(
                                        \begin{array}{ccc}
                                          0.8 & 0 & 0 \\
                                          0 & 1 & 0 \\
                                          0 & 0 & 0.2 \\
                                        \end{array}
                                      \right)
                        \cdot \left(
                                      \begin{array}{ccc}
                                          0.577 & 0.577 & 0.577 \\
                                          0.7071 & 0 & -0.707 \\
                                          0.408 & -0.817 & 0.408 \\
                                        \end{array}
                                      \right) \cdot \left(
                                                      \begin{array}{c}
                                                        x^1_1 \\
                                                        x^1_2 \\
                                                        x^1_3 \\
                                                      \end{array}
                                                    \right)
                                       \nonumber \\
                       \vect{x}^2 &=& \textrm{sign}(\vect{y}^2) \odot \frac{|\vect{y}^2|}{\vect{b} + |\vect{y}^2|} \nonumber
                 \end{eqnarray}
                 }
                 \vspace{-0.3cm}
\end{itemize}
\blue{The biological basis of this simplified model is straightforward:
integration over wavelengths is done using the standard spectral sensitivity function, $V_\lambda$ \cite{Stiles82},
and we assume a simple, point-wise and fixed, exponential relation between luminance and brightness \cite{Stiles82,Fairchild13}.
Regarding spatial pattern detection, we assume frequency-selective linear analyzers \cite{Graham89} in the rows of $F$.
The first sensor (first row) is tuned to the DC component of brightness, the second sensor (second row) to the low frequency component, and the last sensor (third row) to the high frequency.
Each of these linear sensors has different (frequency dependent) gain in the diagonal matrix $G$. This gain is band-pass, i.e. similar to the Contrast Sensitivity Function, CSF \cite{Campbell68}.
Finally, the contrast response undergoes a compressive transform where the interactions between coefficients are neglected as in \cite{Legge80,Legge81}, by using an identity matrix as interaction kernel $H$.}

\blue{As a result, the responses at the $k$-th \emph{photo-receptor} of the first L+NL layer represent the \emph{luminance}, $y^1_k$, and the \emph{brightness}, $x^1_k$, at the $k$-th spatial location.
Given the frequency analysis meaning of $F$, the responses $y^2_1$ and $x^2_1$ are related to the average brightness of the image, while $y^2_k$ and $x^2_k$, with $k>1$, are related to the amplitude or \emph{contrast}
of the low- and high-frequency AC components. With this in mind we will be able to identify the trends of biologically meaningful magnitudes from the proposed expressions in terms of the \emph{luminance} and \emph{contrast} of the images in the stimulus space.}

\vspace{0.25cm}
\blue{The first part of the discussion is focused on examples of the insight into the system that can be obtained from the presented analytical results.}
% \blue{Then, the discussion shows how Result~I and Result~II are useful for the experimental determination of sensible parameters for the 4-layer model outlined above.}
\blue{Then, we show that Result~I is convenient in new psychophysics such as MAximum Differentiation (MAD) \cite{Wang08}; and
Result~II is convenient for parameter estimation in classical experiments.
In fact, MAD and Result~I were used to determine the 2nd and 3rd layers of the illustrative L+NL cascade,
and Result~II was used as alternative to brute-force optimization to maximize correlation with subjective opinion in 1st and 4th layers.}
\blue{The good visual examples of MAD and the goal-optimization curves are practical demonstrations of the correctness of the analytical results.}
Finally, the analytical inversion, Result~III, is compared here with conventional blind decoding techniques \cite{Kay08,Kamitani05,Marre15} used for visual brain decoding.

%
%The obtention of the specific parameters of the model makes extensive use of the results presented here.
%First, following \cite{Malo15}, the layers 2nd and 3rd were determined using MAximum Differentiation (MAD) psychophysics.
%This new psychophysical procedure relies on stimuli generated from the \emph{Jacobian of the response with regard to the image} (i.e. MAD requires Result~I). Then, the layers 1st and 4th were fitted to reproduce subjective image distortion. The conventional brute-force fits used in image distortion models \cite{Watson02,Laparra10a,Malo10,Bertalmio17}, are improved here by applying the \emph{Jacobian of the response with regard to the parameters} for gradient descent. In other words, parameter fitting
%in classical psychophysics improves using Result~II. Finally, the \emph{analytical inversion of the nonlinear stages} (Result~III), which is mandatory in perceptually inspired image compression \cite{Malo06a}, is compared here with conventional blind decoding techniques \cite{Kay08,Kamitani05,Marre15} used for visual brain decoding.
%
%\red{Strictly speaking, the model could have been determined using automatic differentiation as opposed to the use of the explicit expressions (both for MAD and experiments for the reproduction of distortion perception, however this would have prevented the intuition obtained from the expressions.}

\subsection{Physiological, psychophysical and functional trends from the expressions}
\label{insight}

\paragraph*{\blue{Physiology.}} \blue{The \emph{receptive field} of a neuron is the function that describes how the amplitude of the stimulus at different locations
affects its response.
In the simplest (linear) setting, the receptive field of the $k$-th neuron of the $n$-th layer is a vector of weights, $\vect{w}^n_k$, and the variation of the response is given by the dot product of this vector times the variation of the stimulus, $\Delta x^n_k = {\vect{w}^n_k}^\top \cdot \Delta \vect{x}^0$  \cite{Olshausen96,Ringach02}.
In a nonlinear system, $S$, the variation of the response(s) due to the variation of the input is described by the first term of the linear approximation in Eq. \ref{linear_approx}, $\Delta \vect{x}^n = \nabla_{\!\!\vect{x}^0} S \cdot \Delta \vect{x}^0$. Therefore, the receptive fields of the sensors at the $n$-th layer can be thought as the rows of the corresponding Jacobian w.r.t. the stimulus.}

\blue{Using the above receptive field definition based on the Jacobian, a number of interesting qualitative consequences can be extracted from
the analytical Result I (Eq. \ref{deriv_DN}) and the associated chain rule expressions, Eqs. \ref{chain_rule1} and \ref{chain_rule2}.}

\begin{figure}[!t]
	\centering
    \small
    \setlength{\tabcolsep}{2pt}
    \vspace{-2cm}
    \begin{tabular}{c}
    %\hspace{-1.0cm}  \includegraphics[height=0.92\textheight]{general_scheme_bbb.JPG} \\
    \hspace{-5.1cm}  \includegraphics[width=1.35\textwidth]{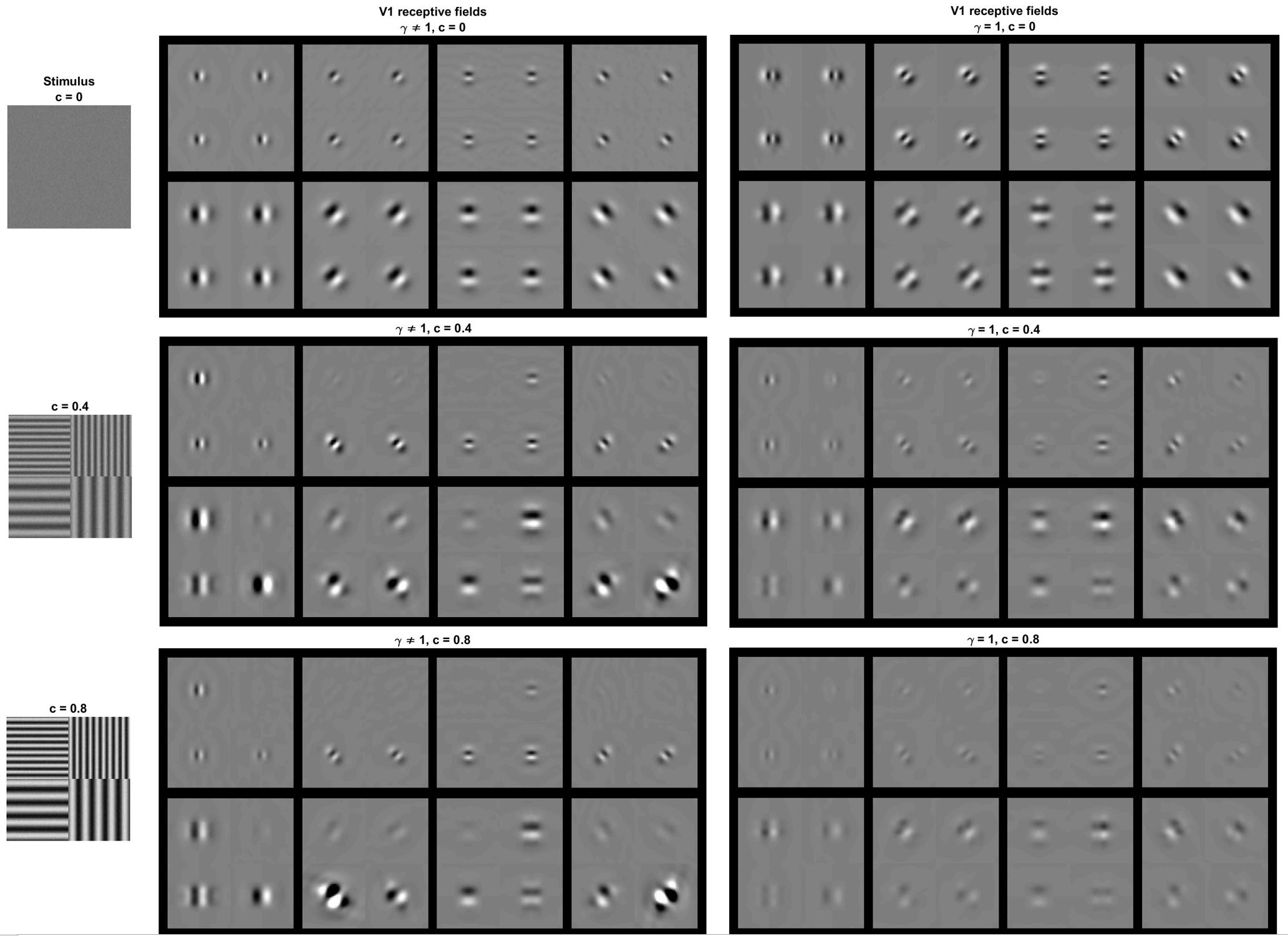} \\
    \vspace{-0.0cm}
    \hspace{-5.2cm}\pbox{1.4\textwidth}{\blue{\textbf{Fig 2.} \scriptsize{\textbf{Insight into physiology~I: Context dependence of receptive fields (from $\nabla_x S$).}
    Comparison of cortical receptive fields tuned to different frequencies, orientations, and locations while adapted to illustrative stimuli of different contrast. The stimuli are shown at the left column. Each row corresponds to the receptive fields induced by the corresponding stimulus.
    The panel at the center shows the receptive fields assuming $\gamma \neq 1$ (as obtained from the experiments). On the contrary, the panel at the right displays the receptive fields setting $\gamma = 1$ at every layer on purpose.
    Note that signal-dependent changes in the receptive fields involve (i) attenuation for sensors stimulated with their preferred signal and (ii) stronger effects on the shape of the receptive field in the left panel when $\gamma \neq 1$, as predicted by the theory.}
    }
    }
    \end{tabular}
    \vspace{-0.5cm}
\end{figure}
%\begin{figure}[!t]
%	\centering
%    \vspace{-0.5cm}
%    \small
%    \setlength{\tabcolsep}{2pt}
%    \begin{tabular}{c}
%    \hspace{-5cm} \includegraphics[width=1.2\textwidth]{receptive_fields.JPG} \\
%    \end{tabular}
%    \vspace{-0.0cm}
%	\hspace{-6cm}\caption{\scriptsize{\textbf{(a.1) Insight in physiology~(I): Context dependence of receptive fields (from $\nabla_x S$).}
%    Trends in Jacobian w.r.t. stimulus.}}\label{MAD_technique}
%    \vspace{-0.15cm}
%\end{figure}

\blue{{\bf First, the shape of the receptive fields at the $i$-th layer is mediated by the functions in the rows of the matrices $L^i$, but it is going to be a signal-dependent combination of these functions.} Note that if the Jacobian is not diagonal, the receptive fields are spatially non-trivial, i.e. not simple \emph{delta} functions.
In the cartoon example this non-diagonal nature comes from the matrix with frequency analyzers, $F$. In the same way, in the cortical layer of our full model (4-th layer), this spatially meaningful part is mediated by a filterbank of wavelet-like linear sensors (in the matrix $L^4$).
According to the chain rule, Eq. \ref{chain_rule2}, these wavelet-like receptive fields will be \emph{modified} by $\nabla_{\vect{y}^4} \mathcal{N}^{(4)}$}.
\blue{This is relevant because, if the Jacobian of the nonlinear part is signal-dependent and nondiagonal, the corresponding linear filters will be recombined in interesting ways leading to variations of the receptive fields. Result I tells us that, in general, this is going to be the case because the diagonal matrices in Eq. \ref{deriv_DN} depend on the signal, and the interaction matrix $H$ is, in principle, non-diagonal. This anticipates that receptive fields at this cortical layer are going to be signal-dependent combinations of wavelet functions.
A closer look at Eq. \ref{deriv_DN} allows to make more specific statements about this adaptive behavior of the receptive fields.}

\blue{ {\bf Second, the fundamental effect of the background signal is reducing the amplitude of the receptive fields (or reducing the global gain of the sensors).}
Note that in Eq. \ref{deriv_DN} this is done in two different ways: a global divisive effect through $\mathds{D}^{-1}_{\mathcal{D}(\vect{e})}$, and a subtractive
effect through $- \mathds{D}_{|\vect{x}|} \cdot H$. In both cases, the bigger the input or output activity (the bigger $\vect{e}$ and $|\vect{x}|$, i.e. \emph{the contrast}), the stronger the attenuation and subtraction. Moreover, this reduction is sensor-specific. Note that left-multiplication by diagonal matrices implies a different factor \emph{per row} (see Eq. \ref{left_diag} in Suppl. Material \ref{matrix_properties}), therefore activity in
the $k$-th sensor is going to reduce the amplitude of the $k$-th row and it is going to increase the subtraction of the linear combination described by the $k$-th row of $H$ (but not of other rows!). % If these were the only terms in Eq. \ref{deriv_DN},
The terms in $\mathds{D}^{-1}_{\mathcal{D}(\vect{e})} \cdot (I - \mathds{D}_{|\vect{x}|} \cdot H)$ mean that $H$ would determine \emph{a fixed} combination of the wavelet filters that would be subtracted from the original filters to a bigger or lower extent depending on the contrast of each wavelet component of the signal. However, there is an extra signal-dependent matrix in the Jacobian: $\mathds{D}_{\left(\gamma |\vect{y}|^{\gamma -1}\right)}$.}

\blue{{\bf Third, the way the linear filters are recombined depends on $H$, but this recombination is not fixed: it may be signal dependent.}
However, if $\gamma = 1$ this dependence vanishes. This effect comes from the extra matrix we mentioned above.
This matrix is right-multiplying the interaction kernel $H$, and hence its effect is substantially different:
it applies a different factor \emph{per column} (see Eq. \ref{right_diag} in Suppl. Material \ref{matrix_properties}).
As a result, the neighbor filters will be combined differently if $\gamma \neq 1$. Otherwise this matrix becomes diagonal and
the combination of neighbors is totally determined by $H$, leading to a contrast dependent attenuation but not a strong change in shape.
}

\blue{In summary, in Result I, Eq. \ref{deriv_DN}, one can identify specific signal-dependent changes in the receptive fields
that involve (i) attenuation for sensors stimulated with their preferred signal and (ii) stronger effects on the shape of the
receptive field depending on the excitation-inhibition exponent $\gamma$.}
\blue{All these effects can be seen in the simulation of Fig.~2, %\ref{AdaptiveRF},
where we compare the receptive fields at the 4-th layer
tuned to different frequencies at different locations of illustrative signals of different contrast. We compare the receptive fields using
$\gamma = 0.65$ (found in the experiments described in the next sections) with those obtained using $\gamma = 1$.}

\vspace{0.25cm}
\blue{Result II is also useful to address physiologically interesting questions.
For instance, \emph{how uncertainty in the synaptic weights affects the response of the sensors?.}
Such question is interesting because the assumptions done to set these filters may be poor,
e.g. selection of a wavelet filterbank in the cortical layer which is not biologically plausible.
Similarly, parameters coming from experimental measurements are noisy. How critical is the
experimental error in terms of the final impact in the response?.
In such situations, the Jacobian w.r.t. the parameters (Result II) that describes the impact of variations
of the parameters in the response has obvious interest.}

\begin{figure}[!t]
	\centering
    \small
    \setlength{\tabcolsep}{2pt}
    \vspace{-0cm}
    \begin{tabular}{c}
    %\hspace{-1.0cm}  \includegraphics[height=0.92\textheight]{general_scheme_bbb.JPG} \\
    %\hspace{-5.15cm}  \includegraphics[width=1.2\textwidth]{impact_parameters.JPG} \\
    %\hspace{-5.15cm} $\times0.5 \, \vect{b}$  & Baseline & $\times2 \, \vect{b}$  & $\times2 \, \vect{b} \,\,\, \times10 \, G$  \\
    \hspace{-5.15cm}  \includegraphics[width=1.3\textwidth]{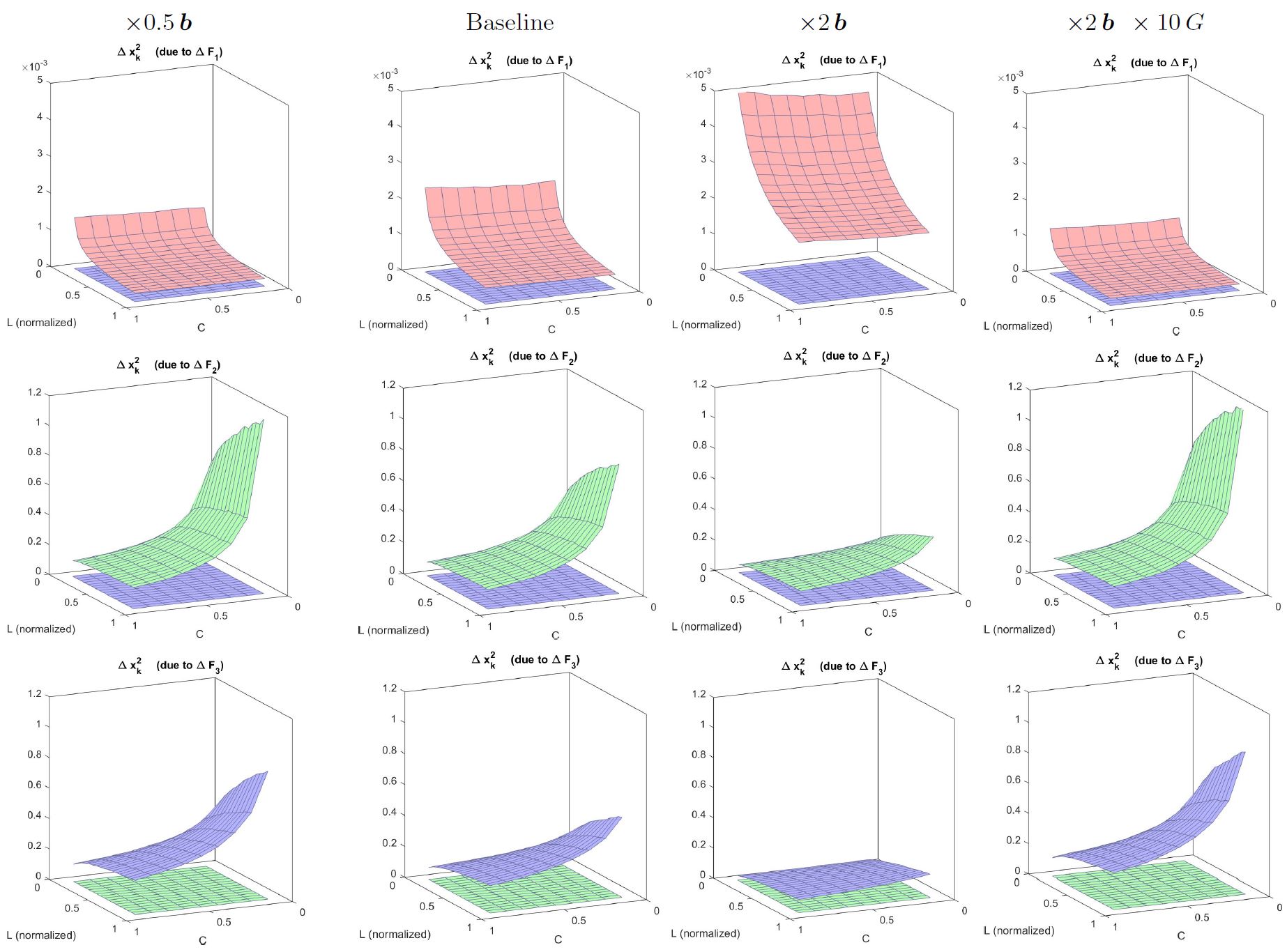} \\
    \vspace{-0.0cm}
    \end{tabular}
    \begin{tabular}{c}
    \hspace{-5.3cm}\pbox{1.4\textwidth}{\blue{\textbf{Fig 3.} \scriptsize{\textbf{Insight into physiology~II: Impact on the response of uncertainty in different filters (from $\nabla_{\theta} S$). }
                         Top: distortions in the zero frequency filter, Middle: distortion in the low-frequency filter, and Bottom: distortion in the high-frequency filter.
                         Variation in the response of the zero, low- , and high-frequency sensors is represented in red, green, and blue respectively.
                         The different columns were computed using variations in the parameters of the simplified model (baseline) to point out the trends seen in Eq. \ref{impact}. Specifically: (i) Impacts in the responses of AC sensors increases with input luminance and gain, and decreases when contrast or semisaturation increase; (ii) Impact of DC filters increases with contrast and strongly decreases with luminance.
                         }
                         }
                         }
                         %\label{FMmodel}
                         \end{tabular}
    \vspace{-0.5cm}
\end{figure}

\blue{Here we show an example of this use of Result II in the simplified three-sensors model outlined above.
In particular, we address how uncertainty in the frequency analyzers (rows of $F$) has an impact on the response of the different
sensors, $x^2_k$, across the stimulus space. In absence of Result~II, we could add random noise to the
filters and empirically check the variation for natural images of different luminance and contrast (see Fig.~3).
However, Result II allows us to anticipate the outcome of such experiment.}
\blue{In this case, using the part of Eq. \ref{elementary_jacobian2} that corresponds to the derivative w.r.t. the filters in the matrix $L^2 = G \cdot F$,
the impact of a variation of the $k$-th filter in $F$, the row vector $\Delta F_k$, is:}
\begin{equation}
      \blue{ \Delta x^2_k = \frac{1 - |x^2_k|}{b_k + |y^2_k|} \,\, G_{kk} \,\,\,\, \Delta F_k \cdot \vect{x}^1 }
      \label{impact}
\end{equation}
\blue{In this equation we can see that different filters have different behavior over the image space.
On the one hand, the impact in the response of AC sensors ($k>1$) will increase with the luminance
of the input due to the direct dependence with $\vect{x}^1$, which is related to brightness.
However, when increasing the contrast of the images the responses $y^2_k$ and $x^2_k$ also increase and then, the subtractive and divisive
effects in the fraction of Eq.~\ref{impact} reduce the impact. Of course, increasing the semisaturation or the gain implies the
corresponding decrease and increase in the impact of the AC filters.}
\blue{On the contrary, the impact of the DC filter behaves quite differently: it increases (a little bit) with the contrast,
through the energy that may be captured by the random variation in $\Delta F_k$. But, more importantly, it strongly decreases
with luminance because of the subtractive and divisive effects caused by increased values in $|x^2_1|$ and $|y^2_1|$, which are increasing functions of brightness.}
\blue{Fig.~3 confirms all these trends.}

\vspace{0.25cm}
\paragraph*{\blue{Psychophysics.}} \blue{The \emph{sensitivity} of the system is characterized by its discrimination ability:
the sensitivity is bigger where the discrimination regions determined by the JNDs are smaller \cite{Graham89,Regan91}.}
\blue{The trends of the sensitivity of the system in the image space for a range of luminance and contrast can be
identified from Result~I, Eq. \ref{deriv_DN}, and the associated expression for the metric, Eq. \ref{distance2}.}
\blue{This is because the volume of the discrimination region at each point of the stimulus space is inversely proportional
to the determinant of the metric. In our simplified model the frequency analysis transform is orthonormal, i.e. $|F|=1$, as a result,
the sensitivity in the space of luminance images depends on these three factors in brackets:}
\begin{equation}
      \hspace{-3cm}\blue{\textrm{sensitivity} = |\nabla_{\vect{y}^1} S|^2 = \gamma_1^{2 d} \left[ \left( \prod_{k=1}^d {(y^1_k)}^{2 (\gamma_1 - 1)} \right) (b_1 + y^2_1)^{-2} (1-x^2_1)^2 \right] \left[  \prod_{k=2}^d (b_k + y^2_k)^{-2} \prod_{k=2}^d (1-x^2_k)^2 \right] \left[ \prod_{k=1}^d G_{kk}^2 \right]}
      \label{sensit}
      %\hspace{-3cm}\blue{\textrm{sensitivity} = |\nabla_{\vect{y}^1} S|^2 = \gamma_1^{2 d} \left[ \left( \prod_{k=1}^d {(y^1_k)}^{2 (\gamma_1 - 1)} \right) (b_1 + y^2_1)^{-2} (1-x^2_1)^2 \right] \left[  \prod_{k=2}^d (b_k + %y^2_k)^{-2} \prod_{k=2}^d (1-x^2_k)^2 \right] \left[ \prod_{k=1}^d G_{kk}^2 \prod_{\lambda = 1}^{d_\lambda} V_{\lambda}^2 \right]}
\end{equation}
\blue{The first factor clearly \emph{decreases with luminance} because its three terms decrease with luminance, brightness and the nonlinear response to brightness (either by division or by subtraction). Note that the first term in this factor is divisive because $\gamma_1 < 1$ (saturating transform), and hence $ 2(\gamma_1-1)<0$. Setting $\gamma_1 = 1$ would reduce the dependence with luminance because the first term in this factor would be 1 for every image.
The second factor \emph{decreases with contrast} because the responses of the AC sensors of the second layer increase with contrast and hence, the two terms of the second factor decrease with contrast (by division and subtraction respectively). Note that increasing the semisaturation factor will reduce the dependence with contrast and luminance.
Finally, the third factor \emph{increases with the area under the CSF-like gain} in the diagonal of the matrix $G$.
In Fig.~4 we compute the inverse of the volumes of the discrimination regions for 3-pixel natural images covering the luminance and contrast range and the above trends are confirmed.}

%\begin{figure}[!t]
%	\centering
%    \small
%    \setlength{\tabcolsep}{2pt}
%    \vspace{-0cm}
%    %\begin{tabular}{c}
%    %%\hspace{-1.0cm}  \includegraphics[height=0.92\textheight]{general_scheme_bbb.JPG} \\
%    %\hspace{-5.15cm}  \includegraphics[width=1.2\textwidth]{Sensitivity_4.JPG} \\
%    %\vspace{-0.0cm}
%    \begin{tabular}{cccc}
%    \hspace{-5.15cm} Baseline  & $\gamma_1 = 1$ &  Wider $G$  & $\times3 \, \vect{b}$ \\
%    \hspace{-5.15cm}  \includegraphics[width=0.3\textwidth]{sensi_1.JPG} & \includegraphics[width=0.3\textwidth]{sensi_2.JPG} & \includegraphics[width=0.3\textwidth]{sensi_3.JPG} & \includegraphics[width=0.3\textwidth]{sensi_4.JPG} \\
%    \vspace{-0.0cm}
%    \end{tabular}
%    \begin{tabular}{c}
%    \hspace{-5.3cm}\pbox{1.4\textwidth}{\blue{\textbf{Fig 4.} \scriptsize{\textbf{Insight into physchophysics: Sensitivities from the volume of the JND regions (related to $\nabla_x S$). }
%                         From left to right: (a) Baseline situation (shows the expected luminance/contrast dependence), (b) Linear luminance-to-brightness response is set to linear ($\gamma_1 = 1$), (c) contrast sensitivity is increased, (d) semisaturation is increased.}
%                         }
%                         }
%                         %\label{FMmodel}
%                         \end{tabular}
%    \vspace{-0.5cm}
%\end{figure}

\begin{figure}[!t]
	\centering
    \small
    \setlength{\tabcolsep}{2pt}
    \vspace{-0cm}
    %\begin{tabular}{c}
    %%\hspace{-1.0cm}  \includegraphics[height=0.92\textheight]{general_scheme_bbb.JPG} \\
    %\hspace{-5.15cm}  \includegraphics[width=1.2\textwidth]{Sensitivity_4.JPG} \\
    %\vspace{-0.0cm}
    \begin{tabular}{c}
    % \hspace{-5.15cm} Baseline  & $\gamma_1 = 1$ &  Wider $G$  & $\times3 \, \vect{b}$ \\
    \hspace{-5.15cm}  \includegraphics[width=1.3\textwidth]{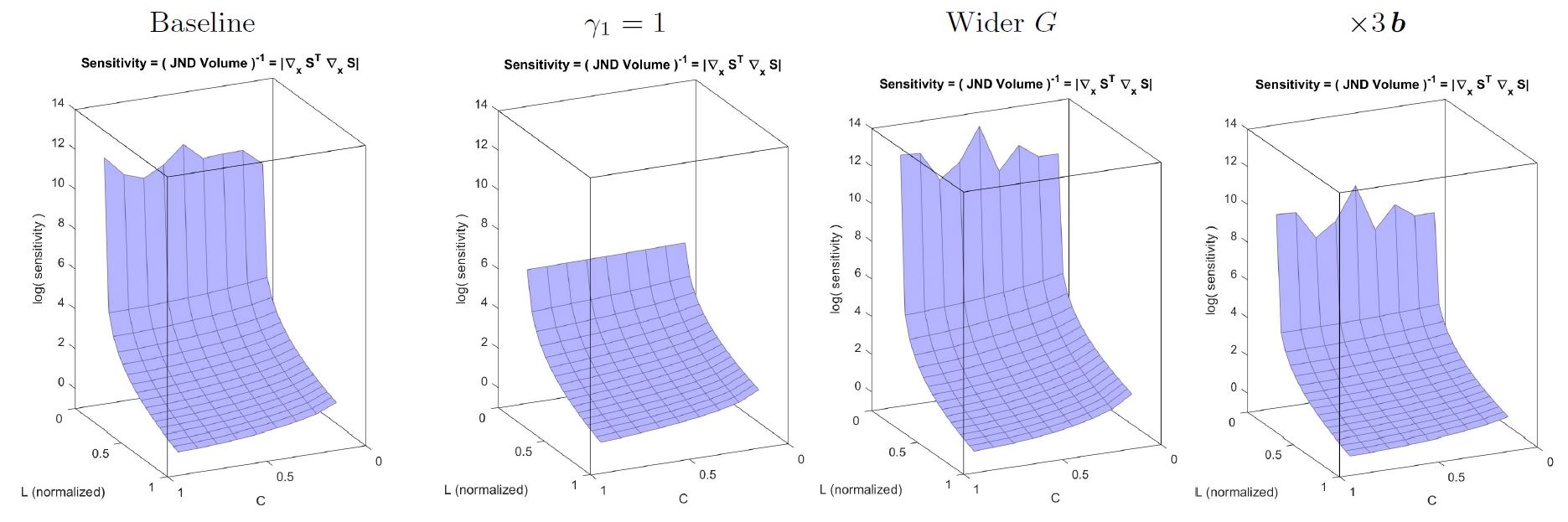} \\
    \vspace{-0.0cm}
    \end{tabular}
    \begin{tabular}{c}
    \hspace{-5.3cm}\pbox{1.4\textwidth}{\blue{\textbf{Fig 4.} \scriptsize{\textbf{Insight into physchophysics: Sensitivities from the volume of the JND regions (related to $\nabla_x S$). }
                         From left to right: (a) Baseline situation (shows the expected luminance/contrast dependence), (b) Linear luminance-to-brightness response is set to linear ($\gamma_1 = 1$), (c) contrast sensitivity is increased, (d) semisaturation is increased.}
                         }
                         }
                         %\label{FMmodel}
                         \end{tabular}
    \vspace{-0.5cm}
\end{figure}

\vspace{0.25cm}
\paragraph*{\blue{Function.}} \blue{It has been argued that one of the basic \emph{functions} of the retina-cortex neural pathway is maximizing
the information about the stimuli transmitted to subsequent areas of the brain \cite{Barlow61,Olshausen96,Barlow01}.
Under certain conditions \cite{Bell97}, the transmitted information is maximized by reducing the redundancy between the coefficients of the
neural representation of the stimulus. Therefore, measuring how redundancy is reduced when facing different kinds of stimuli tells us how
efficient is the system in transmitting the information about them. A very general measure of redundancy in a set of variables is \emph{multi-information}, MI,
which measures the amount of bits shared by them \cite{Studeny98}. Therefore, an appropriate way to assess the efficiency of a system in transmitting the
information about certain stimuli is measuring how the multi-information in the internal representation, MI($\vect{x}^n$), is reduced with regard to the multi-information
in the input space, MI($\vect{x}^0$). Interestingly, this difference depends on the Jacobian of the transform from $\vect{x}^0$ to $\vect{x}^n$ \cite{Studeny98},
and hence our Result~I is helpful here. According to \cite{Studeny98}, the multi-information reduction under a transform, $S$, is:}
\begin{equation}
      \blue{\Delta \textrm{MI} = \textrm{MI}(\vect{x}^0) - \textrm{MI}(\vect{x}^n) = \sum_{k=1}^{d_0} h(x^0_k) - \sum_{k=1}^{d_n} h(x^n_k) + \mathds{E}\left[ log_2 |\nabla_{\vect{x}^0} S| \right]}
      \label{MI_reduct}
\end{equation}
\noindent \blue{where $h(\cdot)$ represents the entropy of the considered scalar variable, which is easy to compute from the corresponding univariate probability density function
of the stimuli, and $\mathds{E}\left[ log_2 |\nabla_{\vect{x}^0} S| \right]$ is the expected value of the $log_2$ of the determinant of the Jacobian over the considered kind of stimuli.
In the case of natural images and vision models with reasonable parameters, the effect of the nonlinearities is performing a sort of PDF equalization \cite{Malo10,Bertalmio14},
therefore, the first term, $\Delta h$, should be fairly independent of the average contrast and luminance, and one would expect that the main dependence of $\Delta \textrm{MI}$
is given by the term that depends on the Jacobian.}
\blue{In our simplified model, the determinant of the Jacobian is the square root of the sensitivity given in Eq. \ref{sensit}.}
\blue{As a result, one would expect that the efficiency shows the same trends as the sensitivity.}

\begin{figure}[!b]
	\centering
    \small
    \setlength{\tabcolsep}{2pt}
    \vspace{-0cm}
    \begin{tabular}{c}
    %\hspace{-5.15cm} Baseline  & $\gamma_1 = 1$ &  Wider $G$  & $\times3 \, \vect{b}$ \\
    \hspace{-5.15cm}  \includegraphics[width=1.3\textwidth]{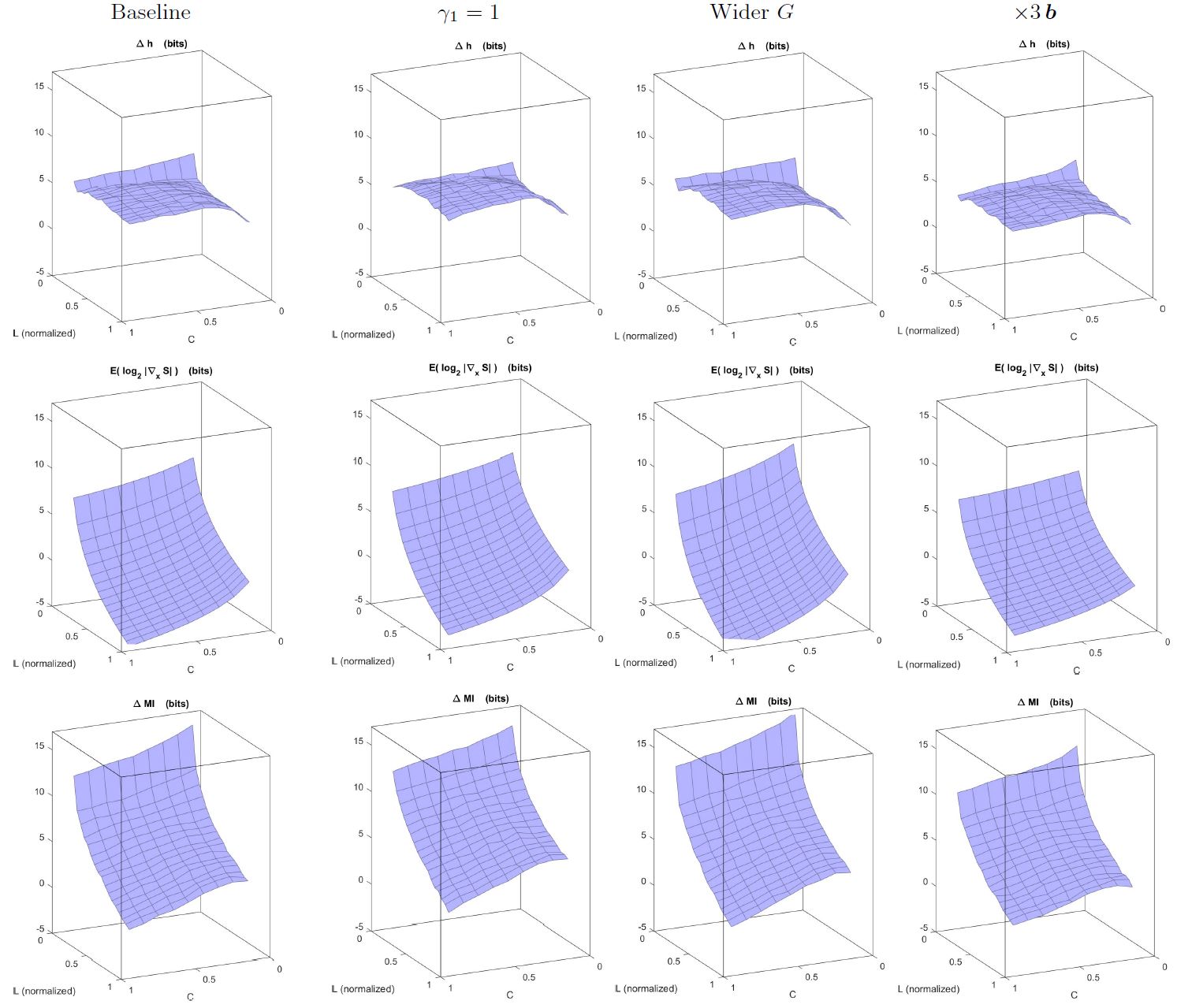}\\
    \vspace{-0.0cm}
    \end{tabular}
    \begin{tabular}{c}
    \hspace{-5.3cm}\pbox{1.4\textwidth}{\blue{\textbf{Fig 5.} \scriptsize{\textbf{Insight into function: Reduction in multi-information from the difference in marginal entropies and the Jacobian of the transform. }
                         Top: differences in marginal entropies. Middle: Term depending on the Jacobian.
                         Bottom: final multi-information reduction.
                         From left to right: (a) Baseline situation (shows the expected luminance/contrast dependence), (b) Linear luminance-to-brightness response, (c) increased contrast sensitivity, (d) increased semisaturation.
                         We can see how $\Delta h$ is fairly constant and the final efficiency follows the trends of the Jacobian: it is large in low-luminance, low-contrast regions of input space. Interestingly, these are the regions more populated by natural images.
                         }
                         }
                         }
                         %\label{FMmodel}
                         \end{tabular}
    \vspace{-0.5cm}
\end{figure}

\blue{In the illustration shown in Fig.~5 we took $10^5$ natural images of 3-pixels and adjusted their average luminance and contrast to get different sets of stimuli over the whole range. For each set of $10^5$ samples we computed the $\Delta \textrm{MI}$ according to Eq. \ref{MI_reduct}. Fig.~5 shows that $\Delta h$ is fairly constant
and that the redundancy reduction follows the trends expected from Eq.~\ref{sensit}. Additionally, \emph{sensitivity} and \emph{efficiency} do follow similar trends: they are bigger in the low-luminance, low-contrast regions of the input space. Interestingly, these are the regions more populated by natural images \cite{HuangMumford99,Simoncelli01,malo2000role}}

\setcounter{figure}{5}

%%%%%%%%%%%%%%%%%%%%%%%%%%%%%%%%%%%%%%%%%%%%%%%%%%%%%%%%%%%%%%%%%%%%%%%%%  END OF 3.1 INSIGTH
%%%%%%%%%%%%%%%%%%%%%%%%%%%%%%%%%%%%%%%%%%%%%%%%%%%%%%%%%%%%%%%%%%%%%%%%%%%%%%%%%%%%%%%%%%%%%%%%%%%%%%%%%%%%%%%%%%%%%%%%%%%%%%%%%%%%%%%%%%%%%%%%%%
%%%%%%%%%%%%%%%%%%%%%%%%%%%%%%%%%%%%%%%%%%%%%%%%%%%%%%%%%%%%%%%%%%%%%%%%%%%%%%%%%%%%%%%%%%%%%%%%%%%%%%%%%%%%%%%%%%%%%%%%%%%%%%%%%%%%%%%%%%%%%%%%%%
%%%%%%%%%%%%%%%%%%%%%%%%%%%%%%%%%%%%%%%%%%%%%%%%%%%%%%%%%%%%%%%%%%%%%%%%%%%%%%%%%%%%%%%%%%%%%%%%%%%%%%%%%%%%%%%%%%%%%%%%%%%%%%%%%%%%%%%%%%%%%%%%%%
%%%%%%%%%%%%%%%%%%%%%%%%%%%%%%%%%%%%%%%%%%%%%%%%%%%%%%%%%%%%%%%%%%%%%%%%%%%%%%%%%%%%%%%%%%%%%%%%%%%%%%%%%%%%%%%%%%%%%%%%%%%%%%%%%%%%%%%%%%%%%%%%%%

\subsection{Jacobian with regard to the image in stimulus synthesis}
\label{MAD_section}
Many times, stimuli design implies that the desired image should fulfill certain properties in the response domain.
Examples include (i) \emph{artistic style transfer} \cite{Gatys16}, in which the response to the synthesized image should be close to the response to the image from which the content is inherited, and should have a covariance structure close to the one in the response to the image from which style is inherited; and (ii) \emph{Maximum Differentiation} \cite{Wang08,Malo15,Malo16}, in which the synthesized images should have maximum/minimum perceptual distance with regard to a certain reference image with a constraint in the energy of the distortion.
In both cases, fulfilling the requirements implies modifying the image so that the response is modified in certain direction. In such situations the Jacobian of the response with regard to the image (Result~I) is critical.

Here we discuss in detail the case of MAximum Differentiation (MAD).
This technique is used to rank competing vision models by using them to solve a simple geometric question and visually assessing which one gave the better solution. While in conventional psychophysics the decision between two models relies on how well they fit thousands of individual measurements (either contrast incremental thresholds or subjective ratings of distortions), in MAD the decision between two models reduces to a single visual experiment.

The geometric question for the perception model in MAD is the following \cite{Wang08}: given a certain reference image, $\vect{x}^0_A$, and the set of distorted images departing a certain amount of energy from the reference image, the sphere with center in $\vect{x}^0_A$ and certain fixed radius (or certain Mean Squared Error); the problem is looking for the images with maximum and minimum perceptual distance on the sphere, lets call them $\vect{x}^0_{\textrm{min}}$ and $\vect{x}^0_{\textrm{max}}$.
If the vision model is meaningful, $\vect{x}^0_{\textrm{min}}$ and $\vect{x}^0_{\textrm{max}}$ should have a very different visual appearance.
The more accurate vision model will be the one leading to the pair of images which are maximally different.
The discriminative power of this visual experiment comes from the fact that the synthesis of these stimuli involves comparing the performance of the models under consideration in every possible direction of the space of images.

\begin{figure}[!b]
	\centering
    \vspace{-0.5cm}
    \small
    \setlength{\tabcolsep}{2pt}
    \begin{tabular}{ccc}
    \hspace{-5cm} (a) & (b) & (c) \\
    \hspace{-5cm} \includegraphics[width=5.5cm,height=5.15cm]{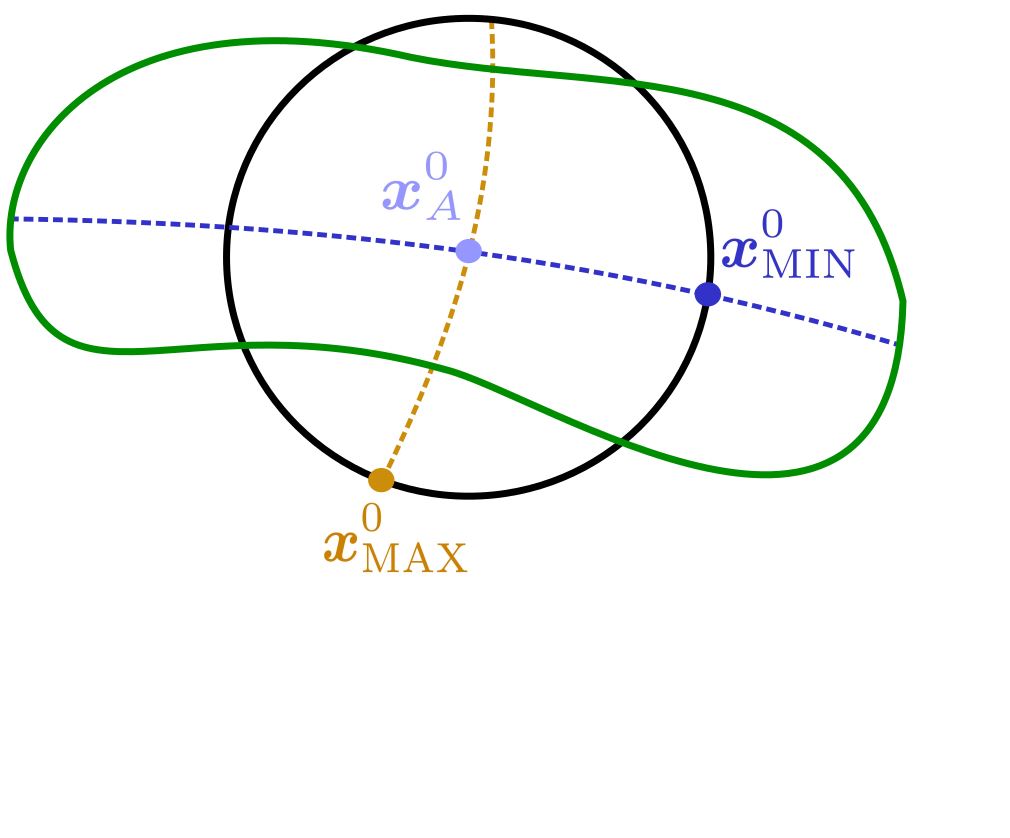} \hspace{0.2cm} & \hspace{0.2cm}
                    \includegraphics[width=5.5cm,height=5.15cm]{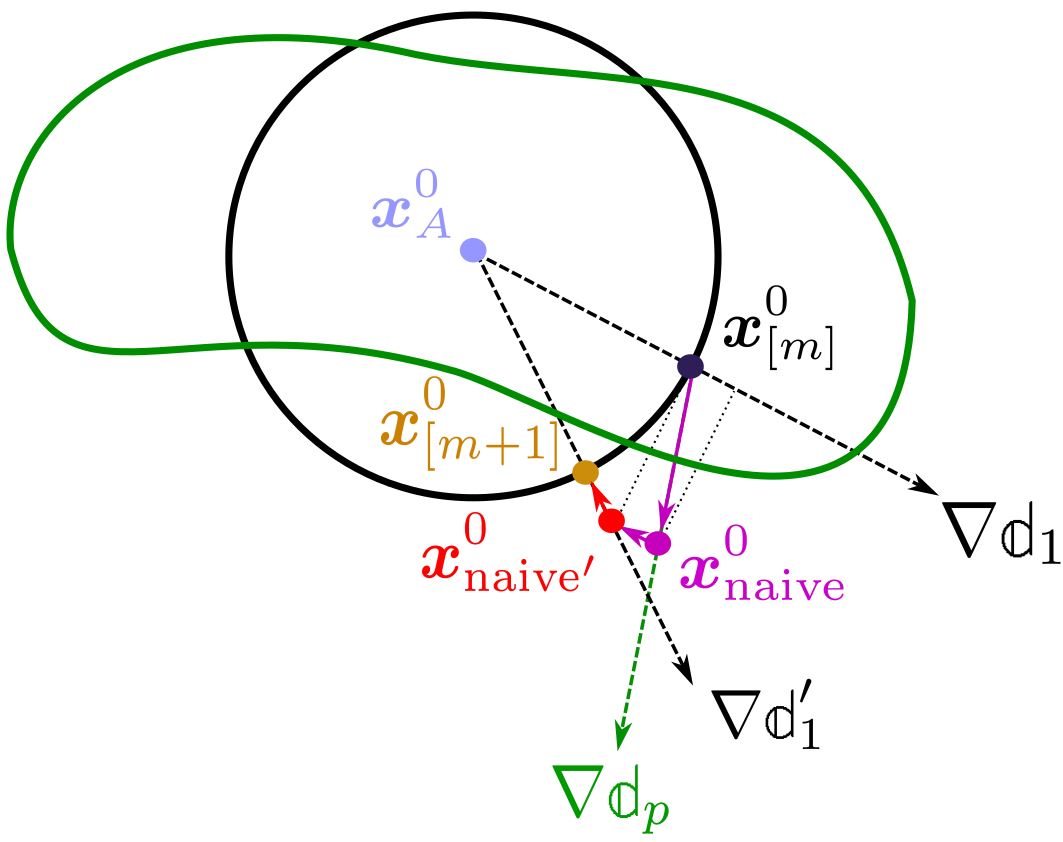} \hspace{0.2cm} & \hspace{0.2cm}  \includegraphics[width=5.5cm,height=5.15cm]{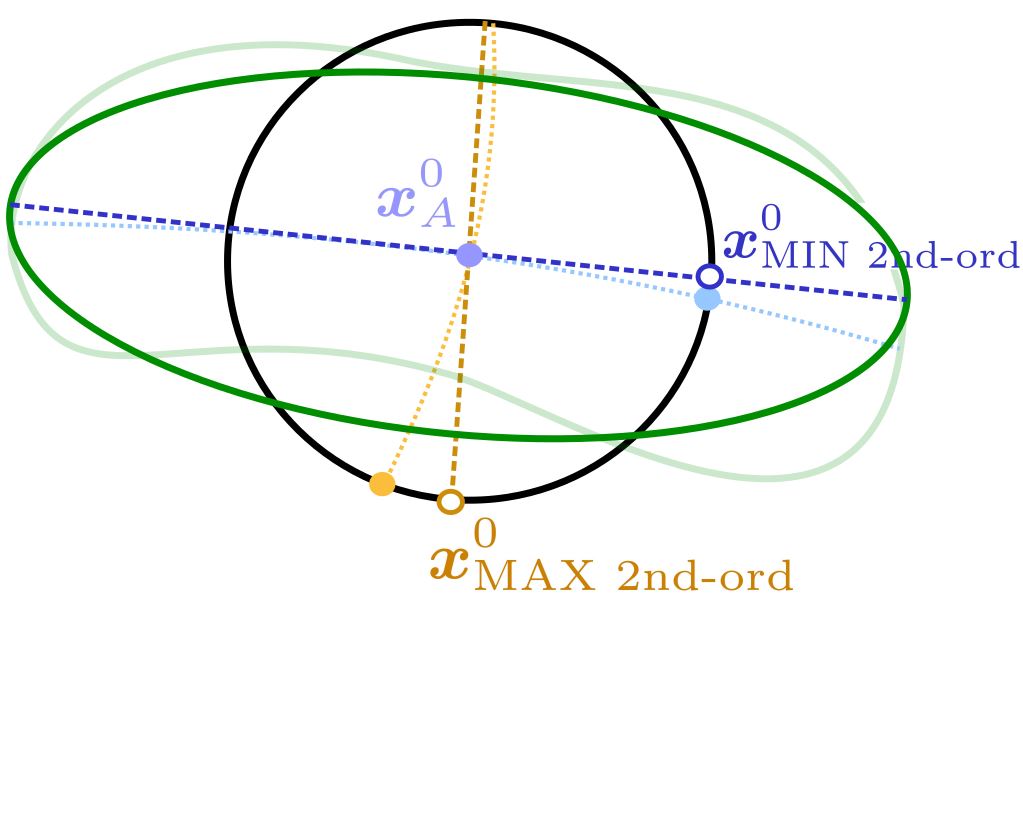} \\
    \end{tabular}
    \vspace{-0.25cm}
	\hspace{-6cm}\caption{\scriptsize{\textbf{Stimulus generation in MAximum Differentiation (MAD).} (a) The MAD concept: given an original image, e.g. the point $\vect{x}^0_A$ in light blue, a perceptual distance measure, $\mathbb{d}_p$, coming from a vision model (leading to the discrimination region in green), and certain fixed Euclidean distance, $\mathbb{d}_1$, (sphere in black); the problem is looking for the best and worst images on the sphere according to the perceptual distance $\mathbb{d}_p$. The solution is given by the images that are in the directions (or paths) leading to the biggest or the lowest perceptual distortions ($\vect{x}^0_{\textrm{MAX}}$ and $\vect{x}^0_{\textrm{MIN}}$, in dark blue and orange respectively).
In the example, the path leading to the biggest perceptual difference (for the same Euclidean length) is the curve in orange because it represents the shortest path from the original image to the discrimination boundary in green. Equivalently, the path leading to the lowest perceptual difference (for the same Euclidean length) is the curve in blue because it represents the longest path to the discrimination boundary in green.
(b) The MAD algorithm: start from a random point at the sphere and modify it to increase (or alternatively decrease) the perceptual distance following $\nabla \mathbb{d}_p$. Note that the naive application of the gradient implies a solution out of the sphere. This has to be projected on the sphere through the appropriate correction: first remove the component parallel to $\nabla \mathbb{d}_1$, and then project in the direction of $\nabla \mathbb{d}_1'$. (c) Second order approximation: approximate the perceptual discrimination regions by ellipsoids (local linear approximation of the vision model). In this way the MAD images are given by the directions of the maximum and minimum eigenvalue of the 2nd order metric matrix.}}\label{MAD_technique}
    \vspace{-0.15cm}
\end{figure}

Fig. \ref{MAD_technique}(a) illustrates the geometric problem in MAD. The following paragraphs show how the different solutions to this geometric problem reduce to the use of Result I.

\paragraph*{General, but numeric, solution to MAD.} In general there is no analytical solution for such problem and hence one has to start from a guess image and modify it according to the direction of the gradient of the perceptual distance, $\nabla_{\!\!\vect{x}^0} \, \mathbb{d}_p$, to maximize or minimize this distance.
Of course, the problem, illustrated in Fig. \ref{MAD_technique}(b), is that, a naive modification of the $m$-th guess, $\vect{x}^0_{[m]}$, in the direction of this gradient puts the solution out of the sphere: note the location of $\vect{x}^0_{\textrm{naive}}$ in pink, in Fig. \ref{MAD_technique}(b).
As proposed in \cite{Wang08}, this departure from the sphere is solved by
(i) subtracting the component parallel to the gradient of the Euclidean distance, $\nabla_{\!\!\vect{x}^0} \, \mathbb{d}_1$ (see the point $\vect{x}^0_{\textrm{naive}'}$ in red), and
(ii) projecting this displaced point back into the sphere (see the point $\vect{x}^0_{[m+1]}$ in orange). In summary, the complete iteration for the stimulus that maximizes/minimizes the distance is as follows \cite{Wang08}:
\begin{equation}
  \vect{x}^0_{[m+1]} = \vect{x}^0_{[m]} \pm \lambda \Bigg( \nabla_{\!\!\vect{x}^0} \, \mathbb{d}_p - \frac{\nabla_{\!\!\vect{x}^0} \, \mathbb{d}_1 \cdot \nabla_{\!\!\vect{x}^0} \, \mathbb{d}_p^\top}{\nabla_{\!\!\vect{x}^0} \, \mathbb{d}_1 \cdot \nabla_{\!\!\vect{x}^0} \, \mathbb{d}_1^\top} \,\, \nabla_{\!\!\vect{x}^0} \, \mathbb{d}_1 \Bigg)^\top + \nu \,\, \nabla_{\!\!\vect{x}^0} \, \mathbb{d}_1'^\top
  \label{MADiterat}
\end{equation}
where, $\lambda$ is the constant that controls the convergence of the gradient descent, and $\nu$ can be computed analytically since the Euclidean distance of the point projected onto the sphere should be $ | \vect{x}^0_{\textrm{naive}'} + \nu \,\, \nabla_{\!\!\vect{x}^0} \, \mathbb{d}_1'^\top - \vect{x}^0_A |^2_2 = r^2$, where, given the gradients, the only unknown is $\nu$.
Note that the gradients of the distances are \emph{row vectors} since they should be applied on the column vectors describing the increments in the images: $\Delta \mathbb{d} = \nabla_{\vect{x}^0} \mathbb{d} \cdot \Delta \vect{x}^0$ (row vector times column vector).
That is why we need to transpose the gradients before adding the modifications to $\vect{x}^0_{[m]}$, and the reason
for the transposes in the scalar products of gradients (as in the projection $\nabla_{\!\!\vect{x}^0} \, \mathbb{d}_1 \cdot \nabla_{\!\!\vect{x}^0} \, \mathbb{d}_p^\top$, row vector times column vector).
Note also that the gradients without prime are computed at $\vect{x}^0_{[m]}$, and the gradient with prime is computed at $\vect{x}^0_{\textrm{naive}'} = \vect{x}^0_{[m]} \pm \lambda \Big( \nabla_{\!\!\vect{x}^0} \, \mathbb{d}_p^\top - \frac{\nabla_{\!\!\vect{x}^0} \, \mathbb{d}_1 \cdot \nabla_{\!\!\vect{x}^0} \, \mathbb{d}_p^\top}{\nabla_{\!\!\vect{x}^0} \, \mathbb{d}_1 \cdot \nabla_{\!\!\vect{x}^0} \, \mathbb{d}_1^\top} \,\, \nabla_{\!\!\vect{x}^0} \, \mathbb{d}_1^\top \Big)$.

Now, lets address the gradients. The Euclidean distance with regard to the reference image evaluated at certain $\vect{x}^0_B$ is $\mathbb{d}_1(\vect{x}^0_B) = \Big( (\vect{x}^0_B - \vect{x}^0_A)^\top \cdot (\vect{x}^0_B - \vect{x}^0_A) \Big)^{1/2}$. Therefore, the gradient of the Euclidean distance with regard to $\vect{x}^0_B$ is:
\begin{equation}
      \nabla_{\!\!\vect{x}^0_B} \, \mathbb{d}_1(\vect{x}^0_B) = \frac{\partial \Big(\mathbb{d}_1(\vect{x}^0_B)^2\Big)^{1/2} }{\partial \mathbb{d}_1(\vect{x}^0_B)^2} \cdot \frac{ \partial \mathbb{d}_1(\vect{x}^0_B)^2 }{ \partial (\vect{x}^0_B - \vect{x}^0_A) } \cdot \frac{\partial (\vect{x}^0_B - \vect{x}^0_A)}{ \partial \vect{x}^0_B } =
      \frac{1}{\mathbb{d}_1(\vect{x}^0_B)} \,\,(\vect{x}^0_B - \vect{x}^0_A)^\top \cdot I
      \nonumber
\end{equation}

More interestingly (since this was not addressed in \cite{Wang08}), the gradient of the perceptual distance in the cascaded setting considered here, which is defined at the response domain, Eq.~\ref{distance}, is,
\begin{equation}
      \nabla_{\!\!\vect{x}^0_B} \, \mathbb{d}_p(\vect{x}^0_B) = \frac{\partial \Big(\mathbb{d}_p(\vect{x}^0_B)^2\Big)^{1/2} }{\partial \mathbb{d}_p(\vect{x}^0_B)^2} \cdot \frac{ \partial \mathbb{d}_p(\vect{x}^0_B)^2 }{ \partial (\vect{x}^n_B - \vect{x}^n_A) } \cdot \frac{\partial (\vect{x}^n_B - \vect{x}^n_A)}{ \partial \vect{x}^0_B } = \frac{1}{\mathbb{d}_p(\vect{x}^0_A,\vect{x}^0_B)} \,\, (\vect{x}^n_B - \vect{x}^n_A)^\top \cdot \nabla_{\!\!\vect{x}^0} S(\vect{x}^0_B)
      \label{grad_distance}
\end{equation}
which depends on the responses for the considered images, $\vect{x}^n = S(\vect{x}^0)$, and on the Jacobian of the response with regard to the input $\nabla_{\!\!\vect{x}^0} S(\vect{x}^0_B)$.

Eq. \ref{grad_distance} together with the auxiliary results on $\nabla_{\!\!\vect{x}^0} S$ (Eqs. \ref{chain_rule1} and \ref{chain_rule2})  imply that the application of MAD in the cascaded setting considered here, reduces to the use of Result~I, i.e. Eq. \ref{deriv_DN} for the canonical nonlinearity, or the equivalent equations for the alternative nonlinearities considered (i.e. Eqs. \ref{Jacobian W-C} and \ref{jacobian_two_gamma}).

%descent constrained to a sphere finally depends on $\nabla_{\!\!\vect{x}^0} \, d_p$ \cite{Wang08}.
%In our setting, Eq. \ref{distance}, this implies,
%\begin{equation}
%  \frac{\partial d^2_p(\vect{x}^0_A, \vect{x}^0_B)}{\partial \vect{x}^0_B} = \,\, 2  (\vect{x}^n_B - \vect{x}^n_A)^\top \cdot \nabla_x S(\vect{x}^0_B)
%  \label{grad_distance}
%\end{equation}

%\paragraph*{Linear approximation of MAD. Gradient descent.} Eq. \ref{grad_distance} can be simplified if we assume that $\Delta x^n = \nabla_x S(x^0_A) \cdot \Delta x^0$. In that case, in the gradient of the distance there is no need to compute the response (only the Jacobian),
%\begin{equation}
%  \frac{\partial d^2_p(\vect{x}^0_A, \vect{x}^0_B)}{\partial \vect{x}^0_B} = \,\, 2  (\vect{x}^0_B - \vect{x}^0_A)^\top \cdot \nabla_x S(\vect{x}^0_B)^\top \cdot \nabla_x S(\vect{x}^0_B)
%  \label{grad_distance2}
%\end{equation}

\paragraph*{Analytic, but approximated, solution to MAD.}
As stated in Section \ref{general_cons} when talking about the distance, Eq. \ref{distance2}, in the local-linear approximation the general discrimination regions
% defined by the inverse of spheres in the response domain
are approximated by ellipsoids. In the illustration of Fig.~\ref{MAD_technique}, the (general) curved region in dark green in Figs.~\ref{MAD_technique}.a and~\ref{MAD_technique}.b
is approximated by the ellipsoid in Fig.~\ref{MAD_technique}.c.

Under this approximation, the minimization/maximization of the perceptual distance on the Euclidean sphere has a clear analytic solution:
the images with maximum and minimum perceptual distance will be those in the directions of the eigenvectors with minimum and maximum eigenvalues of the metric matrix $M(\vect{x}^0_A)$.
These, again, depend on the Jacobian of the response with regard to the stimulus, and hence on Result~I.

The view of the MAD problem in terms of a metric matrix is also useful when breaking large images into smaller patches for computational convenience.
In these patch-wise scenarios the global metric matrix actually has block-diagonal structure
(see the Supplementary Material~\ref{summation_in_mad}).
Therefore, given the properties of block-diagonal matrices \cite{Golub}, the global eigenvectors (and hence the solution to MAD) actually reduce
to the computation of the eigenvectors of the smaller metric matrices for each patch.

\paragraph*{Illustration of the general and the analytic solutions.}
Here we take a reference image and we launch a gradient descent/ascent search in the sphere of constant \blue{Root Mean Square Error} (RMSE) to
look for the best/worst version of this image.

For the same image we compute the Jacobian with regard to the stimulus and we compute the eigenvectors of bigger and lower eigenvalue, i.e. the directions that lead to most/least visible distortions in the 2nd order approximation of the distance (approximated analytic MAD solution).
For computational convenience we take a patch-wise approach considering distinct regions subtending 0.65 deg.
This region-oriented approach certainly generates some artifacts in the block boundaries. However, the moderate visual impact of these edge effects
suggests that for regions of this size (and above) it is fair to assume the block-wise independence of distortion
See additional comments on this computationally convenient assumption in Supplementary Material \ref{summation_in_mad}.
%
% The Supplementary Material \ref{summation_in_mad} shows that the patch-wise approach is good enough for this image size.
% On the positive side, note that the patch-wise approach dramatically
% reduces the memory requirements in the computations.

\begin{figure}[!t]
	\centering
    \vspace{-0.6cm}
    \small
    \setlength{\tabcolsep}{2pt}
    \begin{tabular}{c}
    %\hspace{-1.7cm} \includegraphics[width=5cm,height=5cm]{Figs/figures/mad_concept.JPG} &
    %                \includegraphics[width=6cm,height=5cm]{Figs/figures/mad_algorithm.JPG} \hspace{0.0cm} &
    %\hspace{0.0cm}  \includegraphics[width=2.8cm,height=4.5cm]{Figs/figures/second_order_approx.JPG} \\
    %\hspace{-5cm} \includegraphics[width=18cm,height=9cm]{Figs/MAD_discussion1.pdf} \\
    \hspace{-5cm} \includegraphics[width=18cm,height=9cm]{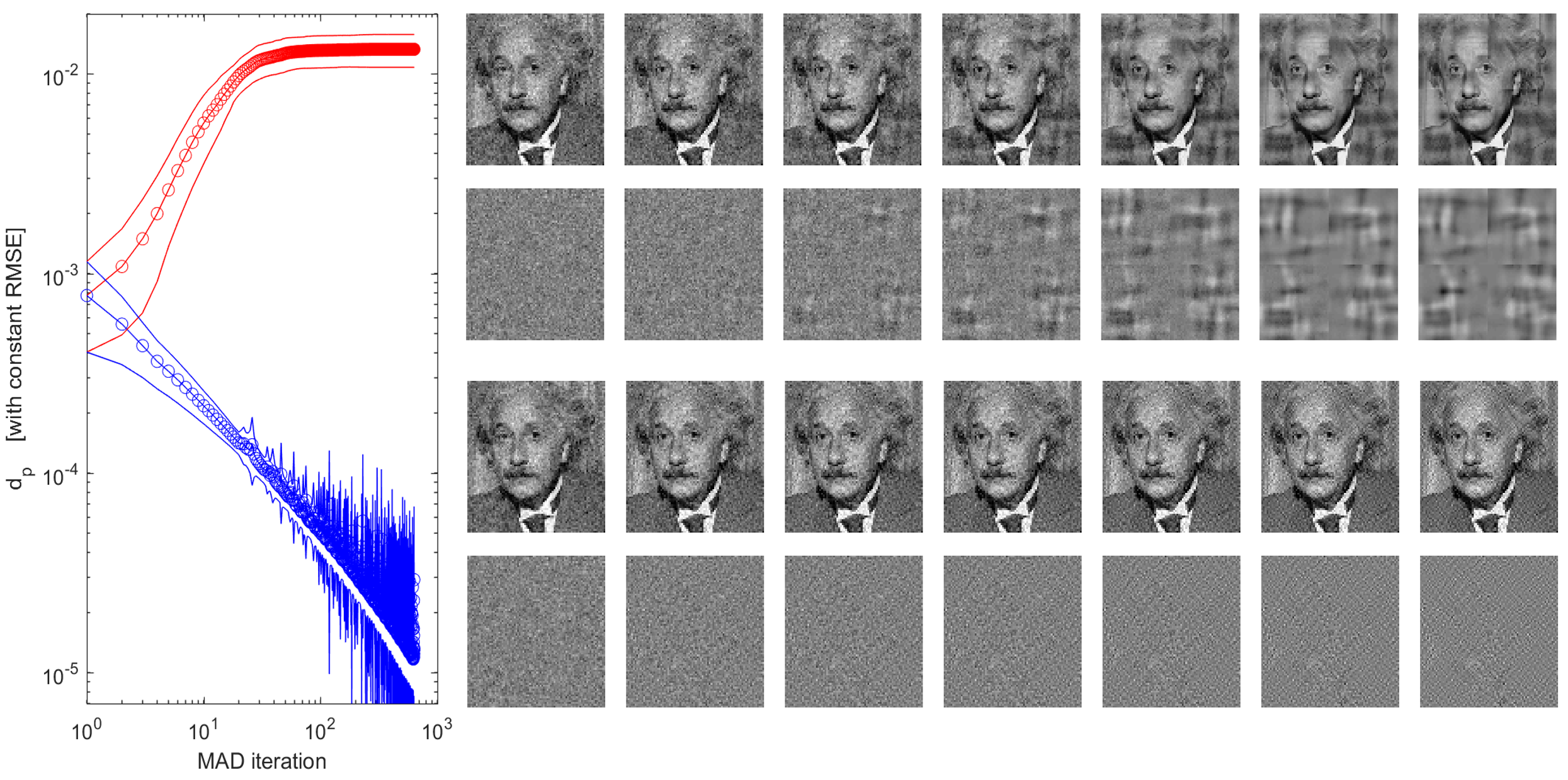} \\
    \end{tabular}
    \vspace{-0.5cm}
	\hspace{-0cm}\caption{\scriptsize{\textbf{Gradient-descent/ascent MAD.} \emph{Left panel:} evolution of the perceptual distance maximized (red curve) or minimized (blue curve) from an initial white-noise distorted image. \emph{Right panel (from left-to-right):} evolution of the intermediate MAD images on the sphere of constant RMSE. The \blue{two rows at the top} show the evolution of progressively-worse images while maximizing the perceptual distance. The \blue{two rows at the bottom} show the equivalent evolution of the progressively-better images while minimizing the perceptual distance.
\blue{In each case (top and bottom) we show the image+distortions and the isolated distortions.
In each case (top and bottom), the first image at the left is the initial randomly selected image in the sphere of constant RMSE.
This image gets progressively worse/better.}
%The smooth increase/decrease of the distance shows that the proposed expressions actually work (derivative of the distance and %Jacobian).
%The progressive subjective degradation/improvement of the images shows that the considered model is accurate.
}
}
\label{evolution}
    \vspace{-0.15cm}
\end{figure}

\begin{figure}[!t]
	\centering
    \vspace{-0.5cm}
    \small
    \setlength{\tabcolsep}{2pt}
    \begin{tabular}{c}
       %\hspace{-4cm} \includegraphics[width=4.6cm,height=4.6cm]{Figs/original.PNG} \\[4mm]
       \hspace{-0cm} \includegraphics[width=12.5cm,height=12.5cm]{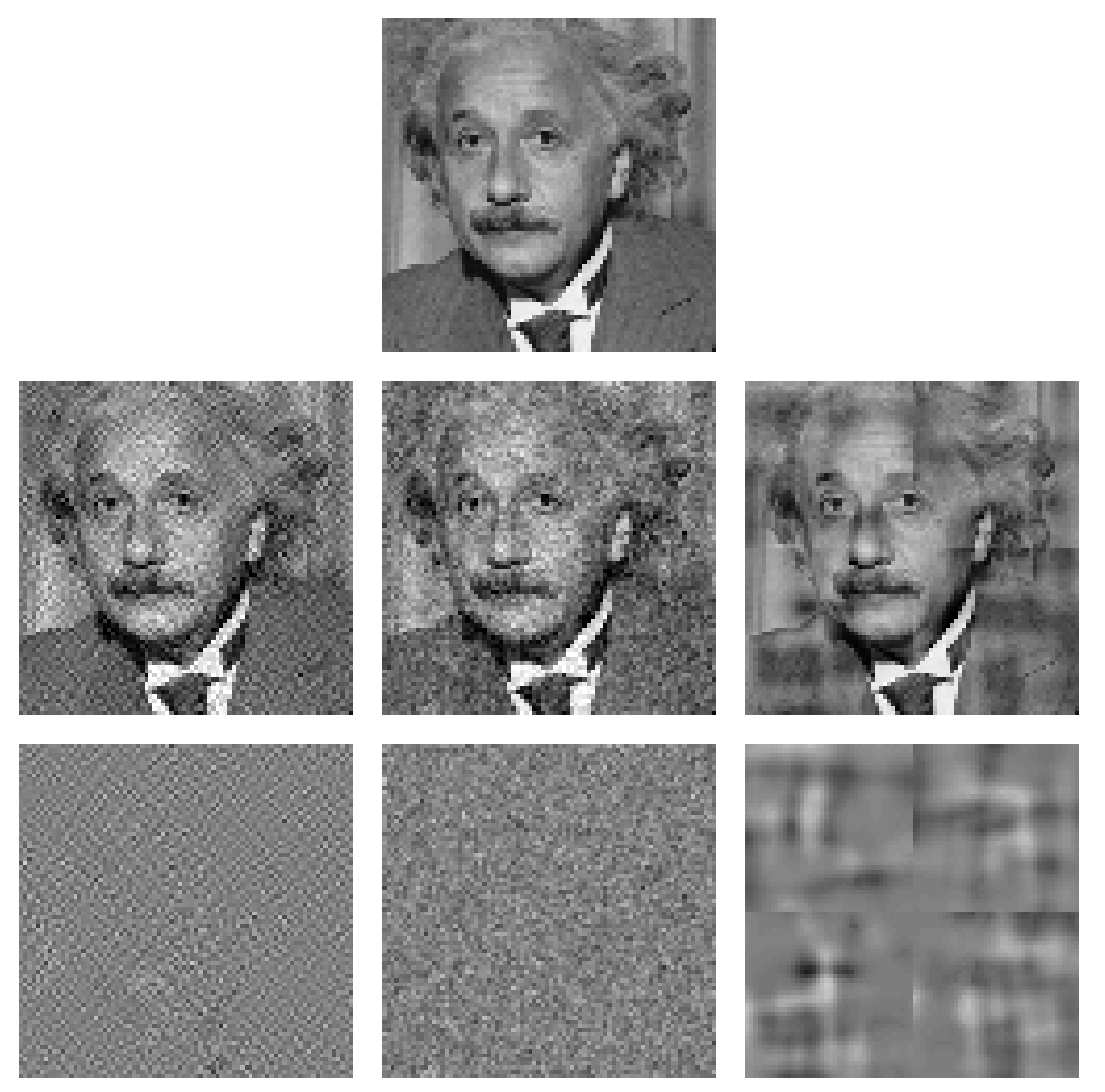} \\
    \end{tabular}
    \vspace{-0.0cm}
	\hspace{-0cm}\caption{\scriptsize{\textbf{General numeric solution.} \emph{Top:} original image \blue{$\vect{x}^0_A$}.
    \emph{Central row:}
    extreme images (MAD best and worst -left and right-) with the same RMSE than the white-noise corrupted image at the center.
    \blue{The central image is the initial random selection in the sphere of constant RMSE}.
    Extreme images were computed using the gradient descent/ascent described in Eq. \ref{MADiterat} (i.e. require Result I). The fact that these extreme images are visually better and worse than the central image indicates that the theory works.
    \emph{Bottom row:} isolated distortions of the same energy: \blue{$\Delta \vect{x}^0 = \vect{x}^0_A - \vect{x}^0_{\textrm{MAD}}$ versus the initial white noise}.
    }}
    \label{MADgral}
    \vspace{-0.15cm}
\end{figure}

\begin{figure}[!t]
	\centering
    \vspace{-0.5cm}
    \small
    \setlength{\tabcolsep}{2pt}
    \begin{tabular}{c}
       %\hspace{-4cm} \includegraphics[width=4.6cm,height=4.6cm]{Figs/original.PNG} \\[4mm]
       %\hspace{-0cm} \includegraphics[width=12.5cm,height=12.5cm]{Figs/mad_analytic_80c.PNG} \\
       \hspace{-0cm} \includegraphics[width=12.3cm,height=12.3cm]{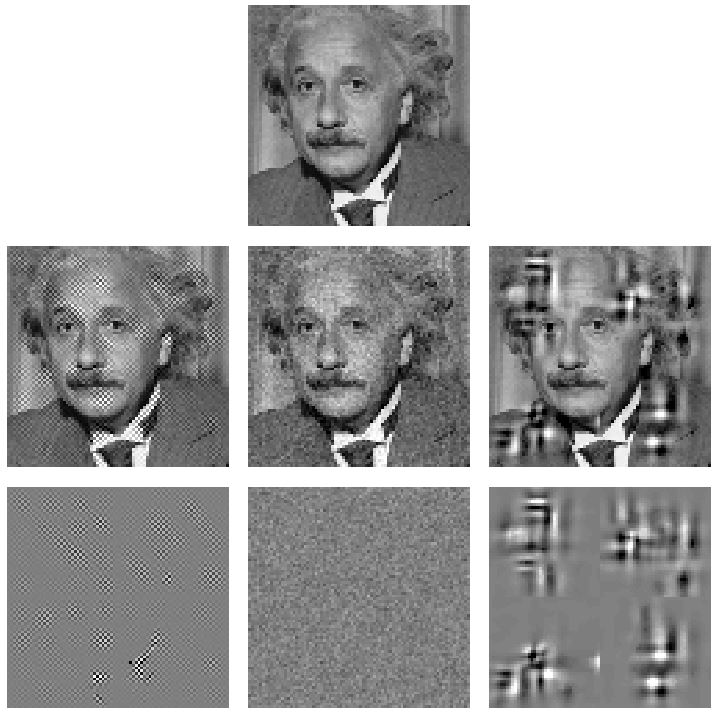} \\
    \end{tabular}
    \vspace{-0.0cm}
	\hspace{-0cm}\caption{\scriptsize{\textbf{Analytic, but approximated, solution.} \emph{Top:} original image \blue{$\vect{x}^0_A$}. \emph{Central row:}
    extreme images (MAD best and worst -left and right-) with the same RMSE than the white-noise corrupted image at the center.
    Extreme images are built from the eigenvectors of the metric matrix, $M(\vect{x}^0_A) = \nabla_x S(\vect{x}^0_A)^\top \cdot \nabla_x S(\vect{x}^0_A)$ (hence requiring Result I). \blue{In this case no search is carried out: the image corrupted with white noise in the center of 2nd row is there only for comparison purposes}. The fact that the extreme images are visually better and worse than the central image indicates that the theory works. \emph{Bottom row:} isolated distortions of the same energy: \blue{$\Delta \vect{x}^0 = \vect{x}^0_A - \vect{x}^0_{\textrm{MAD}}$ versus white noise}.
    }}
    \label{MADanal}
    \vspace{-0.15cm}
\end{figure}

Fig. \ref{evolution} shows the evolution of the general MAD distances and the solutions from the initial guess on the sphere (image corrupted with white-noise). Monotonic increase and decrease in the red and blue distance curves and progressive degradation or improvement in the images indicate both (a) the correctness of Result I, and (b) the accuracy of the parameters of the model used in this illustration.
Figures \ref{MADgral} and \ref{MADanal} show the results of the general MAD search and its analytic approximation respectively.

The main trend is this: the numerical procedure leads to noises of similar visual
nature than the analytic procedure. This means that the iterative search is certainly attracted
to the subspaces with low and high eigenvalues of the 2nd order metric.
\blue{More specifically, in both cases (a) the algorithms tend to allocate high-contrast low-frequency artifacts in low-contrast regions (jacket and background) to increase the visibility of the noise,
and (b) the algorithms tend to allocate high-frequency noise in high-contrast regions (e.g. the tie) to minimize its visibility.
These distortions are completely consistent with the trends identified above in the analytical \emph{sensitivity} and \emph{efficiency} of the system, Eqs. \ref{sensit}-\ref{MI_reduct} and Figs.~4-5:
focus on the low-contrast region and the role played by the CSF-like gain.}

There are differences between the general and the analytic solutions. In this example the visual difference between
the pairs of the analytic (approximated) solution seems bigger than the visual difference between the general (numeric) solution.
In principle, the numeric solution follows more closely the actual geometry induced by perception
(amorphous discrimination region versus approximated ellipsoid). However, the finite length of the gradient descent search
and the eventual trapping in local minima may prevent the practical use of the general technique
(not to speak about the substantially higher computational cost of the search!).

Nevertheless, the qualitative similarities of the solutions (the nature of the distortions and its spatial location)
is more relevant than the small quantitative differences.
In the model considered throughout this Discussion (Fig.~1), the 2nd and 3rd layers (contrast, and energy masking)
were determined using analytic MAD as in \cite{Malo15}. The experimental determination consisted of deciding between
different distorted pairs corresponding to eigenvectors coming from models from different parameters.
The final values found are those referred in the associated code (see Supplementary Materials \ref{example_model} and \ref{toolbox}).
Interestingly, MAD images in this paper (Figs. \ref{evolution}-\ref{MADanal}, and Fig. \ref{demo3} in the demo of the Toolbox)
were computed using two extra layers in the model (1st and 4th, accounting for brightness and wavelet masking respectively).
The important point is that, either numeric or analytic MAD images, they give rise to distinct pairs by putting wavelet-like
localized distortions of the right frequency in the regions with the right contrast or luminance, \blue{as expected
from the \emph{sensitivity} and \emph{efficiency} discussed above}.

Recently proposed visualization techniques to assess the biological plausibility of deep-network architectures \cite{Laparra17}
reduce to our analytic-MAD result originally proposed in \cite{Malo15}.
The relevance of Result~I is that it makes explicit the Jacobian expressions which are hidden in automatic differentiation
techniques used in \cite{Laparra17}. \blue{With the expressions proposed here one may anticipate what kind of patterns and where they should
be located to lead to highly (or hardly) visible distortions.}
% Incidentally, visual judgement of our distorted images and those presented in \cite{Laparra17}
% suggests that the perceptual quality of our model is better since worse-case images are more annoying in our case than in
% the cases presented there.

\blue{MAD may have two different problems (a) problems in identifying the best/worst distortions for a
given image (due to local minima in the distance or noise in the eigenvectors), and (b) substantial change of the directions
of the distortions in the image space.
In our experience we found that, given a reference image, the directions are fairly independent of the initial guess and consistent with the eigenvectors. Therefore, the second problem is more severe than the first because if the dependence is big, multiple experiments would be needed to decide between models \cite{Malo16}.}
\blue{Note that even in this undesirable situation the analytical Jacobian is also useful to assess how poor MAD can be.
This assessment could be done by measuring the variability of the directions defined by the Jacobian.}

%%%%%%%%%%%%%%%%%%%%%%%%%%%%%%%%%%%%%%%%%%%%%%%%%%%%%%%%%%%%%%%%%%%%%%%%%  END OF 3.2 MAD  (use Jacobian wrt stimulus)
%%%%%%%%%%%%%%%%%%%%%%%%%%%%%%%%%%%%%%%%%%%%%%%%%%%%%%%%%%%%%%%%%%%%%%%%%%%%%%%%%%%%%%%%%%%%%%%%%%%%%%%%%%%%%%%%%%%%%%%%%%%%%%%%%%%%%%%%%%%%%%%%%%
%%%%%%%%%%%%%%%%%%%%%%%%%%%%%%%%%%%%%%%%%%%%%%%%%%%%%%%%%%%%%%%%%%%%%%%%%%%%%%%%%%%%%%%%%%%%%%%%%%%%%%%%%%%%%%%%%%%%%%%%%%%%%%%%%%%%%%%%%%%%%%%%%%
%%%%%%%%%%%%%%%%%%%%%%%%%%%%%%%%%%%%%%%%%%%%%%%%%%%%%%%%%%%%%%%%%%%%%%%%%%%%%%%%%%%%%%%%%%%%%%%%%%%%%%%%%%%%%%%%%%%%%%%%%%%%%%%%%%%%%%%%%%%%%%%%%%
%%%%%%%%%%%%%%%%%%%%%%%%%%%%%%%%%%%%%%%%%%%%%%%%%%%%%%%%%%%%%%%%%%%%%%%%%%%%%%%%%%%%%%%%%%%%%%%%%%%%%%%%%%%%%%%%%%%%%%%%%%%%%%%%%%%%%%%%%%%%%%%%%%

\subsection{Jacobian with regard to the parameters in model optimization}
\label{use_jacobian_param}

%The Jacobian of the response with regard to the parameters, %$\nabla_{\!\!\vect{\Theta}} S$, is necessary to set the %parameters of the model
%either from direct and indirect experiments.
%The use of the analytical Jacobian in the optimization is better than the brute force approach used previously (e.g. in \cite{Watson02,Laparra10a,Malo10}). BECAUSE??

The standard methodology to set the free parameters of a model is looking for the values that better reproduce experimental results (either direct physiological recordings or indirect psychophysical results).
Sometimes brute-force exhaustive search (as done in \cite{Winawer13,Watson02,Laparra10a,Bertalmio17}) is good enough given the low dimensionality of the parameter space.
However, when considering thousands of parameters (as may happen in the considered model), brute-force approaches are definitely unfeasible.
In this high-dimensional scenario the Jacobian with regard to the model parameters (i.e. Result~II) may be very convenient to look for the optimal solution, as for instance using gradient descent.

Interestingly, model fitting procedures based on alternative goals (as for instance optimal encoding/decoding performance, as in \cite{LaparraICLR17}) also depend on gradient descent and this Jacobian w.r.t. parameters.
Unfortunately, the use of this Jacobian in similar biological models for optimal encoding/decoding (in  \cite{LaparraICLR17}) or to reproduce psychophysical data (in \cite{Laparra17}), was hidden behind automatic differentiation. On the contrary, here we gave the explicit equations (Result~II) and show their practical performance and correctness in analyzing psychophysical data.

In this section we discuss how to use the presented Result II
(generic Eqs. \ref{global_jacobian2}-\ref{elementary_jacobian2} and specific equations for the Divisive Normalization, Eqs. \ref{jacobian2_global}-\ref{dNdc}), to obtain the model parameters from classical subjective image quality ratings.

\paragraph*{Reproducing direct input-output data.}
In a controlled input-output situation (as in \cite{Winawer13}), it is usual to have a set of experimental physiological responses, $\vect{x}^n_{\textrm{exp}}$, for a given set of known inputs, $\vect{x}^0_{\textrm{exp}}$, and the goal is finding the model that behaves like the recorded data.

%So, the optimal parameters ($\Theta$) will be those that minimize a certain distance between the actual (or experimental) responses, and the responses computed with a model, $S(\vect{x}^0_{\textrm{exp}},\Theta)$.

A popular cost function depending on the parameters is the quadratic norm of the deviation between the theoretical and the experimental responses $\varepsilon(\vect{\Theta}) = \left| \vect{x}^n_{\textrm{exp}} - S(\vect{x}^0_{\textrm{exp}},\Theta) \right|^2_2$.
Minimization of this deviation, requires the derivative of the cost with regard to the parameters,
\begin{equation}
      \frac{\partial \,\,\, \varepsilon(\vect{\Theta})}{\partial \vect{\Theta}} \,\,=\,\, 2 \left( \vect{x}^n_{\textrm{exp}} - S(\vect{x}^0_{\textrm{exp}},\vect{\Theta}) \right)^\top \cdot \nabla_{\!\!\Theta} S(\vect{x}^0_{\textrm{exp}},\vect{\Theta})
      \label{optim_dir}
\end{equation}
which of course depends on the Jacobian w.r.t the parameters (and hence on Result II).

\blue{The analytical inverse, Result~III, has an interesting consequence in terms of the determination of $\vect{\Theta}$ in a controlled
input-output situation.
In general estimation problems the solution is necessarily more accurate if one combines multiple constraints to restrict the range of possible outcomes.
In our particular case, an alternative constraint to the one considered above is the minimization of
the distance between the actual input and the theoretical input that would be obtained from the inverse applied to the actual output.
Assuming that the decoding transform is the inverse of the encoding transform, this implies minimizing $\varepsilon'(\vect{\Theta}) = \left| \vect{x}^0_{\textrm{exp}} - S^{-1}(\vect{x}^n_{\textrm{exp}},\Theta) \right|^2_2$.
This extra constraint would lead to a gradient similar to Eq. \ref{optim_dir}, but involving the inverse function. That is using Result~III, and the Jacobian of the inverse wrt the parameters.
Interestingly, it holds $\nabla_{\!\!\vect{\Theta}} S^{-1} = - \left( \nabla_{\!\!\vect{x}^0} S \right)^{-1} \cdot \nabla_{\!\!\vect{\Theta}} S$. Therefore, this additional constraint reduces to the combined use of Result~III, Result~I, and Result~II.}

\paragraph*{Reproducing indirect data.}
By indirect data we refer to certain behavior that is mediated by the responses of the underlying L+NL mechanisms, but it is different from the actual responses themselves.
This is the conventional situation in psychophysics.
An illustrative example is the subjective assessment of perceived differences in image quality databases.

In this particular image quality situation instead of having a set of physiological responses for a given input, we have a Mean Opinion Score (MOS) of a set of distorted images (which is the usual ground truth in the image quality literature \cite{HandbookImageQuality}), and we want to adjust our model to reproduce this opinion.

In this case, the goal function is the correlation between the experimental subjective distance and the perceptual distance computed using the model explained above.
More specifically, consider a set of $N$ corrupted images, $\vect{z}^0_{[c]} = \vect{x}^0_{[c]} + \Delta \vect{x}^0_{[c]}$, with $c = 1 \ldots N$.
For this set, we assume we know the $N$ mean opinion scores, $\vect{M} = (m^{[1]}, \ldots , m^{[N]})^\top$, and we can compute the $N$ perceptual distortions,
$\vect{D} = (\mathbb{d}^{[1]}_p, \ldots , \mathbb{d}^{[N]}_p)^\top$, using the model (e.g. using Eq. \ref{distance}).

Therefore, the optimal parameters will be those maximizing the alignment between the ground truth, $\vect{M}$ and the model predictions $\vect{D}$.
Using the Pearson correlation, $\varrho$, as alignment measure, we have,
\begin{equation}
      \Theta^\star \,\,\,\,=\,\,\,\, \max_{\Theta} \,\,\, \varrho(\Theta) \,\,\,\,=\,\,\,\, \max_{\Theta} \,\,\, \frac{\vect{M_s}^T \cdot \vect{D_s}(\Theta)}{|\vect{M_s}| |\vect{D_s}(\Theta)|}
\end{equation}
where subindex $s$ stands for subtraction of the mean of the vectors.
%As a result, the parameters can be obtained through gradient descent of $\varrho(\Theta)$, and the update of the parameters at iteration $t$ is,
%\begin{equation}
%      \Theta(t+1) \,\,=\,\, \Theta(t) - \lambda \frac{\partial \varrho(\Theta(t))}{\partial \Theta}
%      \label{optim_indir}
%\end{equation}

The maximization of the correlation, $\varrho$, requires its derivative with regard to the parameters of the model.
Interestingly (see the Supplementary Material \ref{J_correl}), it turns out that the derivative of this goal function also depends on $\nabla_{\!\!\Theta} S$ (i.e. it depends on Result II):
\begin{equation}
\frac{\partial \varrho}{\partial \Theta} = \left(\frac{\vect{M_s}^T}{|\vect{M_s}|\cdot|\vect{D_s}|}-\frac{\vect{M_s}^T\vect{D_s}}{|\vect{M_s}|\cdot|\vect{D_s}|^3}\cdot \vect{D_s}^T \right)\cdot \left(I - \frac{1}{N} \cdot \mathds{1} \right)
 \cdot
 \left[\begin{array}{c}
\frac{ \Delta {\vect{x}^n_{[1]}}^\top}{\left|\Delta \vect{x}^n_{[1]} \right|} \cdot \left[ \nabla_{\!\!\Theta} S(\vect{z}^0_{[1]}) - \nabla_{\!\!\Theta} S(\vect{x}^0_{[1]}) \right] \\
{}\\
\frac{ \Delta {\vect{x}^n_{[2]}}^\top}{\left|\Delta \vect{x}^n_{[2]} \right|} \cdot \left[ \nabla_{\!\!\Theta} S(\vect{z}^0_{[2]}) - \nabla_{\!\!\Theta} S(\vect{x}^0_{[2]}) \right] \\
    \vdots \\
\frac{ \Delta {\vect{x}^n_{[N]}}^\top}{\left|\Delta \vect{x}^n_{[N]} \right|} \cdot \left[ \nabla_{\!\!\Theta} S(\vect{z}^0_{[N]}) - \nabla_{\!\!\Theta} S(\vect{x}^0_{[N]}) \right] \\
  \end{array}  \right]
  \label{deriv_correl}
\end{equation}

\paragraph*{Cascaded L+NL model in image quality.}
The reproduction of image quality ratings is a good way to check the performance of vision models in a variety of observation conditions (variety of natural backgrounds and variety of suprathreshold tests).

The image quality results discussed in this section illustrate three interesting issues:
\begin{itemize}
\item They are a complementary evidence of the quality of the modular model (different from MAD results of the previous section).
\item They point out the benefits of the modular nature of the model since inclusion of extra layers leads to consistent improvements of the performance (either by using canonical L+NL layers or by using alternative formulations such as the two-gamma tone mapping model or the Wilson-Cowan nonlinearity after a linear wavelet stage).
\item They reveal the relevance of Result II in finding the model parameters in high-dimensional scenarios and the correctness of the presented expressions.
\end{itemize}

\begin{figure}[!t]
    \vspace{-2cm}
	\centering
    \small
    \setlength{\tabcolsep}{2pt}
    %\begin{tabular}{cc}
    %  \hspace{-5cm} \includegraphics[width=6cm,height=5cm]{RMSE.PNG} \hspace{0.05cm} & \hspace{0.05cm}
    %                \includegraphics[width=6cm,height=5cm]{SSIM.PNG} \\[-1cm]
    %\end{tabular}
    %\begin{tabular}{ccc}
    %  \hspace{-5cm} \includegraphics[width=6cm,height=5cm]{DN.PNG} \hspace{0.05cm} & \hspace{0.05cm}
    %                \includegraphics[width=6cm,height=5cm]{WC.PNG} \hspace{0.05cm} & \hspace{0.05cm}  \includegraphics[width=6cm,height=5cm]{DNopt.PNG} \\
    %\end{tabular}
     \begin{tabular}{c}
     \hspace{-4cm} \includegraphics[width=18cm]{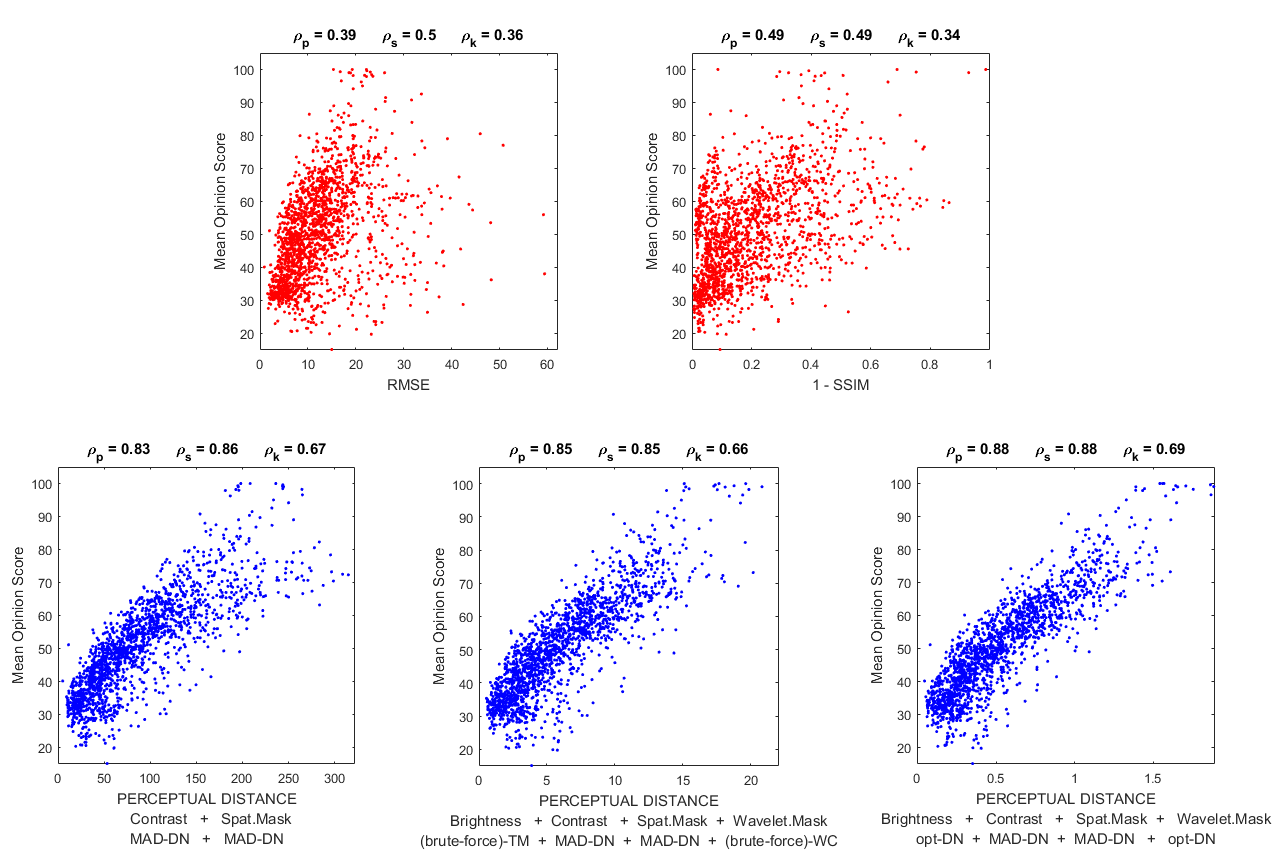} \\
     \end{tabular}
    \vspace{-0.0cm}
%	\hspace{-8cm}\caption{\scriptsize{\textbf{Prediction of subjective distortion.}
%Correlation between predicted subjective distortion (in abscisas) and actual opinion (in ordinates)
%is a good performance measure of models in a variety of observation conditions.
%These results show the Pearson, Spearman and Kendall correlations of different models in reproducing the MOS values of the TID database \cite{Ponomarenko08}.
%Results show:
%(1) the quality of the baseline 2-stage model determined using MAD that accounts for contrast, frequency filtering and masking in the spatial domain %\cite{Malo15};
%(2) improvements in the performance due to the consideration of extra L+NL modules that account for brightness and masking in the wavelet domain, e.g. the %two-gamma tone-mapping (TM) transform, and the Wilson-Cowan (WC) interaction between linear wavelet responses \cite{Bertalmio17}; and
%(3) brute-force approaches can be improved through the appropriate search based on the Jacobian of the response w.r.t the parameters.
%Results in red (Euclidean distance, RMSE; and a widely used model of perceptual distortion, SSIM) have been included for convenient reference.}}
	\hspace{-8cm}\caption{\scriptsize{\textbf{Prediction of subjective distortion.}
Correlation between predicted subjective distortion (in abscisas) and actual opinion (in ordinates)
is a good performance measure of models in a variety of observation conditions.
These results show the Pearson, Spearman and Kendall correlations of different models in reproducing the subjective opinion in the TID database \cite{Ponomarenko08}.
\blue{Scatter plots in red correspond to two simpler models for convenient reference: the Euclidean distance, RMSE; and a widely used model of perceptual distortion, SSIM \cite{wang04}.
Scatter plots in blue correspond to progressive improvements of the baseline 2-layer (L+NL + L+NL) model at the left.
Labels in the abscisas indicate (a) the perceptual phenomena taken into account by the layers of the models, and (b) the structure of the layers and how they were estimated. MAD-DN stands for Divisive Normalization layer with parameters estimated using MAD experiments. Brute-force TM and Brute-force WC stand
for Tone-Mapping and Wilson-Cowan layers estimated through the maximization of the Pearson correlation using exhaustive search in a grid.
Finally, opt-DN stands for Divisive Normalization layers estimated through gradient optimization of Pearson correlation (using Result II).
}
%Results show:
%(1) the quality of the baseline 2-stage model determined using MAD that accounts for contrast, frequency filtering and masking in the spatial domain %\cite{Malo15};
%(2) improvements in the performance due to the consideration of extra L+NL modules that account for brightness and masking in the wavelet domain, e.g. the %two-gamma tone-mapping (TM) transform, and the Wilson-Cowan (WC) interaction between linear wavelet responses \cite{Bertalmio17}; and
%(3) brute-force approaches can be improved through the appropriate search based on the Jacobian of the response w.r.t the parameters.
}}
\label{scater}
    \vspace{-0.15cm}
\end{figure}

Figure \ref{scater} shows the performance of the kind of cascaded L+NL models we are considering here in the reproduction of mean opinion scores. We include two reference models (in red) for convenient comparison. The first reference is just the Euclidean distance between inputs (RMSE). The second reference is the most popular perceptual quality predictor in the image processing community (the Structural SIMilarity index, SSIM \cite{wang04}). Our baseline model is the 2-stage L+NL model whose parameters were tuned using MAximum Differentiation \cite{Malo15}.
Results reported here are better than those reported in \cite{Malo15} (using the same parameters) probably because of two reasons: (1) here we are using bigger patches and hence the patch-independence assumption holds better, and (2) we are now applying luminance calibration to digital values of the TID database.
This baseline model corresponds to the 2nd and 3rd canonical stages of the global model we are considering throughout the discussion section (see model details in the Supplementary Material \ref{example_model}).

Substantial jumps in correlation from the RMSE result indicate the well-known
limitation of naive Euclidean distance \cite{Wang09} but also the potential of MAD to set the parameters
of this 2-stage model \cite{Malo15}.
Note that, assuming the Contrast Sensitivity Function (CSF) of the Standard Spatial Observer \cite{Watson02},
this 2-stage model only has 5 free parameters (something affordable using MAD): the widths of the kernels and the semisaturation for contrast computation, and the width of the kernel, the semisaturation, and the excitation exponent in the masking nonlinearity.

Modularity and interpretability of the model is nice because it allows to propose straightforward
improvements of the baseline model: just introduce extra layers according to the program suggested in \cite{Carandini12}.
% We included extra layers before and after the two canonical layers in the baseline model:
Accordingly, we included a brightness perception layer \emph{before} the contrast computation, and a wavelet interaction model \emph{after} the CSF+spatial masking layer.
To stress the generality of the proposed modular approach our first brightness model was the two-gamma tone mapping operator cited in Section \ref{general_cons}, and our first wavelet masking scheme was
the Wilson-Cowan model cited in Section \ref{general_cons} applied to each subband of a steerable pyramid.
Following \cite{Bertalmio17} the 5 extra parameters of these extra layers were obtained through brute-force search \blue{using $50\%$ of the database. Exhaustive search in 5 dimensions is hard but still feasible}. The resulting model not only improves the \blue{Pearson} correlation in image quality (as expected by construction), but it also has sensible behavior in reproducing contrast masking \cite{Bertalmio17}.

Finally, we explicitly explored the maximization of the correlation using different versions of
the brightness and the wavelet+masking stages. In this final case we used canonical
Divisive Normalization layers.
Note that the joint optimization of the 1st and 4th layers is an interesting way to check
Results I and II at once.
First, relation to the Jacobian w.r.t the parameters in Result II is obvious from Eq. \ref{deriv_correl}.
But, more interestingly, note that the chain rule,  Eq. \ref{chain_rule3}, implies that distortions due to variations in the parameters propagate throughout the network.
Then, the Jacobian w.r.t the stimulus (i.e. Result I of all the layers following the one under optimization) is also required in the joint optimization of 1st and 4th layers.

As a result of the chain rule, this optimization has a great experimental value but also a great computational cost. That is why we split the optimization of this illustration in two separate phases.

\begin{figure}[!b]
	\centering
    \vspace{-0.5cm}
    \small
    \setlength{\tabcolsep}{2pt}
    \begin{tabular}{c}
      \hspace{-1cm}  \includegraphics[width=14cm,height=8cm]{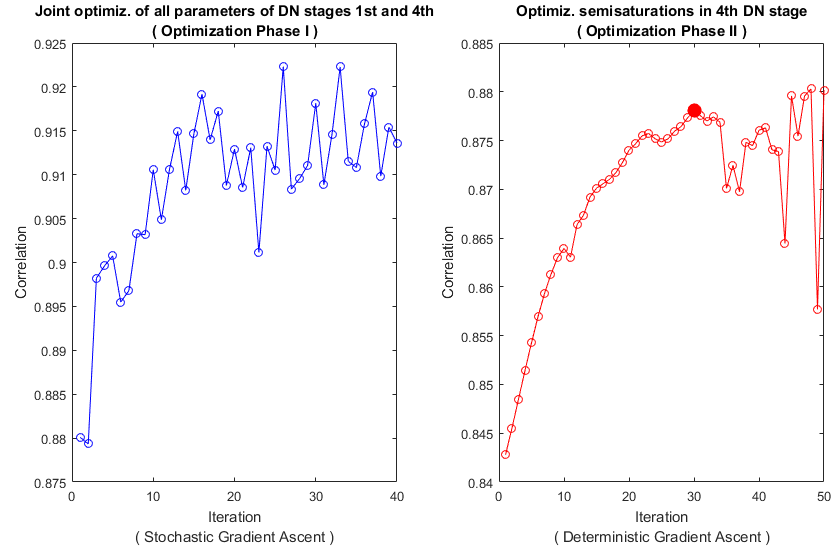}
    \end{tabular}
    \vspace{-0.0cm}
	\hspace{-0cm}\caption{\scriptsize{\textbf{Maximization of correlation with subjective opinion (\emph{phase one} and \emph{phase two}).} \emph{Phase one} (left) involved the joint optimization of the following parameters:
$\beta, \gamma$ of the 1st stage and  $b, \sigma, \gamma, c$ of the 4th stage.
In this case  $\beta_1, \gamma_1, \gamma_4 \in \mathbb{R}^1$ and $\beta_4, \sigma_4, c_4 \in \mathbb{R}^{14} $ one parameter per orientation (with 4 orientation) and scale (with 4 scales) plus both residuals. Note that even though the relatively small number of parameters, the dimensionality is still huge to allow a brute force approach.
Given the computational cost of \emph{phase one} we used stochastic gradient descent on a extremely small training set. Results shown in the curve correspond to the randomly varying training set.
In \emph{phase one} the correlation for the whole database only arrives up to 0.84 because more iteration would be required.
\emph{Phase two} (right) only involved the optimization of $\vect{b}^4$. This is fast enough to use deterministic gradient ascent training with half the database. However, note that if no structure is imposed in $\vect{b}^4$, it has thousands of elements thus brute-force is not possible.
In this case, correlation results shown correspond to the whole database (indicating proper generalization). The parameters leading to the maximum correlation in the test set (peak of the red curve) are those used for the scatter plot of Fig.\ref{scater}.
}}\label{Adam_training}
    \vspace{-0.15cm}
\end{figure}

In the optimization \emph{phase one} we addressed the (highly illustrative but extremely demanding) joint optimization of the 1st and 4th stages. In this \emph{phase one} we used a reduced training set to avoid the computational burden, and we used structured versions of the parameters as discussed after Eq. \ref{dNdc}
to address a relatively low-dimensional problem (but still not affordable through brute force).
Then, in \emph{phase two}, we took the results of \emph{phase one} (which are only a first approximation to the right solution because of the small size of the training set)
and focused on the optimization of a single parameter which is fast to compute but
extremely high-dimensional to point out even more clearly the necessity of using Result II.
Positive results of these first and second learning phases are illustrated in Fig. \ref{Adam_training}.

Computational cost of joint optimization (\emph{phase one}) implies that not that many training points can be used in the gradient ascent.
This implies that in order to generalize the training set has to be stochastically updated.
We show the result of such stochastic maximization of the correlation using only 48 samples of the TID database at a time.
The gradient ascent search in \emph{phase one} (blue curve at the left) certainly increases the correlation for the considered small training set. Note that oscillations come from the random modification of the training set in each iteration.
\blue{The consistent increase of correlation in the stochastic \emph{phase one} points out the correctness of Results I and II. However, in the explored iterations in \emph{phase one}, the correlation in the whole dataset only increased up to 0.84}. This generalization problem means that the training set is too small to avoid overfitting or equivalently,
that extra iterations would be necessary so that this small set could visit
the whole variability in the dataset.

Once Result II (and also Result I) have been checked in the most demanding situation (joint optimization of two layers in \emph{phase one}), we switch to \emph{phase two}.
In the \emph{phase two} only the semisaturation of the 4th stage (only the vector $\vect{b}^4$) was optimized. Note that in this restricted case we do not need the Result I of the intermediate stages anymore. In fact, the computation of the Jacobian w.r.t. this single parameter is so fast that we allowed the search in the full dimensionality of this vector and using deterministic gradient ascent (using a substantial part of the available database). In particular in this \emph{phase two} we trained with 800 randomly chosen points of the database \blue{(about $50\%$)}, as opposed to the reduced number of random regions taken from 48 points used in \emph{phase one}.
In this high-dimensional case (note the huge dimension of $\vect{b}^4$) brute-force is certainly not possible, and hence the gradient ascent (i.e. Result II) is the most sensible way.
In this \emph{phase two}, the correlation on the whole database (in red) consistently increases at the beginning of the search indicating both the correctness of Result II (for this parameter) and the representativeness of the training set.
Finally, as expected in any learning problem using a limited training set, overfitting occurs and the correlation for the test set starts to oscillate. The values found at the (trustable) highlighted point are those used in the final scatter plot of Fig. \ref{scater}.

\blue{Only the \emph{brute-force} and the result of the \emph{phase two} optimization are compared in Fig. \ref{scater}.
These methods used the same amount of training samples but note that the complexity of the models is much bigger in the wavelet-DN case (where brute-force is definitely not possible). Figs \ref{scater} and \ref{Adam_training} (right) report the results for the whole database: of course, as in any learning problem, correlation values in the separated train and test sets are slightly higher and lower respectively.
}

%%%%%%%%%%%%%%%%%%%%%%%%%%%%%%%%%%%%%%%%%%%%%%%%%%%%%%%%%%%%%%%%%%%%%%%%%  END OF 3.3 OPTIMIZATION  (use Jacobian wrt parameters)
%%%%%%%%%%%%%%%%%%%%%%%%%%%%%%%%%%%%%%%%%%%%%%%%%%%%%%%%%%%%%%%%%%%%%%%%%%%%%%%%%%%%%%%%%%%%%%%%%%%%%%%%%%%%%%%%%%%%%%%%%%%%%%%%%%%%%%%%%%%%%%%%%%
%%%%%%%%%%%%%%%%%%%%%%%%%%%%%%%%%%%%%%%%%%%%%%%%%%%%%%%%%%%%%%%%%%%%%%%%%%%%%%%%%%%%%%%%%%%%%%%%%%%%%%%%%%%%%%%%%%%%%%%%%%%%%%%%%%%%%%%%%%%%%%%%%%
%%%%%%%%%%%%%%%%%%%%%%%%%%%%%%%%%%%%%%%%%%%%%%%%%%%%%%%%%%%%%%%%%%%%%%%%%%%%%%%%%%%%%%%%%%%%%%%%%%%%%%%%%%%%%%%%%%%%%%%%%%%%%%%%%%%%%%%%%%%%%%%%%%
%%%%%%%%%%%%%%%%%%%%%%%%%%%%%%%%%%%%%%%%%%%%%%%%%%%%%%%%%%%%%%%%%%%%%%%%%%%%%%%%%%%%%%%%%%%%%%%%%%%%%%%%%%%%%%%%%%%%%%%%%%%%%%%%%%%%%%%%%%%%%%%%%%

\subsection{Analytic inverse in visual brain decoding}
\label{use_inverse}

Visual brain decoding refers to the reconstruction of the input stimulus from physiological recordings of the neural activity (for instance fMRI) \cite{Kay08}. Conventional decoding techniques are based on \emph{learning-through-examples} the response-stimulus relation. First approaches to decoding used plain linear regression \cite{Stanley99}, but now the current practice is using non-linear regression as for instance based on kernel methods (as in \cite{Miyawaki08,Marre15}).
However, given the fact that models of the BOLD signal also have this
cascaded L+NL structure \cite{Winawer13}, the analytic inverse of the transform proposed here (Result III) may have obvious application in decoding the input from the recorded output.

In order to illustrate the eventual benefits of the analytic inverse in the visual decoding problem,
in this discussion we consider a simulation where conventional blind machine-learning techniques (linear regression and \blue{nonlinear kernel-ridge regression as in \cite{Marre15}}) are compared to the analytic inversion.
Here we simulate the recorded neural signal by applying the forward model to noisy inputs and distorting the output. This controlled scenario allows us to generate as many corresponding input-output pairs as necessary to train the machine-learning techniques, as done in the experimental acquisition phase in the brain decoding literature.

\begin{figure}[!t]
	\centering
    \small
    \setlength{\tabcolsep}{2pt}
    \begin{tabular}{c}
       % \hspace{-4cm} \includegraphics[width=4.6cm,height=4.6cm]{original.PNG} \\[4mm]
       \hspace{-0cm} \includegraphics[width=13cm,height=10cm]{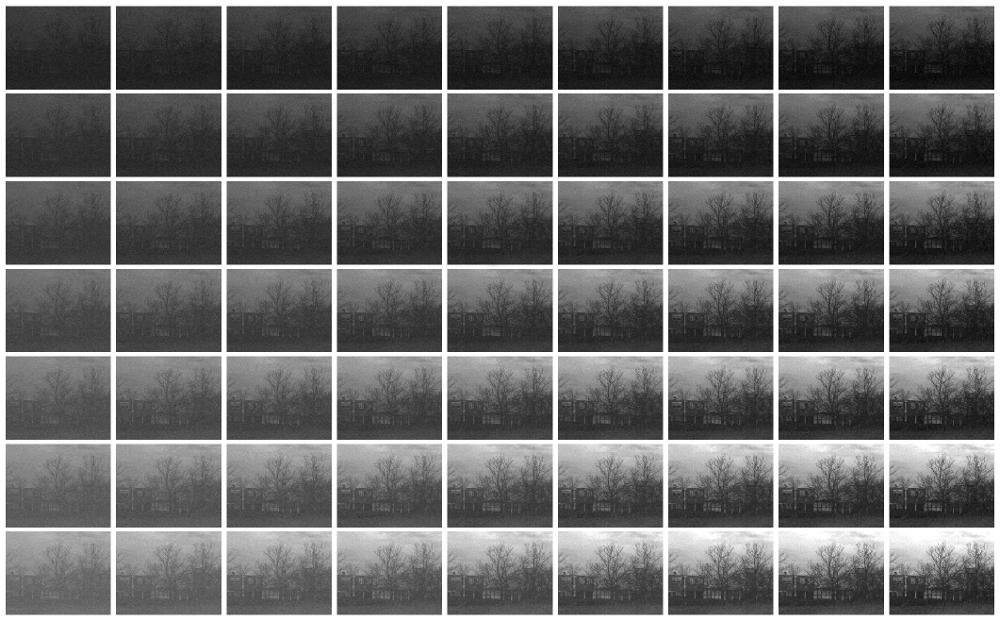} \\
    \end{tabular}
    \vspace{-0.0cm}
	\hspace{-0cm}\caption{\scriptsize{\textbf{Range of illumination conditions.} Average luminance increases from top to bottom and contrast increases from left to right. \blue{Luminance is in the range [0,160] $cd/m^2$. Average luminances are in the range [25, 80] $cd/m^2$, and average contrasts are in the range [0.1, 0.9]}. The learning-based decoders were trained with natural images from the central luminance/contrast condition, under two levels of distortion.}}
\label{range}
    \vspace{-0.15cm}
\end{figure}

In fact, regression techniques depend both on the nature of the input-output pairs and on the nature of the distortions. In our simulation we controlled both:
\begin{itemize}
\item We controlled the nature of the signals by augmenting a calibrated set of
      natural images (Van Hateren database \cite{VanHateren98}) using controlled
      modifications of the illumination conditions.
      Specifically, we linearly modified the images to have different average luminance and contrast.
We considered 7$\times$9 combinations of luminance and contrast covering the range available
in a conventional computer display (see Fig. \ref{range}).
\item Distortion in the signals comes from random variations in the input (e.g. photon-noise at the retina), random variations of the output (e.g. noise in the cortical response), and distortions due to the measurement (e.g. blurring and noise in the BOLD signal).
    There is a debate on the psychophysical relevance of the noise at the input versus the neural noise \cite{Pelli90,Ahumada87,Messe06} that we don't want to address here.
    Just for the sake of the illustration, we controlled these distortions by using uncorrelated noise at the input and blur+noise in the acquisition of the neural signal given by the model.
    The outputs of the model were blurred using a Gaussian kernel with width of 0.05 degrees
    (in visual angle units, in the spatial domain corresponding to each subband).
    We considered two distortion regimes: low-distortion and high distortion.
    The low-distortion regime involved Gaussian noise at the input with $\sigma = 3 cd/m^2$, and Poisson noise at the responses with Fano factor 0.02.
    The high-distortion regime involved the same sources of noise with input deviation $\sigma = 30 cd/m^2$, and internal Fano factor 0.05.

    % Note on the actual values of blur and noise.
    %
    % This experiment was done with the "post_modvis" model optimized brute force.
    % The blur and noise data in the spatial domain is from the files training_n_b_n_A training_n_b_n_B and training_n_b_n_C
    %
    % The noise in the wavelet domain was actually Gaussian, but given the average of the coefficients in X4 may be expressed using Fano factor. The approximate average from data_pre_loop_A is 1e-4. Therefore, since the variance of the noise is 0.25e-5 and 0.5e-5, the fano factor is in the reported range.

\end{itemize}

We trained the machine learning algorithms with images from the central luminance/contrast condition
and responses under the low and high distortion regimes.
We considered 5000 input-output examples in the training. We tested on an image not considered in the training set.
In the test we considered the different illumination conditions (the one used in the training and the other conditions considered in Fig. \ref{range}),
and we applied the two distortion regimes.

In each test example, in which illumination may or may not correspond to the training, we decoded the response with 5 decoders:
(1)~linear decoder trained for the considered distortion,
(2)~linear decoder trained with noise of different nature,
(3)~non-linear decoder trained for the considered distortion,
(4)~non-linear decoder trained with noise of different nature,
(5)~analytical decoder.

\begin{figure}[!t]
	\centering
    \small
    \setlength{\tabcolsep}{2pt}
    \begin{tabular}{cccc}
       \hspace{-6cm} \includegraphics[width=4.5cm,height=7cm]{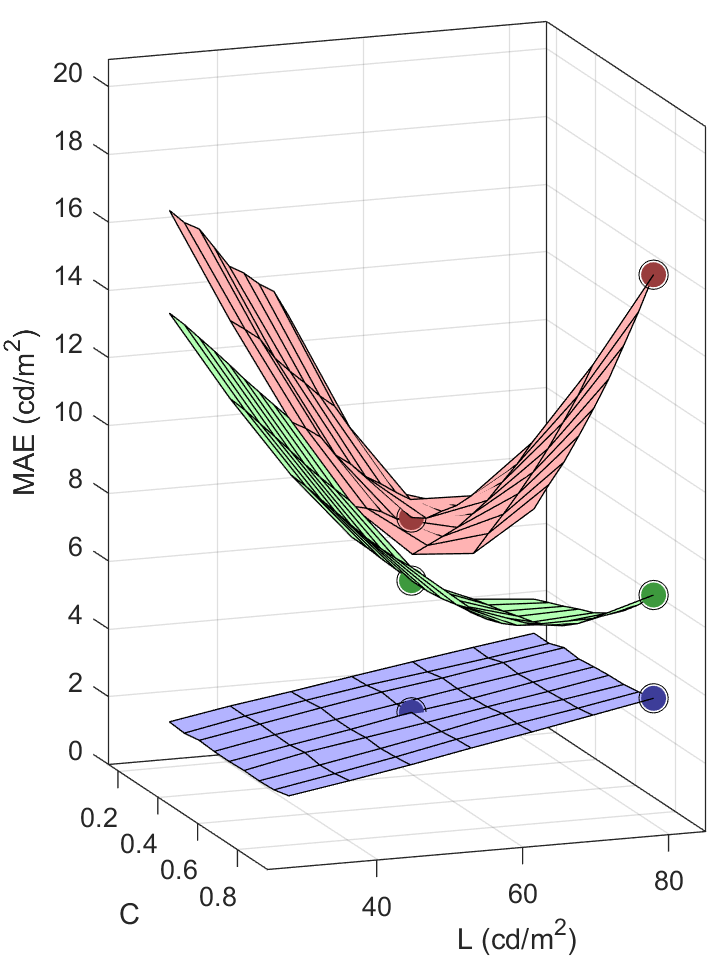} &
       \includegraphics[width=4.5cm,height=7cm]{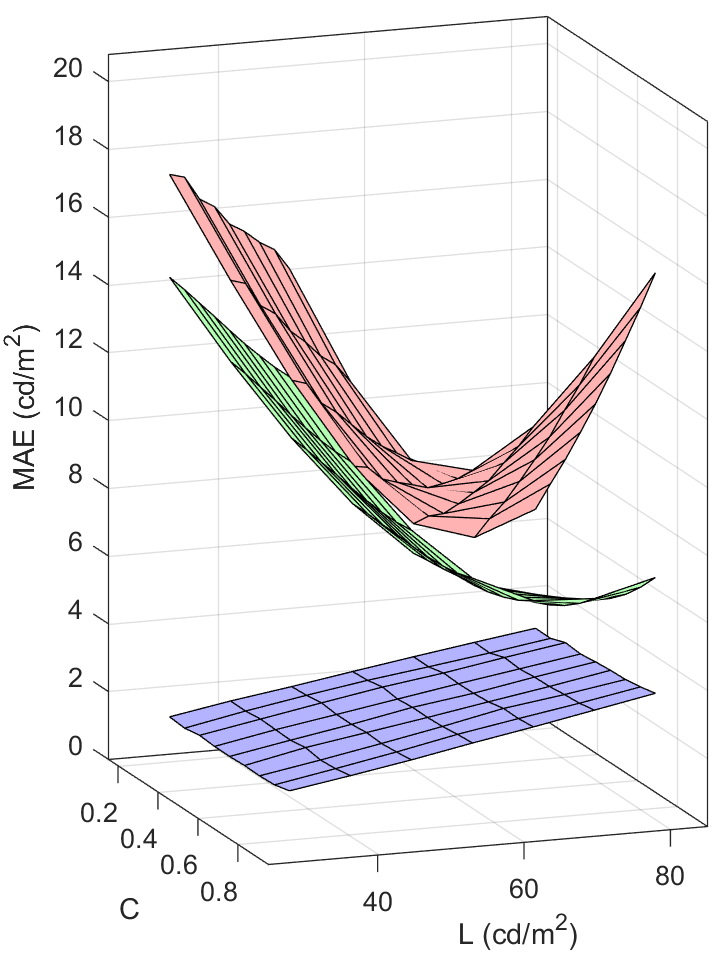} &
       \includegraphics[width=4.5cm,height=7cm]{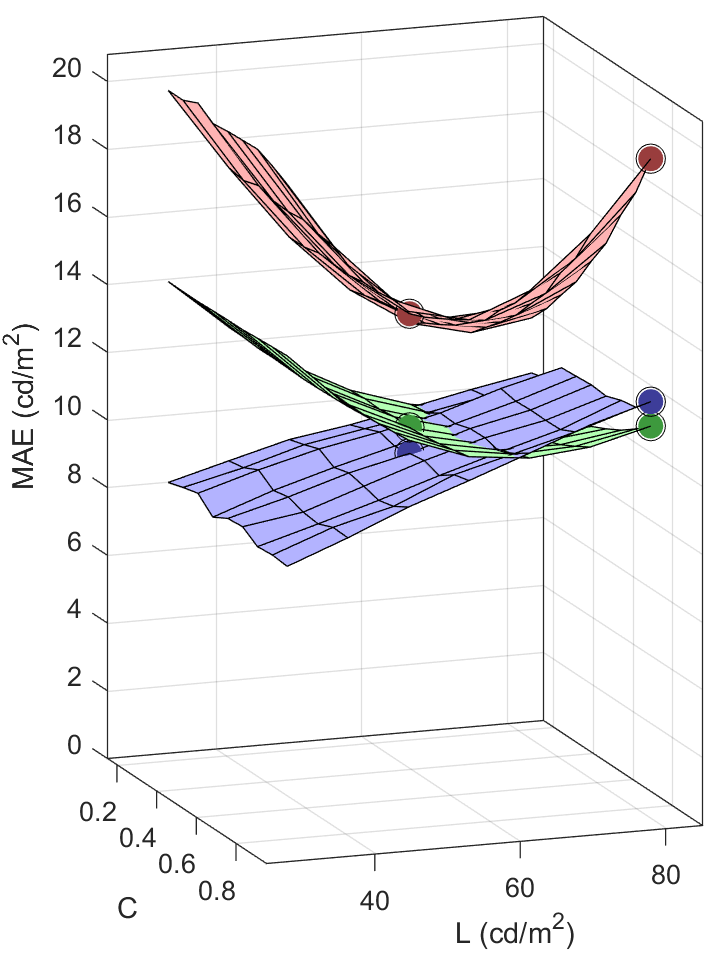} &
       \includegraphics[width=4.5cm,height=7cm]{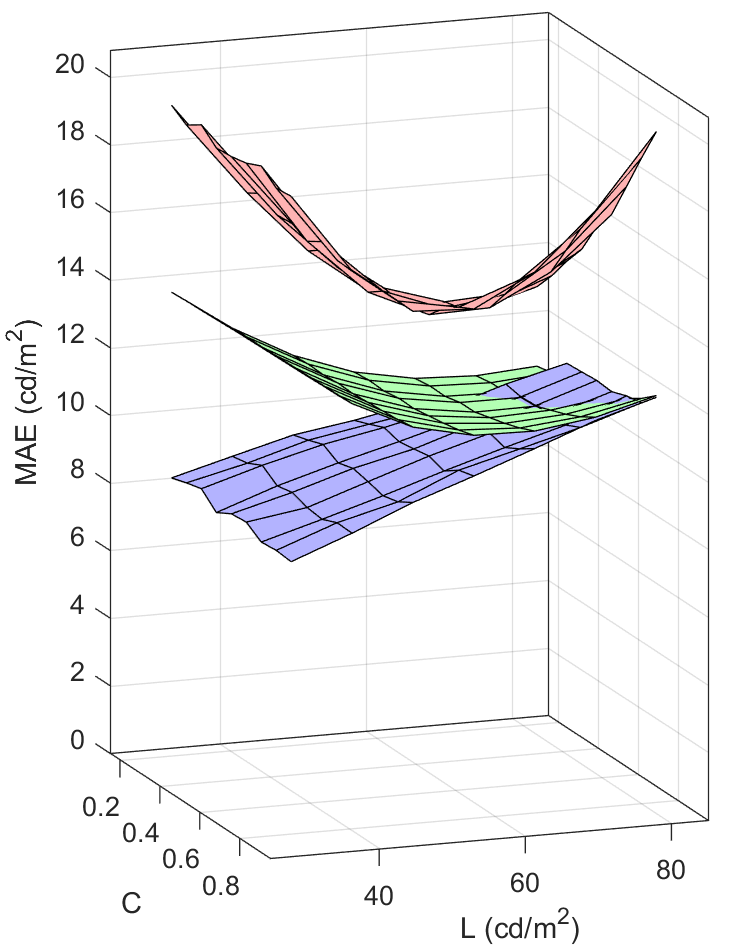}\\
    \end{tabular}
    \vspace{-0.0cm}
	\hspace{-0cm}\caption{\scriptsize{\textbf{Decoding error.} Left panels show the Mean Absolute Error (MAE) in the low-distortion regime and panels of the right show the MAE in the high-distortion regime. \blue{Image luminance is in the range [0,160] $cd/m^2$}. In each case surfaces red, green and blue represent the error of the linear regression, non-linear regression and analytic decoders. In each distortion regime, the panel at the left shows the results of the decoders trained for that regime. The panel of the right represents the case where the signal is reconstructed by decoders trained in a different distortion regime. The highlighted points correspond to the error of the decoded images shown \blue{in Figs. \ref{ReconsA} and \ref{ReconsC}}. Note that the points at the center correspond to the optimum for the learning-based decoders: the training illumination and distortion. Axis label, L means Luminance, and C means Contrast}}
    \label{MAEs}
    \vspace{-0.15cm}
\end{figure}

Distortion in the decoded signals is shown in Fig. \ref{MAEs}. Here we use the Mean Absolute Error in the input domain (in $cd/m^2$ units) as distortion measure, just because it has direct physical interpretation (subjective accuracy will be apparent in the visual examples below).
Results show that the error of the analytic decoder is lower and substantially more independent from the illumination conditions than the error of the machine learning models that depend on the training.
The error surfaces of the data-dependent decoders are curved because they are trained for the central condition in the range. Therefore they have generalization problems in other regions.
For bigger distortions all methods have lower performance (the analytic decoding is affected too), but note that in this case, appropriate training becomes more critical because using decoders trained in other distortion conditions increase the error (see how the green surface goes up in the plot of the right).

\begin{figure}[!t]
	\centering
    \small
    \setlength{\tabcolsep}{2pt}
    \begin{tabular}{c}
       \hspace{-6cm} \includegraphics[width=18cm,height=10.5cm]{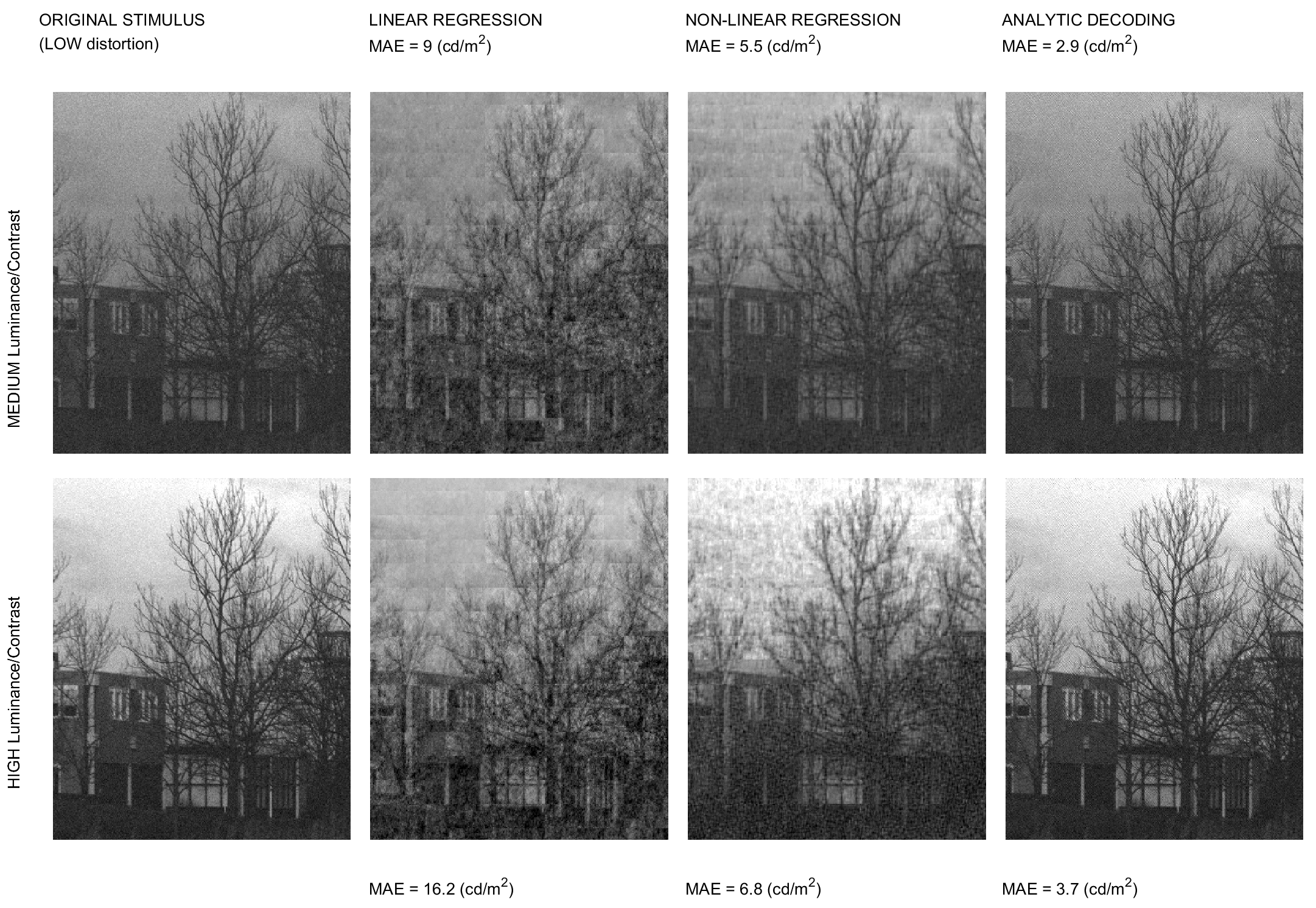}\\
    \end{tabular}
    \vspace{-0.0cm}
	\hspace{-0cm}\caption{\scriptsize{\textbf{Reconstructions in the low-distortion regime.} These six reconstructions correspond to the highlighted dots in the surfaces at the left plot of Fig. \ref{MAEs}}. The analytic decoding clearly overperforms the learning algorithms even in the case that the image has the illumination conditions used in the training (medium luminance/contrast) and the decoders are those trained for the considered distortion.}
    \label{ReconsA}
    \vspace{-0.15cm}
\end{figure}

\begin{figure}[!t]
	\centering
    \small
    \setlength{\tabcolsep}{2pt}
    \begin{tabular}{c}
       \hspace{-6cm} \includegraphics[width=18cm,height=10.5cm]{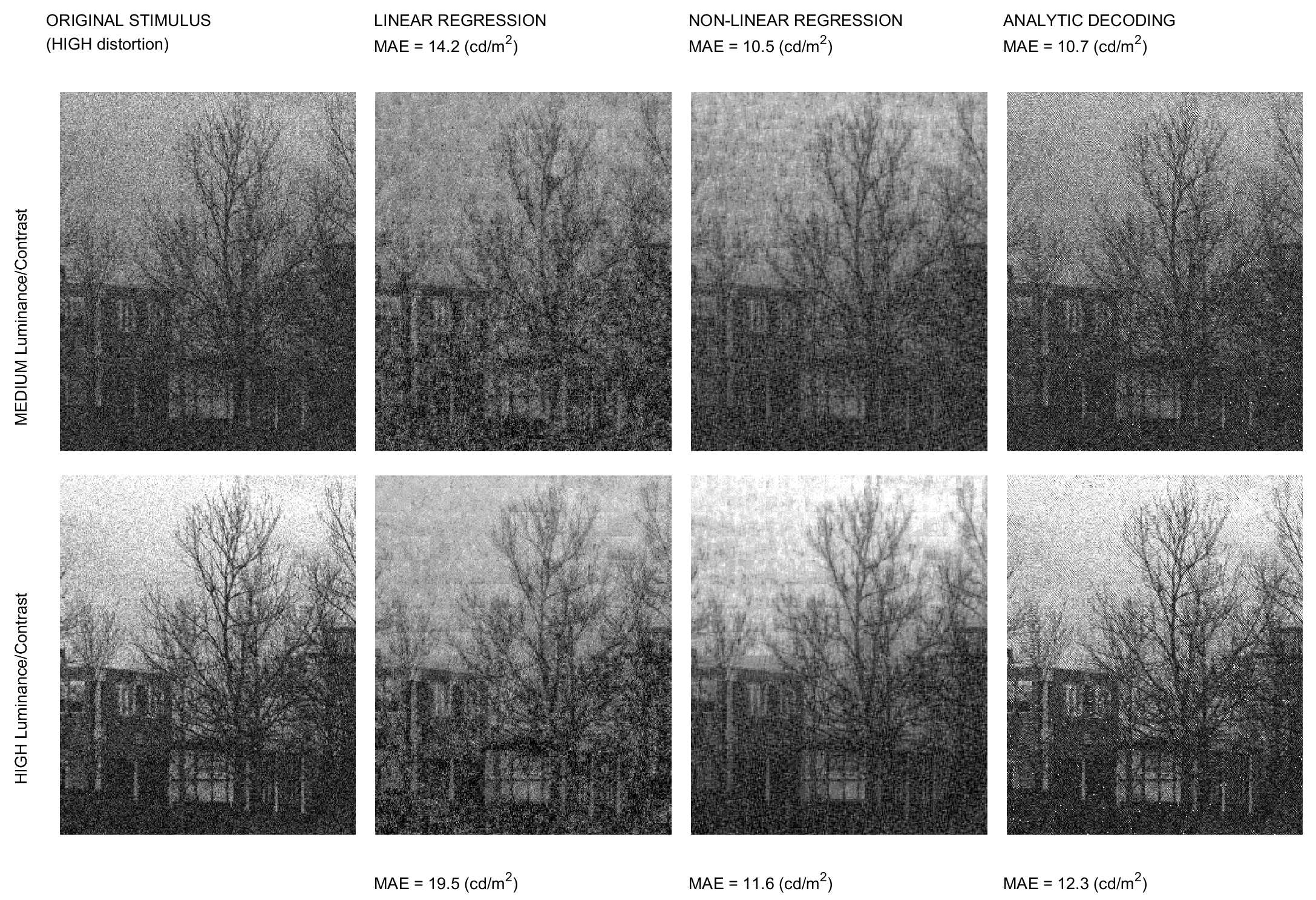}\\
    \end{tabular}
    \vspace{-0.0cm}
	\hspace{-0cm}\caption{\scriptsize{\textbf{Reconstructions in the high-distortion regime.} These six reconstructions correspond to the highlighted dots in the surfaces at the right plot of Fig. \ref{MAEs}}. In this example, with decoders properly trained in the high-distortion regime, the analytic decoding does not have the best MAE values (MAE is not a visually meaningful measure anyway), but it is certainly better in preserving the visual structures in the scene. }
    \label{ReconsC}
    \vspace{-0.15cm}
\end{figure}

Beyond Mean Absolute Error or alternative arbitrary measures of reconstruction accuracy (all of them perceptually arguable), it is worth taking an explicit look at the reconstructed images.
Representative examples of the decoded signals are shown in Figs. \ref{ReconsA} and \ref{ReconsC}.

Note that the set of visual examples include the best case scenario for the learning-based decoders:
training in the same illumination conditions and with distortion of the same nature (top row of Figs. \ref{ReconsA} and \ref{ReconsC}).
Even in this best case scenario, the analytic decoder better reproduces the visual structures even in the high distortion condition.
These visual examples illustrate the advantages of considering the analytic decoding (generalization ability) with regard to
the conventional linear and nonlinear regression.

%\vspace{0.5cm}
%\red{Illustrations of the advantages of the analytic response (generalization ability) with regard to linear and nonlinear regression...}
%\vspace{0.5cm}

%%%%%%%%%%%%%%%%%%%%%%%%%%%%%%%%%%%%%%%%%%%%%%%%%%%%%%%%%%%%%%%%%%%%%%%%%  END OF 3.4 DECODING  (use inverse)
%%%%%%%%%%%%%%%%%%%%%%%%%%%%%%%%%%%%%%%%%%%%%%%%%%%%%%%%%%%%%%%%%%%%%%%%%%%%%%%%%%%%%%%%%%%%%%%%%%%%%%%%%%%%%%%%%%%%%%%%%%%%%%%%%%%%%%%%%%%%%%%%%%
%%%%%%%%%%%%%%%%%%%%%%%%%%%%%%%%%%%%%%%%%%%%%%%%%%%%%%%%%%%%%%%%%%%%%%%%%%%%%%%%%%%%%%%%%%%%%%%%%%%%%%%%%%%%%%%%%%%%%%%%%%%%%%%%%%%%%%%%%%%%%%%%%%
%%%%%%%%%%%%%%%%%%%%%%%%%%%%%%%%%%%%%%%%%%%%%%%%%%%%%%%%%%%%%%%%%%%%%%%%%%%%%%%%%%%%%%%%%%%%%%%%%%%%%%%%%%%%%%%%%%%%%%%%%%%%%%%%%%%%%%%%%%%%%%%%%%
%%%%%%%%%%%%%%%%%%%%%%%%%%%%%%%%%%%%%%%%%%%%%%%%%%%%%%%%%%%%%%%%%%%%%%%%%%%%%%%%%%%%%%%%%%%%%%%%%%%%%%%%%%%%%%%%%%%%%%%%%%%%%%%%%%%%%%%%%%%%%%%%%%

%%--------------------------------------------
\section{Concluding remarks}
%%--------------------------------------------
%\input{3_4_conclusions_2nd.tex}
\label{conclusions}

This paper addressed relevant mathematical details of biologically plausible feed-forward cascades of linear+nonlinear neural models.
These details, namely the Jacobians of the transform (w.r.t. the stimulus and w.r.t. the parameters) and the decoding transform, are usually disregarded in the conventional experimental literature
(e.g. \cite{Carandini12} and cites therein), because it is focused on obtaining the encoding transform in specific experimental settings.

The analytical results presented here show that for the considered L+NL cascades the Jacobian with regard to the stimulus, $\nabla_{\vect{x}^0} S$,
the Jacobian with regard to the parameters, $\nabla_{\vect{\Theta}} S$,
and the inverse, $S^{-1}$, reduce to knowing the corresponding Jacobian and inverse
of the nonlinear part of each layer of the cascade,
namely $\nabla_{\vect{y}^i} \mathcal{N}^{(i)}$,  $\nabla_{\vect{\theta}^i} \mathcal{N}^{(i)}$, and
${\mathcal{N}^{(i)}}^{-1}$.
These necessary elements are explicitly given here: the analytical expressions of $\nabla_{\vect{y}^i} \mathcal{N}^{(i)}$, $\nabla_{\vect{\theta}^i} \mathcal{N}^{(i)}$, and ${\mathcal{N}^{(i)}}^{-1}$ for the case canonical Divisive Normalization.
Equivalent results for alternative nonlinearities such as the
Wilson-Cowan model \cite{Wilson72,Cowan16} and models of brightness perception \cite{Cyriac15} are also given.

\vspace{0.25cm}
\blue{The fundamental reason for this analytical treatment is the analytical insight into the \emph{physiology}, the \emph{psychophysics} and the \emph{function} of the visual system. We explicitly saw that the context-dependent changes in the receptive fields, the impact in the response of uncertainty in the filters (or synaptic weights), the trends of the sensitivity and JNDs, and the efficiency of the system in multi-information terms can be identified in the analytical expressions of the Jacobians $\nabla_{\vect{x}^0} S$ and $\nabla_{\vect{\Theta}} S$.}

\blue{It is true that} the artificial neural networks literature addresses similar cascaded architectures~\cite{Goodfellow16}, and this community has been recently attracted by the applications of their derivatives and inverse (e.g. in image synthesis \cite{Gatys16}, or new visualization methods to assess deep networks using derivatives \cite{Laparra17} and the inverse~\cite{Mahendran14}).
However, this literature doesn't address all the analytic results reported here for biologically plausible nonlinearities such as the Divisive Normalization or the Wilson-Cowan model.
One reason is because the most popular artificial networks use simplified nonlinearities and biological plausibility is not one of the goals when training the artificial models \cite{Goodfellow16}.
But more importantly, given the growing popularity of automatic differentiation methods \cite{Baydin17}, derivatives may be extensively used, but explicit expressions are not given.

\blue{It is important to stress that even in the case of training biologically plausible models, while the use of automatic (implicit) differentiation certainly leads to successful working models, it makes the models more difficult to understand: one should check their behavior empirically because it is not summarized \emph{in equations}}.
\blue{On the contrary, as shown in the examples presented in the first part of the discussion, the explicit expressions allow to identify the basic properties of the system. In this way one may anticipate the kind of results that will emerge from the model (e.g. the nature and location of the distortions computed in MAD) and their relation with the sensitivity or the efficiency of the system because this can be \emph{seen} in the equations.}

\blue{Additional examples of the advantages of the analytical treatment include the following. On the one hand, the analytical expressions may suggest simplifications that enable links between models of different nature.
For instance, the truncation of the expansion of the analytical inverse of the Divisive Normalization is convenient to establish the equivalence between the Divisive Normalization and the Wilson-Cowan model \cite{Malo18}.
On the other hand, the analytical expressions allow sensible comparisons between normal and anomalous versions of a model.
For instance, the corresponding pair procedure to simulate the perception of anomalous observers \cite{Capilla04} could be used in more general models if the analytical inverse is available and it depends on the parameter that describes the anomaly.}

\blue{From the experimental and applied perspective,} we discussed how $\nabla_{\vect{x}^0} S$ can be used in the design of stimuli for novel psychophysics (MAD); how $\nabla_{\vect{\Theta}} S$ can be used to get the parameters of the model using classical psychophysics in image quality ratings; and how visual representation decoding may be benefited from the use of $S^{-1}$.
These illustrations \blue{are a practical demonstration of the correctness of the presented expressions} and suggest that
(i)~the proposed modular model can be easily extended including extra layers that can be fitted without relying in brute-force techniques (hence improving the results in \cite{Watson02,Laparra10a,Bertalmio17}), and
(ii)~the analytic inverse seems an interesting alternative to blind regression techniques \cite{Stanley99,Marre15} previously used in visual decoding.

\blue{Finally, it is important to acknowledge that relevant aspects of the sensory system
were not explicitly considered in the examples shown here, as for instance its dynamics or the
eventual presence of feedback.}
%Dynamics may refer to (a) time-varying inputs and responses,
%and to (b) the adaptation of the parameters.

\blue{Regarding time-varying stimuli, the L+NL cascaded architecture considered here has also
been successfully used to model motion sensitive areas such as V1 and MT~\cite{Simoncelli98}.
Therefore, the spatio-temporal extension of the presented formulation is completely straightforward.}
\blue{Focus on the stationary solution of the system, as done in our consideration of the Wilson-Cowan model,
implies ignoring transients. Nevertheless, networks with divisive
feedback lead to regular divisive normalization-like steady states and the semisaturation depends on the signal~\cite{Wilson93}.
Similarly, when divisive normalization is considered to be equivalent to the
steady state of the Wilson-Cowan model, signal dependence appears in the kernel $H$~\cite{Malo18}.
Therefore, some of the parameters that we assumed to be constant
should actually vary with the environment with specific time constants.}

\blue{In relation with the limitations due to ignoring feedback in the considered models
we would like to stress that we are not advocating for feedforward L+NL cascades as the perfect approach in all situations,
but rather introducing the maths for better using a very popular framework, which will in turn help advance our understanding of how biological vision works.}

\blue{In the same vein, a number of alternative models also assume multiple L+NL stages and adaptive non-linearities \cite{Cui13,Cui16,Antolik16,Bethge17,Simoncelli15,Laparra17}. And they all are successfully fitted via gradient based methods to data. While some of them provide a detailed account on how the parameters are fitted \cite{Cui13,Cui16,Simoncelli15}, others rely on automatic differentiation \cite{Antolik16,Bethge17,Laparra17}.
An interesting example related with the approach proposed here is the model considered in \cite{Wichmann17}, where they explicitly provide not only the Jacobian wrt the parameters to fit the model, but also the Jacobian wrt the input
to synthesize MAD stimuli. Unfortunately, as the other cases, they do not address the inverse either.}

\blue{In summary, given the insight that can be obtained from the explicit expressions}, future modeling efforts should not be restricted to the forward transform, but they should also address the derivatives and the inverse \blue{if the level of abstraction is low enough to invert the model}.

%%%%%%%%%%%%%%%%%%%%%%%%%%%%%%%%%%%%%%%%%%%%%%%%%%%%%%%%%%%%%%%%%%%%%%%%%  END OF 4 CONCLUDING REMARKS
%%%%%%%%%%%%%%%%%%%%%%%%%%%%%%%%%%%%%%%%%%%%%%%%%%%%%%%%%%%%%%%%%%%%%%%%%%%%%%%%%%%%%%%%%%%%%%%%%%%%%%%%%%%%%%%%%%%%%%%%%%%%%%%%%%%%%%%%%%%%%%%%%%
%%%%%%%%%%%%%%%%%%%%%%%%%%%%%%%%%%%%%%%%%%%%%%%%%%%%%%%%%%%%%%%%%%%%%%%%%%%%%%%%%%%%%%%%%%%%%%%%%%%%%%%%%%%%%%%%%%%%%%%%%%%%%%%%%%%%%%%%%%%%%%%%%%
%%%%%%%%%%%%%%%%%%%%%%%%%%%%%%%%%%%%%%%%%%%%%%%%%%%%%%%%%%%%%%%%%%%%%%%%%%%%%%%%%%%%%%%%%%%%%%%%%%%%%%%%%%%%%%%%%%%%%%%%%%%%%%%%%%%%%%%%%%%%%%%%%%
%%%%%%%%%%%%%%%%%%%%%%%%%%%%%%%%%%%%%%%%%%%%%%%%%%%%%%%%%%%%%%%%%%%%%%%%%%%%%%%%%%%%%%%%%%%%%%%%%%%%%%%%%%%%%%%%%%%%%%%%%%%%%%%%%%%%%%%%%%%%%%%%%%

\subsubsection*{Acknowledgments.}
This work was partially funded by the MINECO/FEDER/EU projects TEC2013-50520-EXP
and BFU2014-59776-R, and TIN2015-71537-P;
by the European Research Council, Starting Grant ref. 306337; and by the Icrea Academia Award.

%\newpage
\bibliographystyle{plos2015}
%\bibliography{refs_pinwheels_net}
\bibliography{JesusMaloRef,references_mb,refs_butts}

\newpage

\pagenumbering{roman}

\rfoot{\thepage/xix}

%%--------------------------------------------
\section{Supplementary Materials}
%%--------------------------------------------

\subsection{A cascaded L+NL vision model}
\label{example_model}
%Stage 1: Radiance-Brightness transform
%Stage 2: Center-surround contrast computation
%Stage 3: LGN-like receptive fields or CSF-like filtering and energy masking
%Stage 4: V1-like receptive fields and energy masking

The illustrative model considered in the Discussion was originally intended to provide a psychophysically meaningful alternative to the
\emph{modular concept} in Structural Similarity measures (SSIM). The authors of SSIM suggest a separate consideration of luminance, contrast and structure \cite{Wang04},
which is a sensible approach, but the definition of such factors has no obvious perceptual meaning in SSIM.
The idea for a perceptual alternative to SSIM proposed in \cite{Malo15} was addressing one psychophysical factor at a time
(i.e. also a modular approach), by using a cascade of linear-nonlinear transforms.
In this paper we consider a cascade of 4 L+NL layers, each focused on a different psychophysical factor:
\begin{labeling}{Layer $S^{(4)}$}
\item [Layer $S^{(1)}$] linear spectral integration to compute luminance and opponent tristimulus channels, and nonlinear brightness/color response.
\item [Layer $S^{(2)}$] definition of local contrast by using linear filters and divisive normalization.
\item [Layer $S^{(3)}$] linear LGN-like contrast sensitivity filter and nonlinear local contrast masking in the spatial domain.
\item [Layer $S^{(4)}$] linear V1-like wavelet decomposition and nonlinear divisive normalization to account for
orientation and scale-dependent masking.
\end{labeling}
Here we extend previous results by considering two extra layers (1-st and 4-th layers were not considered in \cite{Malo15}).
%On top of its interpretability, the modular structure simplifies the use of MAD to set the free parameters by determining only one layer at a time.
%As a result, this model substantially improves the performance of SSIM and related models in image quality assessment.
Following the suggestion in \cite{Carandini12}, here we use the canonical Divisive Normalization for each of these layers.
Below we present the expressions of the \emph{forward transforms}, their \emph{derivatives w.r.t. the stimulus}, their \emph{derivatives w.r.t. the parameters}, and their \emph{inverses}.

\hiddensubsubsection{Forward transforms.}

The different layers are almost isomorphic:
while the 3rd and 4th layers follow the divisive normalization expression introduced in the main text,
the 1st and 2nd layers only differ in that they operate on positive signals (luminance and brightness respectively),
and in a global normalization constant in the 1st layer.

\subparagraph*{Layer 1: Brightness from Radiance}
\begin{eqnarray}
  \nonumber \mathcal{L}^{(1)}  \equiv \,\,\,\,\,\,\,\,\,\,\,\,\,\,\,\, \vect{y}^1 & \!\! = \!\!& L^1 \cdot \vect{x}^0 \\
  \mathcal{N}^{(1)}  \equiv \,\,\,\,\,\,\,\,\,\,\,\,\,\,\,\, \vect{x}^1 & \!\! = \!\!& K(\vect{y}^1) \cdot \mathds{D}^{-1}_{\left( \vect{b}^1 + H^1 \cdot {\vect{y}^1}^{\gamma^1} \right)} \cdot {\vect{y}^1}^{\gamma^1} \label{layer1}
\end{eqnarray}
\noindent where, $L^1$ is a matrix with the color matching functions for each spatial location. In particular, restricting ourselves to achromatic information,
the only required color matching function would be the spectral sensitivity $V_\lambda$ \cite{Stiles82,Fairchild05}, leading to the luminance in each spatial location.
The global scaling matrix $K(\vect{y}^1) = \kappa \left( \mathds{D}_{\vect{b}^1} +  \mathds{D}_{\left( \frac{\beta}{d} \mathds{1} \cdot {\vect{y}^1}^{\gamma^1} \right)} + I \right)$, just ensures that the maximum brightness value (for normalized luminance equal to 1) is $\kappa$.
The role of the interaction kernel in the denominator $H^1 = \left( \frac{\beta}{d} \mathds{1} + I \right )$,
where $\mathds{1}$ is the all-ones $d\times d$ matrix, and $I$ is the identity matrix,
is setting the anchor for the brightness adaptation.
With this kernel in the denominator the anchor luminance is related to the average luminance energy $\left( \vect{b}^1 + \frac{\beta}{d} \mathds{1} \cdot {\vect{y}^1}^{\gamma^1} \right )$.
The effect of this nonlinear transform is a Weber-like adaptive saturation \cite{Abrams07}. Similar nonlinear behavior can be assumed for the opponent
chromatic channels \cite{Fairchild05,Stockman11,Laparra12}, but we didnt implemented the color version of the model.

\subparagraph*{Layer 2: Contrast from Brightness}
\begin{eqnarray}
  \nonumber  \mathcal{L}^{(2)}  \equiv \,\,\,\,\,\,\,\,\,\,\,\,\,\,\,\, \vect{y}^2 & \!\! = \!\!& L^2 \cdot \vect{x}^1 \\
  \mathcal{N}^{(2)}  \equiv \,\,\,\,\,\,\,\,\,\,\,\,\,\,\,\, \vect{x}^2 & \!\! = \!\!& \mathds{D}^{-1}_{\left( \vect{b}^2 + H^2 \cdot \vect{y}^2 \right)} \cdot \vect{y}^2  \label{layer2}
\end{eqnarray}
\noindent where the linear stage computes the deviation of point-wise brightness with regard to the local brightness through
$L^2 = I - \mathcal{H}^n$, and this kernel in the \emph{numerator}, $\mathcal{H}^n$, represents the convolution by a two-dimensional Gaussian
(as in Eq. \ref{GaussianH}).
The normalization through $H^2 = \mathcal{H}^d \cdot \left( I - \mathcal{H}^n \right)^{-1}$, where the kernel in the \emph{denominator}, $\mathcal{H}^d$, is another two-dimensional Gaussian kernel, leads to the standard definition of contrast: normalization of the deviation of brightness by the local brightness.

\subparagraph*{Layer 3: Contrast sensitivity and spatial masking}
\begin{eqnarray}
  \nonumber \mathcal{L}^{(3)}  \equiv \,\,\,\,\,\,\,\,\,\,\,\,\,\,\,\, \vect{y}^3 & \!\! = \!\!& L^3 \cdot \vect{x}^2 \\
  \mathcal{N}^{(3)}  \equiv \,\,\,\,\,\,\,\,\,\,\,\,\,\,\,\, \vect{x}^3 & \!\! = \!\!& \mathds{D}_{\textrm{sign}(\vect{y}^3)} \cdot \mathds{D}^{-1}_{\left( \vect{b}^3 + H^3 \cdot |\vect{y}^3|^{\gamma^3} \right)} \cdot |\vect{y}^3|^{\gamma^3} \label{layer3}
\end{eqnarray}
\noindent where $L^3$ is the convolution matrix equivalent to the application of a Contrast Sensitivity Function (CSF) \cite{Campbell68}.
The rows of this matrix consist of displaced versions of center-surround (LGN-like) receptive fields (impulse response of the CSF \cite{InglingUriegas}).
The kernel in the denominator, $H^3$, represents the convolution by another two-dimensional Gaussian that computes the local contrast energy
that masks the responses in high-energy environments.

\subparagraph*{Layer 4: Wavelet analysis and frequency masking}
\begin{eqnarray}
  \nonumber \mathcal{L}^{(4)}  \equiv \,\,\,\,\,\,\,\,\,\,\,\,\,\,\,\, \vect{y}^4 & \!\! = \!\!& L^4 \cdot \vect{x}^3 \\
  \mathcal{N}^{(4)}  \equiv \,\,\,\,\,\,\,\,\,\,\,\,\,\,\,\, \vect{x}^4 & \!\! = \!\!& \mathds{D}_{\textrm{sign}(\vect{y}^4)} \cdot \mathds{D}^{-1}_{\left( \vect{b}^4 + H^4 \cdot |\vect{y}^4|^{\gamma^4} \right)} \cdot |\vect{y}^4|^{\gamma^4} \label{layer4}
\end{eqnarray}
\noindent where $L^4$ is the matrix of Gabor-like receptive fields corresponding to V1-like sensors \cite{SimoncelliWoods90}.
The kernel in the denominator, $H^4$, represents the masking interaction between sensors tuned to different space, frequency and orientation \cite{Watson97}.

\hiddensubsubsection{Derivatives.}
Due to the slight differences in the transforms for the 1st and 2nd layers,
the derivatives and inverses are slightly different from Results I, II and III presented in the main text.
However, here we simply list the corresponding expressions because their derivation
is straightforward using the methods described to get Results I, II and III.
Since the formulation of the 3rd and 4th layers is the one given in the main text, the corresponding expressions will not be repeated here.

\subparagraph*{Derivatives w.r.t stimulus.}
\begin{eqnarray}
      \hspace{-3cm}
      \nabla_{\!\!\vect{y}^1} \mathcal{N}^{(1)} \!\!\!\!&=&\!\!\!\! K(\vect{y}^1) \cdot
      \left[ \mathds{D}^{-1}_{\mathcal{D}^{(1)}(\vect{y}^1)} - \mathds{D}_{\left(\frac{{\vect{y}^1}^{\gamma^1}}{{\mathcal{D}^{(1)}(\vect{y}^1)}^2}\right)} \cdot H^1
      \right]  \cdot \mathds{D}_{\left(\gamma^1 {\vect{y}^1}^{\gamma^1-1}\right)}
      + \frac{\kappa \beta}{d} \mathds{D}_{\left(\frac{{\vect{y}^1}^{\gamma^1}}{{\mathcal{D}^{(1)}(\vect{y}^1)}}\right)} \cdot \mathds{1} \cdot \mathds{D}_{\left(\gamma^1 {\vect{y}^1}^{\gamma^1-1}\right)} \label{dNdy_layer1} \\
      \hspace{-3cm}
      \nabla_{\!\!\vect{y}^2} \mathcal{N}^{(2)} \!\!\!\!&=&\!\!\!\! \,\,\,\,\,\,\,\,\,\,\,\,\,\,\,\,\,\,\,\,\, \left[ \mathds{D}^{-1}_{\mathcal{D}^{(2)}(\vect{y}^2)} - \mathds{D}_{\left(\frac{\vect{y}^2}{{\mathcal{D}^{(2)}(\vect{y}^2)}^2}\right)} \cdot H^2
      \right] \label{dNdy_layer2}
\end{eqnarray}
Where, as in the main text, $\mathcal{D}^{(i)}$, stands for the denominator corresponding to the normalization
in the $i$-th layer.

\subparagraph*{Derivatives w.r.t. the semisaturation $\vect{b}$}
\begin{eqnarray}
      \nabla_{\!\!\vect{b}^1} \mathcal{N}^{(1)} \!\!&=&\!\! - K(\vect{y}^1) \cdot \mathds{D}_{{\vect{y}^1}^{\gamma^1}} \cdot \mathds{D}^{-2}_{\mathcal{D}^{(1)}(\vect{y}^1)} + \kappa \mathds{D}_{{\vect{y}^1}^{\gamma^1}} \cdot \mathds{D}^{-1}_{\mathcal{D}^{(1)}(\vect{y}^1)} \label{jacobian2_b1} \\
      \nabla_{\!\!\vect{b}^2} \mathcal{N}^{(2)} \!\!&=&\!\! - \mathds{D}_{\vect{y}^2} \cdot \mathds{D}^{-2}_{\mathcal{D}^{(2)}(\vect{y}^2)}  \label{jacobian2_b2}
\end{eqnarray}

\subparagraph*{Derivatives w.r.t. the excitation-inhibition exponent $\gamma$}
\begin{equation}
      \hspace{-2.8cm} \nabla_{\!\!\gamma^1} \mathcal{N}^{(1)}  =  \Bigg[ K(\vect{y}^1) \!\cdot\! \mathds{D}^{-1}_{\mathcal{D}^{(1)}(\vect{y}^1)} \!\cdot\! \Bigg[ \mathds{D}_{\textrm{log} \vect{y}^1}
      - \mathds{D}^{-1}_{\mathcal{D}^{(1)}(\vect{y}^1)} \!\cdot\! \mathds{D}_{\left( H^1 \cdot \mathds{D}_{\left({\vect{y}^1}^{\gamma^1}\right)} \cdot \textrm{log} \vect{y}^1 \right)} \Bigg] + \frac{\kappa \beta}{d} \mathds{D}_{\left(\frac{{\vect{y}^1}^{\gamma^1}}{{\mathcal{D}^{(1)}(\vect{y}^1)}}\right)} \!\cdot\! \mathds{1} \!\cdot\! \mathds{D}_{\textrm{log} \vect{y}^1}   \Bigg] \!\cdot\! {\vect{y}^1}^{\gamma^1} \label{jacobian2_gamma1}
\end{equation}

\subparagraph*{Derivatives w.r.t. the global scaling constant $\kappa$}
\begin{equation}
      \hspace{-2.8cm} \nabla_{\!\!\kappa} \, \mathcal{N}^{(1)}  = \left( \mathds{D}_{\vect{b}^1} +  \mathds{D}_{\left( \frac{\beta}{d} \mathds{1} \cdot {\vect{y}^1}^{\gamma^1} \right)} + I \right) \cdot \mathds{D}^{-1}_{\left( \vect{b}^1 + H^1 \cdot {\vect{y}^1}^{\gamma^1} \right)} \cdot {\vect{y}^1}^{\gamma^1} \label{jacobian2_kappa}
\end{equation}

\subparagraph*{Derivatives w.r.t. the parameters of kernels $L$ and $H$.}
The parameter of the normalization kernel of the 1st stage, $H^1$, is the weight, $\beta$, of the adaptation anchor for the luminance. The derivative w.r.t. $\beta$ is:
\begin{eqnarray}
      \nabla_{\!\!\beta} \mathcal{N}^{(1)} \!&=&\! \frac{1}{d} \, \mathds{D}^{-1}_{\mathcal{D}^{(1)}(\vect{y}^1)} \!\cdot\! \Bigg[ \kappa \mathds{D}_{ \left( \mathds{1} \cdot {\vect{y}^1}^{\gamma^1} \right) }
      - K(\vect{y}^1) \cdot \mathds{D}_{\left( \frac{{\vect{y}^1}^{\gamma^1}}{\mathcal{D}^{(1)}(\vect{y}^1)} \right) } \!\cdot\! \mathds{1} \Bigg] \!\cdot\! {\vect{y}^1}^{\gamma^1} \label{jacobian2_beta}
\end{eqnarray}
Contrast computation in the 2nd layer depends on the application of two Gaussian kernels: one in the linear stage, $L^2 = \mathbb{1} - \mathcal{H}^n$, and another in the nonlinear stage,
$H^2 = \mathcal{H}^d \cdot (\mathbb{1} - \mathcal{H}^n)^{-1}$. The rows of each of these kernels depends on the corresponding width and amplitude.

For the set of widths and amplitudes of $\mathcal{H}^n$, $\vect{\sigma}^n$ and $\vect{c}^n$:
\begin{equation}
     \nabla_{\vect{\sigma}^n} \mathcal{L}^{(2)} = - \nabla_{\vect{y}^2} \mathcal{N}^{(2)} \cdot \textrm{diag} \left[
       \left(
       \begin{array}{ccc}
                  \hspace{0.5cm} & {\vect{x}^1}^\top & \hspace{0.5cm} \\
                  \hspace{0.5cm} & {\vect{x}^1}^\top & \hspace{0.5cm} \\
                  \hspace{0.5cm} & \vdots & \hspace{0.5cm} \\
                  \hspace{0.5cm} & {\vect{x}^1}^\top & \hspace{0.5cm}
       \end{array}
     \right) \cdot F^n
     \right]
     \label{dLdsigma}
\end{equation}
where, $F^n_{k k'} = c^n_k \, \frac{dp_{k1} dp_{k2}}{2 \pi \,\, {\sigma^n_k}^5} \,\, \left(  \Delta^2_{k k'} - 2 \, {\sigma^n_k}^2 \right) \,\, e^{-\frac{  \Delta^2_{k k'} }{2 \,\, {\sigma^n_k}^2}}$. In this expression, $\Delta^2_{k k'}$ and $dp_{k1} dp_{k2}$ have the same meaning used in the main text in the context of the Gaussian kernels, Eq. \ref{GaussianH}.

\begin{equation}
     \nabla_{\vect{c}^n} \mathcal{L}^{(2)} = - \nabla_{\vect{y}^2} \mathcal{N}^{(2)} \cdot \textrm{diag} \left[
       \left(
       \begin{array}{ccc}
                  \hspace{0.5cm} & {\vect{x}^1}^\top & \hspace{0.5cm} \\
                  \hspace{0.5cm} & {\vect{x}^1}^\top & \hspace{0.5cm} \\
                  \hspace{0.5cm} & \vdots & \hspace{0.5cm} \\
                  \hspace{0.5cm} & {\vect{x}^1}^\top & \hspace{0.5cm}
       \end{array}
     \right) \cdot G^n
     \right]
     \label{dLdc}
\end{equation}
where, $G^n_{k k'} = \frac{dp_{k1} dp_{k2}}{2 \pi \,\, {\sigma^n_k}^2} \,\, e^{-\frac{  \Delta^2_{k k'} }{2 \,\, {\sigma^n_k}^2}}$.

Similarly, for the set of widths and amplitudes of $\mathcal{H}^d$, $\vect{\sigma}^d$ and $\vect{c}^d$:

\begin{equation}
     \nabla_{\vect{\sigma}^d} \mathcal{N}^{(2)} = -\textrm{diag} \left[ \mathds{D}_{\vect{y}^2} \cdot \mathds{D}^{-2}_{\mathcal{D}^{(2)}(\vect{y}^2)} \cdot
     \left(
       \begin{array}{ccc}
                  \hspace{0.5cm} & \left( (\mathbb{1} - \mathcal{H}^n)^{-1} \cdot \vect{y}^2 \right)^\top & \hspace{0.5cm} \\
                  \hspace{0.5cm} & \left( (\mathbb{1} - \mathcal{H}^n)^{-1} \cdot \vect{y}^2 \right)^\top & \hspace{0.5cm} \\
                  \hspace{0.5cm} & \vdots & \hspace{0.5cm} \\
                  \hspace{0.5cm} & \left( (\mathbb{1} - \mathcal{H}^n)^{-1} \cdot \vect{y}^2 \right)^\top & \hspace{0.5cm}
       \end{array}
     \right) \cdot F^d
     \right]
     \label{dNdsigma2}
\end{equation}
where, $F^d_{k k'} = c^d_k \, \frac{dp_{k1} dp_{k2}}{2 \pi \,\, {\sigma^d_k}^5} \,\, \left(  \Delta^2_{k k'} - 2 \, {\sigma^d_k}^2 \right) \,\, e^{-\frac{  \Delta^2_{k k'} }{2 \,\, {\sigma^d_k}^2}}$.

\begin{equation}
     \nabla_{\vect{c}^d} \mathcal{N}^{(2)} = -\textrm{diag} \left[
     \mathds{D}_{\vect{y}^2} \cdot \mathds{D}^{-2}_{\mathcal{D}^{(2)}(\vect{y}^2)} \cdot
     \left(
       \begin{array}{ccc}
                  \hspace{1cm} & \left( (\mathbb{1} - \mathcal{H}^n)^{-1} \cdot \vect{y}^2 \right)^\top & \hspace{1cm} \\
                  \hspace{1cm} & \left( (\mathbb{1} - \mathcal{H}^n)^{-1} \cdot \vect{y}^2 \right)^\top & \hspace{1cm} \\
                  \hspace{1cm} & \vdots & \hspace{0.5cm} \\
                  \hspace{1cm} & \left( (\mathbb{1} - \mathcal{H}^n)^{-1} \cdot \vect{y}^2 \right)^\top & \hspace{1cm}
       \end{array}
     \right) \cdot G^d
     \right]
     \label{dNdc2}
\end{equation}
where, $G^d_{k k'} = \frac{dp_{k1} dp_{k2}}{2 \pi \,\, {\sigma^d_k}^2} \,\, e^{-\frac{  \Delta^2_{k k'} }{2 \,\, {\sigma^d_k}^2}}$.

\hiddensubsubsection{Inverses.}

The inverse of the 2nd layer is simpler than Result III,
\begin{equation}
      \vect{y}^2 = \left( \mathbb{1} - \mathds{D}_{\vect{x}^2} \cdot H^2 \right)^{-1} \cdot \mathds{D}_{\vect{x}^2} \cdot \vect{b}^2
      \label{inv_layer2}
\end{equation}
\noindent On the contrary, the inverse of the 1st layer may require an iterative process.
As stated above, the saturation in Eq. \ref{layer1} depends on an anchor for the luminance adaptation.
In the forward transform this anchor is computed through the average of the known luminance in the considered image.
But of course, the luminance values are not known when computing the inverse.
If there is no other way to assume certain adaptation state (or average luminance), the solution may be achieved iteratively.
First, assume certain reasonable average luminance to estimate the scaling matrix $K$.
Then, use this estimated $K$ to compute a first estimation of the luminance from known the brightness response and the analytic inverse.
This new estimation of the luminance vector can be used again for a better estimation of $K$, and this process can be iterated.
By using the \emph{energy} notation used in the main text, $\vect{e}^1 = {\vect{y}^1}^{\gamma^1}$, and using $n$ as iteration index,
\begin{eqnarray}
      \nonumber \vect{e}^1_n  & \rightarrow & K_n(\vect{e}^1_n) = \kappa \left( \mathds{D}_{\vect{b}^1} +  \mathds{D}_{\left( \frac{\beta}{d} \mathds{1} \cdot {\vect{e}^1_n} \right)} + I \right) \\
      \vect{e}^1_{n+1}  & = &  \left( \mathbb{1} - \mathds{D}_{ \left( K_n^{-1} \cdot \vect{x}^1 \right)} \cdot H^1 \right)^{-1} \cdot \mathds{D}_{\left( K_n^{-1} \cdot \vect{x}^1 \right)} \cdot \vect{b}^1 \label{inv_layer1}
\end{eqnarray}

\hiddensubsubsection{Note: alternative formulations in certain layers.}

The illustrative model used in the Discussion, described in this supplementary material, and implemented in the toolbox associated to the paper, is based on a cascade of almost isomorphic L+NL transforms (see Eqs. \ref{layer1} - \ref{layer4}).
However, as stated in the main text, this implementation based on isomorphic divisive normalization transforms is not the only possible choice.

Section \ref{general_cons} in the main text cites specific alternatives for the brightness transform (the two-gamma curve \cite{Cyriac15,Cyriac16,KaneBertalmio16})
and for the interaction between V1-like sensors (the Wilson-Cowan model \cite{Wilson72,Cowan16}).
The alternative transforms and their jacobians w.r.t. the stimulus were presented
in the main text in sections \ref{general_cons} and \ref{result1}.
The analytic inverse of the Wilson-Cowan interaction was presented in section \ref{result3}.

The inverse of the two-gamma nonlinearity was not addressed there because, given the
coupling between the input and the exponent (see Eq. \ref{exponente}), Eq. \ref{two-gamma} has no analytical inverse.
Nevertheless, iterative approximations to the actual luminance value can be obtained in the following way.
First one makes an initial guess of the exponent (for instance the average value between the two extremes), and then
one obtains the first guess for the luminance assuming this approximate exponent.
Afterwards, the estimate of the exponent is recomputed from the new luminance estimate, and so on.
At the $n$-th iteration,
%\begin{align}   % Dont ask me why, but align and \linenumbers se pegan!
\begin{eqnarray}
n &=& 0 \,\,
\begin{cases}
\,\, \gamma^1_{0} = \frac{1}{2}(\gamma_{L} + \gamma_{H}) \\[0.2cm] \nonumber
\,\, \vect{y}^1_0 = {\vect{x}^1}^{\frac{1}{\gamma^1_0}}
\end{cases}
\\[0.2cm]
n &>& 0 \,\,
\begin{cases}
\,\, \gamma^1_{n} = \gamma^1(\vect{y}^1_{(n-1)}) \\[0.2cm]
\,\, \vect{y}^1_{n} = {\vect{x}^1}^{\frac{1}{\gamma^1_{n}}}
\end{cases}
\label{inv_layer1_two_gamma}
\end{eqnarray}
%\end{align}
where thee subindex $n$ indicates the iteration and $\gamma^1(\vect{y}^1)$ is computed using Eq. \ref{exponente}.
Note that, given the singularity at the origin of the jacobian of the two-gamma response (see the discussion made after Eq. \ref{jacobian_two_gamma}),
the above only holds for big enough luminance, $\vect{y}^1 \in \lbrack \epsilon ^{\gamma^1(\epsilon)},1 \rbrack$.
For small luminance values, $\vect{y}^1 \in \lbrack 0,\epsilon ^{\gamma^1(\epsilon)} \rbrack$, it holds the
robust regime introduced to solve the mentioned singularity, and this parabolic expression is invertible:
\begin{equation}
\vect{y}^1=\dfrac{1}{\sqrt{a_1}} \sqrt{\vect{x}^1 + \dfrac{a^2_2}{4a_1}}-\dfrac{a_2}{2a_1}
\end{equation}

%%%%%%%%%%%%%%%%%%%%%%%%%%%%%%%%%%%%%%%%%%%%%%%%%%%%%%%%%%%%%%%%%%%%%%%%%  END OF 5.1 MODEL
%%%%%%%%%%%%%%%%%%%%%%%%%%%%%%%%%%%%%%%%%%%%%%%%%%%%%%%%%%%%%%%%%%%%%%%%%%%%%%%%%%%%%%%%%%%%%%%%%%%%%%%%%%%%%%%%%%%%%%%%%%%%%%%%%%%%%%%%%%%%%%%%%%
%%%%%%%%%%%%%%%%%%%%%%%%%%%%%%%%%%%%%%%%%%%%%%%%%%%%%%%%%%%%%%%%%%%%%%%%%%%%%%%%%%%%%%%%%%%%%%%%%%%%%%%%%%%%%%%%%%%%%%%%%%%%%%%%%%%%%%%%%%%%%%%%%%
%%%%%%%%%%%%%%%%%%%%%%%%%%%%%%%%%%%%%%%%%%%%%%%%%%%%%%%%%%%%%%%%%%%%%%%%%%%%%%%%%%%%%%%%%%%%%%%%%%%%%%%%%%%%%%%%%%%%%%%%%%%%%%%%%%%%%%%%%%%%%%%%%%
%%%%%%%%%%%%%%%%%%%%%%%%%%%%%%%%%%%%%%%%%%%%%%%%%%%%%%%%%%%%%%%%%%%%%%%%%%%%%%%%%%%%%%%%%%%%%%%%%%%%%%%%%%%%%%%%%%%%%%%%%%%%%%%%%%%%%%%%%%%%%%%%%%

\subsection{Derivative of a linear function with regard to its parameters}
\label{dLineardL}
The technical result in this section, namely Eq. \ref{deriv_linear_param}, is necessary when deriving one of the analytical results of the paper: the Jacobian with regard to the parameters requires this sort of derivative in Eq. \ref{elementary_jacobian2} (dependence with receptive fields)
and in Eq. \ref{jacobian2_H} (dependence with the masking kernel).

Given a general linear function depending on the rectangular matrix $L^i \in \mathbb{R}^{d_i \times d_{i-1}}$,
\begin{equation}
   \vect{y}^i = \mathcal{L}^{(i)}(\vect{x}^{i-1},L^i) = L^i \cdot \vect{x}^{i-1}
   \nonumber
\end{equation}
here we are interested in a matrix expression for its derivatives with regard to the parameters, $L^i$;
i.e. we address the issue of the matrix form of,
\begin{equation}
    \nabla_{\!\!L^i} \mathcal{L}^{(i)} = \frac{\partial \vect{y}^i}{\partial L^i}
    \nonumber
\end{equation}
In order to do so, go back to the individual dimensions for a moment: remember that we have $d_i$ 1-dimensional functions like this,
\begin{equation}
       y^i_k = \sum_{k'=1}^{d_{i-1}} L^i_{k k'} \,\, x^{i-1}_{k'}
       \nonumber
\end{equation}
where $k = 1, \ldots, d_i$.
Therefore, $\nabla_{\!\!L^i} \vect{y}^i \in \mathbb{R}^{d_i \times (d_i \times d_{i-1})}$ because $\vect{y}^i \in \mathbb{R}^{d_i \times 1}$ and we have $d_i \times d_{i-1}$ elements in $L^i$.
In this setting,
\begin{equation}
      \frac{\partial y^i_k}{\partial L^i_{l m}} = \delta_{k l} \,\, x^{i-1}_m
      \label{deriv_linear1_param}
\end{equation}
With a convenient rearrangement of $L^i$, it is possible to give a single matrix expression that summarizes the multiple element-wise derivatives in Eq. \ref{deriv_linear1_param}.

In particular, here we rearrange the parameters of the linear transform in the column vector $\vect{l}^i = \textrm{vect}({L^i}^\top) \in \mathbb{R}^{(d_i \times d_{i-1}) \times 1}$.
This arrangement is sensible in our context (feed-forward neural model) because of the following considerations.
As stated above, the $k$-th row of the matrix, $L^i_{k \star}$, represents the linear weights of the $k$-th linear sensor in the $i$-th linear stage $\mathcal{L}^i$. In other words,
$L^i_{k \star}$ represents the weighted connectivity or interaction of the $k$-th linear sensor with the previous stage.
Given the scalar-product description of receptive fields \cite{Olshausen96,Ringach02}, one may say that each $L^i_{k \star}$ plays the role of a \emph{receptive field} in the domain $\vect{x}^{i-1}$
because the $k$-th response is computed through these scalar products: $y^i_k = L^i_{k \star} \cdot \vect{x}^{i-1} = {\vect{x}^{i-1}}^\top \cdot {L^i_{k \star}}^\top$.
With the proposed rearrangement, the column vector $\vect{l}^i$ is built by concatenating the transposed \emph{receptive fields} of every linear sensor in the $i$-th layer,
\vspace{-0.0cm}
\begin{equation}
   \vect{l}^i = \textrm{vect}({L^i}^\top) = \left(\begin{array}{c}

                                                  \left(\begin{array}{c}
                                                     L^i_{1 1} \\
                                                     L^i_{1 2} \\
                                                     \vdots \\
                                                     L^i_{1 d_{i-1}} \\
                                                  \end{array}\right)\\
                                                  \left(\begin{array}{c}
                                                     L^i_{2 1} \\
                                                     L^i_{2 2} \\
                                                     \vdots \\
                                                     L^i_{2 d_{i-1}} \\
                                                  \end{array}\right)\\
                                             \vdots \\
                                                  \left(\begin{array}{c}
                                                     L^i_{d_i 1} \\
                                                     L^i_{d_i 2} \\
                                                     \vdots \\
                                                     L^i_{d_i d_{i-1}} \\
                                                  \end{array}\right)\\
                                         \end{array}
                                         \right)
                                         = \left(\begin{array}{c}

                                                  \left(\begin{array}{c}
                                                     \vdots \\
                                                     {L^i_{1 \star}}^\top \\
                                                     \vdots \\
                                                  \end{array}\right)\\

                                                  \left(\begin{array}{c}
                                                     \vdots \\
                                                     {L^i_{2 \star}}^\top \\
                                                     \vdots \\
                                                  \end{array}\right)\\[0.25cm]

                                             \vdots \\[0.25cm]

                                                  \left(\begin{array}{c}
                                                     \vdots \\
                                                     {L^i_{d_i \star}}^\top \\
                                                     \vdots \\
                                                  \end{array}\right)\\

                                         \end{array}
                                         \right)
  \label{vectoriz}
\end{equation}
As a result, the small variations of the linear responses due to perturbations in the parameters of the linear transform (which are $\Delta y^i_k = {\vect{x}^{i-1}}^\top \cdot {\Delta L^i_{k\star}}^\top$), can
be put in matrix form using the rearrangement $\vect{l}^i = \textrm{vect}({L^i}^\top)$:
\begin{equation}
                                         \left(\begin{array}{c}
                                                  \Delta y^i_1\\
                                                  \Delta y^i_2\\[0.25cm]
                                                     \vdots \\[0.25cm]
                                                  \Delta y^i_{d_i}\\
                                         \end{array}
                                         \right)
                                         =
                                         \left(\begin{array}{cccc}
                                                  \left( \cdots {\vect{x}^{i-1}}^\top \cdots \right) &  \left( \cdots \,\,\,\,\,\,\,0\,\,\,\,\,\,\, \cdots \right)              & \cdots & \left( \cdots \,\,\,\,\,\,\,0\,\,\,\,\,\,\, \cdots \right) \\
                                                  \left( \cdots \,\,\,\,\,\,\,0\,\,\,\,\,\,\, \cdots \right)              &  \left( \cdots {\vect{x}^{i-1}}^\top \cdots \right) & \cdots & \left( \cdots \,\,\,\,\,\,\,0\,\,\,\,\,\,\, \cdots \right) \\
                                                               \vdots                         &            \vdots                            & \ddots &    \vdots                      \\
                                                  \left( \cdots \,\,\,\,\,\,\,0\,\,\,\,\,\,\, \cdots \right)              &  \left( \cdots \,\,\,\,\,\,\,0\,\,\,\,\,\,\, \cdots \right)              & \cdots & \left( \cdots {\vect{x}^{i-1}}^\top \cdots \right)\\
                                         \end{array}\right)
                                         \cdot
                                         \Delta \vect{l}^i
  \label{block_diag}
\end{equation}
since this leads to the appropriate scalar products. Identifying terms with the linear approximation in terms of the Jacobian,
\begin{equation}
       \Delta \vect{y}^i = \nabla_{\!\!L^i} \mathcal{L}^{(i)} \cdot \Delta \vect{l}^i = \frac{\partial \vect{y}^i}{\partial L^i} \cdot \Delta \vect{l}^i
       \nonumber
\end{equation}
we see that, assuming the referred rearrangement of the parameters, the Jacobian we are interested in is given by the block diagonal matrix shown above.

In summary, given the linear transform $\vect{y}^i = L^i \cdot \vect{x}^{i-1}$, the Jacobian with regard to its parameters is:
\begin{equation}
       \nabla_{\!\!L^i} \mathcal{L}^{(i)} = \frac{\partial \vect{y}^i}{\partial L^i} = \mathds{B}^{d_i}_{({\vect{x}^{i-1}}^\top)}
      \label{deriv_linear_param}
\end{equation}
where $\mathds{B}^m_{(A)}$ is a \emph{block-diagonal} matrix built by replicating $m$ times the matrix (or vector) $A$ along the diagonal.

%%%%%%%%%%%%%%%%%%%%%%%%%%%%%%%%%%%%%%%%%%%%%%%%%%%%%%%%%%%%%%%%%%%%%%%%%  END OF 5.2 DERIV LINEAL
%%%%%%%%%%%%%%%%%%%%%%%%%%%%%%%%%%%%%%%%%%%%%%%%%%%%%%%%%%%%%%%%%%%%%%%%%%%%%%%%%%%%%%%%%%%%%%%%%%%%%%%%%%%%%%%%%%%%%%%%%%%%%%%%%%%%%%%%%%%%%%%%%%
%%%%%%%%%%%%%%%%%%%%%%%%%%%%%%%%%%%%%%%%%%%%%%%%%%%%%%%%%%%%%%%%%%%%%%%%%%%%%%%%%%%%%%%%%%%%%%%%%%%%%%%%%%%%%%%%%%%%%%%%%%%%%%%%%%%%%%%%%%%%%%%%%%
%%%%%%%%%%%%%%%%%%%%%%%%%%%%%%%%%%%%%%%%%%%%%%%%%%%%%%%%%%%%%%%%%%%%%%%%%%%%%%%%%%%%%%%%%%%%%%%%%%%%%%%%%%%%%%%%%%%%%%%%%%%%%%%%%%%%%%%%%%%%%%%%%%
%%%%%%%%%%%%%%%%%%%%%%%%%%%%%%%%%%%%%%%%%%%%%%%%%%%%%%%%%%%%%%%%%%%%%%%%%%%%%%%%%%%%%%%%%%%%%%%%%%%%%%%%%%%%%%%%%%%%%%%%%%%%%%%%%%%%%%%%%%%%%%%%%%

\subsection{Derivation of the Jacobian with regard to the stimulus}
\label{J_wrt_stimul}
In this section we provide the proofs for (a) Eq. \ref{deriv_DN} for the canonical divisive normalization, and (b) Eq. \ref{jacobian_two_gamma} for the particular two-gamma nonlinearity. Remember the Jacobian of the particular Wilson-Cowan nonlinearity was already derived in the Results section (Eqs. \ref{Jacobian inverse W-C} and \ref{Jacobian W-C}).
% In this technical section is where the benefits of the matrix notation become evident.

\paragraph*{Divisive normalization: proof of Eq. \ref{deriv_DN}.}
Explicitly considering the sign and amplitude terms of the nonlinearity in Eq. \ref{divisive_norm2}, we can write, $\mathcal{N}^{(i)}(\vect{y}^i) = \mathds{D}_{\textrm{sign}(\vect{y}^i)} \cdot N^{(i)}(\vect{e}^i)$. Then, using the chain rule, we have:
\begin{alignat}{2}
      \nonumber
      \nabla_{\!\!\vect{y}^i} \mathcal{N}^{(i)} & = \nabla_{\!\!\vect{y}^i} \, \textrm{sign}(\vect{y}^i) \cdot \mathds{D}_{N^{(i)}(\vect{e}^i)} + \mathds{D}_{\textrm{sign}(\vect{y}^i)} \cdot \nabla_{\!\!\vect{y}^i} N^{(i)}(\vect{e}^i) &   \\ \nonumber
        & \quad \;\;  \text{where,} \\ \nonumber
        & \quad \;\; \nabla_{\!\!\vect{y}^i} \, \textrm{sign}(\vect{y}^i) \cdot \mathds{D}_{N^{(i)}(\vect{e}^i)} = 0, \,\,\, \text{since sign($y^i_{k}$) is constant  $\forall y^i_{k} \neq 0$, and $N^{(i)}(0) = 0$} \\  \nonumber
        & \quad \;\; \nabla_{\!\!\vect{y}^i} N^{(i)}(\vect{e}^i) = \nabla_{\!\!\vect{e}^i} N^{(i)}(\vect{e}^i) \cdot \nabla_{\!\!\vect{y}^i} \vect{e}^i & \\   \nonumber
        & \quad \;\;\;\; \quad \;\;\;\; \quad \;\;\;\;\;\;\;\; \text{where,}
\end{alignat}
\begin{alignat}{2}
      \nonumber
        & \quad \;\;\;\; \quad \;\;\;\; \quad \;\;\;\;\;\;\;\; \nabla_{\!\!\vect{e}^i} N^{(i)}(\vect{e}^i) = \nabla_{\!\!\vect{e}^i} \vect{e}^i \,\, \cdot \mathds{D}_{\left(\frac{1}{\mathcal{D}^{(i)}(\vect{e}^i)} \right)} + \mathds{D}_{\vect{e}^i} \cdot \nabla_{\!\!\vect{e}^i} \, \frac{1}{\mathcal{D}^{(i)}(\vect{e}^i)} \\  \nonumber
        & \quad \;\;\;\; \quad \;\;\;\; \quad \;\;\;\;\;\;\;\; \qquad \qquad \text{where,} \\  \nonumber
        & \quad \;\;\;\; \quad \;\;\;\; \quad \;\;\;\;\;\;\;\; \qquad \qquad \nabla_{\!\!\vect{e}^i} \vect{e}^i = \mathbbm{1} \\  \nonumber
        & \quad \;\;\;\; \quad \;\;\;\; \quad \;\;\;\;\;\;\;\; \qquad \qquad \nabla_{\!\!\vect{e}^i} \, \frac{1}{\mathcal{D}^{(i)}(\vect{e}^i)} = -\mathds{D}_{\frac{1}{{\mathcal{D}^{(i)}(\vect{e}^i)}^2}} \cdot \nabla_{\!\!\vect{e}^i} \mathcal{D}^{(i)}(\vect{e}^i) = -\mathds{D}_{\frac{1}{{\mathcal{D}^{(i)}(\vect{e}^i)}^2}} \cdot H^i  \\  \nonumber
        & \quad \;\;\;\; \quad \;\;\;\; \quad \;\;\;\;\;\;\;\; \qquad \qquad \text{therefore,} \\  \nonumber
        & \quad \;\;\;\; \quad \;\;\;\; \quad \;\;\;\;\;\; \qquad \qquad = \mathds{D}_{\left(\frac{1}{\mathcal{D}^{(i)}(\vect{e}^i)} \right)} - \mathds{D}_{\vect{e}^i} \cdot \mathds{D}_{\frac{1}{{\mathcal{D}^{(i)}(\vect{e}^i)}^2}} \cdot H^i
\end{alignat}
\begin{alignat}{2}
        \nonumber
        & \quad \;\;\;\; \quad \;\;\;\; \quad \;\;\;\;\;\; \nabla_{\!\!\vect{y}^i} \vect{e}^i = \nabla_{\!\!|\vect{y}^i|} \vect{e}^i \cdot \nabla_{\!\!\vect{y}^i} |\vect{y}^i| \\  \nonumber
        & \quad \;\;\;\; \quad \;\;\;\; \quad \;\;\;\;\;\;\;\; \qquad \qquad \text{where,} \\  \nonumber
        & \quad \;\;\;\; \quad \;\;\;\; \quad \;\;\;\;\;\;\;\; \qquad \qquad \nabla_{\!\!|\vect{y}^i|} \vect{e}^i = \mathds{D}_{\gamma^i |\vect{y}^i|^{\gamma^i -1}} \\  \nonumber
       & \quad \;\;\;\; \quad \;\;\;\; \quad \;\;\;\;\;\;\;\; \qquad \qquad \nabla_{\!\!\vect{y}^i} |\vect{y}^i| = \mathds{D}_{\textrm{sign}(\vect{y}^i)}, \,\, \text{since the slope of $|\vect{y}^i|$ is sign($\vect{y}^i$)} \\  \nonumber
        & \quad \;\;\;\; \quad \;\;\;\; \quad \;\;\;\;\;\;\;\; \qquad \qquad \text{therefore,} \\  \nonumber
        & \quad \;\;\;\; \quad \;\;\;\; \quad \;\;\;\;\;\; \qquad \qquad = \mathds{D}_{\gamma^i |\vect{y}^i|^{\gamma^i -1}} \cdot \mathds{D}_{\textrm{sign}(\vect{y}^i)} \\  \nonumber
        & \quad \;\;\;\; \quad \;\;\;\; \quad \; \text{as a result,} \\  \nonumber
        & \quad \;\;\;\; \quad \;\;\;\; \quad \; = \left[ \mathds{D}_{\left(\frac{1}{\mathcal{D}^{(i)}(\vect{e}^i)} \right)} - \mathds{D}_{\frac{\vect{e}^i}{{\mathcal{D}^{(i)}(\vect{e}^i)}^2}} \cdot H^i \right] \cdot \mathds{D}_{\gamma^i |\vect{y}^i|^{\gamma^i-1}} \cdot \mathds{D}_{\textrm{sign}(\vect{y}^i)} \\  \nonumber
        & \text{finally, putting all the pieces together, we have Eq. \ref{deriv_DN}}.
\end{alignat}

\paragraph*{Two-gamma nonlinearity: proof of Eq. \ref{jacobian_two_gamma}.} Given the separation in sign/amplitude, $\vect{x} = \mathds{D}_{\textrm{sign}(\vect{y})} \cdot |\vect{y}|^{\gamma(|\vect{y}|)}$, the derivative is,
\begin{equation}
      \frac{\partial \vect{x}}{\partial \vect{y}} = \frac{\partial \vect{x}}{\partial |\vect{y}|} \cdot \frac{\partial |\vect{y}|}{\partial \vect{y}}
      = \Big[  \frac{\partial \, \textrm{sign}(\vect{y})}{\partial |\vect{y}|} \cdot \mathds{D}_{|\vect{y}|^{\gamma(|\vect{y}|)}} + \mathds{D}_{\textrm{sign}(\vect{y})} \cdot \frac{\partial \, |\vect{y}|^{\gamma(|\vect{y}|)}}{\partial |\vect{y}|} \Big] \cdot \frac{\partial |\vect{y}|}{\partial \vect{y}}
      \nonumber
\end{equation}
where the first term in the parenthesis cancels for the same reasons stated in the previous proof, and the slope the magnitude is the sign, therefore,
\begin{equation}
      \frac{\partial \vect{x}}{\partial \vect{y}} = \mathds{D}_{\textrm{sign}(\vect{y})} \cdot \frac{\partial \, |\vect{y}|^{\gamma(|\vect{y}|)}}{\partial |\vect{y}|}  \cdot \mathds{D}_{\textrm{sign}(\vect{y})} = \frac{\partial \, |\vect{y}|^{\gamma(|\vect{y}|)}}{\partial |\vect{y}|}
      \nonumber
\end{equation}
where signs cancel out because all the matrices are diagonal and hence the product is commutative. Now, by calling $f = |\vect{y}|^{\gamma(|\vect{y}|)}$, our problem is computing $\frac{\partial f}{\partial |\vect{y}|}$.
This notation is convenient since, taking the element-wise logarithm,
\begin{equation}
      \log f = \gamma(|\vect{y}|) \odot \log |\vect{y}|
      \label{logarit}
\end{equation}
and hence, applying the chain rule in the derivative of $\log f$ we have,
\begin{equation}
      \frac{\partial \log f}{\partial |\vect{y}|} = \frac{\partial\log f }{\partial f} \cdot \frac{\partial f}{\partial |\vect{y}|} = \mathds{D}_f^{-1} \cdot \frac{\partial f}{\partial |\vect{y}|} \,\,\,\,\,\,\Rightarrow\,\,\,\,\,\, \frac{\partial f}{\partial |\vect{y}|} = \mathds{D}_f \cdot \frac{\partial \log f}{\partial |\vect{y}|}
      \label{deriv_log_f}
\end{equation}
because the derivative matrices are diagonal, and hence all the products and quotients are Hadamard.

On the other hand, direct derivation of Eq. \ref{logarit}, leads to,
\begin{equation}
      \frac{\partial \log f}{\partial |\vect{y}|} = \frac{\partial \gamma(|\vect{y}|)}{\partial |\vect{y}|} \cdot \mathds{D}_{\log |\vect{y}|} + \mathds{D}_{\gamma(|\vect{y}|)} \cdot \frac{\partial \log |\vect{y}|}{\partial |\vect{y}|}
      \label{deriv_f}
\end{equation}
where, given the expression of the exponent $\gamma$, Eq. \ref{exponente}, its straightforward derivative:
\begin{equation}
      \dfrac{\partial \gamma(|\vect{y}|)}{\partial |\vect{y}|} = \mathds{D}_{\Big( (\gamma_H-\gamma_L)\cdot \dfrac{m \, |\vect{y}|^{ (m-1) }\cdot  \mu_1^m}{ (\mu_1^m + |\vect{y}|^{m})^2 } \Big)}
      \nonumber
\end{equation}
Then, plugging this derivative into Eq. \ref{deriv_f}, and the result into Eq. \ref{deriv_log_f}, we have Eq. \ref{jacobian_two_gamma}.
%\begin{equation}
%      \frac{\partial \vect{x}}{\partial \vect{y}} = \mathds{D}_{|\vect{y}|^{\gamma(|\vect{y}|)}} \cdot \Big[  \frac{\partial \gamma(|\vect{y}|)}{\partial |\vect{y}|} \cdot \mathds{D}_{\log |\vect{y}|} + \mathds{D}_{\frac{\gamma(|\vect{y}|)}{|\vect{y}|}} \Big]
%\end{equation}
%which leads to Eq. \ref{jacobian_two_gamma}.

%%%%%%%%%%%%%%%%%%%%%%%%%%%%%%%%%%%%%%%%%%%%%%%%%%%%%%%%%%%%%%%%%%%%%%%%%  END OF 5.3 JACOB WRT STIMULUS
%%%%%%%%%%%%%%%%%%%%%%%%%%%%%%%%%%%%%%%%%%%%%%%%%%%%%%%%%%%%%%%%%%%%%%%%%%%%%%%%%%%%%%%%%%%%%%%%%%%%%%%%%%%%%%%%%%%%%%%%%%%%%%%%%%%%%%%%%%%%%%%%%%
%%%%%%%%%%%%%%%%%%%%%%%%%%%%%%%%%%%%%%%%%%%%%%%%%%%%%%%%%%%%%%%%%%%%%%%%%%%%%%%%%%%%%%%%%%%%%%%%%%%%%%%%%%%%%%%%%%%%%%%%%%%%%%%%%%%%%%%%%%%%%%%%%%
%%%%%%%%%%%%%%%%%%%%%%%%%%%%%%%%%%%%%%%%%%%%%%%%%%%%%%%%%%%%%%%%%%%%%%%%%%%%%%%%%%%%%%%%%%%%%%%%%%%%%%%%%%%%%%%%%%%%%%%%%%%%%%%%%%%%%%%%%%%%%%%%%%
%%%%%%%%%%%%%%%%%%%%%%%%%%%%%%%%%%%%%%%%%%%%%%%%%%%%%%%%%%%%%%%%%%%%%%%%%%%%%%%%%%%%%%%%%%%%%%%%%%%%%%%%%%%%%%%%%%%%%%%%%%%%%%%%%%%%%%%%%%%%%%%%%%

\subsection{Derivation of the Jacobian with regard to the parameters}
\label{J_wrt_param}
In this section we provide the proofs for the Jacobian of the canonical divisive normalization, namely for the Eqs. \ref{jacobian2_gamma} - \ref{dNdsigma}.
%In this technical section the benefits of the matrix notation of divisive normalization,
%$\mathcal{N}^{(i)} = \mathds{D}_{\textrm{sign}(\vect{y}^i)} \cdot \mathds{D}_{\mathcal{D}^{(i)}(\vect{e}^i)}^{-1} \cdot \vect{e}^i$, Eq. \ref{divisive_norm2}, become evident.

%\paragraph*{Dependence on $K^i$: Proof of Eq. \ref{jacobian2_K}.} In Eq. \ref{divisive_norm2}, $\vect{y}^i$ and the denominator $D^{(i)}(\vect{e}^i)$ are independent on $K^i$, therefore the only term to take into account in deriving with regard to the matrix, $K^i$, is the vector $K^i \cdot \vect{e}^i$,
%\begin{equation}
%      \nabla_{\!\!K^i} \mathcal{N}^{(i)} = \mathds{D}_{\textrm{sign}(\vect{y}^i)} \cdot \mathds{D}^{-1}_{\mathcal{D}^{(i)}(\vect{e}^i)} \cdot
%      \frac{\partial K^i \cdot \vect{e}^i}{\partial K^i}
%      \nonumber
%\end{equation}
%Now, using the derivative of a linear function with regard to the matrix (the result in Eq. \ref{deriv_linear_param}),
%we have $\nabla_{\!\!K^i} K^i \cdot \vect{e}^i = \mathds{B}^{d_i}_{({\vect{e}^i}^\top)}$, and hence we get Eq. \ref{jacobian2_K}.
%Note that $\nabla_{\!\!K^i} \mathcal{N}^{(i)} \in \mathbb{R}^{d_i \times (d_i \times d_i)}$, as it should be, since there are $(d_i \times d_i)$ parameters in $K^i$.
%Note also that small perturbations of the response from perturbations in $K^i$ should be computed as: $\Delta \vect{x}^i = \nabla_{\!\!K^i} \mathcal{N}^{(i)} \cdot \textrm{vect}({\Delta K^i}^\top)$, where $\textrm{vect}(\cdot)$ is the column-wise rearrangement used in Eq. \ref{vectoriz}.

\paragraph*{Dependence on $\gamma^i$: Proof of Eq. \ref{jacobian2_gamma}.} The divisive normalization, Eq. \ref{divisive_norm2}, reproduced above depends on the exponent $\gamma^i$ through the vector of energies $\vect{e}^i = |\vect{y}^i|^{\gamma^i}$. Therefore,
\begin{equation}
      \nabla_{\!\!\gamma^i} \mathcal{N}^{(i)} = - \mathds{D}_{\textrm{sign}(\vect{y}^i)} \cdot \mathds{D}^{-2}_{\mathcal{D}^{(i)}(\vect{e}^i)} \cdot
      \frac{\partial \mathds{D}_{\mathcal{D}^{(i)}(\vect{e}^i)}}{\partial \gamma^i} \cdot \vect{e}^i + \mathds{D}_{\textrm{sign}(\vect{y}^i)} \cdot \mathds{D}^{-1}_{\mathcal{D}^{(i)}(\vect{e}^i)} \cdot \frac{\partial \vect{e}^i}{\partial \gamma^i}
      \label{eq_gamma_1}
\end{equation}
where, if we perturb $\gamma$, the vector in the diagonal gets perturbed, so,
\begin{equation}
      \frac{\partial \mathds{D}_{\mathcal{D}^{(i)}(\vect{e}^i)}}{\partial \gamma^i} = \mathds{D}_{\frac{\partial \mathcal{D}^{(i)}(\vect{e}^i)}{\partial \gamma^i}} = \mathds{D}_{\frac{\partial \mathcal{D}^{(i)}(\vect{e}^i)}{\partial \vect{e}^i} \cdot \frac{\partial \vect{e}^i}{\partial \gamma^i}}
      \label{eq_gamma_2}
\end{equation}
where the chain rule was also applied. Now, lets address the two derivatives in Eq. \ref{eq_gamma_2}.
First, given that $\mathcal{D}^{(i)}(\vect{e}^i) = \vect{b}^i + H^i \cdot \vect{e}^i$, the first term is,
\begin{equation}
      \frac{\partial \mathcal{D}^{(i)}(\vect{e}^i)}{\partial \vect{e}^i} = H^i
      \label{eq_gamma_3}
\end{equation}
Regarding the second term in Eq. \ref{eq_gamma_2}, $\nabla_{\!\!\gamma^i} \vect{e}^i$, by taking element-wise logarithms,
\begin{equation}
      \log \vect{e}^i = \gamma^i \log |\vect{y}^i|
      \label{eq_gamma_4}
\end{equation}
this is convenient because, on the one hand, following a reasoning similar to the one used in Eq. \ref{deriv_log_f}, namely chain rule and direct derivation, we have:
\begin{equation}
      \frac{\partial \vect{e}^i}{\partial \gamma^i} = \mathds{D}_{\vect{e}^i} \cdot \frac{\partial \log \vect{e}^i}{\partial \gamma^i} =  \mathds{D}_{\vect{e}^i} \cdot \log |\vect{y}^i|
      \label{eq_gamma_5}
\end{equation}
Now, plugging Eqs. \ref{eq_gamma_3} and \ref{eq_gamma_5} into Eq. \ref{eq_gamma_2} and this in Eq. \ref{eq_gamma_1}, we get Eq. \ref{jacobian2_gamma}.
Note that $\nabla_{\!\!\gamma^i} \mathcal{N}^{(i)} \in \mathbb{R}^{d_i \times 1}$ as it should be since $\gamma^i$ is scalar (a single parameter to be perturbed which affects to the set of $d_i$ responses in $\vect{x}^i$).

% Democracia matematica!
% Comparacion del segundo t�rmino de las 3 demos que hay en los apuntes.
% La segunda y la tercera coinciden (o estan bien, o estan igual de mal).
%
%DD = diag(randn(3,1));
%H = randn(3,3);
%K = randn(3,3);
%e = randn(3,1);
%ly = randn(3,1);
%
%% 1    MAL!
%DD*H*diag(K*e)*diag(ly)*e
%
%% 2    BIEN!
%diag(H*diag(e)*ly)*DD*K*e
%%DD*diag(H*diag(e)*ly)*K*e
%
%% 3    BIEN!
%DD*diag(K*e)*H*diag(e)*ly
%
% Escribo en Latex la segunda

\paragraph*{Dependence on $\vect{b}^i$: Proof of Eq. \ref{jacobian2_b}.} Before deriving with regard to $\vect{b}^i$, it is convenient to rearrange the terms in Eq. \ref{divisive_norm2} to put the only parameter depending on $\vect{b}^i$ (the denominator) at the end. As the Hadamard product is commutative we can rearrange terms, and using the diagonal matrix notation for the terms independent of $\vect{b}^i$,
\begin{equation}
      \mathcal{N}^{(i)} = \textrm{sign}(\vect{y}^i) \odot \vect{e^i} \odot \frac{1}{\mathcal{D}^{(i)}(\vect{e}^i)} = \mathds{D}_{\textrm{sign}(\vect{y}^i)} \cdot \mathds{D}_{\vect{e}^i} \cdot \frac{1}{\mathcal{D}^{(i)}(\vect{e}^i)}
      \label{rearranged_DN}
\end{equation}
Now, we derive the vector at the end (the Hadamard quotient with the denominator),
\begin{equation}
      \nabla_{\!\!\vect{b}^i} \mathcal{N}^{(i)} = \mathds{D}_{\textrm{sign}(\vect{y}^i)} \cdot \mathds{D}_{\vect{e}^i} \cdot \frac{\partial (\vect{b}^i + H^i \cdot \vect{e}^i)^{-1}}{\partial \vect{b}^i} = - \mathds{D}_{\textrm{sign}(\vect{y}^i)} \cdot \mathds{D}_{\vect{e}^i} \cdot \mathds{D}_{\mathcal{D}^{(i)}(\vect{e}^i)}^{-2} \cdot \mathbbm{1}
\end{equation}
which is Eq. \ref{jacobian2_b}. This Jacobian is a $d_i \times d_i$ matrix consistently with the dimensions of the output and of the vector $\vect{b}^i$.

\paragraph*{Dependence on general $H^i$: Proof of Eq. \ref{jacobian2_H}.} With the same rearrangement done in Eq. \ref{rearranged_DN} to leave the denominator in the end, we take derivatives with regard to $H^i$,
\begin{equation}
      \nabla_{\!\!H^i} \mathcal{N}^{(i)} = \mathds{D}_{\textrm{sign}(\vect{y}^i)} \cdot \mathds{D}_{\vect{e}^i} \cdot \frac{\partial (\vect{b}^i + H^i \cdot \vect{e}^i)^{-1}}{\partial H^i} = - \mathds{D}_{\textrm{sign}(\vect{y}^i)} \cdot \mathds{D}_{\vect{e}^i} \cdot \mathds{D}_{\mathcal{D}^{(i)}(\vect{e}^i)}^{-2} \cdot \frac{\partial H^i \cdot \vect{e}^i}{\partial H^i}
\end{equation}
where, again, we apply the block diagonal result for the derivative of a linear function with regard to the parameters, Eq.\ref{deriv_linear_param}, leading to $\mathds{B}^{d_i}_{{\vect{e}^i}^\top}$ and hence obtaining Eq.\ref{jacobian2_H}.
Given this final block-diagonal term this Jacobian is a $d_i \times (d_i \times d_i)$ matrix, consistently with the number of parameters than may vary in $H^i$.
Note also that
%, similarly to the case of $K^i$,
small perturbations of the response from perturbations in $H^i$ should be computed using is the column-wise rearrangement for $H^i$ defined in Eq. \ref{vectoriz}.

\paragraph*{Dependence on parametric kernels: Proof of Eqs. \ref{dNdsigma} and \ref{dNdc}.} %and \ref{jacobian2_K_gauss}.}
% Lets start with the parametric $H^i$ since the same kind of reasoning applies to $K^i$.
Considering the Gaussian parametrization, Eq. \ref{GaussianH}, the derivative with regard to the width of the $k$-th row is,
\begin{equation}
      \frac{\partial \mathcal{N}^{(i)}}{\partial \sigma^i_k} = \frac{\partial \mathcal{N}^{(i)}}{\partial H^i} \cdot \frac{\partial H^i}{\partial H^i_{k \star}} \cdot \frac{\partial H^i_{k \star}}{\partial \sigma^i_k}
\end{equation}
%The size of the Jacobian reduces dramatically because $\frac{\partial H^i}{\partial H^i_{k \star}}$ selects a single submatrix in the block-diagonal matrix,
%$\mathds{B}^{d_i}_{\vect{e}^i}$, that is in $\frac{\partial \mathcal{N}^{(i)}}{\partial H^i}$.
%This is because
The matrix that transmits the variations in a single row of the kernel into the whole vector-rearranged kernel,
$\Delta \, \textrm{vect}({H^i}^\top) = \frac{\partial H^i}{\partial H^i_{k \star}} \cdot \Delta {H^i_{k \star}}^\top$,
has to be formed by $d_i-1$ replicas of the zero matrix, with the identity matrix, $\mathbbm{1}$, in the $k$-th location,
\begin{equation}
       \frac{\partial H^i}{\partial H^i_{k \star}} = \left(\begin{array}{c}
                  \emptyset_{d_i \times d_i} \\
                  \emptyset_{d_i \times d_i} \\
                  \vdots \\
                  \emptyset_{d_i \times d_i} \\[0.2cm]
                  \mathbbm{1}_{d_i \times d_i} \\[0.2cm]
                  \emptyset_{d_i \times d_i} \\
                  \vdots \\
                  \emptyset_{d_i \times d_i}
                  \end{array}
                  \right)\\
                  \,\,\,
                  \begin{array}{l}
                  {\textrm{\tiny{1st submatrix}}} \\
                  {\textrm{\tiny{2nd submatrix}}} \\
                  \vdots \\
                  %\small{\textrm{\tiny{\emph{(k-1)}-th submatrix}}}
                  \\[0.2cm]
                  {\textrm{\tiny{\emph{k}-th submatrix}}} \\[0.2cm]
                  %\small{\textrm{\tiny{\emph{(k+1)}-th submatrix}}} \\
                  \\
                  \vdots \\
                  \textrm{\tiny{$d_i$-th submatrix}}
                  \end{array}
       \label{replica_zeros}
\end{equation}
This derivative, $\frac{\partial H^i}{\partial H^i_{k \star}}$, selects a single submatrix in the huge block-diagonal matrix,
$\mathds{B}^{d_i}_{\vect{e}^i}$, that is in $\frac{\partial \mathcal{N}^{(i)}}{\partial H^i}$, Eq. \ref{jacobian2_H}.
As a result, this product leads to a square matrix where the only nonzero row is in the $k$-th location,
\begin{equation}
       \mathds{B}^{d_i}_{\vect{e}^i} \cdot \frac{\partial H^i}{\partial H^i_{k \star}} = \left(\begin{array}{c}
                  \emptyset_{1 \times d_i} \\
                  \emptyset_{1 \times d_i} \\
                  \vdots \\
                  \emptyset_{1 \times d_i} \\[0.1cm]
                  {\vect{e}^i}^\top \\[0.1cm]
                  \emptyset_{1 \times d_i} \\
                  \vdots \\
                  \emptyset_{1 \times d_i}
                  \end{array}
                  \right)\\
                  \,\,\,
                  \begin{array}{l}
                  {\textrm{\tiny{1st row}}} \\
                  {\textrm{\tiny{2nd row}}} \\
                  \vdots \\
                  %\small{\textrm{\tiny{\emph{(k-1)}-th submatrix}}}
                  \\[0.2cm]
                  {\textrm{\tiny{\emph{k}-th row}}} \\[0.2cm]
                  %\small{\textrm{\tiny{\emph{(k+1)}-th submatrix}}} \\
                  \\
                  \vdots \\
                  \textrm{\tiny{$d_i$-th row}}
                  \end{array}
       \label{producto}
\end{equation}
When multiplying the above matrix by the extra diagonal matrix in
$\frac{\partial \mathcal{N}^{(i)}}{\partial H^i}$, Eq. \ref{jacobian2_H},
we have the only nonzero row scaled by the $k$-th component of the diagonal of
$- \mathds{D}_{\textrm{sign}(\vect{y}^i)} \cdot  \mathds{D}_{\vect{e}^i} \cdot \mathds{D}^{-2}_{\mathcal{D}^{(i)}(\vect{e}^i)}$, i.e.
\begin{equation}
       \frac{\partial \mathcal{N}^{(i)}}{\partial \sigma^i_k} =
       \left(
       \begin{array}{c}
                  \emptyset_{1 \times d_i} \\
                  \emptyset_{1 \times d_i} \\
                  \vdots \\
                  \emptyset_{1 \times d_i} \\[0.1cm]
                  - \textrm{sign}(y^i_k) \,\, e^i_k \,\, {\mathcal{D}^{(i)}}^{-2}_k  \,\,\,\, {\vect{e}^i}^\top \\[0.1cm]
                  \emptyset_{1 \times d_i} \\
                  \vdots \\
                  \emptyset_{1 \times d_i}
       \end{array}
       \right)
                  \cdot
                  \frac{\partial H^i_{k \star}}{\partial \sigma^i_k}
       \label{single_row}
\end{equation}
Now, taking into account the straightforward derivative of a normalized 2D Gaussian with regard to its width,
\begin{equation}
     \frac{\partial}{\partial \sigma} \left[ c \frac{dp^2}{2 \pi \, \sigma^2} \, \exp\left(-\frac{\Delta^2_{k k'}}{2 \, \sigma^2}\right)
     \right]
     \,\,\,=\,\,\,
     c \frac{dp^2}{2 \pi \, \sigma^5} \left(\Delta^2_{k k'} - 2 \, \sigma^2 \right) \, \exp\left(-\frac{\Delta^2_{k k'}}{2 \, \sigma^2}\right)
     \nonumber
\end{equation}
%\begin{equation}
%     \frac{\partial}{\partial \sigma} \Big[ \frac{1}{\sigma \sqrt{\pi}} \, \exp\left(-\frac{(k - j)^2}{\sigma^2}\right)
%     \Big]
%     \,\,\,=\,\,\,
%     \frac{2 \,\, (k - j)^2 - \sigma^2}{\sigma^4 \,\, \sqrt{\pi}} \, \exp\left(-\frac{(k - j)^2}{\sigma^2}\right)
%     \nonumber
%\end{equation}
and considering that this holds $\forall k' \in \textrm{subband} \, k$, the considered $\frac{\partial H^i_{k \star}}{\partial \sigma^i_k}$ is a column vector:
\begin{equation}
       \nabla_{\!\!\sigma^i_k} \mathcal{N}^{(i)} =
       \left(
       \begin{array}{c}
                  \emptyset_{1 \times d_i} \\
                  \emptyset_{1 \times d_i} \\
                  \vdots \\
                  \emptyset_{1 \times d_i} \\[0.1cm]
                  - \textrm{sign}(y^i_k) \,\, e^i_k \,\, {\mathcal{D}^{(i)}}^{-2}_k  \,\,\,\, {\vect{e}^i}^\top \\[0.1cm]
                  \emptyset_{1 \times d_i} \\
                  \vdots \\[0.25cm]
                  \emptyset_{1 \times d_i}
       \end{array}
       \right)
                  \cdot
       \left(
       \begin{array}{c}
                  F^i_{k 1} \\
                  F^i_{k 2}  \\
                  F^i_{k 3}  \\[0.9cm]
                  \vdots \\[0.9cm]
                  F^i_{k d_i}
       \end{array}
       \right)
      \label{single_element}
\end{equation}
where
\begin{equation}
F^i_{k k'} = \left \{
\begin{array}{lc}
0 & \,\, \forall  \,\, k' \notin \textrm{subband} \,\,\, k \\
\\
c^i_k \, \frac{dp_{k1} dp_{k2}}{2 \pi \,\, {\sigma^i_k}^5} \,\, \left(  \Delta^2_{k k'} - 2 \, {\sigma^i_k}^2 \right) \,\, e^{-\frac{  \Delta^2_{k k'} }{2 \,\, {\sigma^i_k}^2}} & \forall \,\,  k' \in \textrm{subband} \,\,\, k
\end{array}
\right.
\nonumber
\end{equation}

Note that the matrix-on-vector product in Eq. \ref{single_element} is a vector with a single nonzero element (the one in the $k$-th location).
Therefore, the derivatives with regard to all the widths can be expressed in a single matrix expression by replicating the
transposed vector ${\vect{e}^i}^\top$ in $d_i$ rows, and by stacking the different column vectors $F^i_{k k'}$ $\forall \, k$. Then, the elements we are looking for are in the diagonal of the resulting matrix.
This leads to the diagonal matrix in Eq. \ref{dNdsigma}.

In the case of the derivative with regard to the amplitudes of the Gaussians, Eq. \ref{single_row} also holds $\forall \, k$ with the corresponding change of variables.
Then, taking into account that the remaining derivative with regard to the amplitude is simply the Gaussian,
and the consideration of all coefficients $k$ stacked leads to Eq. \ref{dNdc}.

%%%%%%%%%%%%%%%%%%%%%%%%%%%%%%%%%%%%%%%%%%%%%%%%%%%%%%%%%%%%%%%%%%%%%%%%%  END OF 5.4 JACOB WRT PARAMETERS
%%%%%%%%%%%%%%%%%%%%%%%%%%%%%%%%%%%%%%%%%%%%%%%%%%%%%%%%%%%%%%%%%%%%%%%%%%%%%%%%%%%%%%%%%%%%%%%%%%%%%%%%%%%%%%%%%%%%%%%%%%%%%%%%%%%%%%%%%%%%%%%%%%
%%%%%%%%%%%%%%%%%%%%%%%%%%%%%%%%%%%%%%%%%%%%%%%%%%%%%%%%%%%%%%%%%%%%%%%%%%%%%%%%%%%%%%%%%%%%%%%%%%%%%%%%%%%%%%%%%%%%%%%%%%%%%%%%%%%%%%%%%%%%%%%%%%
%%%%%%%%%%%%%%%%%%%%%%%%%%%%%%%%%%%%%%%%%%%%%%%%%%%%%%%%%%%%%%%%%%%%%%%%%%%%%%%%%%%%%%%%%%%%%%%%%%%%%%%%%%%%%%%%%%%%%%%%%%%%%%%%%%%%%%%%%%%%%%%%%%
%%%%%%%%%%%%%%%%%%%%%%%%%%%%%%%%%%%%%%%%%%%%%%%%%%%%%%%%%%%%%%%%%%%%%%%%%%%%%%%%%%%%%%%%%%%%%%%%%%%%%%%%%%%%%%%%%%%%%%%%%%%%%%%%%%%%%%%%%%%%%%%%%%

\subsection{Derivation of the inverse}
\label{proof_inverse}
%The Hadamard quotient in the all-positive normalized vector, $N$, in Eq. \ref{divisive_norm1}, implies that the denominator in,
%\begin{equation}
%      |\vect{x}^i| = \frac{\vect{e}^i}{\vect{b}^i + H^{i} \cdot \vect{e}^i} \nonumber
%\end{equation}
%can be written in the left hand side as a Hadamard product,
From Eq. \ref{divisive_norm2}, the vector of absolute values of the responses is $|\vect{x}^i| = \mathds{D}^{-1}_{\left( \vect{b}^i + H^i \cdot \vect{e}^i \right)} \cdot \vect{e}^i$. Therefore, inverting the matrix, it holds,
\begin{equation}
       \mathds{D}_{\left(\vect{b}^i + H^{i} \cdot \vect{e}^i \right)} \cdot |\vect{x}^i| = \vect{e}^i \nonumber
\end{equation}
which can be written using the Hadamard product,
\begin{equation}
       (\vect{b}^i + H^{i} \cdot \vect{e}^i) \odot |\vect{x}^i| = \vect{e}^i \nonumber
\end{equation}
Now, using the diagonal matrix notation of Hadamard products and the fact that $\mathds{D}_{\vect{a}} \cdot \vect{b} = \mathds{D}_{\vect{b}} \cdot \vect{a}$, we have:
\begin{gather*}
      \mathds{D}_{\vect{b}^i} \cdot |\vect{x}^i| + \mathds{D}_{|\vect{x}^i|} \cdot H^{i} \cdot \vect{e}^i = \vect{e}^i \\
      \mathds{D}_{\vect{b}^i} \cdot |\vect{x}^i| = \left( \mathbbm{1} - \mathds{D}_{|\vect{x}^i|} \cdot H^{i} \right) \cdot \vect{e}^i \\
      \vect{e}^i = \left( \mathbbm{1} -  \mathds{D}_{|\vect{x}^i|} \cdot H^{i} \right)^{-1} \cdot D_{\vect{b}^i} \cdot |\vect{x}^i|
\end{gather*}
and considering that $\vect{e}^i = |\vect{y}^i|^{\gamma^i}$, and that $\vect{y}^i$ inherits the sign from $\vect{y}^i$, it follows Eq. \ref{inv_DN}.

%%%%%%%%%%%%%%%%%%%%%%%%%%%%%%%%%%%%%%%%%%%%%%%%%%%%%%%%%%%%%%%%%%%%%%%%%  END OF 5.5 INVERSE
%%%%%%%%%%%%%%%%%%%%%%%%%%%%%%%%%%%%%%%%%%%%%%%%%%%%%%%%%%%%%%%%%%%%%%%%%%%%%%%%%%%%%%%%%%%%%%%%%%%%%%%%%%%%%%%%%%%%%%%%%%%%%%%%%%%%%%%%%%%%%%%%%%
%%%%%%%%%%%%%%%%%%%%%%%%%%%%%%%%%%%%%%%%%%%%%%%%%%%%%%%%%%%%%%%%%%%%%%%%%%%%%%%%%%%%%%%%%%%%%%%%%%%%%%%%%%%%%%%%%%%%%%%%%%%%%%%%%%%%%%%%%%%%%%%%%%
%%%%%%%%%%%%%%%%%%%%%%%%%%%%%%%%%%%%%%%%%%%%%%%%%%%%%%%%%%%%%%%%%%%%%%%%%%%%%%%%%%%%%%%%%%%%%%%%%%%%%%%%%%%%%%%%%%%%%%%%%%%%%%%%%%%%%%%%%%%%%%%%%%
%%%%%%%%%%%%%%%%%%%%%%%%%%%%%%%%%%%%%%%%%%%%%%%%%%%%%%%%%%%%%%%%%%%%%%%%%%%%%%%%%%%%%%%%%%%%%%%%%%%%%%%%%%%%%%%%%%%%%%%%%%%%%%%%%%%%%%%%%%%%%%%%%%

\subsection{Region-based approach to MAximum Differentiation}
\label{summation_in_mad}

Analytic MAximum Differentiation involves computing the eigenvectors of the metric matrix based on the Jacobian (Eq. \ref{distance2}). This matrix is huge even for moderate size images and one would like to break this problem into smaller pieces. Here we consider how a region-based strategy may affect MAD.

%In this section we consider the issue of breaking the hard problem of processing large images (e.g. subtending several degrees)
%into easier subproblems consisting of processing smaller regions of the images (e.g. each subtending about one degree).
%Note that dimensionality is a major problem. A one-degree image sampled at the rate suggested in Section \ref{general_cons} (80 cycles/degree) implies that the dimension of luminance vectors is $d_0=6400$. Moreover, assuming
%that the last representation for V1-like sensors is a reasonable redundant wavelet (e.g. steerable filters with 3 scales and 4 orientations \cite{Simoncelli92}) the dimension escalates up to $d_n=40100$. This implies dealing with \emph{huge} kernels $\in \mathbb{R}^{40100 \times 40100}$ for input images of $80 \times 80$ pixels. Real images require breaking the problem into smaller pieces.

The use of patch-wise strategies implies assuming that perception of a patch is independent of the content of neighbor patches. Caution has to be taken with such an assumption.
In fact, the main reason to include the nonlinear stages is accounting for the masking effect of the neighbor sensors tuned to surrounding locations.
The scale of the local interactions between the sensors
(e.g. the width of the connections in $L^i$ and the width of the interaction in $H^i$)
will induce edge effects if patch size is limited.
Therefore, this is the length scale that must be taken into account in order to break the problem into pieces.
For example, in the model used in the Discussion the wider interaction has a length scale of 0.2 degrees
(see the specific values in the Toolbox, Supplem. Material 8).

Here, we point out that assuming region-independence in perception implies certain structure in $\nabla_{\vect{x}^0} S$.
This structure has consequences on the metric matrix, which leads to simplified solutions of MAD,
which are also consistent with the original region-independence assumption and with the quadratic summation assumption.

Imagine that large images are formed by $N$ distinct spatial blocks. In this case, data corresponding to the difference between
two images can be arranged in a column vector by stacking the vectors corresponding to the $N$ distinct regions:
$\Delta \vect{x}^0 = \left( {\Delta \vect{x}^0_{[1]}}^\top  \, \Delta {\vect{x}^0_{[2]}}^\top \, \cdots \Delta {\vect{x}^0_{[N]}}^\top \right)^\top$.
Assuming that these distinct regions are perceptually independent, implies a block-diagonal structure in $\nabla_{\vect{x}^0} S$, because the variation of the responses corresponding to the $i$-th region should not depend on the variation of the inputs for the $j$-th region:
\begin{equation}
                                         \left(\begin{array}{c}
                                                  \Delta \vect{x}^n_{[1]}\\[0.1cm]
                                                  \Delta \vect{x}^n_{[2]}\\[0.2cm]
                                                     \vdots \\[0.2cm]
                                                  \Delta \vect{x}^n_{[N]}\\
                                         \end{array}
                                         \right)
                                         =
                                         \left(\begin{array}{cccc}
                                                  \nabla_{\vect{x}^0_{[1]}} S(\vect{x}^0_{[1]}) &         \emptyset_{d_n \times d_0}    & \cdots &   \emptyset_{d_n \times d_0}    \\[0.1cm]
                                                  \emptyset_{d_n \times d_0}            & \nabla_{\vect{x}^0_{[2]}} S(\vect{x}^0_{[2]}) & \cdots &   \emptyset_{d_n \times d_0}    \\[0.2cm]
                                                               \vdots                   &            \vdots                     & \ddots &    \vdots \\[0.2cm]
                                                  \emptyset_{d_n \times d_0}            &       \emptyset_{d_n \times d_0}      & \cdots & \nabla_{\vect{x}^0_{[N]}} S(\vect{x}^0_{[N]})\\
                                         \end{array}\right)
                                         \cdot
                                          \left(\begin{array}{c}
                                                  \Delta \vect{x}^0_{[1]}\\[0.1cm]
                                                  \Delta \vect{x}^0_{[2]}\\[0.2cm]
                                                     \vdots \\[0.2cm]
                                                  \Delta \vect{x}^0_{[N]}\\
                                         \end{array}
                                         \right)
  \label{stakazo}
\end{equation}
where each (relatively small) rectangular sub-matrix, $\nabla_{\vect{x}^0_{[i]}} S(\vect{x}^0_{[i]})$, describes the behavior for the $i$-th region.

In this situation, in the 2nd-order (or local-linear) approximation, Eq. \ref{distance2}, the perceptual difference induced by the large $\Delta \vect{x}^0$ would be given by $\mathbb{d}^2_p = \Delta {\vect{x}^0}^\top \cdot \nabla_{\vect{x}^0} S(\vect{x}^0)^\top \cdot \nabla_{\vect{x}^0} S(\vect{x}^0) \cdot \Delta {\vect{x}^0}$, and using the block diagonal structure in Eq. \ref{stakazo}, one has:
\begin{equation}
    \hspace{-1cm}
    \mathbb{d}^2_p   = \begin{array}{c}
                                    \left( \Delta {\vect{x}^0_{[1]}}^\top \, \Delta {\vect{x}^0_{[2]}}^\top \cdots \Delta {\vect{x}^0_{[2]}}^\top \right) \cdot \\[1.75cm]
                            \end{array}
                            \left(\begin{array}{cccc}
                                                  M(\vect{x}^0_{[1]}) &         \emptyset_{d_0 \times d_0}    & \cdots &   \emptyset_{d_0 \times d_0}    \\[0.1cm]
                                                  \emptyset_{d_0 \times d_0}            & M(\vect{x}^0_{[2]}) & \cdots &   \emptyset_{d_0 \times d_0}    \\[0.2cm]
                                                               \vdots                   &            \vdots                     & \ddots &    \vdots \\[0.2cm]
                                                  \emptyset_{d_0 \times d_0}            &       \emptyset_{d_0 \times d_0}      & \cdots & M(\vect{x}^0_{[N]})\\
                                         \end{array}\right)
                                         \cdot
                                          \left(\begin{array}{c}
                                                  \Delta \vect{x}^0_{[1]}\\[0.1cm]
                                                  \Delta \vect{x}^0_{[2]}\\[0.2cm]
                                                     \vdots \\[0.2cm]
                                                  \Delta \vect{x}^0_{[N]}\\
                                         \end{array}
                                         \right)
  \label{metrica_stakazo}
\end{equation}
where each (relatively small) square sub-matrix $M(\vect{x}^0_{[i]}) = \nabla_{\vect{x}^0_{[i]}} S(\vect{x}^0_{[i]})^\top \cdot \nabla_{\vect{x}^0_{[i]}} S(\vect{x}^0_{[i]})$ is the metric matrix corresponding to the $i$-th region.

\begin{figure}[!t]
    \vspace{-0.5cm}
	\centering
    \small
    \setlength{\tabcolsep}{2pt}
    \begin{tabular}{cc}
       %\hspace{-4cm} \includegraphics[width=4.6cm,height=4.6cm]{Figs/original.PNG} \\[4mm]
       %\hspace{-0cm} \includegraphics[width=12.5cm,height=12.5cm]{Figs/mad_analytic_80c.PNG} \\
       \hspace{-3cm} \includegraphics[width=7.5cm,height=7.5cm]{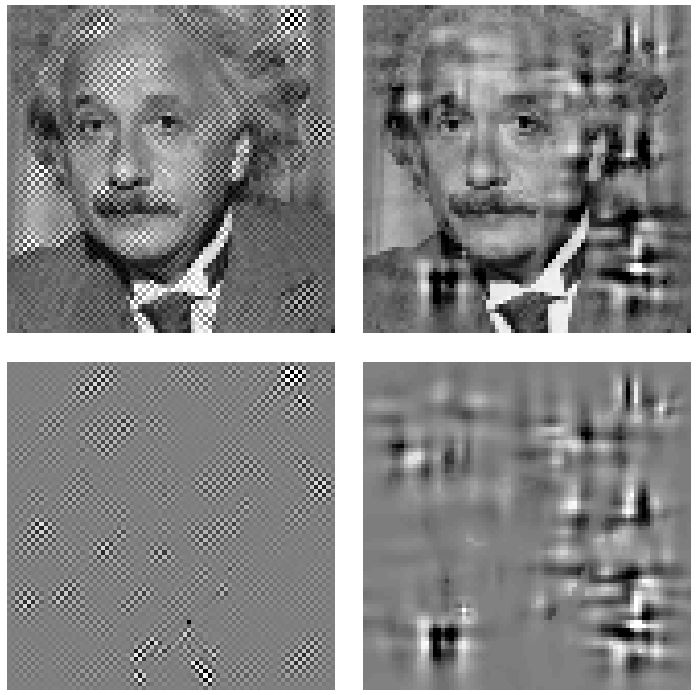} \hspace{0.25cm} & \hspace{0.25cm}
       \includegraphics[width=7.5cm,height=7.5cm]{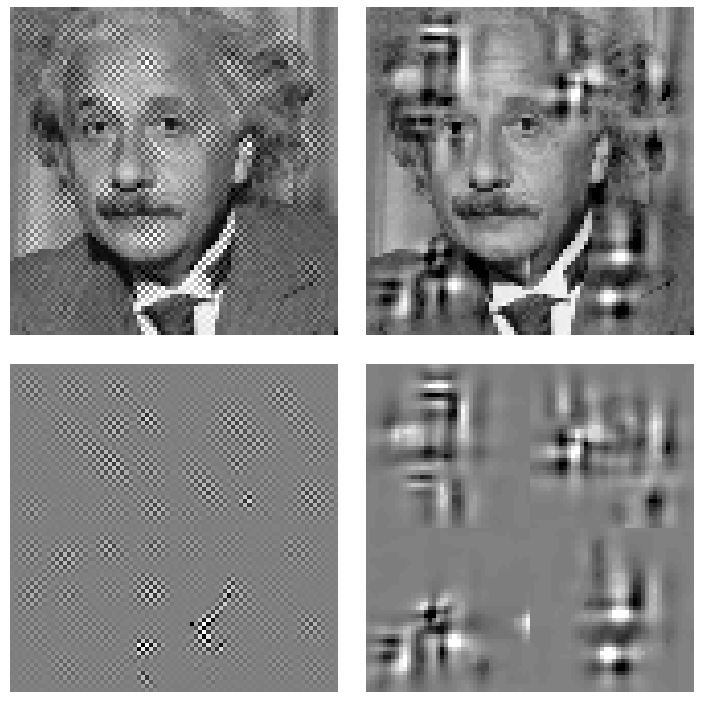} \\
%       \hspace{-3cm} (a)  \hspace{0.25cm} & \hspace{0.25cm} (b)
    \end{tabular}
    \vspace{-0.0cm}
	\hspace{-0cm}\caption{\scriptsize{\textbf{Global vs region-based MAD computation.} \emph{Left:} best and worst MAD-analytic images computed from an image subtending 1.25 deg. \emph{Right:} best and worst MAD-analytic images computed from 4 subimages subtending 0.63 deg each.}}
    \label{MADblocks}
    \vspace{-0.15cm}
\end{figure}

Eq. \ref{metrica_stakazo} has one interesting consequence for its use in Maximum Differentiation:
since the eigenvectors of a (large) block-diagonal matrix can be computed from the eigenvectors of the
(smaller) blocks in the diagonal \cite{Golub}, there is no need to explicitly build and work with the huge matrices
in Eqs. \ref{stakazo} and \ref{metrica_stakazo}.
This relieving mathematical result is completely consistent with intuition:
if perceptual independence is assumed (and responses are computed block-wise), then,
large MAD images can be also computed block-wise from the eigenvectors of the smaller metric matrices corresponding to each image block.

Figure \ref{MADblocks} shows an example of the comparison of block-wise versus global approaches to compute MAD images.
In both best and worst images the energy is focused roughly in the same spatial regions with the same frequency content. Qualitative behavior is similar meaning that edge effects are negligible. %Therefore, it is computationally sensible to take a block-wise approach.
\blue{The above warning about assuming region-independence in perception implies that,
strictly speaking, one cannot ensure that patch-wise computation of eigendistortions will not modify the judgement and the interpretation of MAD. However, as suggested by the example in Fig. \ref{MADblocks}, the response to the MAD question is fairly independent of block size given \emph{big-enough} regions. How big is this is an experimental issue that would probably depend on the specific length-scale of the parameter that one is addressing in the MAD experiment.
Qualitative check of edge effects at different sizes may be necessary before launching the actual experiment to choose a small, yet safe, block size.}

A lateral consequence of Eq. \ref{metrica_stakazo} is related to the summation of distortions across the visual field.
Note that the total distortion is the quadratic sum of individual distortions in the regions of the image,
$\mathbb{d}^2_p = \sum_{i=1}^{N} \Delta {\vect{x}^0_{[i]}}^\top \cdot M(\vect{x}^0_{[i]}) \cdot \Delta {\vect{x}^0_{[i]}} = \sum_{i=1}^{N} {\mathbb{d}^{[i]}_p}^2$. This quadratic summation is consistent with the quadratic norm chosen for the summation over individual response elements.

%%%%%%%%%%%%%%%%%%%%%%%%%%%%%%%%%%%%%%%%%%%%%%%%%%%%%%%%%%%%%%%%%%%%%%%%%  END OF 5.6 REGION-BASED MAD
%%%%%%%%%%%%%%%%%%%%%%%%%%%%%%%%%%%%%%%%%%%%%%%%%%%%%%%%%%%%%%%%%%%%%%%%%%%%%%%%%%%%%%%%%%%%%%%%%%%%%%%%%%%%%%%%%%%%%%%%%%%%%%%%%%%%%%%%%%%%%%%%%%
%%%%%%%%%%%%%%%%%%%%%%%%%%%%%%%%%%%%%%%%%%%%%%%%%%%%%%%%%%%%%%%%%%%%%%%%%%%%%%%%%%%%%%%%%%%%%%%%%%%%%%%%%%%%%%%%%%%%%%%%%%%%%%%%%%%%%%%%%%%%%%%%%%
%%%%%%%%%%%%%%%%%%%%%%%%%%%%%%%%%%%%%%%%%%%%%%%%%%%%%%%%%%%%%%%%%%%%%%%%%%%%%%%%%%%%%%%%%%%%%%%%%%%%%%%%%%%%%%%%%%%%%%%%%%%%%%%%%%%%%%%%%%%%%%%%%%
%%%%%%%%%%%%%%%%%%%%%%%%%%%%%%%%%%%%%%%%%%%%%%%%%%%%%%%%%%%%%%%%%%%%%%%%%%%%%%%%%%%%%%%%%%%%%%%%%%%%%%%%%%%%%%%%%%%%%%%%%%%%%%%%%%%%%%%%%%%%%%%%%%

\subsection{Maximization of correlation with subjective opinion}
\label{J_correl}
The linear correlation between the ground truth, $\vect{M}$ and the model predictions $\vect{D}$ is,
\begin{equation}
      \varrho = \frac{E(  (\vect{M} - \vect{\bar{M}}) (\vect{D}-\vect{\bar{D}})  ) }{\sigma(\vect{M}) \sigma(\vect{D})}
      \label{correlation}
\end{equation}
where $E(\cdot)$ stands for expected value,  $\vect{\bar{v}}$  stands for the average of vector $\vect{v}$, and $\sigma(\cdot)$ stands for standard deviation.

However, there is a more convenient expression to apply the chain rule in the derivatives with regard to the parameters.
Note that the sums in the average to compute $E(\cdot)$ and $\sigma(\cdot)$ can be written as dot products and norms,
and subtraction of the mean can be written as $\vect{M_s}=\vect{M}-\frac{1}{N}\cdot\mathds{1}\cdot\vect{M}$, where $\mathds{1}$ is an all-ones matrix.
As a result, the linear correlation can be written as,
\begin{equation}
      \varrho = \frac{\vect{M_s}^T \cdot \vect{D_s}}{|\vect{M_s}| |\vect{D_s}|}
      \label{correlation2}
\end{equation}

In this way, the derivatives of $\varrho(\vect{\vect{\Theta}})$ with regard to $\vect{\vect{\Theta}}$:
\begin{equation}
\frac{d \varrho}{d \vect{\vect{\Theta}}} =\underset{(1)}{\frac{d \varrho}{d \vect{D_s}}} \cdot\underset{(2)}{ \frac{d \vect{D_s}}{d\vect{\vect{\Theta}}_{}}} \;\;\;\;\;\;\;\;\;\;\;\; \text{where} \;\; \left\{\begin{array}{c}
  \frac{d \varrho}{d \vect{D_s}}  \in \mathbb{R}^{1 \times N}\\
  \;\;\;\\
 \frac{d \vect{D_s}}{d \vect{\vect{\Theta}}}  \in \mathbb{R}^{N \times m}\\
    \end{array}  \right.
    \label{deri_rho}
\end{equation}

Developing the term $(1)$

\begin{align}
\frac{d \varrho}{d \vect{D_s}} =
&\frac{d}{d \vect{D_s}} \left[ \frac{\vect{M_s}^T \vect{D_s}}{|\vect{M_s}|\cdot|\vect{D_s}|} \right] \nonumber \\ \nonumber
=& \frac{\vect{M_s}^T}{|\vect{M_s}|\cdot|\vect{D_s}|} + \frac{\vect{M_s}^T \vect{D_s}}{|\vect{M_s}|}
\cdot \frac{d}{d \vect{D_s}} (\vect{D_s}^T \cdot \vect{D_s})^{-1/2} \\ \nonumber
=& \frac{\vect{M_s}^T}{|\vect{M_s}|\cdot|\vect{D_s}|}- \frac{1}{2}\frac{\vect{M_s}^T \vect{D_s}}{|\vect{M_s}|}
\cdot  (\vect{D_s}^T \cdot \vect{D_s})^{-3/2}  \cdot \frac{d }{d \vect{D_s}}  (\vect{D_s}^T \cdot \vect{D_s})\\ \nonumber
=& \frac{\vect{M_s}^T}{|\vect{M_s}|\cdot|\vect{D_s}|}-\frac{\vect{M_s}^T \vect{D_s}}{|\vect{M_s}|}
\cdot  \frac{1 }{|\vect{D_s}|^3} \cdot \vect{D_s}^T\\
=& \frac{\vect{M_s}^T}{|\vect{M_s}|\cdot|\vect{D_s}|}-\frac{\vect{M_s}^T\vect{D_s}}{|\vect{M_s}|\cdot|\vect{D_s}|^3}\cdot \vect{D_s}^T
\label{termino1}
\end{align}

\vspace{0.2cm}
Developing the term $(2)$
\begin{align}
 \frac{d \vect{D_s}}{d\vect{\vect{\Theta}}}= &
  \frac{d}{ d \vect{\vect{\Theta}}} \left( \vect{D}- \frac{1}{N}\mathds{1} \vect{D} \right)
  =\frac{d \vect{D}}{ d \vect{\vect{\Theta}}}  - \frac{1}{N} \mathds{1} \frac{d \vect{D}}{d\vect{\vect{\Theta}}} \label{termino21}\\ \nonumber
  {}&{} \\\nonumber
\text{where},&{}\\ \nonumber
&\frac{d \vect{D}}{d\vect{\vect{\Theta}}}=\left(\begin{array}{c}
   \;\;\; \frac{\partial \mathbb{d}_p^{[1]}}{\partial \vect{\vect{\Theta}}} \;\;\;\\[0.2cm]
    \;\;\; \frac{\partial \mathbb{d}_p^{[2]}}{\partial \vect{\vect{\Theta}}} \;\;\;\\
    \vdots \\[0.2cm]
    \;\;\;  \frac{\partial \mathbb{d}_p^{[N]}}{\partial \vect{\vect{\Theta}}}  \;\;\;\\
  \end{array}  \right), \\ \nonumber
\end{align}

\begin{align}
% {}&{} \\ \nonumber
\text{where, for the i-th image},&{} \nonumber \\  \nonumber
 {}&{} \\  \nonumber
\frac{\partial \mathbb{d}_p^{[i]}}{\partial \vect{\vect{\Theta}}}=&  \frac{\partial \left(\Delta {\vect{x}^n_{[i]}}^\top  \cdot \Delta \vect{x}^n_{[i]} \right)^{1/2} }{\partial \vect{\Theta}}= \frac{1}{2 \mathbb{d}_p^{[i]}} \cdot \frac{\partial \left(\Delta {\vect{x}^n_{[i]}}^\top \cdot  \Delta \vect{x}^n_{[i]} \right) }{\partial \vect{\Theta}} \\
=&\frac{1}{2 \mathbb{d}_p^{[i]}} 2 \cdot\Delta {\vect{x}^n_{[i]}}^\top   \cdot\frac{\partial \Delta \vect{x}^n_{[i]} }{\partial \vect{\Theta}} \nonumber \\ \nonumber
=&\frac{1}{\mathbb{d}_p^{[i]}}  \cdot\Delta {\vect{x}^n_{[i]}}^\top   \cdot\frac{\partial (\vect{z}^n_{[i]}-\vect{x}^n_{[i]}) }{\partial \vect{\Theta}}\\ \nonumber=&\frac{1}{\mathbb{d}_p^{[i]}}  \cdot \Delta {\vect{x}^n_{[i]}}^\top \cdot \left( \frac{\partial \vect{z}^n_{[i]}}{\partial \vect{\Theta}}- \frac{\partial \vect{x}^n_{[i]}}{\partial \vect{\Theta}} \right)\\
=&\frac{1}{\mathbb{d}_p^{[i]}}  \cdot\Delta {\vect{x}^n_{[i]}}^\top \cdot \left[ \nabla_{\vect{\Theta}} S(\vect{z}^0_{[i]}) - \nabla_{\vect{\Theta}} S(\vect{x}^0_{[i]}) \right]
\label{termino22}
\end{align}
Plugging Eqs. \ref{termino1}-\ref{termino22} into Eq. \ref{deri_rho}, one gets Eq. \ref{deriv_correl}.

%%%%%%%%%%%%%%%%%%%%%%%%%%%%%%%%%%%%%%%%%%%%%%%%%%%%%%%%%%%%%%%%%%%%%%%%%  END OF 5.7 OPTIMIZATION
%%%%%%%%%%%%%%%%%%%%%%%%%%%%%%%%%%%%%%%%%%%%%%%%%%%%%%%%%%%%%%%%%%%%%%%%%%%%%%%%%%%%%%%%%%%%%%%%%%%%%%%%%%%%%%%%%%%%%%%%%%%%%%%%%%%%%%%%%%%%%%%%%%
%%%%%%%%%%%%%%%%%%%%%%%%%%%%%%%%%%%%%%%%%%%%%%%%%%%%%%%%%%%%%%%%%%%%%%%%%%%%%%%%%%%%%%%%%%%%%%%%%%%%%%%%%%%%%%%%%%%%%%%%%%%%%%%%%%%%%%%%%%%%%%%%%%
%%%%%%%%%%%%%%%%%%%%%%%%%%%%%%%%%%%%%%%%%%%%%%%%%%%%%%%%%%%%%%%%%%%%%%%%%%%%%%%%%%%%%%%%%%%%%%%%%%%%%%%%%%%%%%%%%%%%%%%%%%%%%%%%%%%%%%%%%%%%%%%%%%
%%%%%%%%%%%%%%%%%%%%%%%%%%%%%%%%%%%%%%%%%%%%%%%%%%%%%%%%%%%%%%%%%%%%%%%%%%%%%%%%%%%%%%%%%%%%%%%%%%%%%%%%%%%%%%%%%%%%%%%%%%%%%%%%%%%%%%%%%%%%%%%%%%

%%
%%\subsection{Dummy section to generate text for figures}
%%\input{4_x_text_figures.tex}
%%
%%%--------------------------------------------
%%%\section{Supplementary Materials}
%%%--------------------------------------------
%%

\subsection{The \texttt{BioMultiLayer-L+NL} Toolbox}
\label{toolbox}

This \verb"Matlab" toolbox implements all the maths related to the kind of
multilayer feedforward L+NL neural models considered in the paper.
Specifically, it implements the vision model detailed in the Supplementary Material \ref{example_model},
which consists of a casacade of isomorphic linear+nonlinear transforms based on linear receptive fields
and canonical Divisive Normalization nonlinearities).

The \verb"BioMultiLayer-L+NL" toolbox includes the forward transform, the inverse, and all the derivatives
(w.r.t. the signal and w.r.t. the parameters).
The derivatives allow (i) Novel MAximum Differentiation (MAD) psychophysics
and (ii) fitting the model from classical psychophysics. The inverse allows
improved decoding of neural signals.

For convenience, this release includes other public-domain toolboxes
(please cite these sources as well):
\begin{itemize}
   \item CSF of the Standard Spatial Observer (\verb"SSO" folder) by J. Malo and A.B. Watson \cite{Watson02}.
         % [Watson $\&$ Malo IEEE ICIP 2002].
   \item Steerable wavelet pyramid (\verb"matlabPyrTools" folder) by E.P. Simoncelli \cite{Simoncelli92}.
         % [Simoncelli et al. Shiftable Multi-Scale Transforms IEEE Trans.Inf.Theory, 1992].
         % Compilation of mex files of this folder is convenient before using the toolbox.
\end{itemize}

\hiddensubsubsection{Installing the toolbox}
(1) Download the toolbox from \verb"http://isp.uv.es/docs/BioMultiLayer_L_NL.zip".
(2) Decompress all the contents of the file.
(3) Compile the mex files of the \verb"matlabPyrTools" toolbox.
(4) Include the folders in the Matlab path.

Once the toolbox is installed you may enter \verb"help BioMultiLayer_L_NL" at the Matlab prompt
to get an overview of all the routines in the toolbox.
It is convenient to take a look at the demo scripts included in the toolbox for worked-out examples.

\begin{figure}
  \vspace{-0.5cm}
  \centering
   \hspace{-5cm}\includegraphics[width=1.35\textwidth]{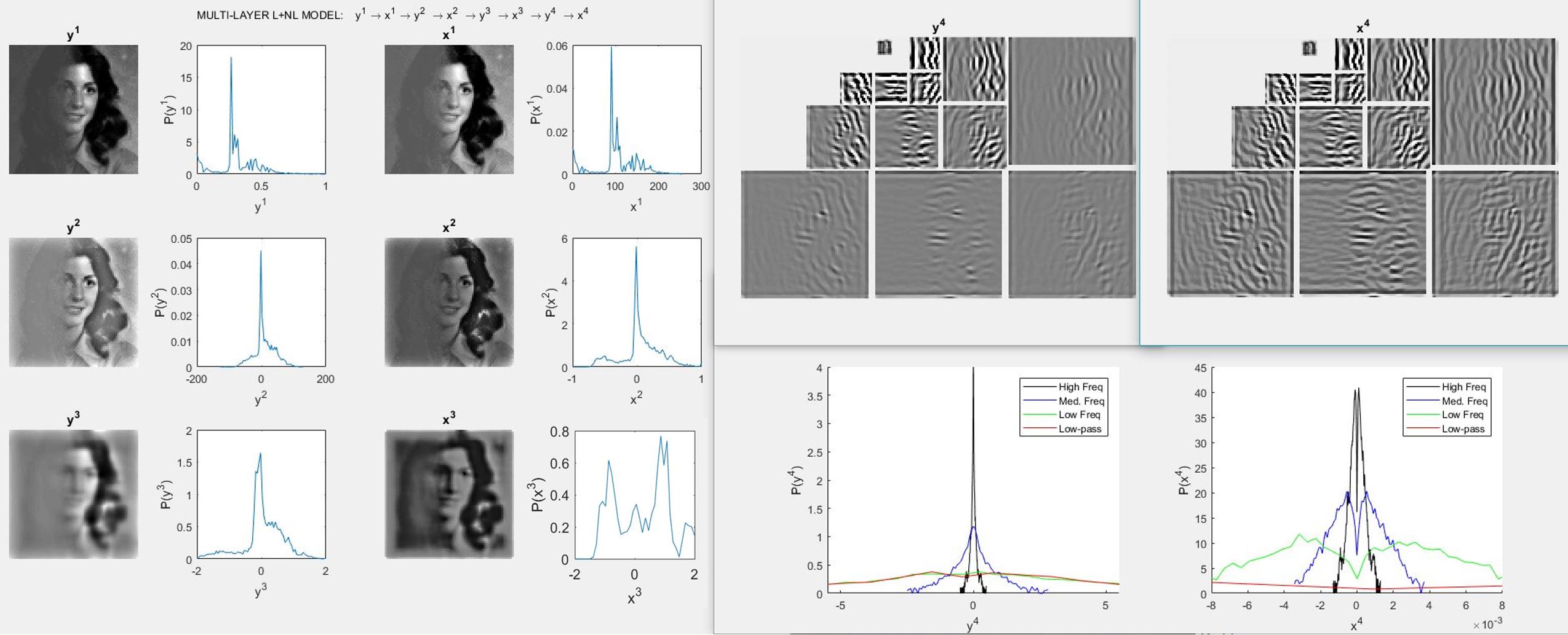}
  \caption{\scriptsize{\textbf{Forward Transform (in the demo script \texttt{demo\_deep\_DN\_iso.m}).} Responses and marginal PDFs along the layers of the network. Note that the bimodal PDFs of $\vect{x}^3$ and $\vect{x}^4$} are consistent with the predictive behavior reported in \cite{Malo10}.
  Equalization behavior at $\vect{x}^1$ reported at \cite{KaneBertalmio16} is not that evident in this example because this is a low-dynamic-range image.}\label{demo1}
\end{figure}

\hiddensubsubsection{Basic routines: responses, derivatives and inverse}

The basic use of the \verb"BioMultiLayer-L+NL" toolbox reduces to two routines (the forward and the inverse transform):
\begin{itemize}
      \item Given an image, \blue{\texttt{\textbf{deep\_model\_DN\_isomorph.m}}} computes the responses and the Jacobians of the vision model made of isomorphic Linear+Nonlinear layers with Divisive Normalization.
          This function sequentially calls \verb"stage_L_NL.m" that computes the response and the Jacobians of a single layer in the network.

      \item Given a response vector, \blue{\texttt{\textbf{inv\_deep\_model\_DN\_isomorph.m}}} decodes the response and reconstructs the input image. This function sequentially calls the function \verb"inv_stage_L_NL.m" that computes the inverse of a single layer in the network.
\end{itemize}
The user is referred to the \verb"help" of these functions for details on how to use them.
Additionally, step-by-step examples on how to use the above functions is given in the demo script \verb"demo_deep_DN_iso.m".
Particularly relevant is how to select which Jacobian(s) should be computed.
This is controlled with a structure described in the \verb"help" of \verb"stage_L_NL.m".

Initialization note: before using the above utilities, (a) images have to be prepared for the toolbox, and (b) the parameters
of the model have to be set.

\begin{figure}[b!]
  \centering
  \hspace{-0.5cm}\includegraphics[width=0.94\textwidth]{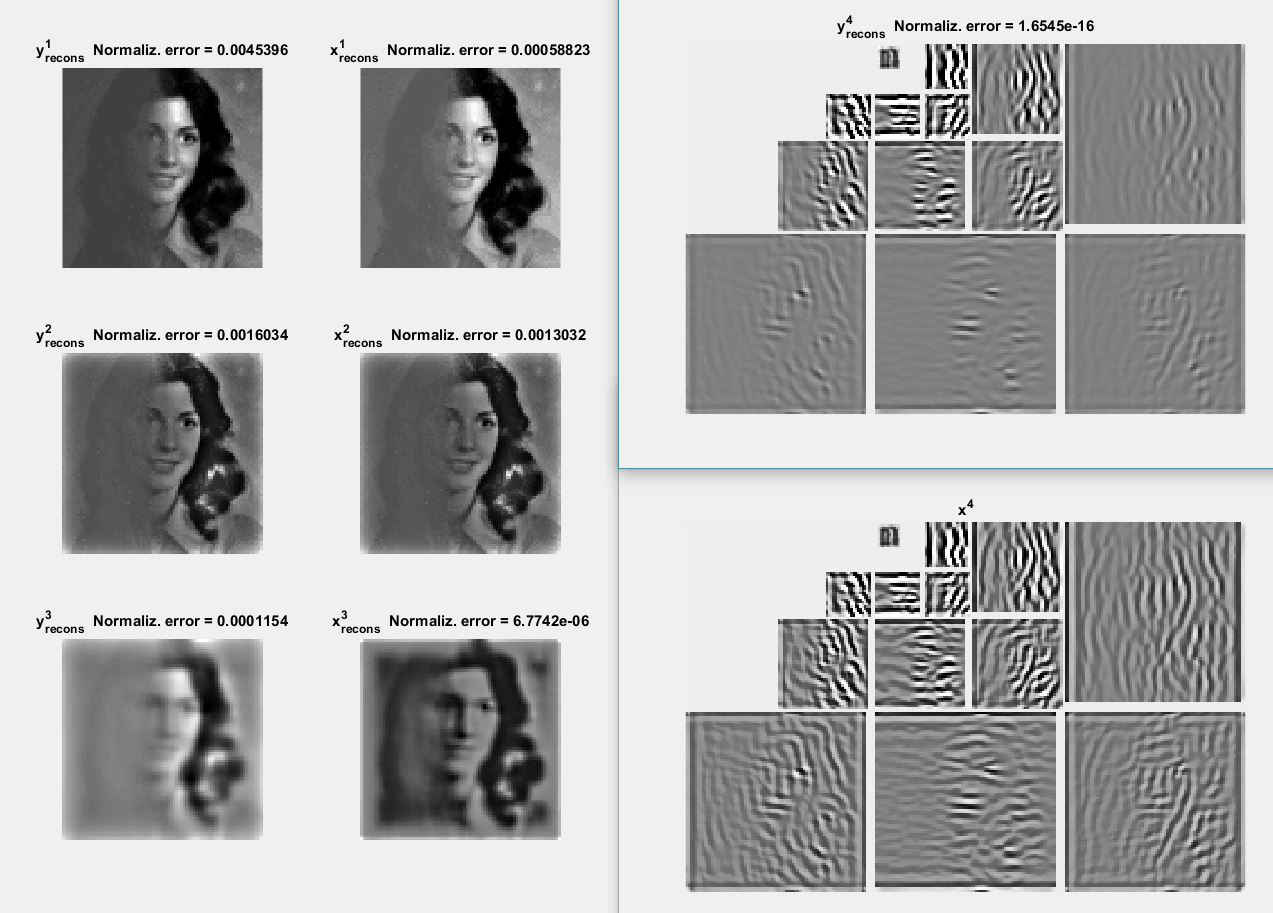}
  \caption{\scriptsize{\textbf{Inverse Transform (in the demo script \texttt{demo\_deep\_DN\_iso.m}).} Decoded signals from the response at the last layer of the network.}}\label{demo2}
\end{figure}

\begin{figure}[t]
  \centering
  \vspace{-0.5cm}
  \begin{tabular}{cc}
   \hspace{-1cm}\includegraphics[width=0.51\textwidth]{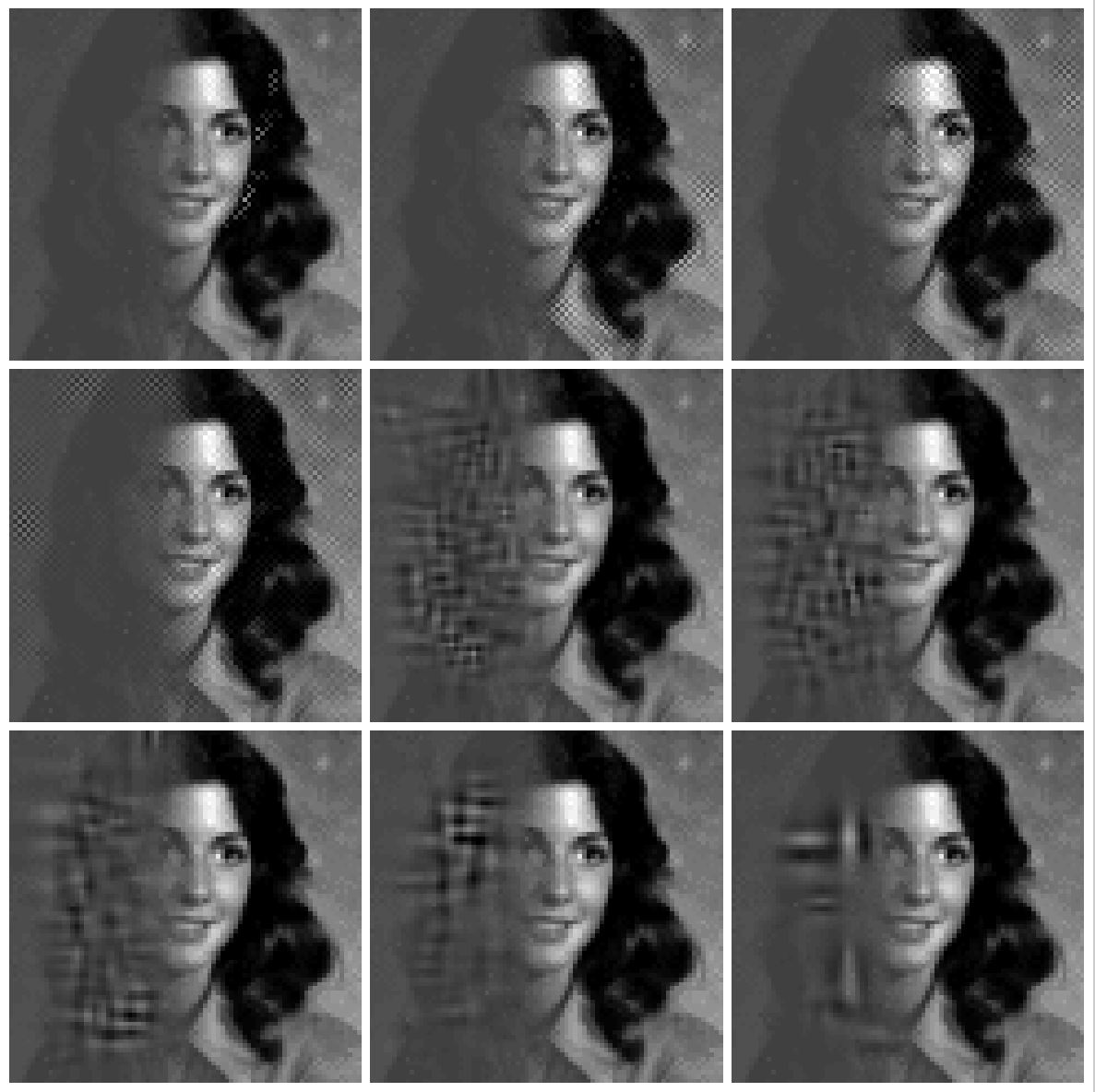} &
  \includegraphics[width=0.51\textwidth]{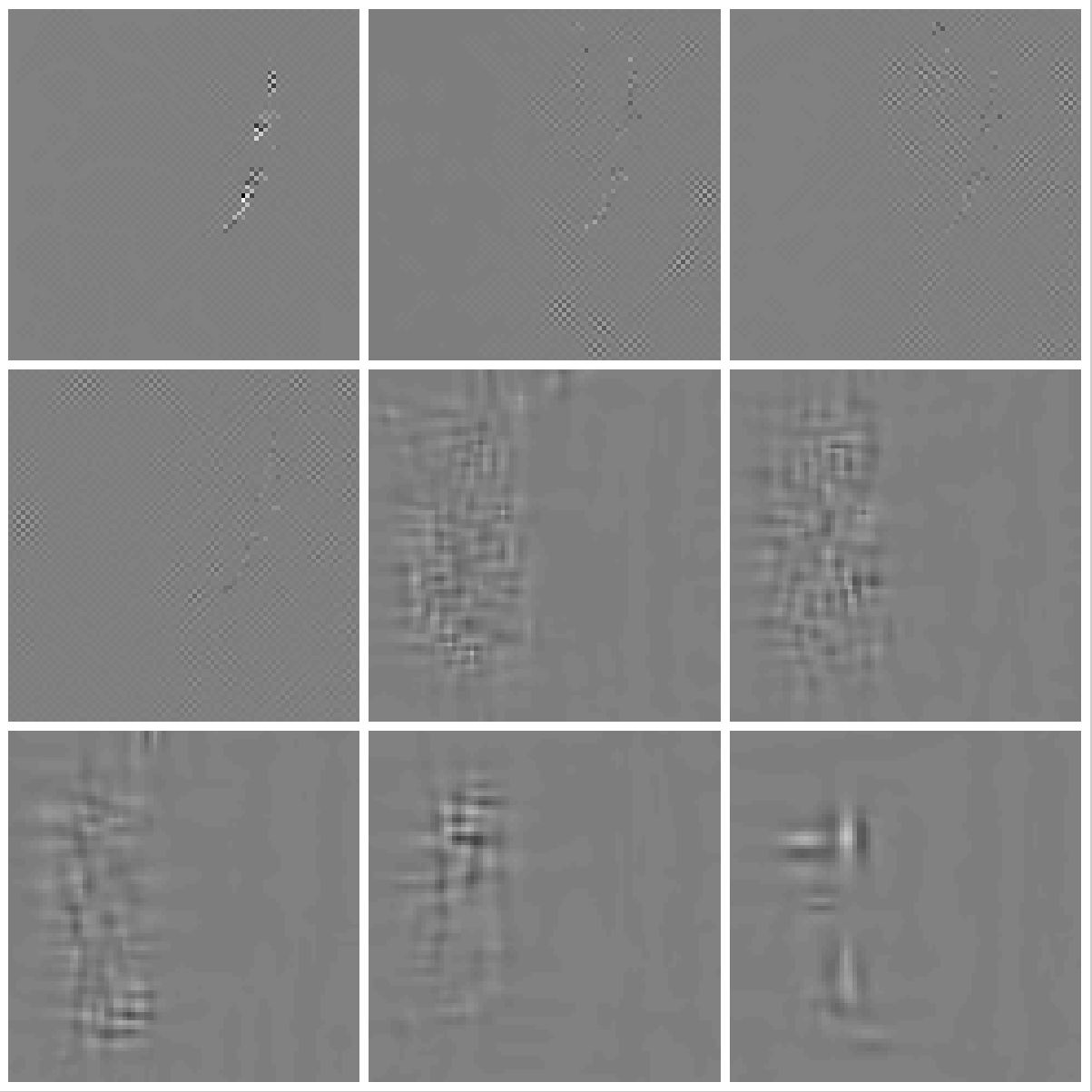}
  \end{tabular}
  \caption{\scriptsize{\textbf{Derivatives w.r.t. the signal: Analytic MAD (in the demo script \texttt{demo\_deep\_DN\_iso.m})} Left panel: Distorted images in the directions of low and high eigenvalues of the 2nd order metric matrix. Low eigendistortions correspond to the upper-left images where distortions are not noticeable, and high eigendistortions correspond to the lower-right images where distortions are highly noticeable. Right panel: isolated distortions. All the distortions have the same energy (induce the same MSE), but they are visually very different. Note that highly visible distortions are concentrated in the low contrast part of the image (and the other way around for highly visible distortions), which is consistent with the masking phenomenon and indicates the quality of the model.}}\label{demo3}
\end{figure}

\paragraph*{(a) Input stimuli, image normalization, and image arrangement}
The current version of the \verb"BioMultiLayer-L+NL" toolbox operates on \emph{achromatic images} in normalized luminance units.
This means that in the current implementation the model starts from luminance vectors, $\vect{y}^0$, and not from hyperspectral stimuli, $\vect{x}^0$.
In other words, the spectral and the chromatic elements of the first stage in Fig. 1, or the linear integration of the spectrum, $L^{1}$,
in Eq. \ref{layer1}, are not included in this release.
This missing linear stage, $L^1$, can be easily implemented using the \verb"Colorlab" toolbox \cite{Colorlab}.
Specifically, the \verb"Colorlab" functions \verb"spect2tri.m" and \verb"xyz2atd.m" compute CIE XYZ tristimulus values from radiances,
and transform the CIE XYZ result to convenient opponent chromatic representations (luminance, red-green and yellow-blue) \cite{Stockman11,Fairchild05}.

Normalized luminance refers to division by \emph{what is assumed to be the maximum luminance in the considered class of scenes}.
This implies that input values in $\vect{y}^0$ are mainly in the [0, 1] range except for highlights that may be over 1.
See the comment on proper image normalization and luminance calibration in \verb"deep_model_DN_isomorph.m".
More accurate transforms from conventional digital images to tristimulus images can be done using
the \verb"Colorlab" \cite{Colorlab} or the \verb"Psychtoolbox" \cite{Psychtoolbox}.

If your input luminance image is a matrix, \verb"y0", vector arrangement according to the \emph{last-dimension-first}
convention cited in Eq. \ref{the_image}, is simply obtained using \verb"y0(:)", and large images can be patch-wise vectorized using \verb"im2col.m".
The \texttt{BasicVideoTools} toolbox \cite{BasicVideoTools} has convenient generalizations of these vectorization
functions to be applied in spatio-spectral (or spatio-temporal) arrays (namely \texttt{im2colcube.m} and \texttt{col2imcube.m}).
These could be applied to extend the current toolbox to start from the hyperspectral stimulus $\vect{x}^0$ instead of
starting from $\vect{y}^0$.

\paragraph*{(b) Model parameters}
The considered network has many parameters.
The Jacobians implemented in this toolbox enable experimental methods to determine the parameters.
In fact, a side result of this paper is the specific set of values of the parameters obtained from MAD psychophysics and image quality optimization.
These parameters have to be passed to different functions of the \verb"BioMultiLayer-L+NL" toolbox in a specific \verb"struct" variable.
In order to simplify the construction of the parameter variable, it is generated in the script
\verb"parameters_DN_isomorph.m". An example on how to set the parameters and call the script is given in \verb"demo_deep_DN_iso.m".
If you are happy with certain set of parameters, you may store them in a \verb"*.mat" file so that you will not need to generate them again.

\begin{figure}[t]
  \centering
  \vspace{-0.5cm}
  \begin{tabular}{c}
  \hspace{-1.5cm}\includegraphics[width=1.05\textwidth]{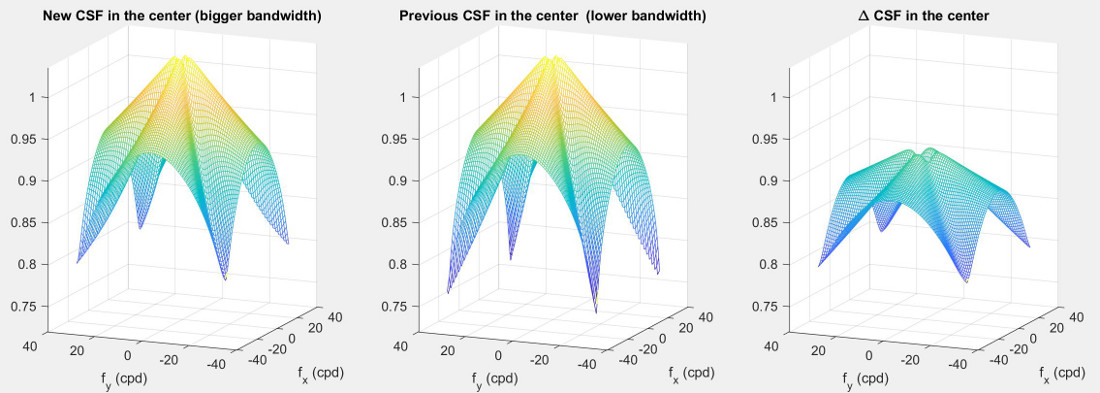}\\
  \hspace{-1.5cm}\includegraphics[width=1.05\textwidth]{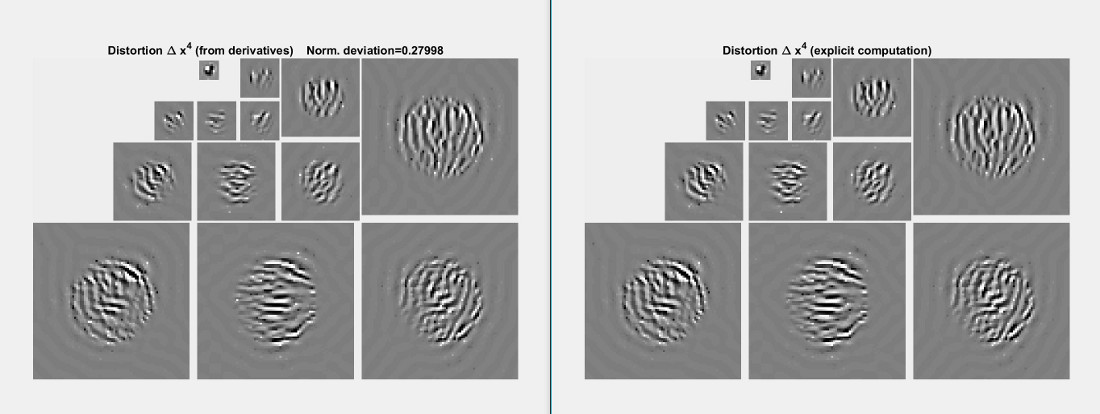}
  \end{tabular}
  \caption{\scriptsize{\textbf{Derivatives w.r.t. the parameters: Propagation of the perturbation in the CSF (in the demo script \texttt{demo\_deep\_DN\_iso.m})} Imagine a modification in the CSF filter of stage 3. For instance, increasing the bandwidth of the filter.
  How would it affect the response?. This can be done in two different ways: (1) the exact one (compute the response with and without the modification and subtract), and (2) using the first order approximation based on the Jacobian of the response w.r.t. the parameters (linearly propagate the effect of the modification).
  The top panel shows the different filter functions used in the center of the visual field, and the bottom panel shows the correponding perturbation in the response computed exactly (right) and according to the first order approximation (left).}}\label{demo4}
\end{figure}

\begin{figure}[h!]
  \centering
  \vspace{0.2cm}
  \begin{tabular}{c}
  \hspace{-2cm}\includegraphics[width=1.1\textwidth]{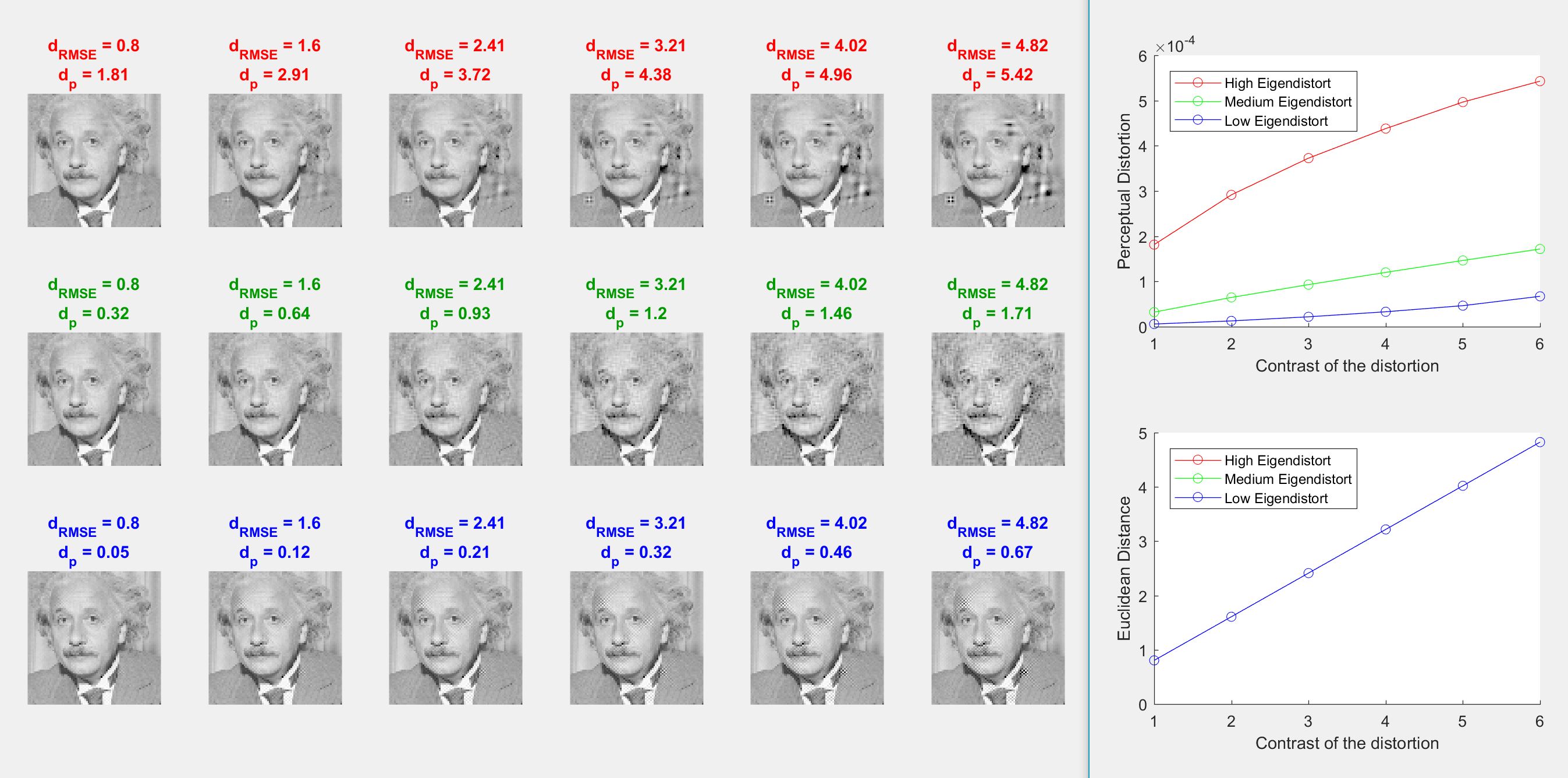}
  \end{tabular}
  \caption{\scriptsize{\textbf{Perceptual distance (demo script \texttt{demo\_metric\_DN\_iso.m})} Distortions of different nature (eigen distortions of the metric of low, medium and high eigenvalue in the top, medium and bottom rows). The distortion linearly increases the RMSE contrast from left to right columns. The images in each column have the same RMSE (Euclidean distance) with regard to the original image. The numbers and the plots at the right display the Euclidean and the perceptual distances.}}\label{demo5}
\end{figure}

\hiddensubsubsection{Advanced routines}

The flexible vision model in the \verb"BioMultiLayer-L+NL" toolbox may be used to
(a) compute perceptual distances between images,
(b) generate stimuli for MAximum Differentiation (MAD), and
(c) look for a specific model that maximizes the correlation with subjective opinion in image quality ratings.
These advanced applications are available using:
\begin{itemize}
      \item \blue{\texttt{\textbf{metric\_deep\_DN\_iso.m}}} computes the perceptual distance between an original image and a distorted image according to the multi-layer vision model. The use of this function is illustrated in the script \verb"demo_metric_deep_DN_iso.m".
      \item \blue{\texttt{\textbf{mad\_deep\_DN\_iso.m}}} performs the general MAximum Differentiation (MAD) search for an image and a multi-layer vision model defined by its parameters. The use of this function and the equivalent analytic 2nd order result using eigenvectors of the metric is illustrated in the script \verb"demo_mad_DN_iso.m".
      \item \blue{\texttt{\textbf{maximize\_correlation\_DN\_iso.slurm}}} launches stochastic gradient descent search in the parameter space to look for the model that maximizes the correlation with the mean opinion score of subjectively rated image quality databases.
          A \verb"*.slurm" function is invoked because this search is computationally expensive and the optimization has to be parallelized over multiple cores in a cluster. The folder \verb"Corr_max_TID" contains an illustrative example of such parallelization.
\end{itemize}

\hiddensubsubsection{Demos}

\paragraph*{Basic use of the toolbox: \blue{\texttt{\textbf{demo\_deep\_DN\_iso.m}}}}
This script illustrates the use of the toolbox to compute:
    (1) the response at the different layers
    (2) the inverse (decoded signal)
    (3) the Jacobians (w.r.t. the signal and w.r.t. the parameters).

  Responses (see Fig. \ref{demo1}) and inverses (see Fig. \ref{demo2})
  are shown for an illustrative natural image with nonstationary contrast.
  Marginal PDFs of the responses are also shown in Fig. \ref{demo1} to illustrate the
  predictive effect of Divisive Normalization.
  The Jacobian with regard to the stimuli is used to compute the 2nd order metric for that image.
  We compute eigen-distortions corresponding to such metric matrix and they make perceptual sense (see Fig. \ref{demo3}).
  The Jacobian with regard to the parameters is used to estimate the effect of perturbations of the model (see Fig. \ref{demo4}).

\paragraph*{Perceptual distances: \blue{\texttt{\textbf{demo\_metric\_deep\_DN\_iso.m}}}}
This script shows how the Euclidean distance in the response domain is
   substantially different from the Euclidean distance in the input space,
   and it is better correlated to subjective opinion.
   In this script, distortions of different nature are linearly scaled in
   contrast to lead to different distorted images.
   Euclidean and perceptual distances are computed and displayed together
   with the images (see Fig. \ref{demo5}).

\paragraph*{MAximum Differentiation search: \blue{\texttt{\textbf{demo\_mad\_DN\_iso.m}}}}
This script compares the three methods to generate maximally different images
   (1) General MAximum Differentiation search
   (2) Simplified MAximum Differentiation search based on the 2nd order approximation of distance
   (3) Analytic result based on the 2nd order approximation of distance.
The procedure in this demo was used to compute the results in the Discussion section \ref{MAD_section}.

\paragraph*{Checking the analytical expressions: \blue{\texttt{\textbf{check\_results.m}}}}
   This script shows that the analytic results and the implementation are correct.

   Analytic derivatives are compared to derivatives computed through finite differences.
   The analytic inverses and the inverses based on expansions are compared to the actual input.
   Derivatives and inverses are computed using patches from natural images
   (new patches are randomly selected in each realization of this script).
   Reasonable values for the parameters of the model are assumed.
   Agreement between the compared quantities is graphically and numerically assessed.
   \emph{Graphic assessment:} Values of the analytic results are plotted versus
   the equivalent numerical results. Good alignment along the unit-slope diagonal
   means good agreement.
   \emph{Numeric assessment:} Deviations between the compared results are expressed
   as ratios between the norm of the deviation over the norm of the analytic result.
   small values of this ratio mean good agreement.

   A summary of the numeric results computed on 50 patches of natural images is given in Table 1.

\begin{figure}
	\centering
    \small
    \setlength{\tabcolsep}{2pt}
    \vspace{0.0cm}
    \begin{tabular}{c}
    \hspace{-5.15cm}  \includegraphics[width=1.2\textwidth]{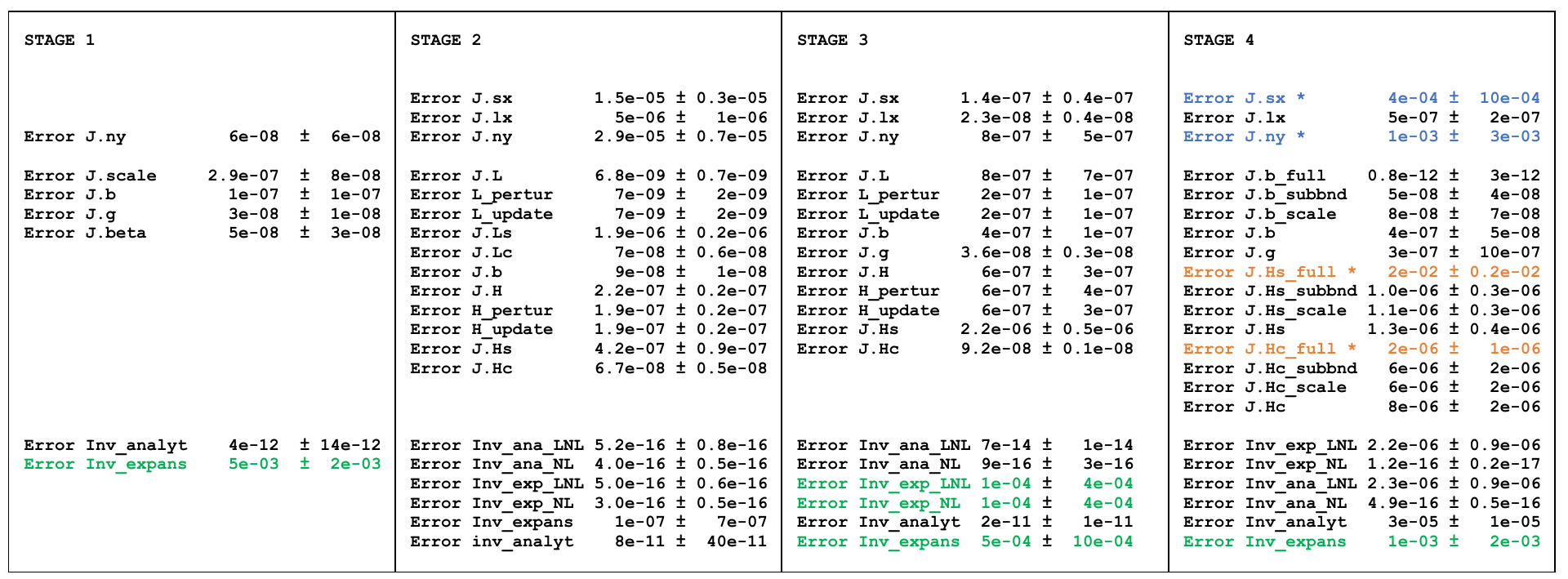} \\
    \vspace{-0.0cm}
    \hspace{-5.3cm}\pbox{1.25\textwidth}{\textbf{Table 1.} \scriptsize{\textbf{Numerical check of analytical Jacobians and Inverses.}
                         Data displays the normalized error. Normalized error for the Jacobians stands for the ratio of the 2-norm of the numerical-vs-analytical deviation over the 2-norm of the analytical result.
                         For the inverses the normalized error stands for the ratio of the 2-norm of the deviation of the reconstructed
                         signal over the 2-norm of the input signal.
                         The different columns contain the errors at the different stages.
                         The data-blocks from top to bottom are:
                         (1) errors of the derivatives w.r.t. the signal (check of Result I),
                         (2) errors of the derivatives w.r.t. the parameters (check of Result II), and
                         (3) errors of the inverses (isolated in each stage, and propagated through the stages).
                         These numerical experiments were carried out using \texttt{\textbf{check\_results.m}}.
                         Small errors (errors are always orders of magnitude lower than the actual value) show that expressions and implementation are correct.
                         We highlighted in color the larger errors. Errors of different color indicate different nature in
                         the source of the deviation.
                         Errors in green are obtained at the inverse using the expansion method.
                         As expected, these errors decrease by increasing the number of terms in the expansion
                         (i.e. it is only a matter of increasing the cpu time).
                         Errors in blue come from checking the analytical derivative with finite differences.
                         The finite nature of the increment induces an error. These errors decrease by decreasing
                         the size of the finite difference step. Not a theoretical problem either. Finally, errors
                         in orange correspond to the case where too small increments lead to noisy variations
                         of functions (imagine the differences induced in a Gaussian kernel due to a very
                         small difference in its width). This introduces extra errors that accumulate when
                         considering huge kernels (as is the case in the 4th stage). These deviations reduce when
                         reducing the size of the finite difference in the numerical derivative.}
                         }
                         %\label{tabla}
                         \end{tabular}
    \vspace{-0.5cm}
\end{figure}

%%%%%%%%%%%%%%%%%%%%%%%%%%%%%%%%%%%%%%%%%%%%%%%%%%%%%%%%%%%%%%%%%%%%%%%%%  END OF 5.8 TOOLBOX
%%%%%%%%%%%%%%%%%%%%%%%%%%%%%%%%%%%%%%%%%%%%%%%%%%%%%%%%%%%%%%%%%%%%%%%%%%%%%%%%%%%%%%%%%%%%%%%%%%%%%%%%%%%%%%%%%%%%%%%%%%%%%%%%%%%%%%%%%%%%%%%%%%
%%%%%%%%%%%%%%%%%%%%%%%%%%%%%%%%%%%%%%%%%%%%%%%%%%%%%%%%%%%%%%%%%%%%%%%%%%%%%%%%%%%%%%%%%%%%%%%%%%%%%%%%%%%%%%%%%%%%%%%%%%%%%%%%%%%%%%%%%%%%%%%%%%
%%%%%%%%%%%%%%%%%%%%%%%%%%%%%%%%%%%%%%%%%%%%%%%%%%%%%%%%%%%%%%%%%%%%%%%%%%%%%%%%%%%%%%%%%%%%%%%%%%%%%%%%%%%%%%%%%%%%%%%%%%%%%%%%%%%%%%%%%%%%%%%%%%
%%%%%%%%%%%%%%%%%%%%%%%%%%%%%%%%%%%%%%%%%%%%%%%%%%%%%%%%%%%%%%%%%%%%%%%%%%%%%%%%%%%%%%%%%%%%%%%%%%%%%%%%%%%%%%%%%%%%%%%%%%%%%%%%%%%%%%%%%%%%%%%%%%

\subsection{Toolbox-oriented matrix properties}
\label{matrix_properties}

These properties are useful in the implementation of the model for large images or multiple image patches. These properties are particularly convenient in \verb"Matlab" since it is not very efficient in building large diagonal matrices.

\hiddensubsubsection{Products}

\paragraph*{Hadamard product and diagonal matrices.} Useful to define divisive normalization (e.g. in Eq. \ref{divisive_norm2})
\begin{equation}
           \verb"a.*b" \,\, = \,\,
           \vect{a} \odot \vect{b} \,\, = \,\,  \vect{b} \odot \vect{a} \,\, = \,\, \mathds{D}_{\vect{a}} \cdot \vect{b} \,\, = \,\, \mathds{D}_{\vect{b}} \cdot \vect{a} \,\,\,\,\, \textrm{with} \,\,\,\,\, \vect{a},\vect{b} \in \mathbb{R}^{d \times 1}, \,\,\,\,\, \mathds{D}_{\vect{a}}, \mathds{D}_{\vect{b}} \in \mathbb{R}^{d \times d}
      % \nonumber
      \label{haddamard_prod}
\end{equation}

\paragraph*{Kronecker product and matrix replication.} This non-commutative product is useful to avoid the large diagonal matrices below.
\begin{equation}
           \verb"repmat(A,m,n)" \,\, = \,\,
           \verb"kron(ones(m,n),A)" \,\, = \,\,
           \mathds{1}_{m \times n} \otimes A
      % \nonumber
      \label{kronecker_prod}
\end{equation}

\hiddensubsubsection{Large matrices (single image patch)}

\paragraph*{Left-multiplication by diagonal:} weight each \emph{row} of $A$ by the corresponding $v_i$ (e.g. in Result I, Eq. \ref{deriv_DN}):
\begin{equation}
      \mathds{D}_{\vect{v}} \cdot A \,\, = \,\, \left( \mathds{1}_{1 \times d} \otimes \vect{v} \right) \odot A
      \,\, = \,\, \verb"repmat(v,1,d).*A"
      % \nonumber
      \label{left_diag}
\end{equation}

\paragraph*{Right-multiplication by diagonal:} weight each \emph{column} of $A$ by the corresponding $v_i$:
\begin{equation}
      A \cdot \mathds{D}_{\vect{v}} \,\, = \,\, \left( \mathds{1}_{d \times 1} \otimes \vect{v}^\top \right) \odot A
      \,\, = \,\, \verb"repmat(v',d,1).*A"
      % \nonumber
      \label{right_diag}
\end{equation}

\paragraph*{Large block-diagonal matrices in $\nabla_H S$:} When dealing with derivatives w.r.t. nonparametric kernels in Result II (either to optimize the kernel or to compute the effect of perturbations) one find products involving huge block-diagonal matrices. These can be avoided:
\begin{itemize}
      \item In optimization (use Eq. \ref{jacobian2_H} in \ref{deriv_correl}), \verb"deltaS_times_blk_diagJ.m":
          \begin{equation}
                \vect{a}^\top \cdot \mathds{B}^d_{\vect{v}^\top} = \left( \mathds{1}_{1 \times d_v} \otimes \vect{a} \right) \odot \left( \mathds{1}_{d \times 1} \otimes \vect{v}^\top \right)
                \label{deltaS_times_blk_diagJ}
          \end{equation}
      \item In perturbations due to $\Delta H$ (apply  \ref{jacobian2_H}), \verb"blk_diagJ_times_deltaH.m":
          \begin{equation}
                \mathds{B}^d_{\vect{v}^\top} \cdot \textrm{vect}(\Delta H^\top) = diag\left( \left( \mathds{1}_{d \times 1} \otimes \vect{v}^\top \right) \cdot \Delta H^\top \right)
                \label{blk_diagJ_times_deltaH}
          \end{equation}
\end{itemize}

\hiddensubsubsection{Even larger matrices (multiple image patches)}

\begin{tabular}{c}
\parbox{\textwidth}{
This is how different expressions change when working with
$N$ image vectors at the same time (stacked in a single matrix of size $d \times N$) as done by \texttt{im2col.m}.}\\[0.4cm]
%\begin{table}
%\caption{Table Title} \label{stakazo_tabla}
\begin{tabular}{c}
\hspace{-0.5cm}
\begin{tabular}{lcl}
  \hline
  & & \\
  Single vector & \hspace{-0.5cm} $\xrightarrow{\textrm{stack N vectors}}$ \hspace{-1.5cm} & \hspace{1.7cm} Multiple vectors \\
  & & \\\hline
  & & \\
  $\vect{v} \in \mathbb{R}^{d \times 1}$ & \hspace{-0.9cm} $\longrightarrow$ \hspace{-0.9cm} & $\vect{v} = \left( \begin{array}{cccc}
                                                                                           \vdots &     \vdots     &        & \vdots \\
                                                                                   \vect{v}^{[1]} & \vect{v}^{[2]} & \cdots & \vect{v}^{[N]} \\
                                                                                          \vdots &     \vdots     &        & \vdots \\
                                                                                 \end{array}
                                                                               \right) \in \mathbb{R}^{d \times N}$ \\[1cm]
                      &   &                                                         and $\,\,\,\,\,\,\,\,\,\,\,\, \textrm{vect}(\vect{v}) = \left(
                                                                                                                                                     \begin{array}{c}
                                                                                                                                                       \vect{v}^{[1]} \\
                                                                                                                                                       \vect{v}^{[2]} \\
                                                                                                                                                       \vdots \\
                                                                                                                                                       \vect{v}^{[N]} \\
                                                                                                                                                     \end{array}
                                                                                                                                                   \right) \in \mathbb{R}^{(N d)\times 1} $\\
                                                                                                                                                   & & \\\hline
                                                                                                                                                   & & \\
  $\Delta \vect{x}^i = \nabla_{\vect{x}^{i-1}} S^{(i)} \cdot \Delta \vect{x}^{i-1}$ & \hspace{-0.9cm} $\longrightarrow$ \hspace{-0.9cm} & $\textrm{vect}\left( \Delta \vect{x}^i \right) = \nabla_{\vect{x}^{i-1}} S^{(i)} \cdot \textrm{vect}\left( \Delta \vect{x}^{i-1} \right)$ \\
  & & \\
  where $ \nabla_{\vect{x}^{i-1}} S^{(i)} \in \mathbb{R}^{d_i \times d_{i-1}}$ &                   & where $\nabla_{\vect{x}^{i-1}} S^{(i)} \in \mathbb{R}^{N d_i \times N d_{i-1}} $ is block-diag. \\
   & & but stored by convenience as $\in \mathbb{R}^{(N d_i) \times d_{i-1}}$ \\
    &  & \\\hline
    &  &  \\
  $\Delta \vect{x}^i = \nabla_{\vect{\Theta}^{i}} S^{(i)} \cdot \Delta \vect{\Theta}^{i}$  & \hspace{-0.9cm} $\longrightarrow$ \hspace{-0.9cm} & $\textrm{vect}\left( \Delta \vect{x}^i \right) = \nabla_{\vect{\Theta}^{i}} S^{(i)} \cdot \Delta \vect{\Theta}^{i} $ \\
    &                   &  \\
   where $ \nabla_{\vect{\Theta}^{i}} S^{(i)} \in \mathbb{R}^{d_i \times d_{\Theta^i}}$ &                   & where $ \nabla_{\vect{\Theta}^{i}} S^{(i)} \in \mathbb{R}^{(N d_i) \times d_{\Theta^i}}$ \\
    &                   &  \\
  \hline
    &  &  \\
  $\mathds{D}_{\vect{v}} \cdot A$ & \hspace{-0.9cm} $\longrightarrow$ \hspace{-0.9cm} & $\left( \mathds{1}_{1 \times d} \otimes \textrm{vect}(\vect{v}) \right) \odot \left( \mathds{1}_{N \times 1} \otimes A \right) $ \\
    &                   &  \\
  \hline
    &  &  \\
  $A \cdot \mathds{D}_{\vect{v}}$ & \hspace{-0.9cm} $\longrightarrow$ \hspace{-0.9cm} & $\left( \vect{v}^\top \otimes \mathds{1}_{d \times 1}  \right) \odot \left( \mathds{1}_{N \times 1} \otimes A \right) $ \\
    &                   &  \\
  \hline
\end{tabular}
\end{tabular}
%\end{table}
\end{tabular}

%%%%%%%%%%%%%%%%%%%%%%%%%%%%%%%%%%%%%%%%%%%%%%%%%%%%%%%%%%%%%%%%%%%%%%%%%  END OF 5.9 MATRICES
%%%%%%%%%%%%%%%%%%%%%%%%%%%%%%%%%%%%%%%%%%%%%%%%%%%%%%%%%%%%%%%%%%%%%%%%%%%%%%%%%%%%%%%%%%%%%%%%%%%%%%%%%%%%%%%%%%%%%%%%%%%%%%%%%%%%%%%%%%%%%%%%%%
%%%%%%%%%%%%%%%%%%%%%%%%%%%%%%%%%%%%%%%%%%%%%%%%%%%%%%%%%%%%%%%%%%%%%%%%%%%%%%%%%%%%%%%%%%%%%%%%%%%%%%%%%%%%%%%%%%%%%%%%%%%%%%%%%%%%%%%%%%%%%%%%%%
%%%%%%%%%%%%%%%%%%%%%%%%%%%%%%%%%%%%%%%%%%%%%%%%%%%%%%%%%%%%%%%%%%%%%%%%%%%%%%%%%%%%%%%%%%%%%%%%%%%%%%%%%%%%%%%%%%%%%%%%%%%%%%%%%%%%%%%%%%%%%%%%%%
%%%%%%%%%%%%%%%%%%%%%%%%%%%%%%%%%%%%%%%%%%%%%%%%%%%%%%%%%%%%%%%%%%%%%%%%%%%%%%%%%%%%%%%%%%%%%%%%%%%%%%%%%%%%%%%%%%%%%%%%%%%%%%%%%%%%%%%%%%%%%%%%%%

\end{document}